\documentclass[usegraphicx,usenatbib,useAMS,onecolumn]{mn2e}
\usepackage{epsfig}
\usepackage{graphicx}
\usepackage{amsmath,bm}
\usepackage{color}
\usepackage{amssymb}

\newcommand{\bk}{{\bf k}}

\newcommand{\bx}{{\bf x}}

\title[The power spectrum and  bispectrum of SDSS DR11 BOSS galaxies I: bias and gravity]{The power spectrum and  bispectrum of SDSS DR11 BOSS galaxies I: bias and gravity}

\author[H. Gil-Mar\'in et al.]{H\'ector Gil-Mar\'in$^{1}$\thanks{hector.gil@port.ac.uk}, Jorge Nore\~na$^{2,3}$, Licia Verde$^{4,2,5}$, Will J. Percival$^{1}$, \and Christian Wagner$^{6}$, Marc Manera$^7$, Donald P. Schneider$^{8,9}$\\
 $^{1}$ Institute of Cosmology \& Gravitation, University of Portsmouth, Dennis Sciama Building, Portsmouth PO1 3FX, UK\\
 $^{2}$ Institut de Ci\`encies del Cosmos, Universitat de Barcelona, IEEC-UB, Mart\'i i Franqu\`es 1, 08028, Barcelona, Spain\\
  $^{3}$ Department of Theoretical Physics and Center for Astroparticle Physics (CAP), 24 quai E. Ansermet, CH-1211 Geneva 4, CH \\
    $^{4}$ ICREA (Instituci\'o Catalana de Recerca i  Estudis Avan\c{c}ats), Passeig Llu\'is Companys, 23 08010 Barcelona - Spain\\
 $^{5}$ Institute of Theoretical Astrophysics, University of Oslo, Norway\\
 $^{6}$ Max-Planck-Institut f\"ur Astrophysik, Karl-Schwarzschild Str. 1, 85741 Garching, Germany\\
  $^{7}$ University College London, Gower Street, London WC1E 6BT, UK\\
  $^8$ Department of Astronomy and Astrophysics, The Pennsylvania State University, University Park, PA 16802, USA\\
$^9$ Institute for Gravitation and the Cosmos, The Pennsylvania State University, University Park, PA 16802, USA\\
}
\def\gs{\mathrel{\raise1.16pt\hbox{$>$}\kern-7.0pt
\lower3.06pt\hbox{{$\scriptstyle \sim$}}}}         
\def\ls{\mathrel{\raise1.16pt\hbox{$<$}\kern-7.0pt 
\lower3.06pt\hbox{{$\scriptstyle \sim$}}}}         

\begin{document}
\maketitle

\begin{abstract}
We analyse the anisotropic clustering of the Baryon Oscillation Spectroscopic Survey (BOSS) CMASS Data Release 11 sample, which consists of $690 827$ galaxies in the redshift range $0.43 < z < 0.70$ and has a sky coverage of $8 498$ deg$^2$ corresponding to an effective volume of $\sim6\,\rm{Gpc}^3$. We fit the Fourier space statistics,  the power spectrum and bispectrum monopoles to measure the linear and quadratic bias parameters, $b_1$ and $b_2$, for a non-linear non-local bias model, the growth of structure parameter $f$ and the amplitude of dark matter density  fluctuations parametrised by $\sigma_8$.  We obtain  $b_1(z_{\rm eff})^{1.40}\sigma_8(z_{\rm eff})=1.672\pm 0.060$ and  $b_2^{0.30}(z_{\rm eff})\sigma_8(z_{\rm eff})=0.579\pm0.082$ at the effective redshift of  the survey, $z_{\rm eff}=0.57$.  The main cosmological result is the constraint on the combination $f^{0.43}(z_{\rm eff})\sigma_8(z_{\rm eff})=0.582\pm0.084$,  which is complementary to $f\sigma_8$ constraints obtained from 2-point redshift space distortion analyses. A less conservative analysis yields $f^{0.43}(z_{\rm eff})\sigma_8(z_{\rm eff})=0.584\pm0.051$.    We ensure that our result is robust by performing detailed   systematic tests using a large suite of  survey galaxy mock catalogs and  N-body simulations.  The constraints on $f^{0.43}\sigma_8$ are useful for setting additional constrains on neutrino mass, gravity, curvature as well as the number of neutrino species from galaxy surveys analyses (as presented in a  companion paper).
\end{abstract}


\section{Introduction}\label{sec:intro}

The  small inflationary  primordial density fluctuations are believed to be close to those of a Gaussian random  field, thus their statistical properties are fully described by the power spectrum. Gravitational instability amplifies the initial perturbations but the growth eventually becomes non-linear. In this case the three-point correlation function and its counterpart in Fourier space, the bispectrum, are intrinsically second-order quantities, and  the lowest-order statistics sensitive to non-linearities. 
These three-point statistics can not only be used to test the gravitational instability paradigm but also to probe galaxy biasing and thus break the degeneracy between linear bias and the matter density parameter present in power spectrum measurements.  Pioneering work on measuring  the three-point statistics in a cosmological context are \cite{PeeblesGroth:75, GP77} and \cite{FrySeldner:82}. The interpretation of these measurements had to wait for the development of non-linear cosmological perturbation theory, which showed how non-Gaussianity, and in particular the bispectrum, is generated by gravity and how (galaxy) biasing  affects the bispectrum \citep{Fry:1994}. This advance started with the pioneering work of \citet{Goroff:1986} and \citet{Fry84},  and most of the  theory was developed by the early 2000s (e.g., see  \cite{Bernardeauetal:2002} for a review).  Before the bispectrum could be used to probe galaxy bias from galaxy redshift surveys, a full treatment of the redshift-space bispectrum for galaxies had to be developed \citep{MVH97,Scoccetal98,HMV98,VHMM98, Scoccetal99,Scocc00}.  Starting around the year 2000, the golden era of cosmology started producing galaxies redshift surveys covering unprecedented volumes. Despite the number of power spectra analyses performed, the bispectrum work, especially with the goal of  extracting cosmological information, from it, has been much  less extensive \citep{FFFS01,SFFF2001,Verdeetal:2002,JingBorner2004, GS2005,2004MNRAS.353..287W,Marin:2011,2013MNRAS.432.2654M}.
To date, bispectra analyses were performed out with the aim of constraining the bias parameters adopting a simple quadratic, local bias prescription. To the extract cosmological information these constraints had to be combined with e.g., the measurement of $\beta=f/b$ ---where $f$ is the linear growth rate and $b$ the linear galaxy bias--- from redshift space distortions of the power spectrum.

In this paper we consider the galaxy bispectrum and power spectrum  monopole of the CMASS galaxy sample of Sloan Digital Sky Survey III Baryon Oscillation Spectroscopic Survey (BOSS) data release 11 (DR11).  By using jointly the power spectrum and bispectrum we can constrain not only the bias parameters, but also  the gravitational growth of clustering and in particular the combination $f^{0.43}\sigma_8$, where $\sigma_8$ denotes the linear {\it rms} of the  dark matter density perturbations on scales of 8~$h^{-1}{\rm Mpc}$.  This quantity is particularly interesting as it may be used to probe directly the nature of gravity. In fact  in general relativity (GR) the linear growth rate of perturbations is uniquely given by the expansion history. Therefore for a specified expansion history (such as the one measured by Baryon Acoustic Oscillations or by supernovae data), GR predicts the redshift evolution of $\sigma_8$ and $f$.  Most of the tests of gravity on cosmological scales rely on the measurement of the anisotropic power spectrum  in redshift space to constrain the combination $f\sigma_8$. In this paper we offer a different constraint that arises from the combination of 2- and 3-point statistics. 
The fact that the  $f$-$\sigma_8$ combinations of these two approaches differ offers the possibility of measuring both quantities from a combined analysis. We also present constraints on the relation between the clustering  of   mass and that of galaxies in the form of the combinations $b_1^{1.40}\sigma_8$ and $b_2^{0.30}\sigma_8$, where $b_1$ and $b_2$  are two bias parameters for an Eulerian non-local non-linear  bias model, which we assume local in Lagrangian space (\citealt{McDonaldRoy:2009,Baldaufetal:2012,Chan12} and \citealt{Saitoetal:2014}).  These constraints make possible the use both the shape and amplitude of the measured galaxy power spectrum in the mildly non-linear regime to constrain cosmological parameters. 
 This paper is the first of a series of related works. In \citet{HGMetal:inprep} we present the adopted model of the redshift space bispectrum in the mildly non-linear regime.  The full analysis of the survey is presented in  two companion papers. In this paper, we present the details of the measurement of the power spectrum and bispectrum of the CMASS DR11 galaxy sample and all the systematic tests that evaluate the validity of the measurement. In the companion paper \citep{paper2} we focus on the cosmological interpretation of the constraints obtained in combination with  other datasets such as Cosmic Microwave Background (CMB) data.

The rest of this  paper is organised  as follows. In \S~\ref{section_data} we present a description of the CMASS DR11 data and of the resources used for calibrating and testing the theoretical models. In \S~\ref{section_method} we describe  our methodology which includes the estimator used to measure the power spectrum and bispectrum from the galaxy catalogue, the modelling of mildly non-linear power spectra and bispectra for biased tracers  in redshift space  and  the statistical method used to extract cosmological information from the  measurements. In \S~\ref{sec:results} we present the results including the set of best fit parameters a nether errors, and \S~\ref{section:systematics} contains all the systematic tests that we have performed.
Finally, in \S~\ref{sec:conclusions} we summarise the conclusions  and anticipate  future avenues of research.

\section{The  real and SYNTHETIC data}\label{section_data}
\label{sec:dataandsims}
Our analysis of the BOSS galaxy sample relies heavily on  calibration, testing  and  performance assessment using simulated and mock data. Here we describe the real data we use  along with their real-world effects, and  the synthetic data in the form of mock surveys and N-body simulations.
\subsection{The SDSSIII BOSS data}  
\label{sec:data}

As part of the Sloan Digital Sky Survey III (SDSSIII,  \citealt{EISetal:2011}) the Baryon Oscillations Spectroscopic Survey (BOSS) \citep{Dawsonetal:2012,Smeeetal:2013,Boltonetal:2012} 
has measured 
the spectroscopic redshifts of about 1.2 million galaxies (and over 200000 quasars).  The galaxies are selected from multi-colour SDSS imaging \citep{Fukugitaetal:1996,Gunnetal:1998,Smithetal:2002,Gunnetal:2006,Doietal:2010} covering a redshift range of $ 0.15 <z < 0.70$. The survey targeted two samples called LOWZ ($0.15 \le z\le 0.43$) and CMASS ($0.43<z\le 0.70$). In this work  we use only  the CMASS sample. The BOSS survey is  optimised for the measurement of the baryon  acoustic oscillation (BAO)  scale  from the galaxy power spectrum/correlation function and hence covers a large cosmic volume $V_{\rm eff} \simeq6$\, Gpc$^3$ with a number density of galaxies   $n \sim 3 \times 10^{-4}\,[h{\rm Mpc}^{-1}]^3$ to ensure that shot noise is not dominant at BAO scales \citep{Whiteetal:2011}.  

Most of CMASS galaxies are red with a prominent break in their spectral energy distribution at 4000\AA\,, making the sample highly biased ($b \sim 2$). While this choice boosts the clustering signal at BAO scales, it renders the sample not optimal for bispectrum studies: the clustering  boost   comes at the expense of making the bias deviate from the simple linear, local, deterministic, Eulerian bias prescription. The bispectrum is much more sensitive than the power spectrum to these effects.
The CMASS-DR11 sample covers  $8498$ deg$^2$ divided in a northern Galactic cap (NGC) with $6 391$ deg$^2$ and a  southern Galactic cap (SGC) with $2 107$ deg$^2$. Our sample includes 520 806 galaxies in the north and 170 021 galaxies in the south.  The effective redshift of the dataset has been determined to be $z_{\rm eff}=0.57$ in previous works \citep{Andersonetal:2012}.

In order to correct several shortcomings of the CMASS dataset \citep{Rossetal:2013,Andersonetal:2013}, three different incompleteness weights have been included: a redshift failure weight, $w_{\rm rf}$, a fibre collision weight, $w_{\rm fc}$ and a systematics weight, $w_{\rm sys}$, which combines a seeing condition weight and a stellar weight. Thus, each galaxy target is counted as,
\begin{equation}
\label{completness_weights}w_{\rm c}=(w_{\rm rf}+w_{\rm fc}-1)w_{\rm sys}.
\end{equation}

The redshift failure and fibre collision weights account for those galaxies that have been observed, but whose redshift has not been measured. This could be due to several reasons: two galaxies are too close to each other ($< 62''$) to put two fibre detectors (fibre collision), or because the process of measuring of the redshift has simply failed. In  both cases these galaxies are still included in the catalogue by double counting the nearest galaxy, which is assumed to be statistically indistinguishable from the missing galaxy (see \citealt{Rossetal:2013} for details). The systematic weights account for fluctuations in the target density caused by changes in observational efficiency.  The CMASS sample presents correlations between the galaxy density and the seeing in the imaging data used for targeting, as well as the proximity to a star. In order to correct for such effects, the systematic weights are designed to correct these variations giving an isotropic weighted field.

An additional weight that ensures the condition of minimum variance can be set \citep{FKP, Beutleretal:2013},

\begin{equation}
\label{fkp_weights}w_{\rm FKP}({\bf r})= \frac{w_{\rm sys}({\bf r})}{w_{\rm sys}({\bf r})+w_{\rm c}({\bf r})n({\bf r})P_0}
\end{equation}

where $n$ is the mean number density of galaxies and $P_0$ is the amplitude of the galaxy power spectrum at the scale where the error is minimised,  $k\sim0.10\,h{\rm Mpc}^{-1}$.
The effects of the inclusion of the weights in the shot noise term are discussed in Appendix~\ref{appendix_shot_noise}.
Although the weighting scheme could in principle be improved for a population of differently biased tracers \citep{PVP}, the homogeneity of the CMASS galaxy population used here does not warrant the extra complication.

\subsection{The mock survey  catalogs and N-body simulations}
  \label{sec:sims-mocks}
  In order to test the validity of some approximations and the systematic errors of   the adopted modelling and approach,
 we use the following set of simulations.
  \begin{enumerate}
  \item A set of 50 \textsc{PThalos} realisations in periodic boxes. These are halo catalogues created using the  2nd-order Lagrangian Perturbation Theory (2LPT) matter field method by \cite{Maneraetal:2013} with flat LCDM cosmology.  The box-size is 2.4~Gpc$h^{-1}$. The  minimum mass of the 2LPT haloes is $m_p= 5.0\times 10^{12}\,M_\odot h^{-1}$. In order to extract the halo field, a Cloud-in-Cell (hereafter CiC) prescription has also been used with $512^3$ grid cells, whose size is $4.69\, {\rm Mpc}h^{-1}$. These realisations do not have any observational features such as the survey geometry or galaxy weights.
  
  \item A set of 50  \textsc{PThalos} realisations with the survey geometry of  the NGC CMASS  sample from Data Release 10 (DR10) \citep{Ahnetal:2013}. Both DR10 and DR11 have a similar radial selection function, but DR11 has a more uniform angular survey mask than DR10. Thus, DR10 should present stronger mask effects  than DR11. We therefore use DR10 to test the mask corrections we apply to the DR11 sample.  This set has been constructed from the catalogue (i) applying the  CMASS NGC DR10 survey mask. These catalogues are embedded in a box of $3500\,{\rm Mpc}h^{-1}$\footnote{See fig. 11 of \cite{Maneraetal:2013} to see why we need a larger box than in (i)}. CiC prescription has been applied with $512^3$ grid cells, which corresponds to a cell resolution of $6.84\,{\rm Mpc}h^{-1}$.
  
  \item A set of five realisations of dark matter and 20 realisations of N-body haloes  based on N-body dark matter particles simulations with box size  $L_B =1.5\, {\rm Gpc}h^{-1}$ with periodic boundary conditions. The original mass of the dark matter particles is $m_p=7.6 \times 10^{10} M_\odot h^{-1}$, and the minimum halo mass has been selected to be $7.8 \times 10^{12} M_\odot h^{-1}$, which corresponds a bias of $b\sim2$. The halo catalogues are generated by the Friends of Friends algorithm \citep{Davisetal:1985} with a linking length of 0.168 times the mean inter-particle spacing.  In order to extract the dark matter and halo field, a CiC prescription has also been used with $512^3$ grid cells, whose size is $2.93\, {\rm Mpc}h^{-1}$.  No observational features, such as survey geometry or galaxy weights, are incorporated.

  \item A set of 600 + 600 realisations of mocks galaxies with the CMASS DR11 NGC and SGC survey geometry, respectively. This is the galaxy catalogue presented in \cite{Maneraetal:2013} based on \textsc{PThalos}. Galaxies have been added using a Halo Occupation Distribution  (HOD) prescription (see \citealt{Maneraetal:2013}  for details). These catalogues contain both survey geometry and galaxy weights.
  
  \end{enumerate}
Realisations (i) to (iv) are based on  $\Lambda$CDM cosmology with matter density $\Omega_m=0.274$,  cosmological constant $\Omega_\Lambda = 0.726$,  baryon density $\Omega_b = 0.04$,  reduced Hubble constant $h=0.7$, matter density fluctuations characterised by an $\sigma_8=0.8$ and power law primordial power spectrum with spectral slope  $n_s =0.95$ (as used in \citealt{Andersonetal:2012}). All snapshots are at a redshift $z_{\rm sim}=0.55$, which is very close to the effective redshift of the CMASS data $z_{\rm eff}=0.57$. Under the assumption that  general relativity is the correct   description for  gravity, the logarithmic growth factor at this epoch is $f(z_{\rm eff})=0.744$  and $\sigma_8({\rm z_{eff}})=0.6096$.
\begin{enumerate}
  \setcounter{enumi}{4}
\item An additional set of dark matter N-body simulations is used in \S\ref{test_dark_matter} only. They consist of an N-body dark matter particles simulation with flat $\Lambda$CDM cosmology slightly different from the (i) - (iv). The box size is $L_B = 2.4\,{\rm Gpc}h^{-1}$ with periodic boundary conditions and the number of particles is $N_p = 768^3$, with 60 independent runs. The cosmology used is the dark energy density, $\Omega_\Lambda=0.73$, matter density, $\Omega_m=0.27$, Hubble parameter, $h=0.7$, baryon density, $\Omega_bh^2=0.023$, spectral index $n_s = 0.95$ and the amplitude of the primordial power spectrum at $z = 0$, $\sigma_8= 0.7913$. Taking only the gravitational interaction into account, the simulation was performed with GADGET-2 code \citep{Gadget2}. The snapshot used in this paper is at $z=0$. In order to obtain the dark matter field from particles we have applied the CiC prescription using $512^3$ grid cells. Thus the size of the grid cells is $4.68\, {\rm Mpc}h^{-1}$.
\end{enumerate}

\section{Method}
\label{section_method}
In this section we describe the methodology used to extract the measurements of bias parameters and the growth of structure. The performance of our methodology, and the tests performed to quantify any possible systematic errors, are reported  in \S ~\ref{section:systematics}.

\subsection{Definitions}   

The power spectrum $P$ and bispectrum $B$ are the two- and three-point functions in Fourier space. For a cosmological over-density  field $\delta$, they are defined as,
 \begin{eqnarray}
\label{pspmass}\langle \delta_{\bk}\delta_{\bk'}\rangle &\equiv&
(2\pi)^3 P(\bk) \delta^{\rm D}(\bk+\bk'),\\
\label{bispmass}\langle \delta_{\bk_1}\delta_{\bk_2}\delta_{\bk_3}\rangle &\equiv&
\label{bispectrum_definition}(2\pi)^3 B(\bk_1,\bk_2) \delta^{\rm D}(\bk_1+\bk_2+\bk_3),
\end{eqnarray}
where $\delta^{\rm D}$ is the Dirac delta distribution, $\delta_{\bk} \equiv \int d^3\bx\,\delta(\bx)\exp(-i\bk\cdot\bx)$ is the Fourier transform of the overdensity  $\delta(\bx) \equiv \rho(\bx)/\overline{\rho}-1$, where $\rho$ is the dark matter density and $\bar{\rho}$ its mean value. Eq.~\ref{bispectrum_definition} shows that bispectrum can be non-zero only if the $\bk$-vectors close to form a triangle.

In order to compute the galaxy power spectrum and bispectrum, we make use of the Feldman-Kaiser-Peacock estimator (FKP-estimator \citealt{FKP}), which has been used in previous analysis of bispectrum of galaxy surveys \citep{Scoccimarroetal:2001,Verdeetal:2002}.  
The FKP galaxy fluctuation field is defined,
\begin{eqnarray}
F_i({\bf r})\equiv w_{\rm FKP}({\bf r})\lambda_i\left[ w_c({\bf r}) n({\bf r}) - \alpha n_s({\bf r})\right],
\label{fkp_estimator}
\end{eqnarray}
where $n$ and $n_s$ are, respectively, the observed number density of galaxies and the number density of a random catalogue, 
which is a synthetic catalog Poisson sampled  with the same mask and selection function as the survey but otherwise no intrinsic (cosmological) correlations; $w_c$ and $w_{\rm FKP}$ were defined in Eqs. \ref{completness_weights} and \ref{fkp_weights} respectively; $\alpha$ is the ratio between the weighted number of observed galaxies over the random catalogue galaxies, $\alpha\equiv{\sum_i^{N_{\rm gal}}w_c}/{N_s}$ where $N_s$ denotes the number of objects in the synthetic catalog and $N_{\rm gal}$ the  number of galaxies in the  (real) catalog. The pre-factor defined as $\lambda_i$ is a normalisation to be chosen to make the power spectrum (for index $i=2$) and bispectrum  (for index $i=3$) estimators unbiased with respect to their definitions in Eq.~\ref{pspmass}-\ref{bispmass}.
It is convenient to define the coefficients,
\begin{eqnarray}
\label{I_factors}I_{i}\equiv\int d^3{\bf r}\, w_{\rm FKP}^i({\bf r}){\langle {n w_c}\rangle}^i({\bf r}).
\end{eqnarray}
These factors play a key role in the normalisation as shown below. 

\subsection{Estimating the Power spectrum}\label{sec:estimator_ps}
The normalisation for the power spectrum can be conveniently chosen, $\lambda_2\equiv I_{2}^{-1/2}$, to match the theoretical power spectrum when $\bar{n}$ has no  dependence  on position. Thus, the galaxy power spectrum estimator used in this work is,
\begin{eqnarray}
\label{fkp_ps}F_2({\bf r})\equiv I_{2}^{-1/2}w_{\rm FKP}({\bf r})\left[w_c({\bf r})n({\bf r}) - \alpha n_s({\bf r})\right].
\end{eqnarray}
From this expression we obtain,
\begin{eqnarray}
\langle |F_2({\bf k})|^2\rangle=\int \frac{d^3{\bf k}'}{(2\pi)^3}\, P_{\rm gal}({\bf k}')|W_2({\bf k}-{\bf k}')|^2+P_{\rm noise},
\label{power_spectrum_FKP}
\end{eqnarray}
where $P_{\rm gal}$ is the theoretical prediction for the galaxy (or tracer) power spectrum in the absence of any observational effect, $P_{\rm noise}$ is the shot noise term (see Appendix~\ref{appendix_shot_noise} for the model used and \S~\ref{sec:methodshotnoise}) and $W_2$ is the window function, which is defined as,
\begin{eqnarray}
W_2({\bf k})\equiv I_{2}^{-1/2}\int d^3{\bf r}\, w_{\rm FKP}({\bf r}){\langle w_c n\rangle}({\bf r})e^{+i{\bf k}\cdot{\bf r}}.
\label{G_2_definition}
\end{eqnarray}
The random catalogue satisfies the expression $\langle w_c n \rangle({\bf r}) = \alpha \langle n_s \rangle({\bf r})$, and it can be therefore used to generate the window function. We do not consider correcting Eq. \ref{power_spectrum_FKP} by the integral constraint, because its effect it is only relevant at larger scales that the ones considered in this paper.

We will designate the left hand side of  Eq.~\ref{power_spectrum_FKP}  $P^{\rm meas.}$ when $F_2$ is extracted from any of the catalogs (real or simulated) of \S~\ref{sec:sims-mocks}. In \S~\ref{sec:38} we will provide the details of the computation of $F_2$.

For any model $P(k)$ the convolution of  Eq.~\ref{power_spectrum_FKP} can be performed numerically in Fourier space in a minutes-time scale on a single processor for a reasonably large number of grid-cells (such as $512^3$ or $1024^3$) using fftw\footnote{Fastest Fourier Transform in the West: http://fftw.org}. An alternative option, which we do not adopt, would be to reduce the integral of Eq.~\ref{power_spectrum_FKP} to a 1-dimensional integral \citep{Rossetal:2013}, defining a spherically-averaged window function,
and making the assumption that the power spectrum input is an isotropic function, although numerical results demonstrate that this is a good approximation. 
 The model for  $P_{\rm noise}$ in the presence of completeness weight and other real-world effects is presented in Appendix~\ref{appendix_shot_noise}. This derivation assumes that the shot noise follows  Poisson statistics and the accuracy of the error estimation relies on the mocks having the same statistical properties for the shot noise as the data. For our final analysis of the data, we will  treat the shot noise amplitude as a nuisance parameter and marginalise over it. This approach accounts for  possible deviations from Poisson statistics as well as limitations of the mocks.

For the BOSS CMASS DR11 survey $W_2({\bm \epsilon})$ is a rapidly decreasing function with a width of $1/L_{\rm svy.}$, where $L_{\rm svy.}$ characterises the typical size of the survey. Provided that $P_{\rm gal}({\bf k})$ is smooth at small scales, the value of the integral in Eq.~\ref{power_spectrum_FKP} tends to $P_{\rm gal}$ for large values of $\bf k$. 

One of the FKP-estimator limitations is that the line-of-sight vector cannot be easily included in this formalism. This estimator is consequently only suitable for calculating monopole statistics (both power spectrum and bispectrum). Except for narrow angle surveys  \citep{Blakeetal:2013}, higher order multipoles, such as the quadrupole or hexadecapole, require a more complex estimator, such as described by \cite{Yamamotoetal:2006}, as  is implemented in \cite{Beutleretal:2013} for the CMASS DR11 galaxy sample. In what follows we will denote the monopole (angle average) of the right hand side of Eq.~\ref{power_spectrum_FKP} $P^{\rm model}(k)$, when $P_{\rm gal}(k)$ is  the monopole (angle average) of Eq.~\ref{Psg} in \S~\ref{method:powerspectrum}.  

\subsection{Estimating  the Bispectrum}
\label{sec:bispectrumestimator}
As for the power spectrum, we can define a FKP-style estimator for the bispectrum. In general, for the $N$-point correlation function, $\lambda_N$ should be set to $I_{N}^{1/N}$ to provide an unbiased relation between $\langle F^N\rangle$ and the $N$-point statistical moment. Therefore we  set the normalisation factor to $\lambda_3\equiv I_{3}^{-1/3}$ and   the galaxy field estimator for the bispectrum is,
\begin{eqnarray}
\label{fkp_bis}F_3({\bf r})\equiv I_{3}^{-1/3} w_{\rm FKP}({\bf r})\left[ w_c({\bf r})n({\bf r}) - \alpha n_s({\bf r})\right].
\end{eqnarray}
With this estimator, we can write,
\begin{equation}
 \label{Bispectrum_FKP}\langle F_3({\bf k}_1) F_3({\bf k}_2) F_3({\bf k}_3)\rangle=\int \frac{d^3{\bf k'}}{(2\pi)^3} \frac{d^3{\bf k''}}{(2\pi)^3}\, B_{\rm gal}({\bf k}',{\bf k}'')
 W_3({\bf k}_1-{\bf k}',{\bf k}_2-{\bf k}'')+B_{\rm noise}({\bf k}_1,{\bf k}_2),
\end{equation}
where we always assume ${\bf k}_3\equiv -{\bf k}_1-{\bf k}_2$, that ensures that the 3 $k$-vectors form a triangle. As for the power spectrum, the expression for the shot noise, $B_{\rm noise}$, is derived in Appendix~\ref{appendix_shot_noise}  and further discussed in \S~\ref{sec:methodshotnoise}. The window function $W_3$ can be written in terms of the window function of the power spectrum, 
\begin{eqnarray}
\label{window_bispectrum}W_3({\bf k}_A,{\bf k}_B)\equiv \frac{I_2^{3/2}}{I_{3}}\left[ W_2({\bf k}_A) W_2({\bf k}_B) W_2^*({\bf k}_A+{\bf k}_B)  \right].
\end{eqnarray}
 Eqs.~\ref{Bispectrum_FKP}  and \ref{window_bispectrum}  can be derived from the definition of $F({\bf r})$ in Eq.~\ref{fkp_estimator}. We will designate the left hand side of  Eq.~\ref{Bispectrum_FKP}  $B^{\rm meas.}$ when $F_3$ is extracted from any of the catalogs (real or simulated) of \S~\ref{sec:sims-mocks}. In \S~\ref{sec:38} we provide the details about the computation of $F_3$ from a galaxy distribution.
 
 Performing the double convolution between the window function and the theoretical galaxy bispectrum (Eq.~\ref{Bispectrum_FKP}) can be a challenging computation for a suitable number of grids cells (such as $512^3$ or $1024^3$).
In this work we perform an approximation that we have found to work reasonably well, which introduces biases that are negligible compared to the statistical errors  of this survey.
 It consists of assuming that the input theoretical bispectrum is of the form $B_{\rm gal}(k_1,k_2,k_3)\sim P(k_1)P(k_2) {\cal Q}(k_1,k_2,k_3) + {\rm cyc}$, where ${\cal Q}$ can be any function of the 3 $k$-vectors. 
 Then, ignoring the effect of the window function on ${\cal Q}$, the integral of Eq.~\ref{Bispectrum_FKP} is separable. As a consequence, we can simply write,
\begin{eqnarray}
\label{bispectrum_approximation}\int \frac{d^3{\bf k'}}{(2\pi)^3} \frac{d^3{\bf k''}}{(2\pi)^3}\, B_{\rm gal}({\bf k}',{\bf k}'')
 W_3({\bf k}_1-{\bf k}',{\bf k}_2-{\bf k}'')\!&\!\!=\!\!&\int \frac{d^3{\bf k'}}{(2\pi)^3} \frac{d^3{\bf k''}}{(2\pi)^3}\, P(k') P(k'') {\cal Q}(k',k'',|{\bf k}'+{\bf k}''|)W_3({\bf k}_1-{\bf k}',{\bf k}_2-{\bf k}'')\\
\nonumber&\simeq&[P\otimes W_2](k_1)	\times[P\otimes W_2](k_2) \times {\cal Q}(k_1,k_2,k_3),
\end{eqnarray}
where we have defined,
 \begin{eqnarray}
\label{combolucion_window} [P\otimes W_2](k_i)\equiv\int \frac{d^3{\bf k}'}{(2\pi)^3}\, P({\bf k}')|W_2({\bf k}_i-{\bf k}')|^2.
\end{eqnarray}
This approximation works reasonably well  for modes that are not too close to the size of the survey i.e., {\it all} three sides of the $k$-triangle are sufficiently large. The approximation  fails to reproduce accurately the correct bispectrum shape when  (at least) one of the $k_i$ is close to the fundamental frequency, $k_f=2\pi/L$, where $L$ is the typical survey size. In particular for the geometry of CMASS DR11, this limitation only applies to  triangle configurations  where the modulus of one $k$-vector is much shorter than the other two ($k_3\ll k1\sim k2$, the so-called {\it squeezed} configuration) and   the shortest $k$ is $\lesssim0.03\, h{\rm Mpc}^{-1}$. We test the efficiency of this estimator in \S~\ref{5_3_section}.

In what follows we will refer to the right hand side of Eq.~\ref{Bispectrum_FKP} as  $B^{\rm model}$ where we will use the simplification of Eq.~\ref{bispectrum_approximation} and where we consider the  galaxy (or tracer) bispectrum monopole for $P(k)P(k'){\cal Q}(k_1,k_2,k_3)$+cyc.  when the expression for the redshift space galaxy bispectrum  is that  reported in Eq.~\ref{B_gggs} in \S~\ref{sec:bispectrummethod}. 
 
\subsection{The galaxy bias model}
\label{sec:biasmethod}

The galaxy bias is defined as the mapping functional between the dark matter and the galaxy density field. When this relation is assumed to be local and deterministic we can generically write,
\begin{eqnarray}
\label{bias_delta}\delta_{g}({\bf x})={\mathcal B}[\delta({\bf x)}]\delta({\bf x}),
\end{eqnarray}
where all possible non-linearities of the bias are encoded in the functional $\mathcal B$. 
A simple and widely used model for  $\mathcal B$ is a 
 simple Taylor expansion in $\delta$ \citep{FryGazta:1993}, often truncated at the first or second-order (for bispectrum analyses of galaxy catalogs using this bias model see \citealt{Scoccimarroetal:2001,FFFS01,Verdeetal:2002,GS2005} and \citealt{WiggleZ_marin}).  While this model is still widely used in bispectrum forecasts, here we argue that it is insufficient for the precision and bias properties offered by the CMASS sample.  

Recently it has been shown, by both analytical and numerical methods, that the gravitational evolution of the dark matter density field naturally induces non-local bias terms in the halo- (and therefore galaxy-) density field, even when the initial conditions are local (see \citealt{Catelan98} for initial investigations). Some of these non-local bias terms contribute  at mildly non-linear scales and therefore they only introduce non-leading order corrections in the shape and amplitude of the power spectrum and bispectrum. However, other terms contribute at large scales, at the same level as the linear, local bias parameter, $b_1$ \citep{McDonaldRoy:2009,Baldaufetal:2012,Chan12,Saitoetal:2014}. 

In practice, neglecting the non-local bias terms can produce a mis-estimation of the other bias parameters, even  when working only at large, supposedly linear, scales. \cite{FFFS01} were the first to apply a local Lagrangian bias to the IRAS PSCz survey catalogue (Infra-Red Astronomical Satellite Point Source Catalog)\citep{PSCz} and compare it with an Eulerian local bias model. Their results concluded that for that particular galaxy population the local Eulerian bias described better the data that the local Lagrangian bias with a likelihood ratio of ${\mathcal{L}}_E/{\mathcal{L}}_L=1.6$.   However, for  N-body haloes, mock haloes and mock galaxies,  we have checked that the local Eulerian description of the bias produces inconsistent results.  In the  Eulerian local bias model, the power spectrum requires a value of $b_1$ which is significantly higher than the one required by the bispectrum, even at large scales (both in real and in redshift space). In a similar way, the value of $b_2$ required by the halo-halo power spectrum is smaller than the one required by the power spectrum. These discrepancies are reduced when the Lagrangian local model is assumed (see for example Fig. \ref{PS_BS_plot}, where the predictions from the power spectrum and bispectrum actually cross in a region). This result suggests that for the N-body and halo and galaxy catalogues used in this paper, the local Lagrangian bias model provides a better description than the local Eulerian bias model. Of course, we do not know whether for the observed CMASS BOSS LRGs galaxies, this behaviour holds. However, for this point (and many others), we are assuming that the observed galaxy field is qualitatively similar to the simulated galaxy field, and therefore, seems us reasonable to assume the local Lagrangian model instead of local Eulerian for the bias model.

Hence, we use the Eulerian non-linear and non-local bias model proposed by \cite{McDonaldRoy:2009}. The non-local terms  are included through a quadratic term in the tidal tensor $s({\bf x})=s_{ij}({\bf x})s_{ij}({\bf x})$, with $s_{ij}({\bf x})=\partial_i\partial_j\Phi({\bf x})-\delta_{ij}^{\rm Kr}\delta({\bf x})$. Here $\Phi({\bf x})$ is the gravitational potential, $\nabla^2\Phi({\bf x})=\delta({\bf x})$. 
 With this non-local term, our adopted second-order expression for the relation between $\delta_g$ and $\delta$ is:
\begin{eqnarray}
\label{delta_galaxy}\delta_{g}({\bf x}) = b_1 \delta({\bf x}) + \frac{1}{2}b_2 [\delta({\bf x})^{2}-\sigma_2]+\frac{1}{2}b_{s^2}[s({\bf x})^2-\langle s^2 \rangle] + \mbox{higher order terms,}
\end{eqnarray}
where $b_1$ is the linear bias term, $b_2$ is the non-linear bias term and $b_{s^2}$ the non-local bias term. The terms $\sigma_2$ and $\langle s^2\rangle$ ensure the condition $\langle \delta_g\rangle=0$. Most of the third order terms in $\delta_{g}$ contribute to fourth and higher order corrections in the power spectrum and bispectrum and will not be considered in this paper; however, for the power spectrum,  some contributions coming from  these terms  are not negligible  at second order and must be considered (see \citealt{McDonaldRoy:2009} for a full discussion). We refer to this extra bias term as $b_{3\rm nl}$. In Fourier space the Eq.~\ref{delta_galaxy} reads,
\begin{eqnarray}
\delta_{g}({\bf k}) = b_1 \delta({\bf k}) + \frac{1}{2}b_2 \int \frac{d{\bf q}}{(2\pi)^3}\, \delta({\bf q})\delta({\bf k}-{\bf q})+\frac{1}{2}b_{s^2}\int	\frac{d{\bf q}}{(2\pi)^3}\, \delta({\bf q})\delta({\bf k}-{\bf q}) S_2({\bf q},{\bf k}-{\bf q})+ \mbox{higher order terms},
\label{deltah}
\end{eqnarray}
where we ignore the contributions of $\sigma_2$ and $\langle s^2\rangle$ to the $k=0$ mode, which is not observable. $S_2$ is related to the $s_{ij}({\bf x})$ field as, 
\begin{eqnarray}
s^2({\bf k})=\int\frac{d\,{\bf k}'}{(2\pi)^3}\,S_2({\bf k}', {\bf k}-{\bf k}')\delta({\bf k}')\delta({\bf k}-{\bf k}')
\end{eqnarray}
where $s^2({\bf k})$ is just the Fourier transform of $s^2({\bf x})$ field. This relation implies that the $S_2$ kernel is defined as,
\begin{eqnarray}
\label{eq:S2kernel}
S_2({\bf q}_1,{\bf q}_2)\equiv\frac{({\bf q}_1\cdot{\bf q}_2)^2}{(q_1q_2)^2}-\frac{1}{3}.
\end{eqnarray}
The bias model of Eq.~\ref{deltah} depends on four different bias parameters, $b_1$, $b_2$, $b_{s^2}$ (which appear both in the power spectrum and bispectrum) and also $b_{3\rm nl}$ that contributes the second order in the power spectrum. In this paper we assume that, although the galaxy bias is non-local in Eulerian space, is local in Lagrangian space. Under this assumption, the non-local bias terms  can be related at first order to the linear bias term $b_1$\footnote{If we incorporate the pre-factor $1/2$ in the bias parameter $b_{s^2}$, then the relation changes to $b_{s^2}=-\frac{2}{7}(b_1-1)$.}, 
\begin{eqnarray}
b_{s^2}&=&-\frac{4}{7}(b_1-1)\quad \mbox{\citep{Chan12,Baldaufetal:2012}}, \\
b_{3\rm nl}&=&\frac{32}{315}(b_1-1)\quad \mbox{\citep{Beutleretal:2013,Saitoetal:2014}}.
 \end{eqnarray}
With these relations, we are able to express the galaxy biasing as a function of only two free parameters, $b_1$ and $b_2$. Eq.~\ref{deltah} is the starting point for computing the galaxy power spectrum and bispectrum.

 The second order bias parameter, $b_2$ can be quite sensitive to truncation effects. In this sense, $b_2$ should be treated as an effective parameter that absorbs part of the higher order contributions that are not considered when we truncate Eq.~\ref{deltah} at second order. In an other work \citep{HGMetal:inprep} it has been reported that even for dark matter, $b_2$  can present non-zero values due to these sort of effects. We therefore treat $b_2$ as a nuisance parameter, to be marginalised over.

\subsection{The power spectrum model}
\label{method:powerspectrum}
The real-space galaxy power spectrum $P_{{g},\delta\delta}(k)$, can be written as a function of the statistical moments of dark matter using Eq.~\ref{deltah} and perturbation theory as (see \citealt{McDonaldRoy:2009,Beutleretal:2013}),
\begin{eqnarray}
\label{PS_real} P_{{g},\delta\delta}(k)&\!\!\!=\!\!\!&b_1\left [b_1 P_{\delta\delta}(k)+2b_2P_{b2,\delta}(k)+2b_{s^2}P_{bs2,\delta}(k)+2b_{3{\rm nl}}\sigma_3^2(k)P^{\rm lin}(k)\right]   +b_2\Big[     b_2P_{b22}(k)+2b_{s^2}P_{b2s2}(k)\Big]+b_{s^2}^2P_{bs22}(k),
\end{eqnarray}
where $P^{\rm lin}$ and $P_{\delta\delta}$ are the linear and non-linear matter power spectrum, respectively. The other terms correspond to 1-loop corrections due to higher-order bias terms and their explicit form can be found in Appendix~\ref{appendix_power_spectrum}.

The mapping from real space to redshift space quantities involves the power spectrum of the velocity divergence $\theta({\bf k})=[-i {\bf k}\cdot {\bf v}({\bf k})]/[af(a)H]$. We assume that there is no velocity bias between the underling dark matter field and the galaxy field at least on the relatively large scales of interest. According to \cite{Taruyaetal:2010} and \cite{Nishimichietal:2011} (hereafter TNS model), the galaxy power spectrum in redshift space can be approximated as,
\begin{eqnarray}
\label{Psg}P_{g}^{(s)}(k,\mu)=D^P_{\rm FoG}(k,\mu,\sigma_{\rm FoG}^P[z])\left[ P_{g,\delta\delta}(k)+2f\mu^2P_{g,\delta\theta}(k)+f^2\mu^4P_{\theta\theta}(k)+b_1^3A(k,\mu,f/b_1)+b_1^4B(k,\mu,f/b_1)  \right],
\end{eqnarray}
where $\mu$ denotes the cosine of the angle between the $k$-vector and the line of sight, $f$ is the linear growth rate $f=\partial \ln \delta/\partial \ln a$, and $P_{g,\delta\delta}(k)$ is given by Eq.~\ref{PS_real}. 
The quantities $P_{g,\delta\theta}$,  and $P_{\theta\theta}$, are  the non-linear power spectra for the  galaxy density-velocity, and the dark matter velocity-velocity, respectively. The expressions for all these terms are  reported in Appendix~\ref{appendix_power_spectrum}; here it will suffice to say that the model  for the non-linear matter quantities is obtained using resummed perturbation theory (hereafter RPT) at 2-loop as is described in \cite{HGMetal:2012} (hereafter 2L-RPT).

 The factor $D^P_{\rm FoG}$  is often referred to as the Fingers-of-God (hereafter FoG) factor and accounts for the non-linear damping due to the velocity dispersion of satellite galaxies ($\sigma_{\rm FoG}^P[z]$) inside the host halo. However we treat this factor as an effective parameter that enclose our poor understanding of the non-linear redshift space distortions and to be  marginalized over.   The expression adopted  for $D_{\rm FoG}^P$ is also reported in Appendix~\ref{appendix_power_spectrum}.

The angular dependence of the redshift space power spectrum is often expanded in Legendre polynomials (see Appendix~\ref{appendix_power_spectrum} for details). Here we will only  consider the monopole, i.e., the angle-averaged power spectrum. For this reason our analysis is  complementary to and independent of that of \cite{Beutleretal:2013,Chuangetal:2013,Samushiaetal:2013} and \cite{,Sanchezetal:2013}, who use the quadruple to monopole ratio. However, this does not mean that the results presented here and their results can be combined as if they were independent measurements (the survey is the same);  we will explore in future work whether error-bars could be further reduced by combining the two approaches.

\subsection{The bispectrum model}\label{sec:bispectrummethod}
The galaxy-bispectrum  in real space can be written using to the bias model of Eq.~\ref{deltah} as, 
\begin{eqnarray}
\label{B_ggg1}B_{g}(k_1,k_2,k_3)=b_1^3 B(k_1,k_2,k_3) +b_1^2\left[ b_2 P(k_1)P(k_2) + b_{s^2} P(k_1)P(k_2) S_2({\bf k}_1,{\bf k}_2) + {\rm cyc.} \right]
\end{eqnarray}
where $P$ and $B$ are the non-linear matter power spectrum and bispectrum, respectively,  and  we have neglected  terms proportional to $b_2^2$, $b_{s^2}^2$, which  are of higher order. Using the 2L-RPT model for the matter bispectrum proposed by \cite{HGMetal:2011}, we can express the
real space galaxy bispectrum as a function of the non-linear matter power spectrum and the effective kernel, ${\cal F}^{\rm eff}_2({\bf k}_1,{\bf k}_2)$ (see \citealt{HGMetal:2011} and Appendix~\ref{appendix_bispectrum}),
\begin{eqnarray}
\label{B_ggg2}B_{g}(k_1,k_2,k_3)= 2P(k_1)P(k_2)\left[ b_1^3 {\cal F}^{\rm eff}_2({\bf k}_1,{\bf k}_2) + \frac{b_1^2b_2}{2}  + \frac{b_1^2b_{s^2}}{2}S_2({\bf k}_1,{\bf k}_2) \right] + {\rm cyc.}\,.
\end{eqnarray}
The non-local bias ($b_{s2}$) contributes to the leading order and introduces a new shape dependence through the kernel $S_2$ (defined in Eq.~\ref{eq:S2kernel}), which was not present in the matter bispectrum. In this case, we do not consider the contribution of $b_{3\rm nl}$ because for the bispectrum (in contrast to the power spectrum) it only appears in fourth and higher order corrections in $\delta_g$.

Redshift space distortions can be  included in this model by introducing an effective kernel $Z_2^{\rm eff}(k_1, k_2,{\bf \Psi})$ (\citealt{HGMetal:inprep} and  Appendix ~\ref{appendix_bispectrum}), where ${\bf \Psi}$ denotes the parameters to be fitted, of which  the ones of interest are $f, b_1, b_2, b_{s^2}$.
With this the galaxy bispectrum in redshift space as a function of the non-linear real-space matter power spectrum is \citep{HGMetal:inprep}: 
\begin{equation}
\label{B_gggs}B_{g}^{(s)}({\bf k}_1,{\bf k}_2)=D^B_{\rm FoG}(k_1,k_2,k_3,\sigma_{\rm FoG}^B[z])\left[2P(k_1)\,Z_1({\bf k}_1)\,P(k_2)\,Z_1({\bf k}_2)\,Z^{\rm eff}_2({\bf k}_1,{\bf k}_2)+\mbox{cyc.}\right],
\end{equation}
 where  $Z_1$, denotes the redshift space kernel predicted by SPT and the $Z_2^{\rm eff}$ kernel is a phenomenological extension  of the SPT kernel $Z_2$ (for a detailed derivation and explicit expressions, see Appendix~\ref{appendix_bispectrum}).  The  $D_{\rm FoG}^B$ term is a damping factor that aims to describe the Fingers-of-God effect due to velocity dispersion inside virialised structures through the one-free parameter, $\sigma_{\rm FoG}^B$, which we will also marginalise over. Here $\sigma_{\rm FoG}^B$ is a different (nuisance) parameter from $\sigma_{\rm FoG}^P$ in Eq.~\ref{Psg}. In this paper we will treat $\sigma_{\rm FoG}^P$ and $\sigma_{\rm FoG}^B$ as independent parameters although they may be weakly correlated.  The adopted expression for $D_{\rm FoG}^B$ is  reported in Eq.~\ref{D_fog_sc} in Appendix~\ref{appendix_bispectrum}.

As  for the power spectrum, we can expand the redshift space bispectrum in multipoles (see Appendix ~\ref{appendix_bispectrum} for details);  here we will consider only the monopole (i.e., the $\mu$ angle-averaged bispectrum).

  Note that we are truncating the bispectrum description at a different order than the power spectrum. The description of the power spectrum is based on a physical, perturbative,  model;  is very accurate at large scales (few percent) until it dramatically breaks down, and cannot be applied anymore. On the other hand, the bispectrum description  is  phenomenological; has an accuracy of ~5\% at large scales and gradually deviate from the prediction of N-body. Therefore, is natural to expect that these two models present  different ranges of validity when they are applied to biased objects. In this paper we have opted to describe each statistic the best we can, even if this means truncating the power spectrum and bispectrum models at different scales and different orders.

\subsection{Shot noise}
\label{sec:methodshotnoise}
Discreteness introduces extra spurious power to both the power spectrum and bispectrum.
In this paper we consider that the (additive) shot noise contribution may be modified from that of a pure Poisson sampling. We parametrise this deviation through a free parameter, $A_{\rm noise}$,
\begin{eqnarray}
\label{AP}P_{\rm noise}&=&(1-A_{\rm noise})P_{\rm Poisson},\\
\label{AB}B_{\rm noise}({\bf k}_1,{\bf k}_2)&=&(1-A_{\rm noise})B_{\rm Poisson}({\bf k}_1,{\bf k}_2),
\end{eqnarray}
where the terms  $P_{\rm Poisson}$ and $B_{\rm Poisson}({\bf k}_1,{\bf k}_2)$ are the Poisson predictions for the shot noise; their expression can be found in Appendix~\ref{appendix_shot_noise}. For $A_{\rm noise}=0$ we recover the Poisson prediction, whereas when $A_{\rm noise}>0$ we obtain a sub-Poisson shot noise term and $A_{\rm noise}<0$ a super-Poisson noise term. The extreme case of $A_{\rm noise}=1$ corresponds to a sub-Poissonian noise that is null; $A_{\rm noise}=-1$ correspond to a super-Poissonian noise that doubles the Poisson prediction. We  expect that the observed noise is always contained between these two extreme cases, so we constrain the $A_{\rm noise}$ parameter to be, $-1\leq A_{\rm noise} \leq +1$.

\subsection{Measuring power spectrum and bispectrum of  CMASS galaxies from the BOSS survey}\label{sec:38}

In order to compute the power spectrum and bispectrum from a set of galaxies, we need to compute the   suitably weighted field $F_i({\bf x})$ described in \S~\ref{sec:bispectrumestimator}. We use a random catalogue of number density of $\bar{n}_s({\bf r})=\alpha^{-1} \bar{n}({\bf r})$ with $\alpha\simeq0.00255$, and therefore $\alpha^{-1}\simeq400$. In order to do so we place the NGC and SGC galaxy samples in boxes  which we discretise  in grid-cells, using a box with side of  $3500\,h^{-1}{\rm Mpc}$  to fit the NGC galaxies and of $3100\,h^{-1}{\rm Mpc}$ for the SGC galaxies. 

 The number of grid cells used for the analysis is $512^3$. This corresponds to a grid-cell resolution of $6.84\,h^{-1}{\rm Mpc}$ for NGC and $6.05\,h^{-1}{\rm Mpc}$ for SGC. The fundamental wave-lengths are $k_f=1.795\cdot10^{-3}\,h{\rm Mpc}^{-1}$ and $k_f=2.027\cdot10^{-3}\,h{\rm Mpc}^{-1}$ for the NGC and SGC boxes, respectively.  We have checked that for $k\leq0.25\,h{\rm Mpc}^{-1}$, doubling the number of grid-cells per side, from $512^3$ to $1024^3$, produces a negligible change in the power spectrum. This result indicates that using $512^3$ grid-cells provides sufficient resolution at the scales of interest.

We apply the CiC method to associate galaxies to grid-cells to obtain the quantity $F_i({\bf r})$ of Eq.~\ref{fkp_estimator} on the grid.

To obtain $P^{\rm meas.}({\bf k})=\langle|F_2({\bf k})|^2\rangle$,
we bin the power spectrum $k-$modes in 60 bins between the fundamental frequency $k_f$ and the maximum frequency for a given grid-size  with width  $\Delta \log_{10} k=\left[ \log_{10}(k_{\rm M})-\log_{10}(k_f) \right]/60$, where $k_{\rm M}\equiv \sqrt{3} k_f N_{\rm grid}/2$ is the maximum frequency  and $N_{\rm grid}$ is the number of grid-cells per side, in this case 512.

We use the real part of $\langle F_{{\bf k}_1}F_{{\bf k}_2}F_{{\bf k}_3}\rangle$ as our data for the bispectrum, for triangles in $\bk$-space (i.e. where $\bk_1+\bk_2+\bk_3={\bf 0}$). Therefore we have   $B^{\rm meas.}({\bf k}_1,{\bf k}_2,{\bf k}_3)={\rm Re}\left[\langle F_3({\bf k}_1) F_3({\bf k}_2) F_3({\bf k}_3)\rangle\right]$. There is clearly a huge number of possible triangular shapes to investigate; it is not feasible in practice to  consider them all. However, is not necessary to consider all possible triplets as their bispectra are highly correlated.  As shown in \cite{MVH97}, triangles with one $k$-vector in common are correlated, through cross-terms in the 6-point function. In addition, the survey window function induces mode coupling which correlates different triplets  further. In particular, in this paper we focus on those triangles with $k_2/k_1=1$ and $2$, allowing $k_3$ to vary from $|{\bf k}_1-{\bf k}_2|$ to $|{\bf k}_1+{\bf k}_2|$.

We choose to bin $k_1$ and $k_3$ in fundamental $k$-bins of $\Delta k_1=\Delta k_3=k_f$.  Additionally, $k_2$ is binned in fundamental $k$-bins when $k_1=k_2$. However, for those triangles with $k_2/k_1=2$ we bin $k_2$ in $k$-bins of $2k_f$ in order to cover all the available $k$-space. Thus, generically we can write $\Delta k_2=(k_2/k_1) \Delta k_1$.   We have checked that changing the bin-size has a negligible impact on the best fit parameters as well as on their error. We present results in the plots using the bin size adopted in the analysis.

The measurement of the bispectrum is  performed with an approach similar to that described in Appendix A of \cite{HGMetal:2011}. Given fixed $k_1$, $k_2$ and $k_3$, and a $k_i-$bin, defined by $\Delta k_1$, $\Delta k_2$ and $\Delta k_3$, we define the region that satisfies, $k_i-\Delta_{k_i}/2 \le q_i  \le k_i+\Delta_{k_i}/2$. There are a limited number of {\it fundamental} triangles in this $k$-space region, with the number depending on,
\begin{equation}
V_B(k_1,k_2,k_3)=\int_{\mathcal R} \,d{\bf q}_1\,d{\bf q}_2,\,d{\bf q}_3\, \delta^D({\bf q}_1, {\bf q}_2, {\bf q}_3)\simeq8\pi^2k_1k_2k_3\Delta_{k_1}\Delta_{k_2}\Delta_{k_3},
\end{equation}
where the $\simeq$ becomes an equality when $\Delta_{k_i}\ll k_i$. The value of the bispectrum is defined as the mean value of these fundamental triangles. Instead of trying to find these triangles, we cover this ${\mathcal R}$-region with $k$-triangles randomly-orientated in the $k$-space. The mean value of these random triangles tends to the mean value of the fundamental triangles when the number of random triangles is sufficiently large.  We have empirically found that the number of random triangles that we must generate to produce convergence to the mean value of the bispectrum is $\sim5V_B(k_1,k_2,k_3)/k_f^6$, where $k_f\equiv2\pi/L_B$ is the fundamental wavelength, and $L_B$ the size of the box.
 For each choice of $k_i, \Delta k_i\,, \{i=1,2,3\}$ provides us an estimate of what we call a single bispectrum {\it mode}.

When we perform the fitting process to the data set, we need to specify the minimum and maximum scales to consider. The largest scale we use for the fitting process is $0.03\,h{\rm Mpc}^{-1}$. This large-scale limit is caused by the survey geometry of the bispectrum (see \S~\ref{5_3_section} for details). The smaller the minimum  scale, the more $k$-modes are used and therefore the smaller the statistical errors. On the other hand, small scales are poorly modeled in comparison to large scales, such that we expect the systematic errors to grow as the minimum scale decreases. Therefore, we empirically find a compromise between these two effects such that the statistical and systematic errors are comparable. To do so, we perform different best fit analysis for different minimum scales and find the corresponding maximum $k$ by identifying changes on the best fit parameters that are larger than the statistical errors as we increase the minimum scale. 

In the following, when we report a $k_{\rm max}$ value, this means that none of the $k_{1,\,2,\,3}$ of the bispectrum triangles can exceed this value. In addition, our triangle catalogue is always limited by $k_1\leq0.1\,h{\rm Mpc}^{-1}$ when $k_2/k_1=2$ and $k_1\leq0.15\,h{\rm Mpc}^{-1}$ when $k_2/k_1=1$, because of computational reasons. 

The number of  modes  used is typically $\sim 5000$. If we wanted to use the mock catalogs to estimate the full covariance of both quantities (power spectrum and bispectrum), we would need to drastically reduce the number of bins (and modes), so that the total number of (covariance) matrix elements is much smaller than the number of mocks (currently 600 CMASS mocks are available).
This  could be achieved by increasing the $k$-bin size, but with the drawback of a significant loss of shape information. For this reason we will only estimate from the mock catalogs the diagonal elements of the  covariance ($\sigma_P^2[k]$, $\sigma_B^2[k_1,k_2,k_3]$), and use these as described in the next section.

\subsection{Parameter estimation}

Both the  power spectrum  and bispectrum in redshift space depend on cosmologically interesting parameters, the bias parameters as well as nuisance parameters.  
 The dependence is described in details in the above subsections. 
 
In total, for the full model,  we have seven free parameters ${\bf \Psi}=\{b_1, b_2, f, \sigma_8,A_{\rm noise},\sigma_{\rm FoG}^P, \sigma_{\rm FoG}^B \}$:
\begin{itemize}
\item Two parameters constrain the bias $b_1$ and $b_2$. Under the assumption of local Lagrangian bias, $b_1$ determines the value for $b_{s^2}$ and $b_{\rm nl3}$.
\item Two Fingers-of-God, redshift space distortion, parameters  $\sigma_{\rm FoG}^P$ and $\sigma_{\rm FoG}^B$.
\item A shot noise amplitude parameter $A_{\rm noise}$.
\item The logarithmic growth factor parameter $f$. This parameter can be predicted  for a given cosmological model (in particular if $\Omega_m$ is known) if we assume a theory for gravity. However, in this paper we consider this parameter free in order to test possible deviations from GR or, if assuming GR,  for not using a prior on $\Omega_m$.
\item The amplitude of the primordial dark matter power spectrum, $\sigma_8$.
\end{itemize}

 The other cosmological parameters, including $\Omega_m$, the spectral index $n_s$, and the  Hubble parameter $h$ are assumed fixed  to their fiducial values in the fitting process.
 In most cases they are set to the best fit values obtained by the Planck mission based on the cosmic microwave background (CMB) analysis  \citep{Planck_cosmology} in a flat $\Lambda$CDM model. We refer to these set of parameters as Planck13; they are listed in Table~\ref{cosmology_table}.  In selected occasions we will change this set of fiducial parameters to assess how our analysis depends on this assumption. The dependence on $\Omega_m$ is largely absorbed   by having $f$ as  a fitted parameter. 
 
 The probability  distribution for the bispectrum in the mildly non-linear regime is not known (although some progress  are being made see e.g., \citealt{Matsubara2007}); even if one invokes the central limit theorem  and model the distribution of bispectrum modes as a multi-variate Gaussian,   the evaluation of its covariance would be challenging (see e.g., eq. 38--42 of \citealt{MVH97}, appendix A of \citealt{VHMM98} and discussion above). In addition we want to analyse jointly the power spectrum and the bispectrum whose joint distribution is not known. Another approach is therefore needed.
We opt for the approach proposed in \cite{Verdeetal:2002}, which consists of introducing a suboptimal but unbiased estimator. Given an underlying cosmological model, ${\bf \Omega}$, and a set of free parameters to be fitted, 
 $\bf \Psi$, the power spectrum and bispectrum can be written as,
\begin{equation}
P^{\rm model}(k)=P^{\rm model}(k,{\bf \Psi}; {\bf \Omega})\,\, {\rm and}\,\,\,\,
B^{\rm model}(k_1,k_2,k_3)=B^{\rm model}(k_1,k_2,k_3,{\bf \Psi}; {\bf \Omega}).
\end{equation}

We then construct the $\chi^2_{\rm diag.}$-function as,
\begin{equation}
\label{chi2_formula}\chi^2_{\rm diag.}({\bf\Psi})=\sum_{k-{\rm bins}} \frac{\left[ P_{(i)}^{\rm meas.}(k)-P^{\rm model}(k,{\bf \Psi}; {\bf \Omega}) \right]^2}{\sigma_P(k)^{2}}+\sum_{{\rm triangles}} \frac{\left[ B^{\rm meas.}_{(i)}(k_1,k_2,k_3)-B^{\rm model}(k_1,k_2,k_3,{\bf \Psi}; {\bf \Omega}) \right]^2}{\sigma_B(k_1,k_2,k_3)^{2}},
\end{equation}
where we have ignored the contribution from off-diagonal terms, and we take into account only the diagonal terms, whose errors are given by $\sigma_P$ and $\sigma_B$, which are obtained directly from the mock catalogs.

We use a  Nelder-Mead based-method of minimization \citep{NR}. We impose some mild priors: $b_1>0$, $f>0$ and, in some cases, we also require $b_2>0$. As will be clear in \S~\ref{sec:constraining_gravity},  the $b_2>0$ prior has no effect on the results but it makes it easier to find the minimum for some of the mocks realisations.

 We obtain a set of parameters that minimizes $\chi^2_{\rm diag.}$ for a given realisation, $i$, namely ${\bf \Psi}_{(i)}$. By ignoring the off diagonal terms of the covariance matrix  (and  the full  shape of the  likelihood), we  do not have a have  maximum likelihood estimator which is necessarily minimum variance, optimal or  unbiased. However, we will demonstrate with tests on N-body simulations that this approximation does not bias the estimator

 Therefore,  {\it a)} the particular  value of the $\chi^2_{\rm diag.}$ at its minimum  is meaningless  and should not be used to estimate  a goodness of fit and {\it b)} the errors on the parameters cannot be estimated by standard $\chi^2_{\rm diag.}$ differences. The key property  of this method is that  $\langle{\bf\Psi}_{(i)}\rangle$ is an unbiased estimator of the true set ${\bf\Psi}_{{\rm true}}$ and that the dispersion of ${\bf\Psi}_{(i)}$ is an unbiased estimator of the error:  ${\bf\Psi}_{{\rm true}}$ should belong to the interval $\langle {\bf\Psi}_{{\rm i}}\rangle \pm \sqrt{\langle {\bf\Psi}_{{\rm i }}^2\rangle-{\langle {\bf\Phi}_{{\rm i}}\rangle}^2}$ with roughly $68\%$ confidence\footnote{The estimate of the confidence can only be approximate for three reasons a) the  error distribution is estimated from a finite number of realisations b) the realisations might not have the same statistical properties of the real Universe and the errors might slightly depend on that c) the distribution could be non-Gaussian.}.  
 
 We can demonstrate the sub-optimality  analytically as follows. The Cramer-Rao bound says that  the error for any unbiased estimator is always greater or equal to the  square root of the inverse of the Fisher information matrix. The maximum likelihood estimator is asymptotically  the best unbiased estimator that saturates the Cramer-Rao bound (i.e. you cannot do better than a maximum-likelihood estimator). Using the full covariance would correspond to do a maximum likelihood estimator in the region around the maximum, or otherwise said,  using the Laplace approximation. This would be the best unbiased estimator saturating the Cramer-Rao bound. Using only  the diagonal elements  therefore gives a sub-optimal estimator. Always in the  limit of the Laplace approximation, this estimator will still be unbiased.  In practice the maximum likelihood estimator might not be strictly unbiased (it is only asymptotically  and we have made the Laplace approximation to arrive to the above conclusion). Therefore we have  checked   that effectively the estimator is unbiased empirically: applying it to a case where the bias parameters are known, such as CDM simulations. As it was included in the text, this technique was used in \cite{Verdeetal:2002}, and it has been recently applied successfully in \cite{HGMetal:inprep}.

We will follow this procedure, using the   600 mock galaxy surveys from \cite{Maneraetal:2013}, we  estimate the errors from the CMASS DR11 data set in \S~\ref{sec:results}.
Since the realisations are independent, the dispersion on each parameter provides the associated error for a single realisation. 
This is true for the NGC and SGC alone, but not for the combined sample NGC+SGC. Both NGC and SGC catalogues  were created from the same set of 600 boxes of size 2400~$h^{-1}{\rm Mpc}$, just sampling a subsection of galaxies of these boxes to match the geometry of the survey. For the DR11 BOSS CMASS galaxy sample, it was not possible to sample NGC and SGC from the same box without overlap, as in for previous releases such as DR9 \citep{Ahnetal:2012}. In particular, for DR11 the full southern area is contained in the NGC (see  \S 6.1 of \citealt{Percivaletal:2013} for more details). Thus,  to compute the errors of the combined NGC+SGC sample one must use different boxes for the northern and southern components. We estimate the errors simply sampling the NGC from one subset of 300 realisations and combine them with the samples of the SGC from the other subset. In the same manner we can make another estimation sampling the NGC and SGC from the other subset of 300, respectively. We simply combine both predictions taking their mean value. Although we know that the error-bars must somewhat  depend on the assumed cosmology (and bias) in the mocks, in this work we consider this dependence negligible. 

 Note that since we are using 300 realizations to estimate the errors on a larger amount of $k$-bins (around 5000), the errors obtained may present inaccuracies respect to their expected value. A check  on the  performance of this approximation, accuracy of the  estimated errors and effects on the recovered parameters, is presented in the Appendix A of \cite{HGMetal:2011} for dark matter in real space.  There, using 40 realizations, the errors are estimated from the dispersion among realizations  and compared with the (Gaussian) analytic predictions. The result is that the errors estimated from the 40 realizations agree to a $\sim30\%$ accuracy with the analytic predictions up to $k\sim0.2\,h{\rm Mpc}^{-1}$. However, we want to stress that the methodology considered here is not very sensitive to the accuracy of how the errors are estimated. If the errors were overestimated by a constant factor, the best fit values of ${\bf\Psi}_{(i)}$ would be unaffected, and the variance among ${\bf\Psi}_{(i)}$ will be unchanged,  as it is estimated {\it \`a la} Monte-Carlo. If the errors were mis-estimated by shape-dependent factors, the estimator would be less-optimal, but still unbiased. Therefore, the validity of the methodology does not rely on the accuracy of the error-estimation, only its optimality.

\section{Results}\label{sec:results}

We begin by presenting the measured power spectrum and bispectrum and  later discuss the best fit model and the  constraints on the parameters of interest.
\begin{figure*}
\centering
\includegraphics[scale=0.4]{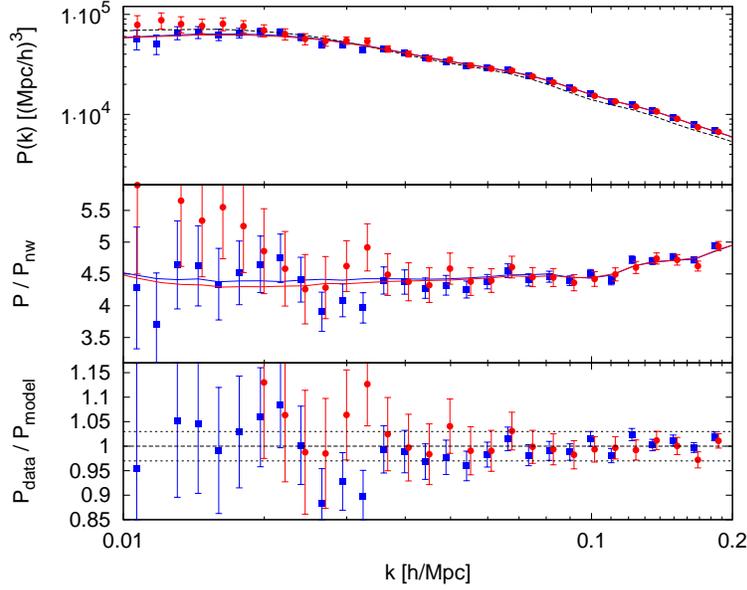}
\caption{Power spectrum data for the NGC (blue squares) and the SGC (red circles) versions and the best fit model prediction (red and blue lines) according to NGC+SGC Planck13  (Table~\ref{tab:bestfitparams}). Blue lines take into account the NGC mask and red lines the SGC mask. The top panel shows the power spectrum, middle panel the power spectrum normalised by a non-wiggle linear power spectrum for clarity, and the bottom panel the relative deviation of the data  from the model. The black dotted lines in the bottom panel mark the 3\% deviation respect to the model. In the top panel the average mocks power spectrum is indicated by the black dashed line. The model and the data show an excellent agreement within $3\%$ accuracy for the entire $k$-range displayed.}
\label{dataPS}
\end{figure*}
The top panel of  Fig.~\ref{dataPS} presents the power spectrum monopole of CMASS DR11 data measurements for NGC (blue squares) and SGC (red circles) galaxy samples. The model prediction using the best fit parameters corresponding to NGC + SGC is also shown and the best fit parameters values are reported in Table~\ref{tab:bestfitparams}. The  blue solid line includes the  NGC mask effect and  red solid line the SGC mask. We also show for reference the averaged value of the 600 realisations of the NGC galaxy sample mocks (black dashed line). 

In the middle panel we display the power spectrum normalised by a  linear power spectrum where the baryon acoustic oscillations have been smoothed (the red and blue lines are as in the top panel).
\begin{table*}
\begin{center}
\begin{tabular}{|c|c|c|c|c|c|c|c}
$k_{\rm max}=0.17\,h{\rm Mpc}^{-1}$ & ${b_1}$  & ${b_2}$  &  $f(z_{\rm eff})$&$\sigma_8(z_{\rm eff})\,\, (/\sigma_8^{\rm Planck13})$ & $\sigma^P_{\rm FoG}$ & $\sigma^B_{\rm FoG}$ &$A_{\rm noise}$\\
\hline\hline
NGC& $ 2.214$& $1.274 $& $0.991$ &$ 0.544\,\,(0.857)$ & $5.748$ & $17.881$ & $-0.319$   \\
\hline
SGC& $1.838 $ &  $0.677$& $0.517$&  $0.694\,\,(1.094)$ & $4.636$&$8.873$&$0.102 $\\
\hline\hline
NGC + SGC &  $ 2.086 $ & $0.902 $ & $0.763$ & $0.597\,\,(0.941)$ & $5.843$ & $15.397 $ & $-0.214$ \\
  \end{tabular}
\end{center}
\caption{Best fit parameters for the combination of NGC and SGC assuming an underlying ``Planck13" Planck cosmology (see text for details). The maximum $k$-vector used in the analysis is also indicated. For the $\sigma_8(z_{\rm eff})$ measurement, the parenthesis indicate the ratio to the fiducial Planck13 value. The units for  $\sigma_{\rm FoG}$   are  Mpc$h^{-1}$.}
\label{tab:bestfitparams}
\end{table*}

The error-bars correspond to the diagonal elements of the covariance and are estimated from the scatter of the mocks. The errors in the plots are therefore correlated, so a ``$\chi^2$-by-eye'' estimate would be highly misleading.  
 
 In the lower panel, we present the fractional differences between the data and the best fit model. The model is able to reproduce all the data points up to $k\simeq0.20\,h{\rm Mpc}^{-1}$, within $3\%$ accuracy (indicated by the black dotted horizontal lines). The SGC sample presents an excess of power at large scales compared to the NGC sample. This feature has been also observed in different analyses of the same galaxy sample \citep{Beutleretal:2013,Andersonetal:2013}. It is likely that this excess of power arises from targeting systematics in the SGC galaxy catalogue. More details about this feature will be reported in the next and final Data Release of the CMASS catalogue.
 
 The differences between the parameters corresponding to NGC, SGC and NGC+SGC observed in Table~\ref{tab:bestfitparams} are due to degeneracies introduced among the parameters. These degeneracies are fully described in \S~\ref{sec:results2}. We do not display errors on these parameters because we do not consider to estimate them using the mocks, since their distribution is highly non-Gaussian. It is only  when we  use a suitable parameter combination (in Table~\ref{data_table}) that the distribution looks more Gaussian and it makes sense to associate an error-bar to them.

\begin{figure*}
\centering
\includegraphics[clip=false, trim= 40mm 10mm 22mm 35mm,scale=0.26]{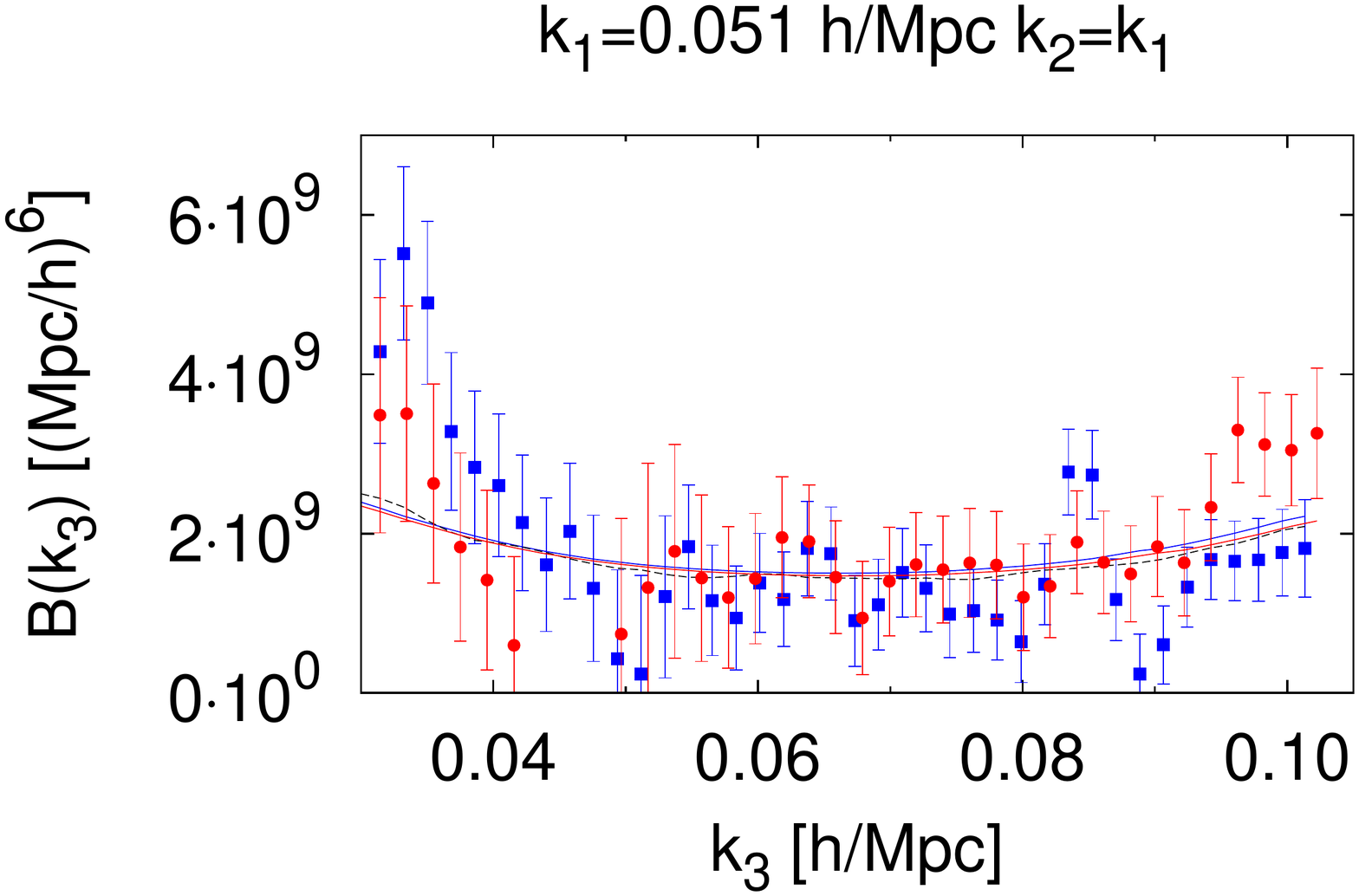}
\includegraphics[clip=false, trim= 40mm 10mm 22mm 35mm,scale=0.26]{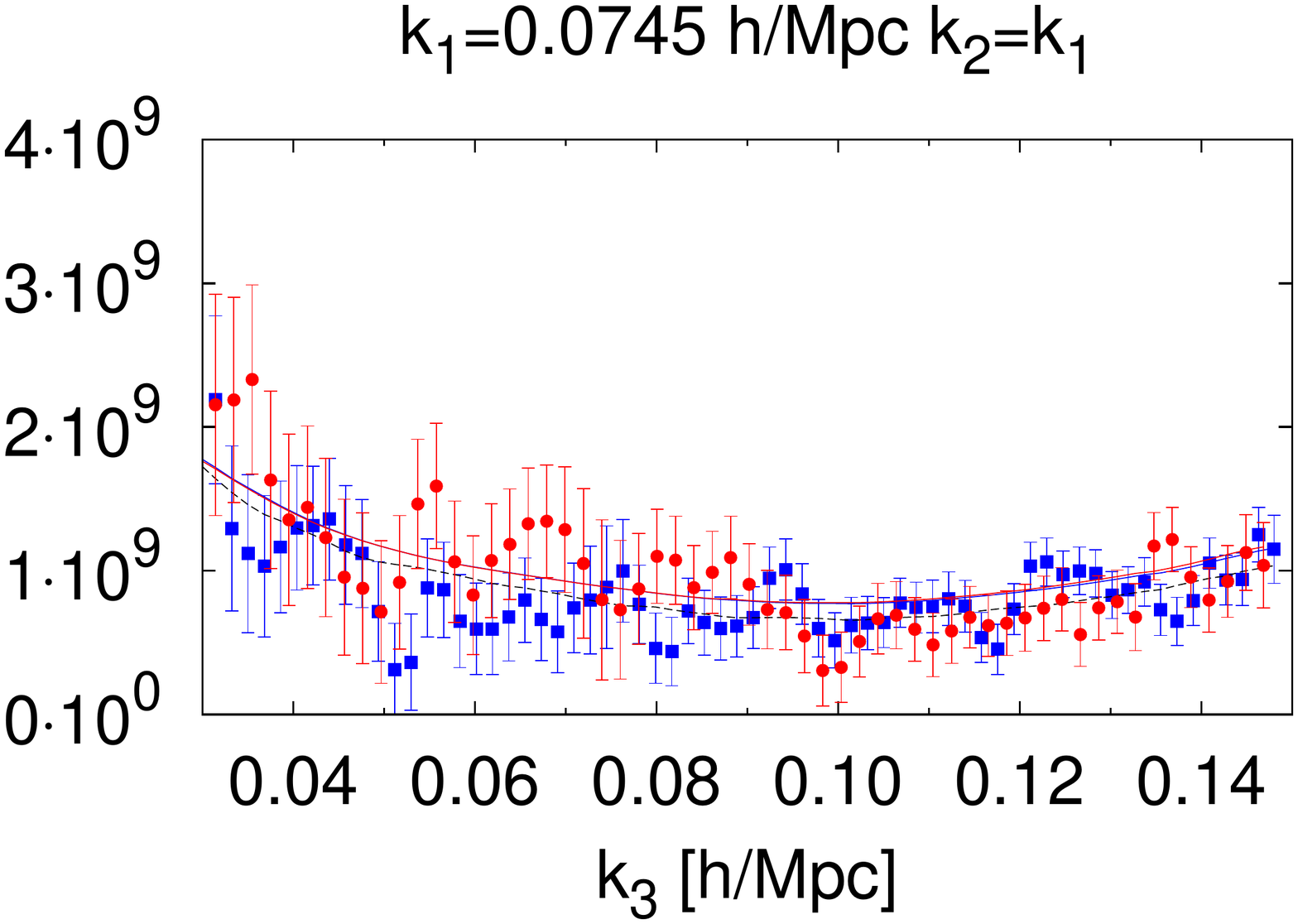}
\includegraphics[clip=false, trim= 40mm 10mm 22mm 35mm,scale=0.26]{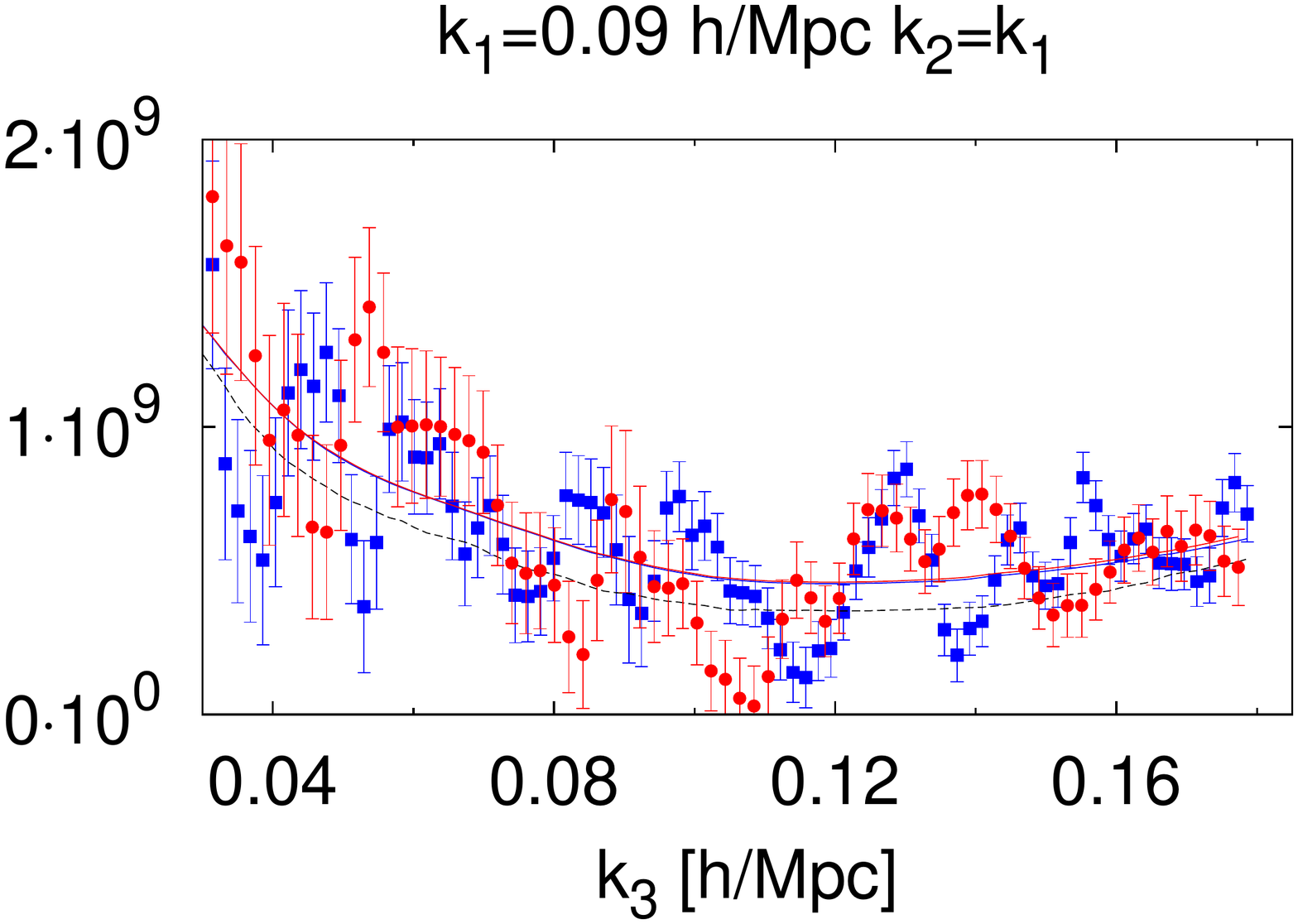}

\includegraphics[clip=false, trim= 40mm 10mm 22mm 35mm,scale=0.26]{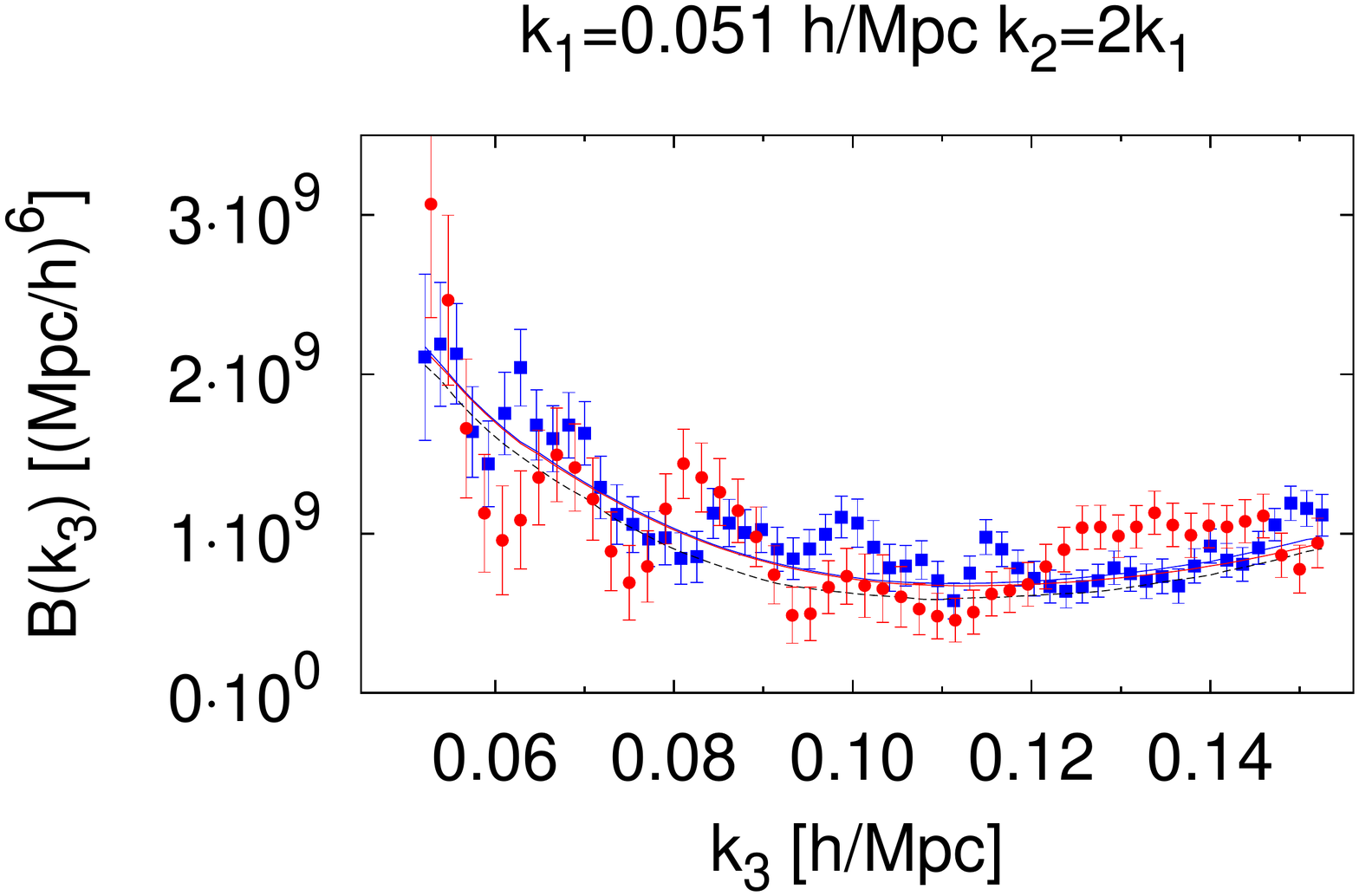}
\includegraphics[clip=false, trim= 40mm 10mm 22mm 35mm,scale=0.26]{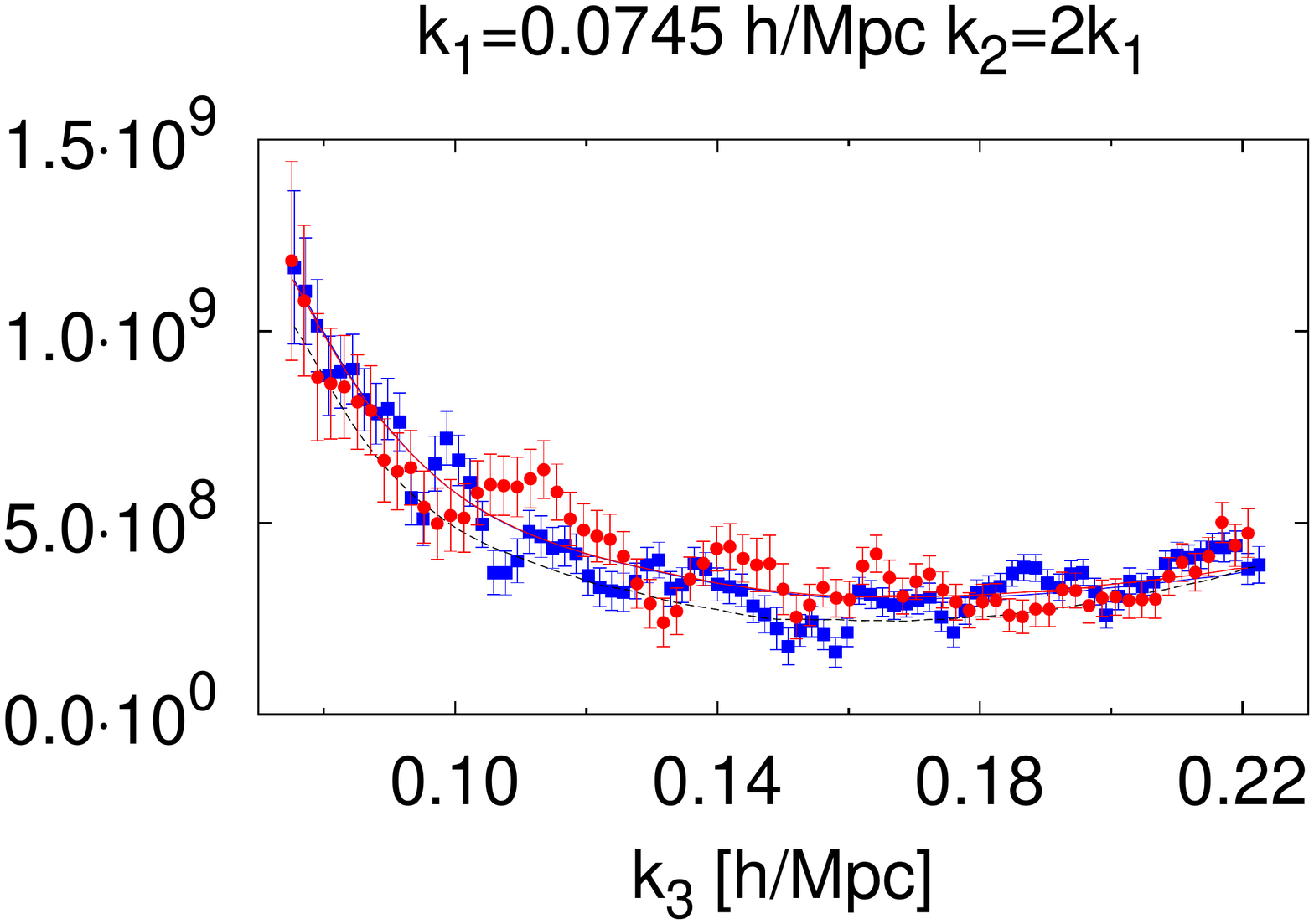}
\includegraphics[clip=false, trim= 40mm 10mm 22mm 35mm,scale=0.26]{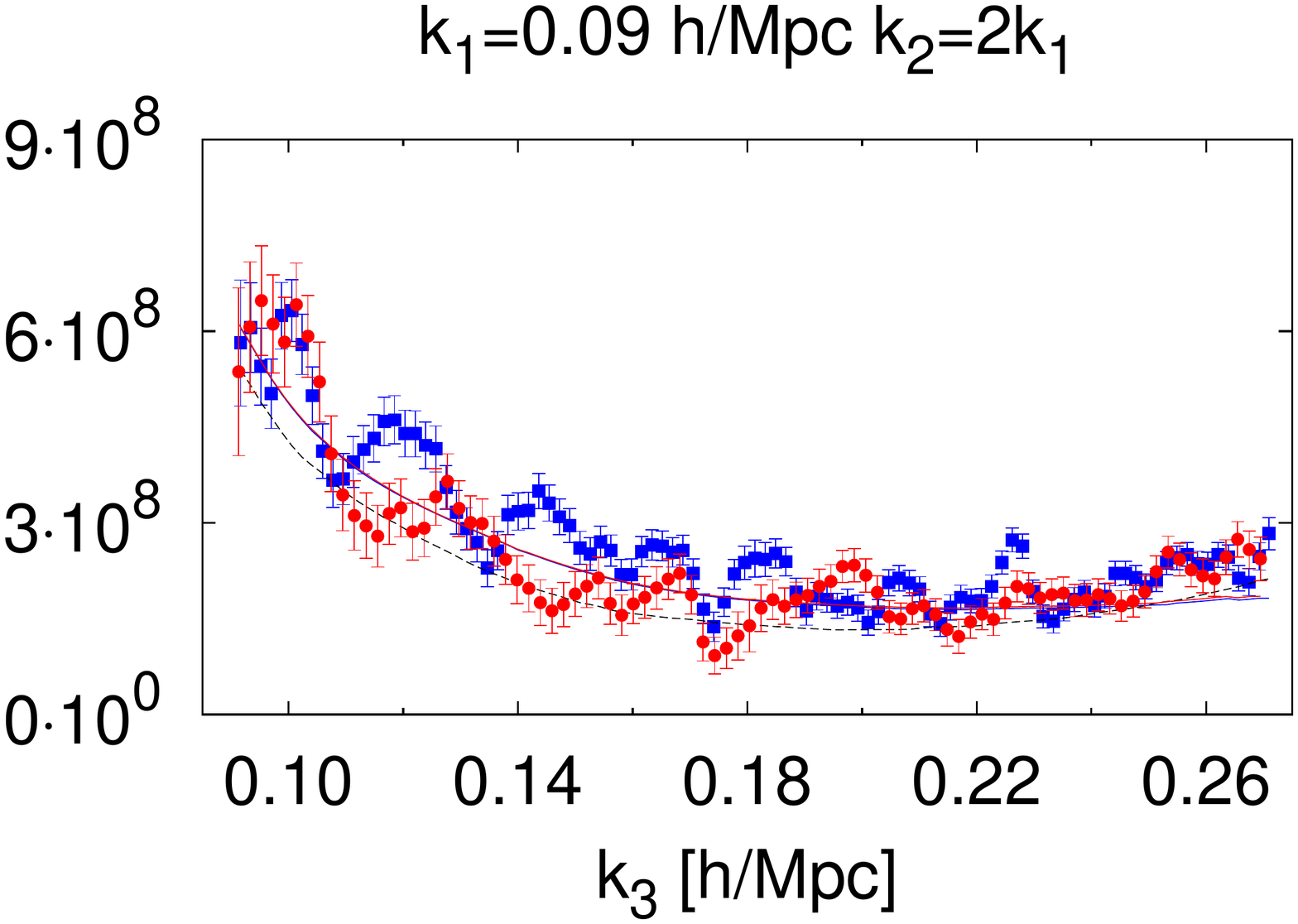}
\caption{Bispectrum data for NGC (blue squares) and SGC (red circles) with the best fit models (red and blue lines) listed in Table~\ref{tab:bestfitparams} as a function of $k_3$ for given $k_1$ and $k_2$. Blue lines take into account the effects of the NGC mask, and red lines for SGC mask. For reference the (mean) bispectrum of  the mock galaxy catalogs are shown by the black dashed lines. Different panels show different scales and shapes. The first row corresponds to triangles with $k_1=k_2$ whereas the second row to $k_1=2k_2$. Left column plots correspond to $k_1=0.051\,h{\rm Mpc}^{-1}$, middle column to $k_1=0.0745 \,h{\rm Mpc}^{-1}$ and the right column to $k_1=0.09 \,h{\rm Mpc}^{-1}$. The model is able to describe the observed bispectrum for $k_3\lesssim0.20\,h{\rm Mpc}^{-1}$.}
\label{dataBS}
\end{figure*}

The six panels of Fig.~\ref{dataBS} show  the measured CMASS DR11 bispectrum for different scales and shapes for the NGC  (blue) and SGC (red) galaxy samples. The best fit model  to the NGC+SGC of Table~\ref{tab:bestfitparams} (also used in Fig.~\ref{dataPS}), is indicated  with the same colour notation. The average  of the 600 NGC galaxy mocks is shown by the  black dashed line. It is not surprising that the mocks   are a worse fit to the bispectrum than the analytic prescription for the best fit parameters; in fact the mocks  have a  slightly different cosmology  and bias parameters compared to the best fit to the data.

Errors and  data-points  are highly correlated, especially those for modes with triangles that share two sides. Consequently, the oscillations observed in the different bispectra panels are entirely due to the sample variance effect;  in fact there is no correspondence for the location of these features between NGC and SGC.

\begin{figure*}
\centering
\includegraphics[clip=false, trim= 40mm 10mm 22mm 35mm,scale=0.26]{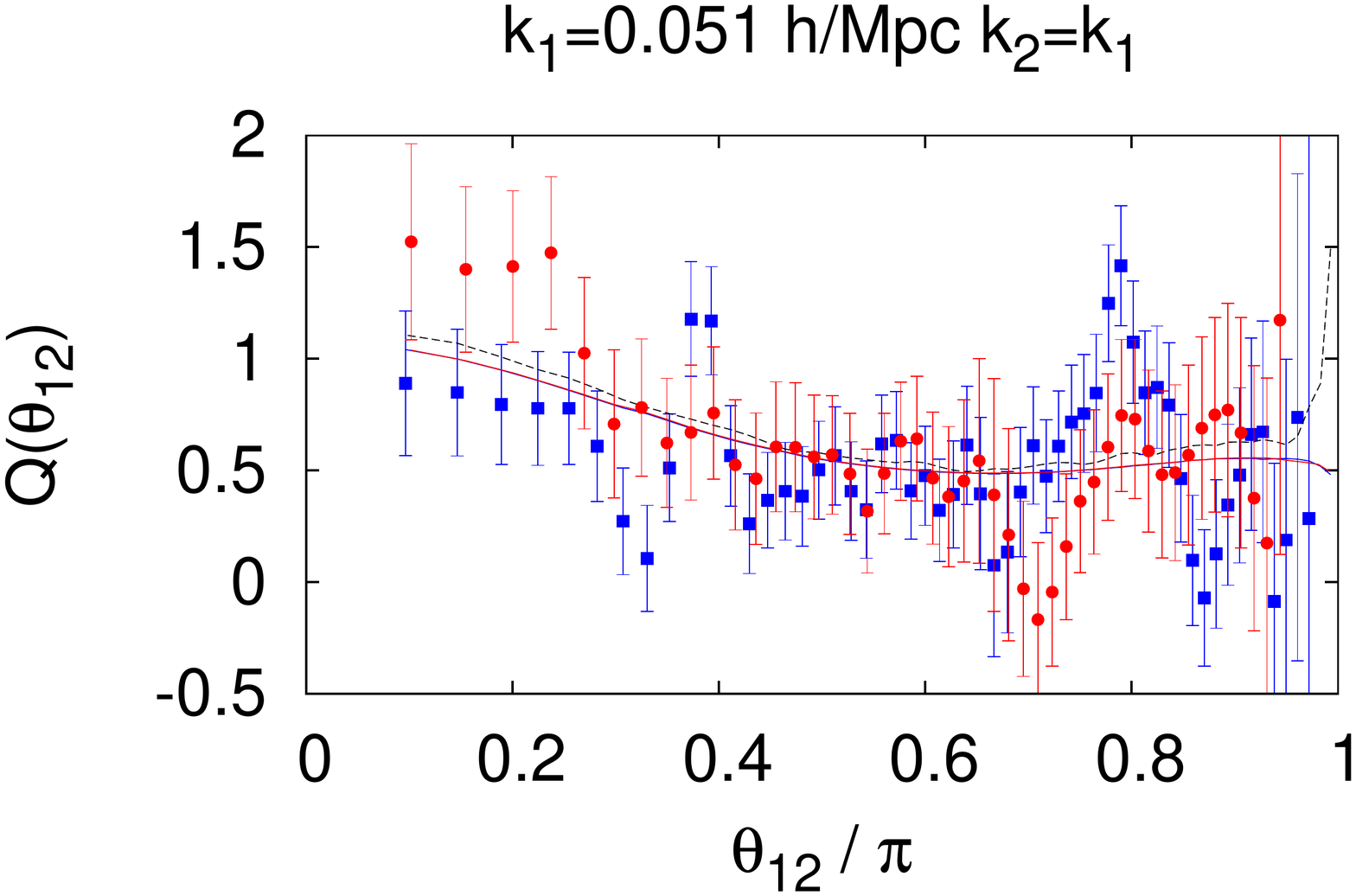}
\includegraphics[clip=false, trim= 40mm 10mm 22mm 35mm,scale=0.26]{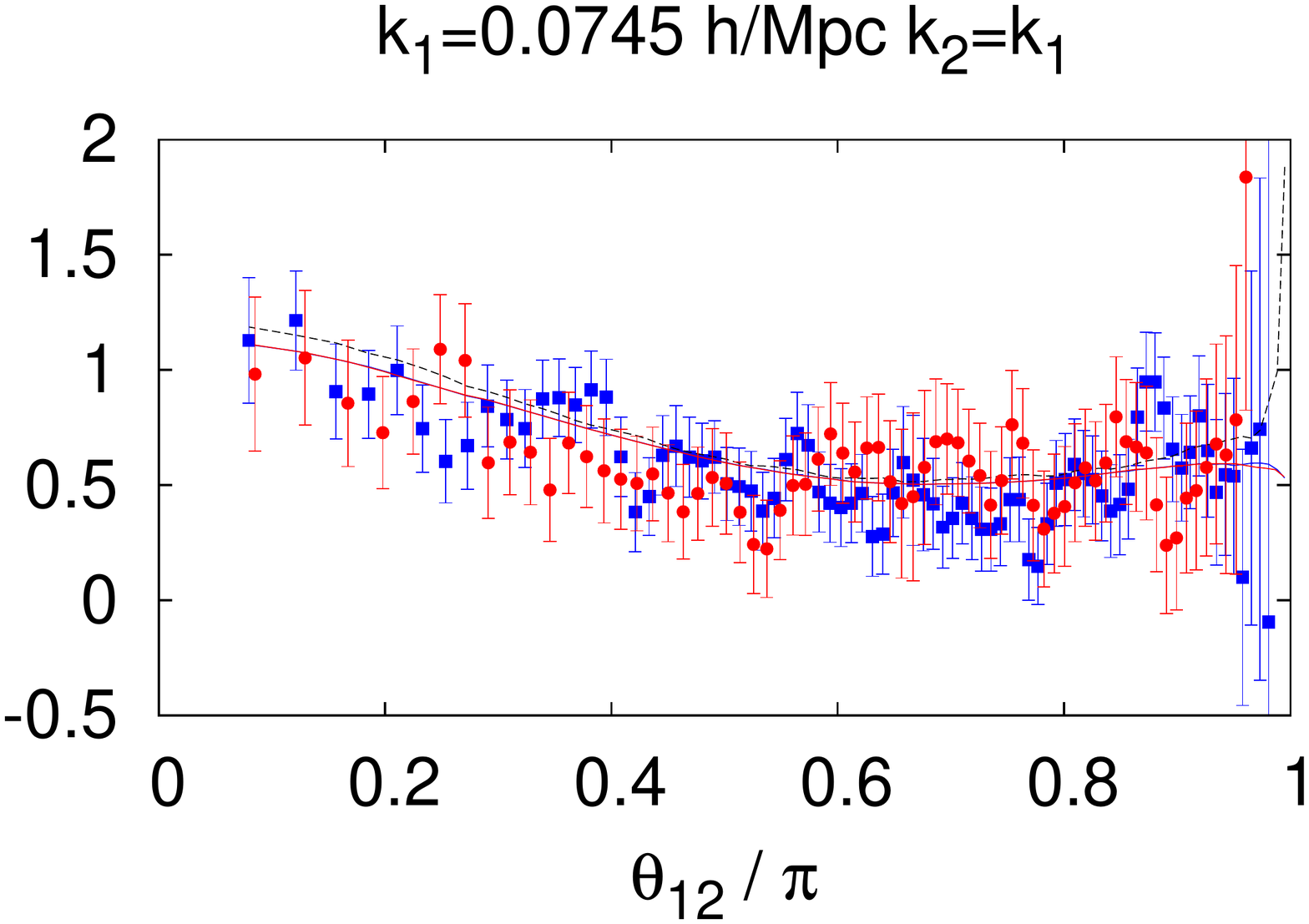}
\includegraphics[clip=false, trim= 40mm 10mm 22mm 35mm,scale=0.26]{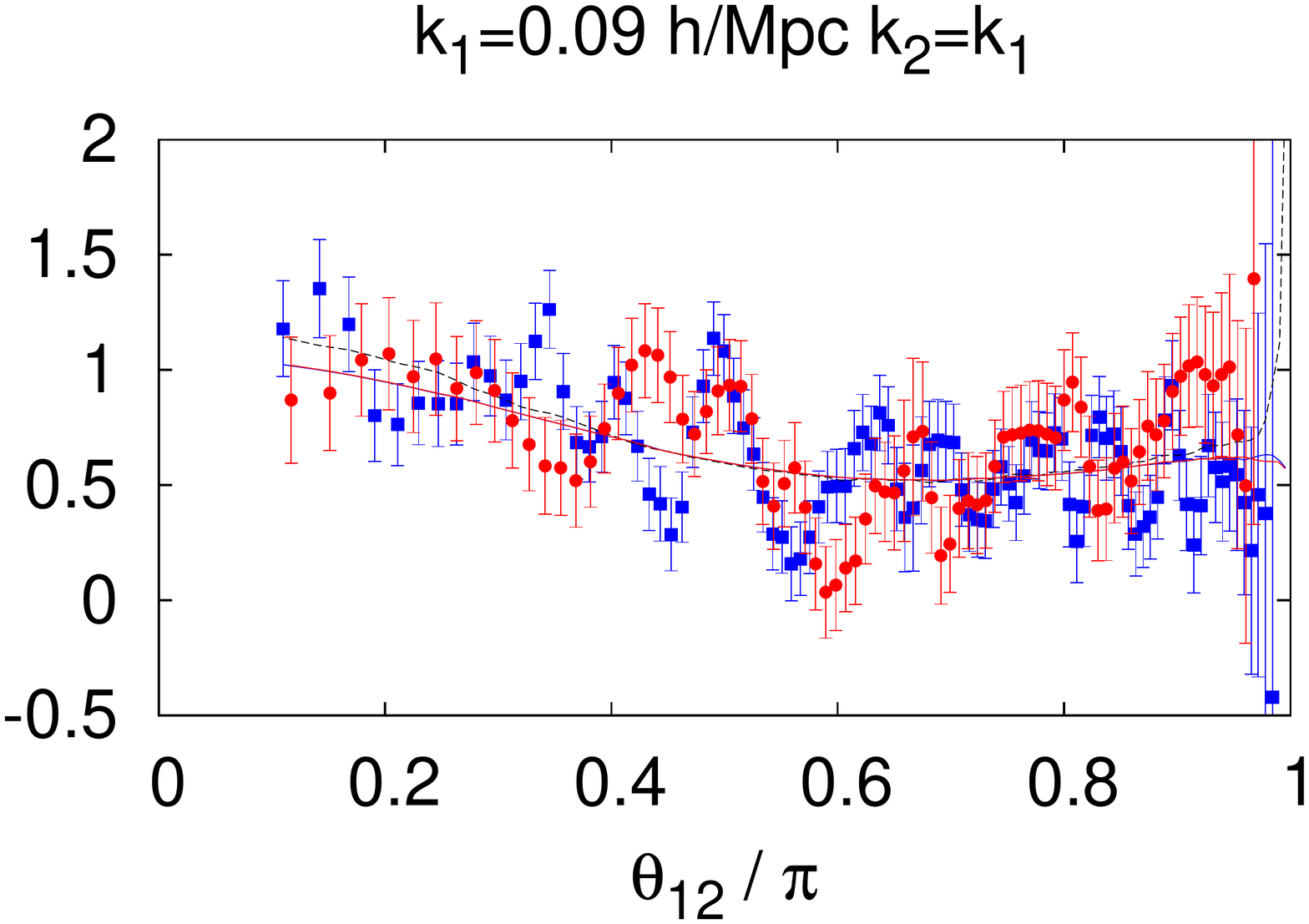}

\includegraphics[clip=false, trim= 40mm 10mm 22mm 35mm,scale=0.26]{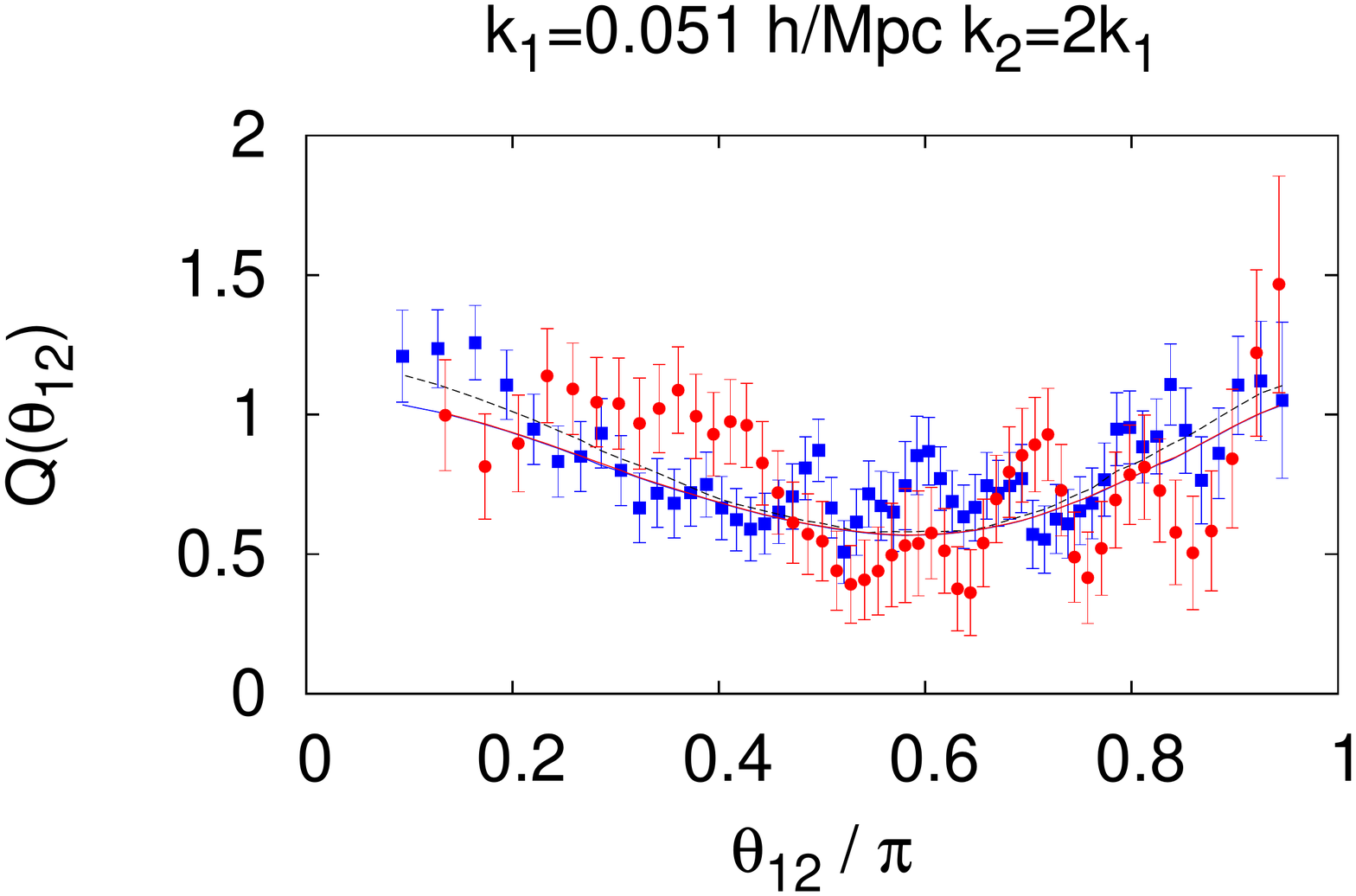}
\includegraphics[clip=false, trim= 40mm 10mm 22mm 35mm,scale=0.26]{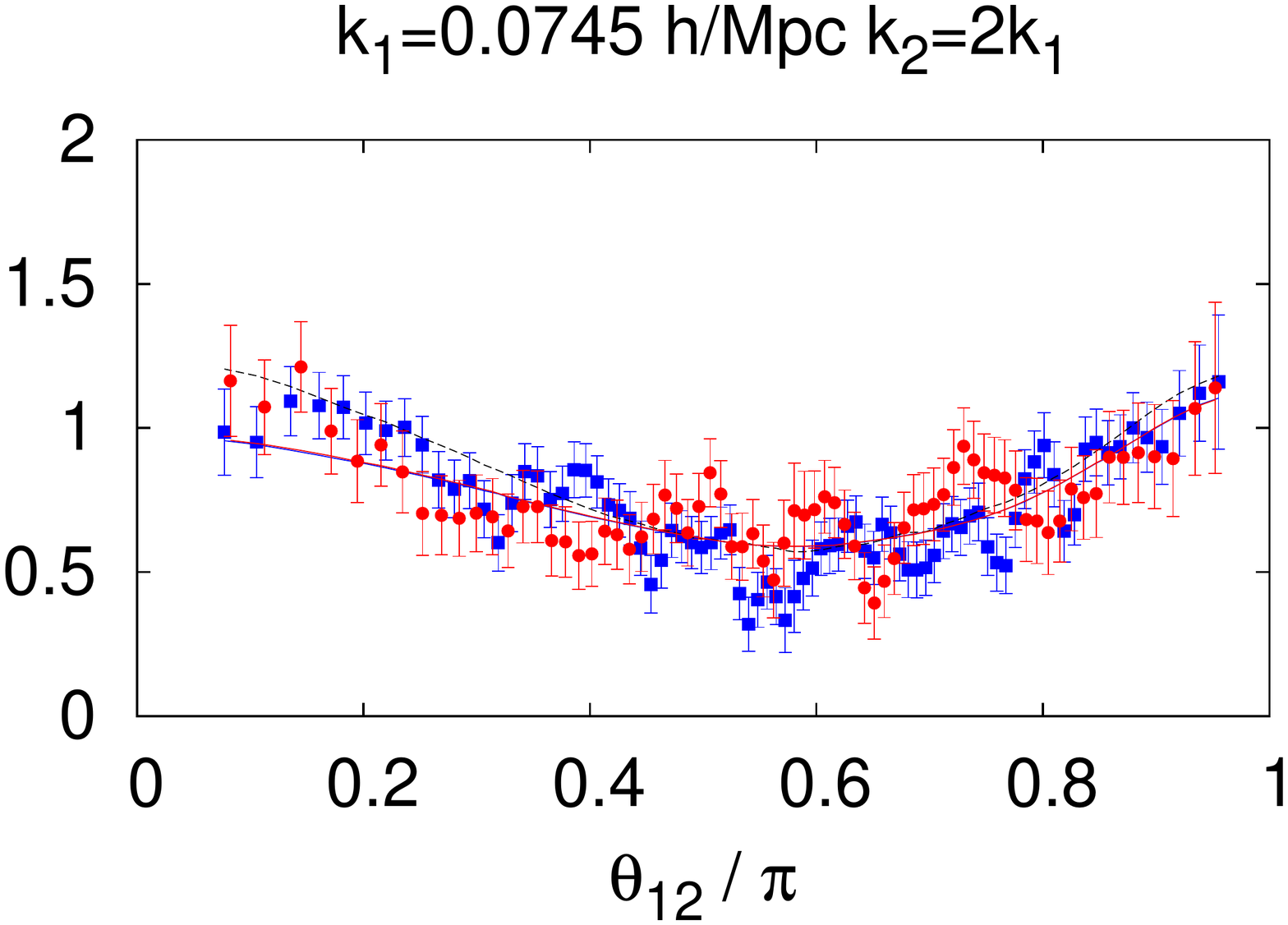}
\includegraphics[clip=false, trim= 40mm 10mm 22mm 35mm,scale=0.26]{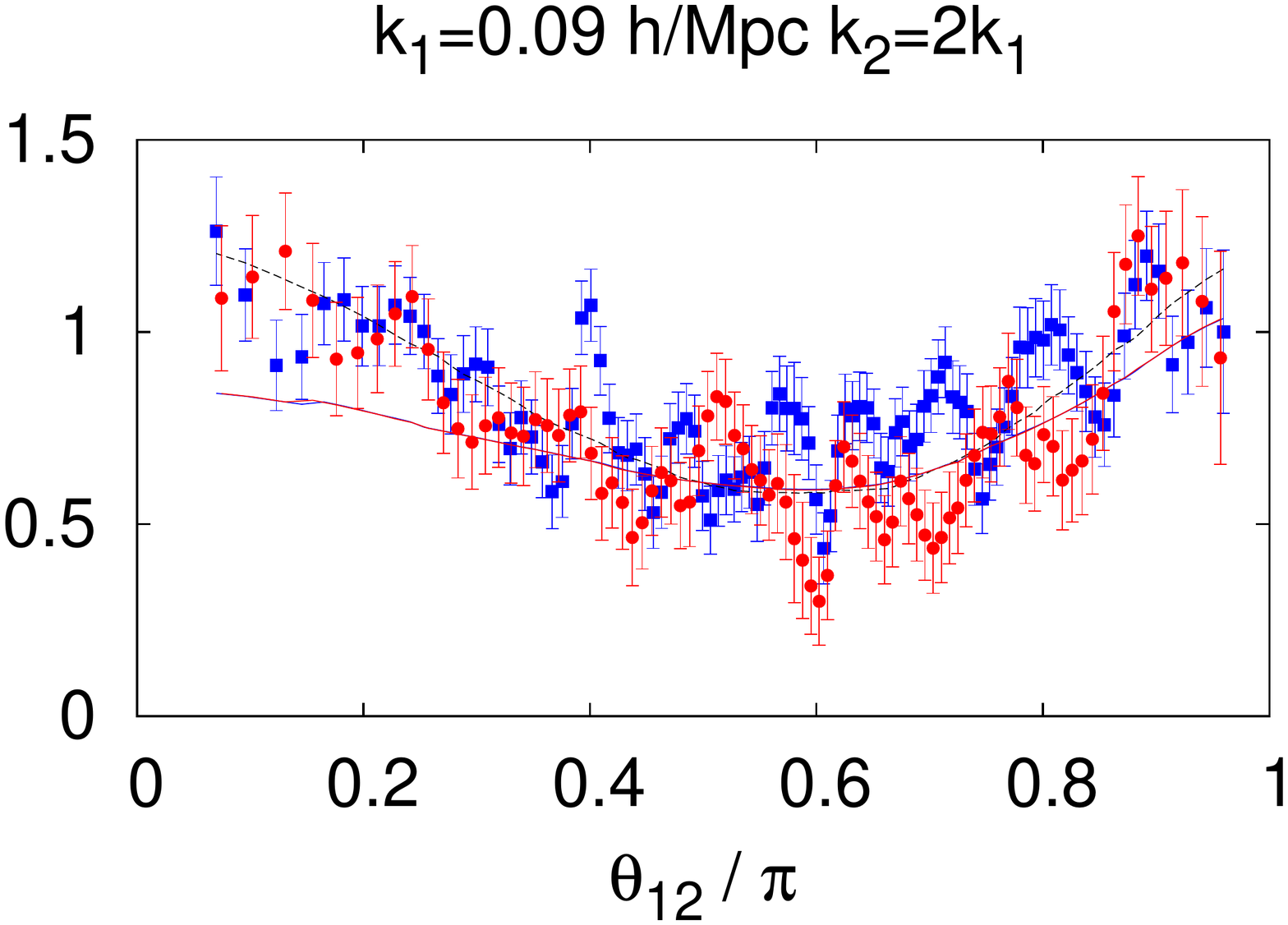}
\caption{Reduced bispectrum for DR11 CMASS data (symbols with errors) and the corresponding model (red and blue lines) for different scales and shapes. Same notation to that in Fig.~\ref{dataBS}. The model is able to describe the characteristic ``U-shape" for scales where $k_i\lesssim0.20\,h{\rm Mpc}^{-1}$.}
\label{dataBSQ}
\end{figure*}

Historically  the bispectrum has been plotted as  the hierarchical amplitude $Q(\theta)$  given a ratio $k_1/k_2$ (see e.g., \citealt{Fry:1994}) defined as \begin{equation}\label{eq:Q}
Q(\theta_{12}|k_1/k_2)=\frac{B(k_1,k_2,k_3)}{P(k_1)P(k_2)+P(k_2)P(k_3)+P(k_1)P(k_3)},
\end{equation}
where $\theta_{12}$ is the angle between the two k-vectors ${\bf k}_1$ and ${\bf k}_2$.  In tree-level perturbation theory  and for a power law power spectrum this quantity is independent of overall scale $k$ and of time\footnote{We are working with monopole quantities, so the  bispectra and power spectra in Eq.~\ref{eq:Q} are the  corresponding monopoles $B^0$ and $P^0$.}. In practice this is not the case (the power spectrum is not a power law and the the leading order  description in perturbation theory must be enhanced even to work at scales $k\lesssim 0.2$).  For ease of comparison with previous literature present a figure of $Q(\theta)$  in Fig.~\ref{dataBSQ}.
This figure does not have any information not contained in Fig.~\ref{dataBS}.

Gravitational instability predicts a characteristic ``U-shape" for  $Q(\theta)$  when $k_i/k_j=2$, but  non-linear evolution and  non-linear bias erase this dependence on configuration. Fig. \ref{dataBS} and \ref{dataBSQ} possess the characteristic  shape at high statistical significance. It is also interesting  that for large $k$ (in particular large  $k_1$ and $k_2/k_1=2$ and $\theta_{12}$ small, therefore $k_3$ nearing $k_1+k_2$) we see the breakdown of our prescription. The theoretical predictions  that produce the blue and red lines, the  power spectra in the denominator of $Q(\theta_{12})$ are computed using 2L-RPT and the prescription of \S~\ref{method:powerspectrum}.  The average of the mocks is a closer match (despite the different cosmology) because non-linearities are better captured.

\subsection{Bias and growth factor measurements}\label{sec:results2}

\begin{figure*}
\centering
\includegraphics[scale=0.3]{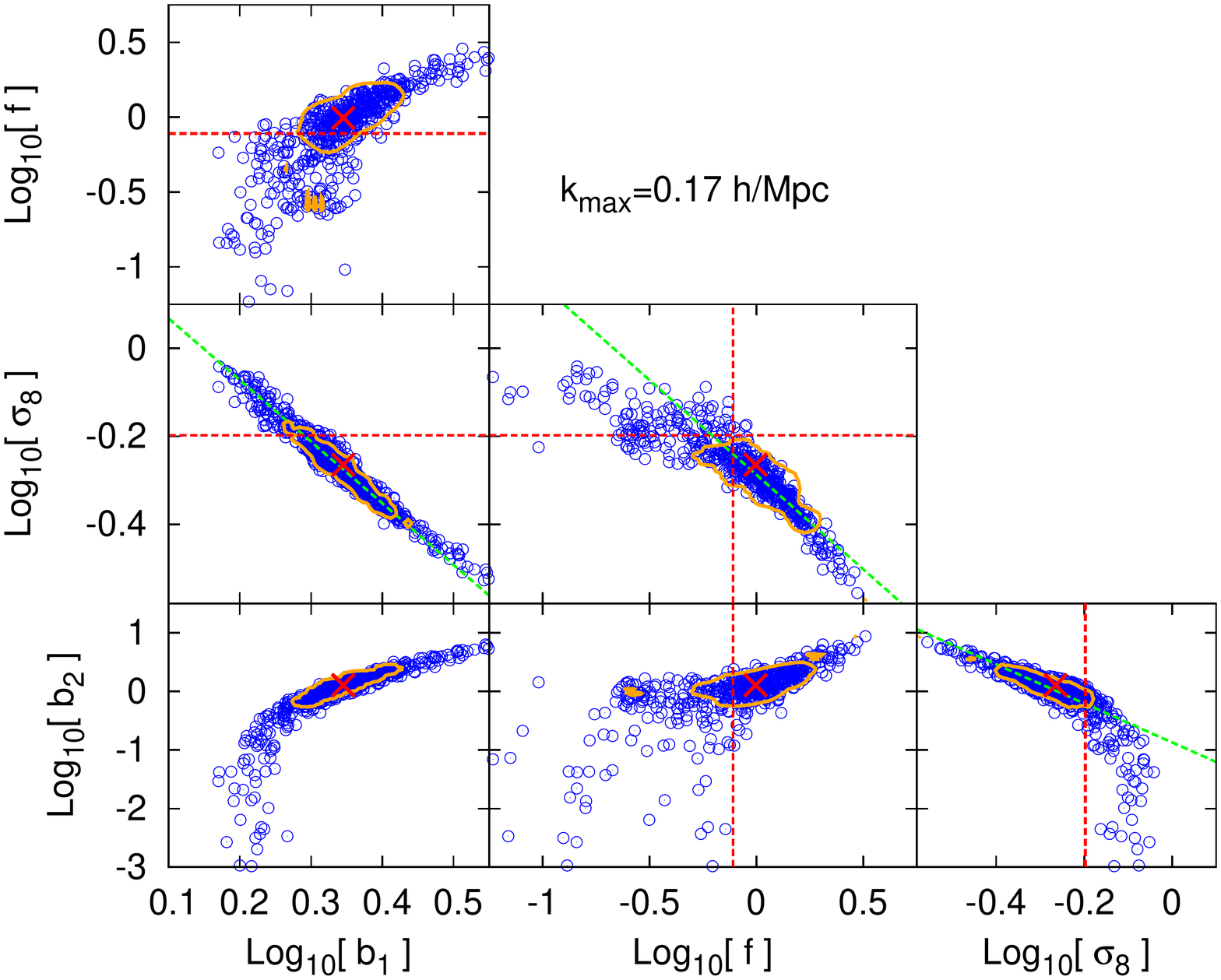}
\includegraphics[scale=0.3]{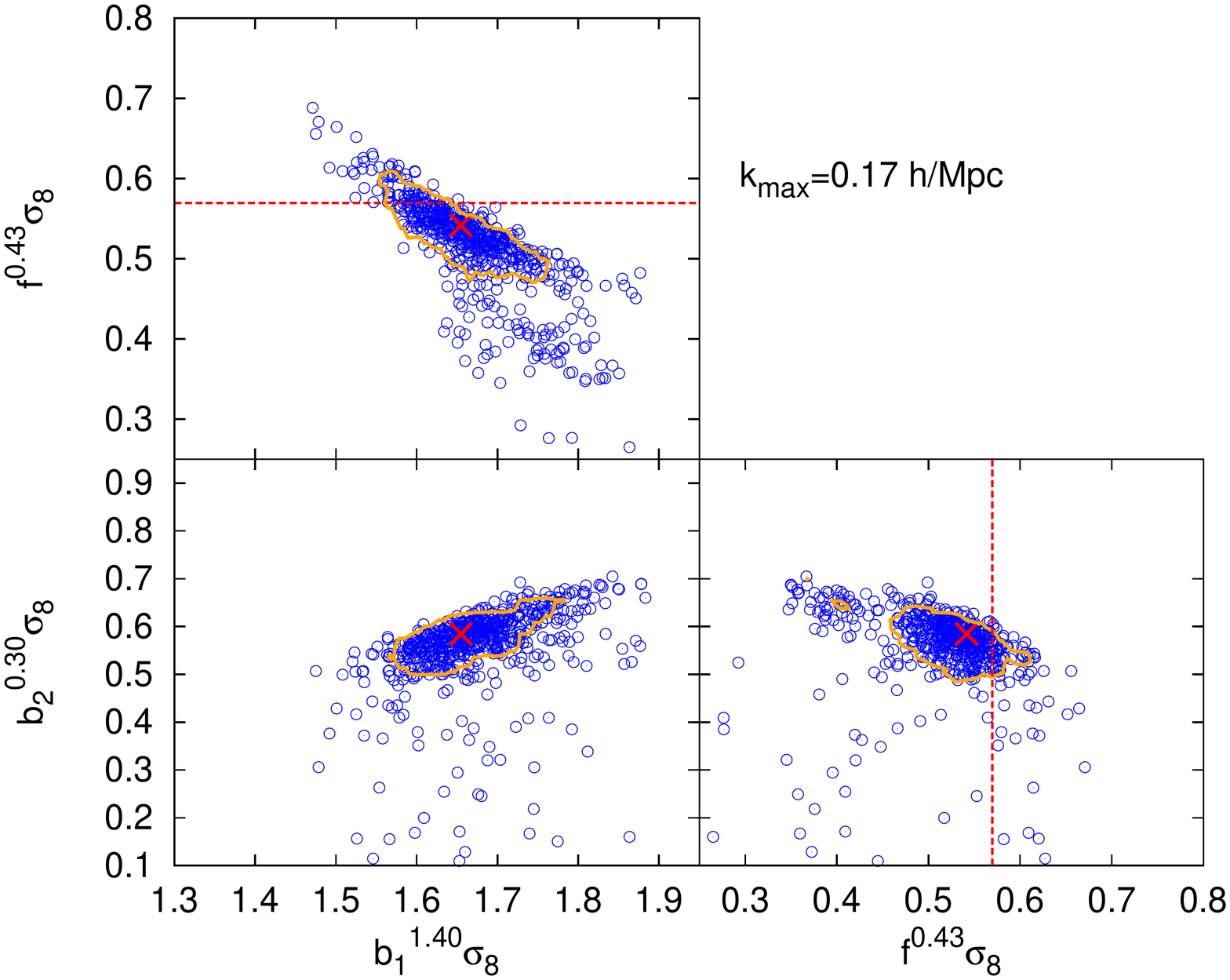}
\caption{ Two dimensional distributions of the parameters of (cosmological) interest. Left panels: We use  $\log_{10} b_1,  \log_{10} b_2, \log_{10} f, \log_{10} \sigma_8$ to obtain simpler degeneracies. The blue points represent the best fit of the 600 NGC mock catalogs and the red cross is the best fit from the data. The mocks distributions of points have been displaced in the $\log_{10}$ space to be centered on the best fit for the NGC data.  If we consider the distribution of the mocks as a sample of the posterior distribution of the parameters, the orange contour lines enclose 68\% of the marginalised posterior. The green dashed lines represent the linearised direction of the degeneracy in parameter space in the region around the maximum of the distribution. The dashed red lines indicate the Planck13 cosmology. Right panels: same notation as the left panels but for the  best constrained combination of parameters. The distributions appear more Gaussian than in the original variables.}
\label{fig:degener}
\end{figure*}

Despite the model depending on  four  cosmological parameters, the data can only constrain three (cosmologically interesting) quantities; there are large degeneracies among these parameters, in particular involving $\sigma_8$. Under the reasonable assumption that  the distribution of the best fit parameters from each of  the 600 mocks is a good approximation to the likelihood surface, there are non-linear degeneracies in the parameters space of $b_1$, $b_2$, $f$ and $\sigma_8$ as shown in the left panel of Fig.~\ref{fig:degener} (and also in Fig.~\ref{plot_degenerations}). These non-linear degeneracies can be reduced (i.e., the parameter degeneracies can be made as similar as possible to a multivariate Gaussian distribution) by a simple re-parametrization. In particular we will use  $\log_{10} b_1,  \log_{10} b_2, \log_{10} f, \log_{10} \sigma_8$, which, when computing marginalised confidence intervals on the parameters, is equivalent to assuming uniform priors on these parameters. Conveniently, this coincides with  Jeffrey's non-informative prior. We can adopt this procedure because $b_1$, $\sigma_8$ and $f$ are positive definite quantities and $b_2$ is positive for CMASS galaxies and for the mocks.  This issue is explored in detail in \S~\ref{sec:constraining_gravity}. Because of these degeneracies,  we combine the four cosmological  parameters into three new variables:  $b_1^{1.40}\sigma_8$, $b_2^{0.30}\sigma_8$ and $f^{0.43}\sigma_8$ (indicated by the dashed green lines in Fig.~\ref{fig:degener}).
This combination is formed {\it after} the fitting process and therefore  the (multi-dimensional) best fit values for $b_1$, $b_2$, $f$ and $\sigma_8$ are not affected by the definition of the new variables. In the new variables the parameter distribution is more Gaussian and the errors can be easily estimated  from the mocks.

 In the left panel of  Fig.~\ref{fig:degener} we show the distribution of CMASS DR11 NGC best fits from the galaxy mocks (blue points) for $\log_{10} b_1$, $\log_{10} b_2$, $\log_{10} f$ and $\log\sigma_8$. The red crosses indicate the best fit values obtained from the CMASS DR11 NGC+SGC data set.  The orange contours enclose  68\% of marginalised posterior when we consider the distribution of mocks as a sample of the posterior distribution of the parameters. The best fit parameters have been displaced in $\log$-space by a constant offset in order to match the centre of the 68\% contour and the measured data points. This allows use of the mocks to see the likely degeneracies around the data best-fit values. Black and red dashed lines show the fiducial values for $f$ and $\sigma_8$ for mocks and data, respectively. The green dashed lines indicate the empirical relation between $\sigma_8$ and the other variables. These empirical relations correspond to power law relations in linear space, and the slope of these lines is not affected by the shift of the mocks, as it is done in $\log$-space. In particular, we have found that these empirical relations correspond to $f^{-0.43}\sim\sigma_8$, $b_1^{-1.40}\sim\sigma_8$ and $b_2^{-0.30}\sim\sigma_8$. In the right panel of Fig.~\ref{fig:degener}, we present the same distribution that in the right panel but for the combined set of variables, $f^{0.43}\sigma_8$, $b_1^{1.40}\sigma_8$ and $b_2^{0.30}\sigma_8$. The distribution of results from the galaxy mocks are closer to a multi-variate Gaussian distribution in these new set of variables than in the original set.
 
 Table~\ref{data_table} lists the best fit values and the errors for these new variables. The data used are always the DR11 CMASS galaxies monopole power spectrum and bispectrum when the Planck13 cosmology is assumed. The first two rows correspond to the NGC and SGC galaxy sample, respectively, whereas in the third row both samples are combined. For the three cases, the maximum scale is  conservatively set to $k_{\rm max}=0.17\,h{\rm Mpc}^{-1}$. A smaller $k_{\rm max}$ would yield too large error-bars, but  at larger $k$ non-linearities become important and  we have evidence that our modelling starts breaking down. This issue is further discussed in \S~\ref{sec:dependenceonkmax}, where we study the dependence of the best fit parameters with $k_{\rm max}$ and  the choice motivated in details in \S~\ref{section:systematics}.  
 
\begin{table*}
\begin{center}
\begin{tabular}{|c|c|c|c|c|c|c}
$k_{\rm max}=0.17\,h{\rm Mpc}^{-1}$ & ${b_1}^{1.40}\sigma_8(z_{\rm eff})$  & ${b_2}^{0.30}\sigma_8(z_{\rm eff})$ & $A_{\rm noise}$ & $\sigma^B_{\rm FoG}$ & $\sigma^P_{\rm FoG}$ & $f^{0.43}(z_{\rm eff})\sigma_8(z_{\rm eff})$\\
 \hline
  NGC & $1.655\pm0.071$ & $0.585\pm0.094$ & $-0.32\pm0.27$ & $17\pm13$ & $5.7\pm1.9$ & $0.541\pm0.092+0.05$  \\
 \hline
 SGC & $1.63\pm0.10$ & $0.62\pm0.15$ & $0.10\pm0.32$ & $8\pm19$ & $4.6\pm3.0$ & $0.52\pm0.12+0.05$ \\
\hline
\hline
NGC + SGC &  $1.672\pm0.060$ & $0.579\pm0.082$ & $-0.21\pm0.24$ & $15\pm12$ & $5.8\pm1.8$ & $0.532\pm0.080+0.05$  \\
  \end{tabular}
\end{center}
\caption{Best fit parameters for NGC, SGC and combination  (NGC+SGC) for Planck13 cosmology. The maximum scale is set to $k_{\rm max}=0.17\,h^{-1}{\rm Mpc}$. The units for  $\sigma_{\rm FoG}$   are  Mpc$h^{-1}$.}
\label{data_table}
\end{table*}

The best-fit $f^{0.43}\sigma_8$ is provided along with  a systematic error-component, in addition to the statistical error. In \S~\ref{sec:systematics} we present a full description of how this systematic error is obtained. In brief,   we have indications that  the model used for describing the power spectrum and bispectrum  of {\it biased} tracers in redshift space  presents a systematic and scale-independent underestimate of $f^{0.43}\sigma_8$ at the level of {\bf $0.05$}. The determination of this systematic error relies on the analysis of N-body haloes as well as mock galaxy catalogs. It is interesting that the systematic correction would cancel if we considered instead the quantity $f\sigma_8$ \citep{HGMetal:inprep}; we will discuss this point in \S~\ref{sec:systematics}.

From the results in Table~\ref{data_table} we do not detect any strong tension between NGC and SGC for any of the parameters. We only observe a non-statistically significant trend $A_{\rm noise}$: the NGC galaxy sample tends to have a  slightly sub-Poisson shot noise, whereas the SGC sample presents a slightly super-Poisson shot noise. However, these differences are not statistically significant and can be explained by a sample variance effect.

 We understand that the parameterization $f^{0.43}\sigma_8$ is  non-standard, although is the one that naturally arises from the shape of the parameter-space. In order to make a connection with the commonly estimated $f\sigma_8$, we can assume a value for $f$ predicted   for a standard $\Lambda$CDM model with parameters set by the  {\it Planck} \citep{Planck_cosmology} $f_{\rm Planck}=0.777$ (as listed in Table \ref{cosmology_table}), we can construct our estimator of $f\sigma_8$ as, $[{f\sigma_8}]_{\rm est.}\equiv[f^{0.43}\sigma_8]f_{\rm Planck}^{0.57}$. From the values of $f^{0.43}\sigma_8$ for NGC+SGC in Table \ref{data_table} we obtain that $[{f\sigma_8}]_{\rm est.}=0.504\pm0.069$. This result is in very good agreement with the prediction from Planck, $[f\sigma_8]_{\rm Planck}=0.493$, with only $2\%$ offset.  

\subsection{Dependence on the maximum $k$}\label{sec:dependenceonkmax}
\begin{figure*}
\centering
\includegraphics[clip=false, trim= 20mm 0mm 12mm 0mm,scale=0.32]{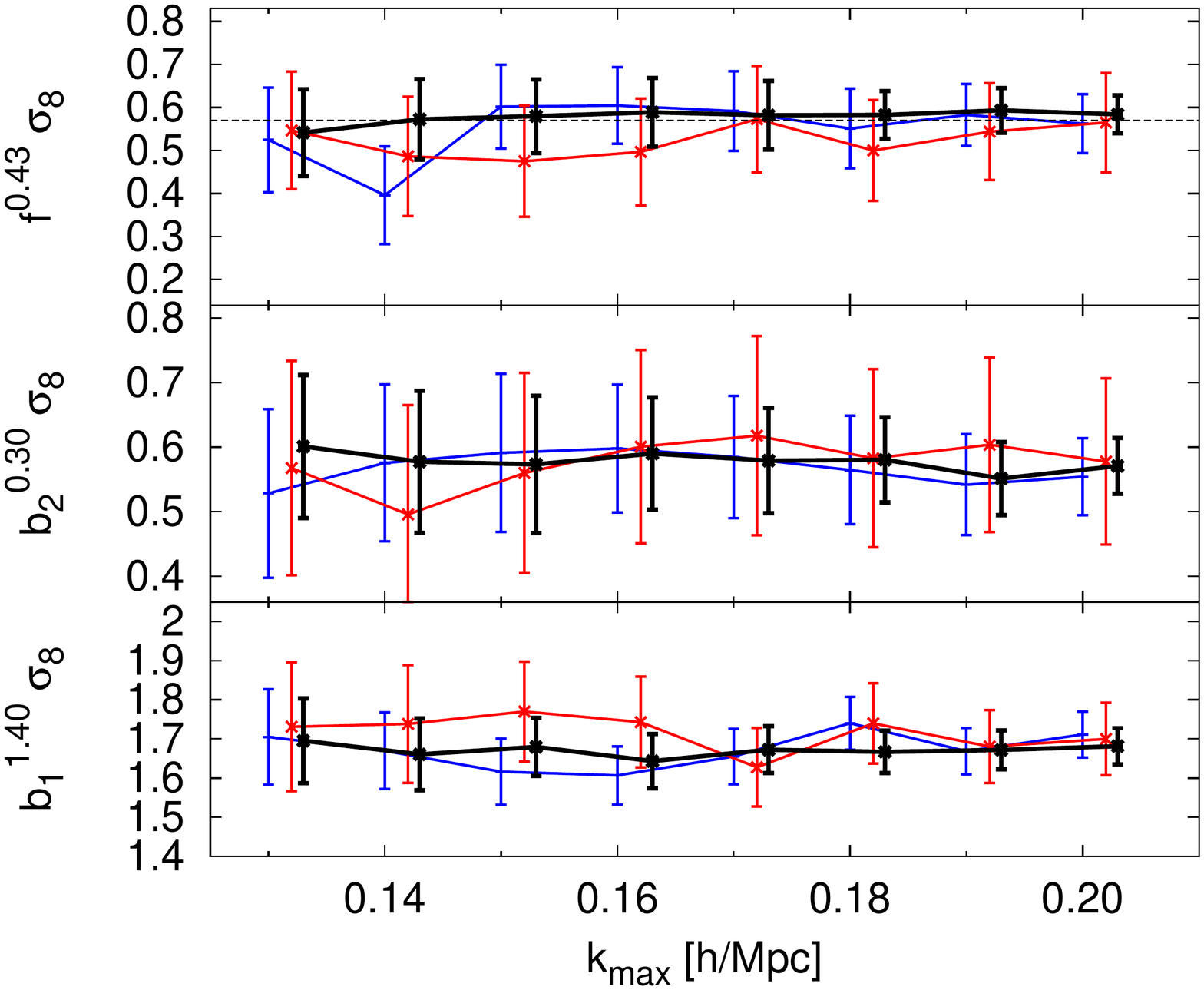}
\includegraphics[clip=false, trim= 20mm 0mm 12mm 0mm,scale=0.32]{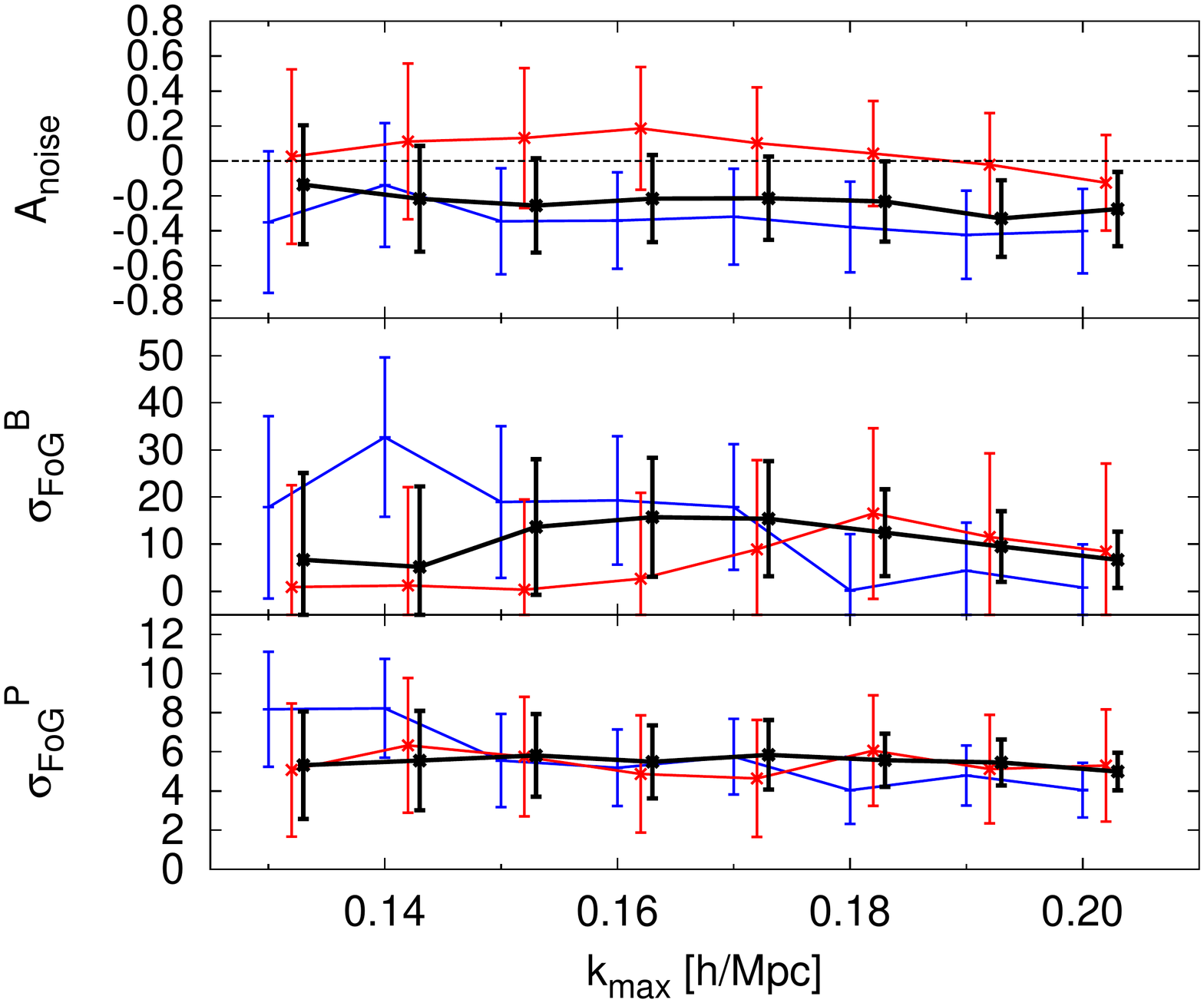}
\label{kmax}
\caption{Best fit parameters as a function of $k_{\rm max}$ for NGC data (blue symbols), SGC data (red symbols) and a combination of both (black symbols) when the Planck13 cosmology is assumed. The quantity $f^{0.43}\sigma_8$ has been corrected by the systematic error as is listed in Table~\ref{data_table}. For the $f^{0.43}\sigma_8$ panel, the corresponding fiducial values for GR are plotted in dashed black line.  In the $A_{\rm noise}$ panel, the dotted line indicates no deviations from Poisson shot noise. The units of $\sigma_{\rm FoG}$ are Mpc$h^{-1}$. There is no apparent dependence with $k_{\rm max}$ for any of the displayed parameters for $k_{\rm max}\leq0.17\,h{\rm Mpc}^{-1}$.}
\label{kmax_data2}
\end{figure*}

In the two panels of Fig.~\ref{kmax_data2}  we  present the effect of varying the maximum $k$ (smallest scale) included, $k_{\rm max}$. The left panel displays the variation of $b_1^{1.40}\sigma_8$, $b_2^{0.30}\sigma_8$ and $f^{0.43}\sigma_8$ as function of $k_{\rm max}$ , while the right panel shows $A_{\rm noise}$, $\sigma_{\rm FoG}^P$ and $\sigma_{\rm FoG}^B$ as function of $k_{\rm max}$. The plotted values for the $f^{0.43}\sigma_8$ quantity have been corrected by the systematic offset of 0.05 as described in \S~\ref{sec:systematics}. The different colour lines correspond to the galaxy catalogue used to perform the analysis: blue lines for the NGC, red lines for the SGC and black lines when the catalogues are combined. 

 The three galaxy samples yield consistent quantities for all values of $k_{\rm max}$; there is no  indication of a breakdown of the model (i.e., abrupt changes in the recovered parameters values when too small scales are included). 
 
Extensive tests (see \S~\ref{section:systematics}) indicate that, at least for  N-body simulations and  mock catalogs, the modelling adopted here starts to break down beyond $k=0.17\,h{\rm Mpc}^{-1}$ for biased tracers in redshift space. However, we have checked that for $0.20\leq k\, [h{\rm Mpc}^{-1}] \leq 0.17$, the modelling is still able to reproduce N-body simulations and mocks catalogs up to a few percent accuracy. Because of this, we adopt a conservative approach, where we stop our analysis at $k_{\rm max}=0.17\,h{\rm Mpc}^{-1}$, and a less conservative approach, where we push the analysis up to $k_{\rm max}=0.20\,h{\rm Mpc}^{-1}$. In both cases we add in quadrature a systematic contribution to the statistical error, $\sigma_{\rm sys}$, which we chose to be 50\% of the systematic shift, $\sigma_{\rm sys}$. Therefore, in both cases the total error is given by $\sigma_{\rm tot}\equiv\sqrt{\sigma^2_{\rm est}+[\sigma_{\rm sys}/2]^2}$. For completeness,  in Table~\ref{data_table_kmax} we report results as function of $k_{\rm max}$ as they are plotted in Fig.~\ref{kmax_data2}.
 
 \begin{table*}
\begin{center}
\begin{tabular}{|c|c|c|c|c|c|c}
$k_{\rm max}\, [\,h{\rm Mpc}^{-1}]$ & ${b_1}^{1.40}\sigma_8(z_{\rm eff})$  & ${b_2}^{0.30}\sigma_8(z_{\rm eff})$ & $A_{\rm noise}$ & $\sigma^B_{\rm FoG}$ & $\sigma^P_{\rm FoG}$ & $f^{0.43}(z_{\rm eff})\sigma_8(z_{\rm eff})\pm\sigma_{\rm est} + \sigma_{\rm sys}\,  (\pm\sigma_{\rm tot})$\\
\hline
0.13 &  $1.69\pm0.11$ & $0.60\pm0.11$ & $-0.14\pm0.34$ & $7\pm18$ & $5.3\pm2.7$ & $0.49\pm0.10+0.05\, (\pm0.10)$  \\
\hline
0.14 &  $1.660\pm0.091$ & $0.58\pm0.11$ & $-0.22\pm0.30$ & $5\pm17$ & $5.6\pm2.5$ & $0.522\pm0.094+0.05\, (\pm0.097)$  \\
\hline
0.15 &  $1.679\pm0.074$ & $0.57\pm0.11$ & $-0.26\pm0.27$ & $14\pm14$ & $5.8\pm2.1$ & $0.529\pm0.086+0.05\, (\pm0.090)$  \\
\hline
0.16 &  $1.643\pm0.069$ & $0.590\pm0.087$ & $-0.22\pm0.25$ & $16\pm13$ & $5.5\pm1.9$ & $0.538\pm0.080+0.05\, (\pm0.084)$  \\
\hline
 0.17 &   $1.672\pm0.060$ & $0.579\pm0.082$ & $-0.21\pm0.24$ & $15\pm12$ & $5.8\pm1.8$ & $0.532\pm0.080+0.05\, (\pm0.084)$  \\
 \hline
0.18 &  $1.667\pm0.054$ & $0.580\pm0.066$ & $-0.23\pm0.23$ & $12.4\pm9.2$ & $5.7\pm1.3$ & $0.532\pm0.055+0.05\, (\pm0.060)$  \\
\hline
0.19 &  $1.672\pm0.049$ & $0.551\pm0.057$ & $-0.33\pm0.22$ & $9.5\pm7.5$ & $5.4\pm1.2$ & $0.543\pm0.052+0.05\, (\pm0.058)$  \\
\hline
0.20 &  $1.681\pm0.046$ & $0.571\pm0.043$ & $-0.28\pm0.21$ & $6.7\pm6.0$ & $4.99\pm0.96$ & $0.534\pm0.044+0.05\, (\pm0.051)$  \\ 
  \end{tabular}
\end{center}
\caption{Best fit parameters for (NGC+SGC) for Planck13 cosmology for different $k_{\rm max}$. This table corresponds to the black line of Fig. \ref{kmax_data2}.  The units for  $\sigma_{\rm FoG}$   are  Mpc$h^{-1}$. In the last column, a total error is given by $\sigma_{\rm tot}\equiv\sqrt{\sigma^2_{\rm est}+[\sigma_{\rm sys}/2]^2}$}
\label{data_table_kmax}
\end{table*}

\subsection{Dependence on the assumed cosmology}
\label{sec:cosmodependence}
In the analysis of the CMASS DR11 data in the above section we have assumed the Planck cosmology (Planck13). This assumption is necessary to obtain the linear power spectrum which is the starting point for the galaxy power spectrum and bispectrum theoretical models. Since the results presented in Table~\ref{data_table} and Fig.~\ref{kmax_data2} may be sensitive to the assumed cosmological parameters, in this section we repeat the analysis for the NGC galaxy sample assuming  two variations of the Planck13 cosmology. We aim at quantifying how sensitive the parameter set $\{b_1^{1.40}\sigma_8,\, b_2^{0.30}\sigma_8,\, f^{0.43}\sigma_8,\,A_{\rm noise},\,\sigma_{\rm FoG}^P,\, \sigma_{\rm FoG}^B \}$ is to the cosmological model assumed.

Table~\ref{cosmology_table} presents the cosmological parameters for the Planck13 cosmology, assumed in \S~\ref{sec:results2}, and  present two additional Planck-like cosmologies sets, namely L-Planck13 and H-Planck13. These sets of parameters are generated using the uncertainties of Planck13 parameters reported in \cite{Planck_cosmology}.  The L-Planck13 cosmology has most  parameters lowered by $1\sigma$ respect to Planck13, whereas for the H-Planck13 cosmology most of the parameters have been increased  by $1\sigma$. These cosmologies would be highly disfavoured by Planck data. We also include the cosmology of the mocks for comparison reasons. The definition of the parameters listed on Table~\ref{cosmology_table} can be found in table 1 of \cite{Planck_cosmology}. The parameters $\Omega_b h^2$, $\Omega_c h^2$, $\tau$, $A_s$, $n_s$ and $h$ are the ``input parameters", whereas $\sigma_8$, $D_+$, $f$, $\Omega_m$ and $f^{0.43}\sigma_8$ are derived from those. We use the CAMB software \citep{CAMB_paper} to generate the linear dark matter power spectrum, $P_{\rm lin}$, from each cosmological parameter set.

\begin{table*}
\begin{center}
\begin{tabular}{|c|c|c|c|c|}
 & Mocks & Planck13 & H-Planck13 & L-Planck13 \\
 \hline
$\Omega_bh^2$ & 0.0196  & 0.022068 & 0.0224 & 0.02174 \\
\hline
$\Omega_ch^2$ & 0.11466 & 0.12029 & 0.1165 & 0.1227 \\
\hline 
$\tau$ & 0.09123 & 0.0925 & 0.135 & 0.059 \\
\hline
$10^9A_s$ & 1.9946 & 2.215 & 2.39 & 2.07 \\
\hline 
$n_s$ & 0.95 & 0.9624 & 0.971 & 0.9522 \\
\hline
$h$ & 0.70 & 0.6711 & 0.688 & 0.660 \\
\hline
\hline
$\sigma_8(z=0)$ & 0.80 & 0.8475 & 0.8680  & 0.8252 \\
\hline
$\sigma_8(z_{\rm eff})$ & 0.6096 & 0.6348  & 0.6564   & 0.6149  \\
\hline
$f(z_{\rm eff})$ & 0.744 & 0.777 & 0.760 & 0.788 \\
\hline
$\Omega_m$ & 0.274  & 0.316  & 0.293  & 0.332  \\
\hline
$f^{0.43}(z_{\rm eff})\sigma_8(z_{\rm eff})$ & 0.537 & 0.570 & 0.583 & 0.555 \\
\end{tabular}
\end{center}
\caption{Parameters for the different cosmology models tested in this paper for the analysis of CMASS data: Planck13, L-Planck13 and H-Planck13. The mocks cosmology is shown as a reference.}
\label{cosmology_table}
\end{table*}%

\begin{figure*}
\centering
\includegraphics[scale=0.4]{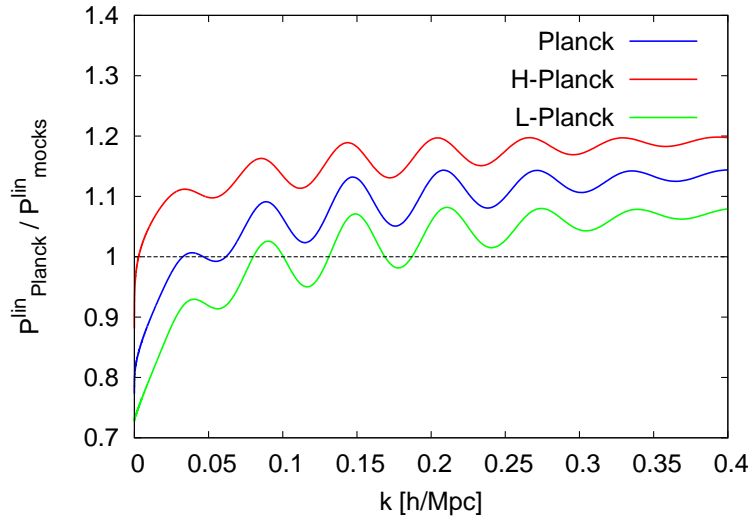}
\caption{Linear power spectrum of Planck13 cosmology (blue line), H-Planck13 cosmology (red line) and L-Planck13 cosmology (green line). All the power spectra have been normalised by the mock linear power spectrum for clarity. The main difference between the Planck cosmologies relies on the amplitude, whereas for the mocks cosmology the BAO oscillations also present a different pattern.  The details of these different cosmologies can be found in Table~\ref{cosmology_table}.}
\label{Plin_plot}
\end{figure*}

Fig.~\ref{Plin_plot} displays the linear dark matter power spectrum of the Planck13, H-Planck13 and L-Planck13 cosmologies normalised by the  power spectrum for  the mocks cosmology in order to visualise the differences. The main changes  are due to  the parameter $A_s$, which regulates the amplitude of the linear power spectrum. However, since in the analysis of the data we always recover the parameters in combination with  $\sigma_8$, we do not expect the results to depend on the choice of  $A_s$. We also observe that the differences in the wiggles pattern among the Planck cosmologies are small. On the range of scales considered for our analysis the effect of other parameters, which change the broadband shape of the power spectrum such as  such as $n_s$, is small.

Table~\ref{data_table_cosmo} lists the best fit parameters obtained from analysing the power spectrum  and bispectrum monopoles from the DR11 CMASS NGC galaxy sample when four different cosmologies are assumed: Planck13, H-Planck, L-Planck and Mocks. As  in Table~\ref{data_table}, the maximum scale for the fit has been set to $0.17\,h{\rm Mpc}^{-1}$.
\begin{table*}
\begin{center}
\begin{tabular}{|c|c|c|c|c|c|c}
$k_{\rm max}=0.17\,h{\rm Mpc}^{-1}$ & ${b_1}^{1.40}\sigma_8(z_{\rm eff})$  & ${b_2}^{0.30}\sigma_8(z_{\rm eff})$ & $A_{\rm noise}$ & $\sigma^B_{\rm FoG}$ & $\sigma^P_{\rm FoG}$ & $f^{0.43}(z_{\rm eff})\sigma_8(z_{\rm eff})$\\
 \hline
  Planck13 & $1.655\pm0.071$ & $0.585\pm0.094$ & $-0.32\pm0.27$ & $17\pm13$ & $5.7\pm1.9$ & $0.541\pm0.092+0.05$  \\
\hline
 H-Planck13 & $ 1.805\pm0.071$ & $0.579\pm0.095$ & $-0.41\pm0.27$ & $9\pm13$ & $3.9\pm1.9$ & $0.526\pm0.092+0.05$  \\
 \hline
 L-Planck13 &  $1.572\pm0.071$ & $0.560\pm0.095$ & $-0.33\pm0.27$ & $18\pm13$ & $5.7\pm1.9$ & $0.529\pm0.092+0.05$  \\
 \hline 
 Mocks & $1.708\pm0.071$ & $0.533\pm0.095$ & $-0.50\pm0.27$ & $8\pm13$ & $3.9\pm1.9$ & $0.493\pm0.092+0.05$ \\
  \end{tabular}
\end{center}
\caption{Best fit parameters to CMASS DR11 NGC galaxy sample for four different underlying cosmologies: Planck13, L-Planck13, H-Planck13 and Mocks. The maximum scale is set to $k_{\rm max}=0.17\,h{\rm Mpc}^{-1}$. The units for $\sigma_{\rm FoG}^{(i)}$   are  Mpc$h^{-1}$.}
\label{data_table_cosmo}
\end{table*}
Considering the relatively large changes in the input cosmological parameters, we do not observe any significant variation  for most of the estimated parameters (shifts compared to the fiducial cosmology are typically $\lesssim 0.5 \sigma$).  The most sensitive parameter to the cosmology is $b_1^{1.40}\sigma_8$, which changes $\simeq1\sigma$ at $k_{\rm max}\leq0.17\,h{\rm Mpc}^{-1}$. On the other hand, the $f^{0.43}\sigma_8$ parameter does not present any significant trend within the cosmologies explored in this paper.
Since we assume that the errors do not depend with cosmology, they are the same for all three cosmologies. 

\begin{figure*}
\centering
\includegraphics[clip=false, trim= 20mm 0mm 12mm 0mm,scale=0.32]{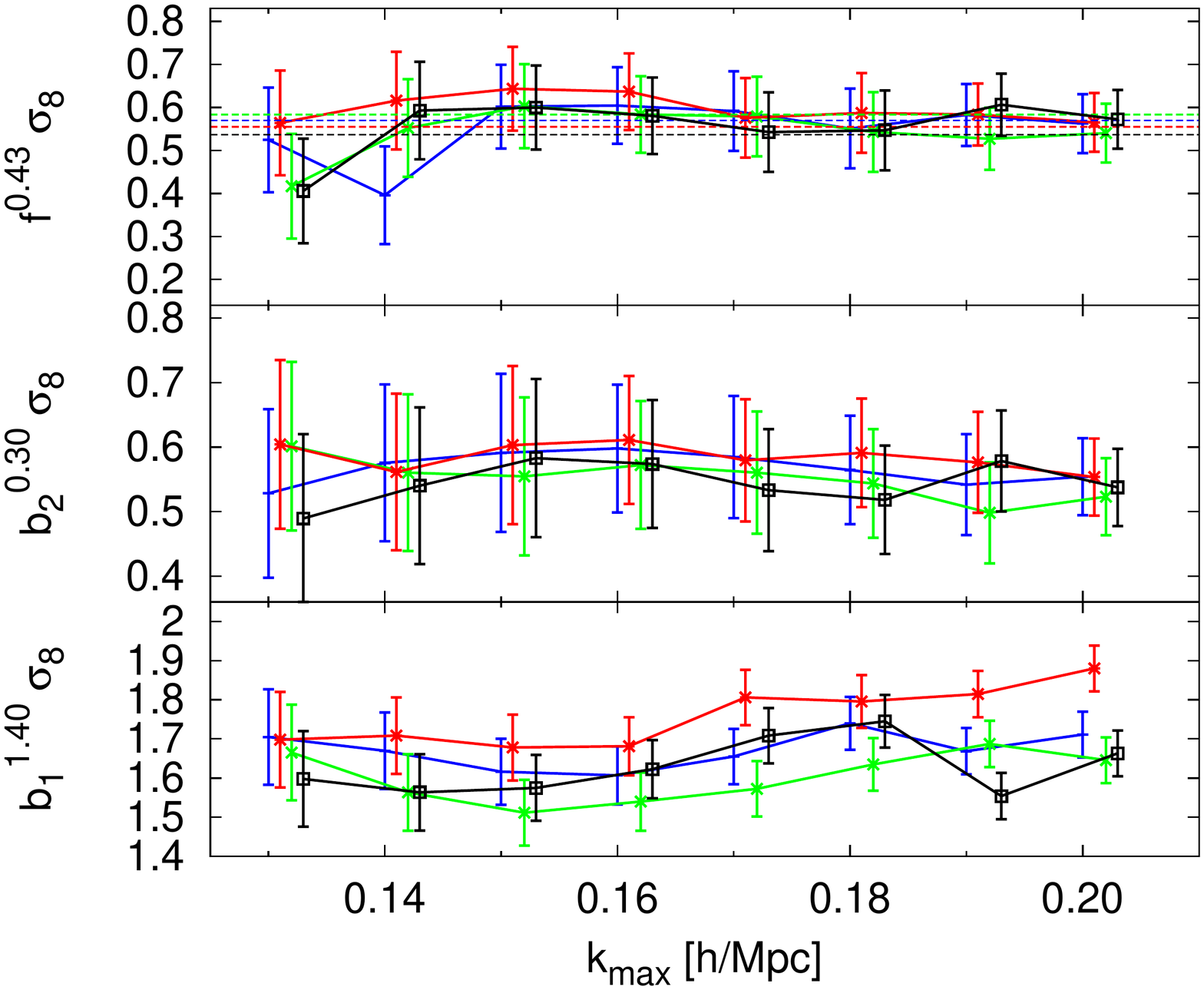}
\includegraphics[clip=false, trim= 20mm 0mm 12mm 0mm,scale=0.32]{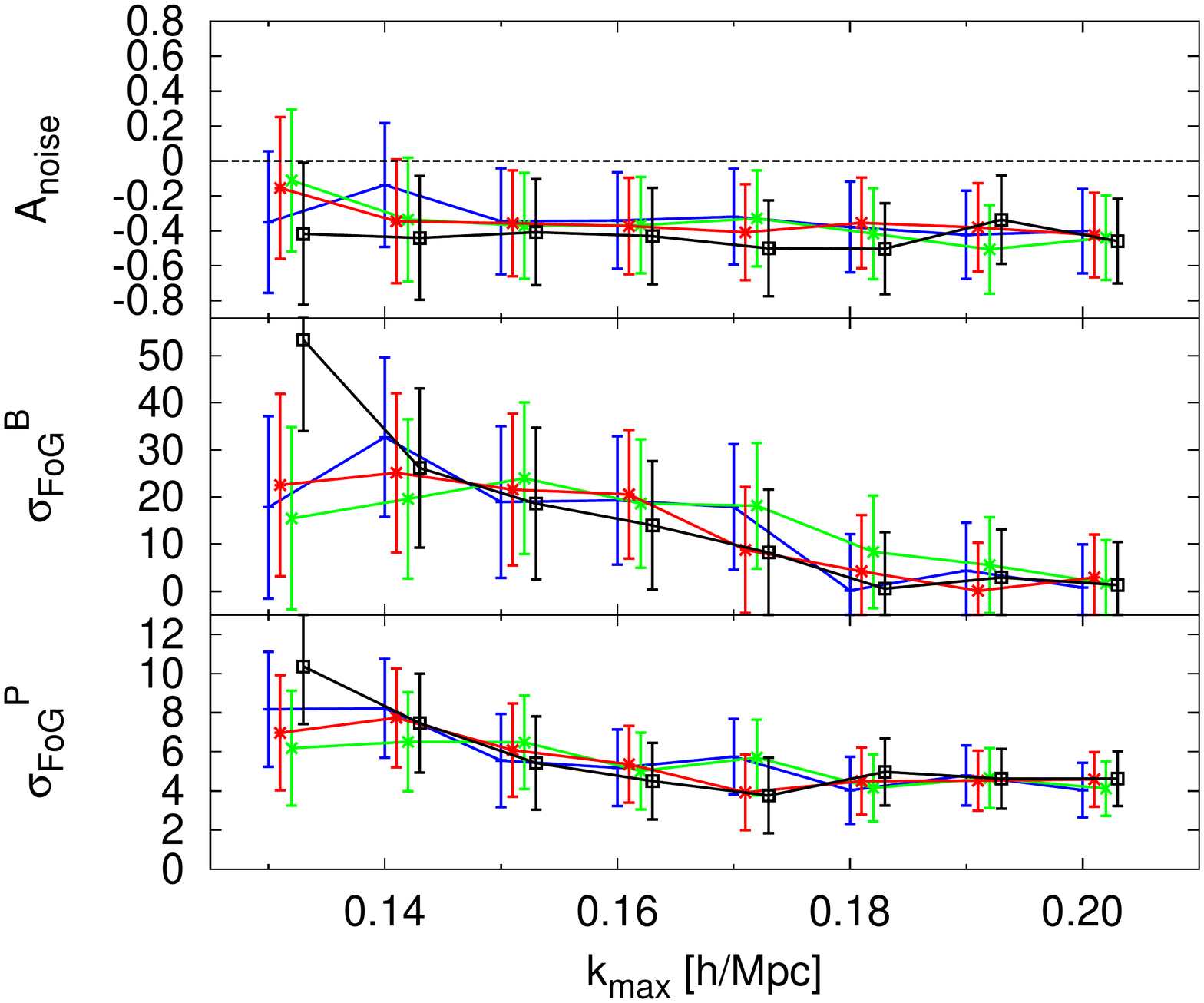}
\label{kmax2}
\caption{Best fit parameters as a function of $k_{\rm max}$ for NGC data assuming different cosmologies (listed in Table~\ref{cosmology_table}): Planck13 (blue symbols), L-Planck13 (green symbols), H-Planck13 (red symbols) and Mocks (black symbols). The quantity $f^{0.43}\sigma_8$ has been corrected by the systematic error as is listed in Table~\ref{data_table}. For the $f^{0.43}\sigma_8$ panel, the corresponding fiducial values for GR are shown by dashed lines for the corresponding cosmology model.  There is no apparent dependence with $k_{\rm max}$ for any of the displayed parameters for $k_{\rm max}\leq0.17\,h{\rm Mpc}^{-1}$.}
\label{kmax_data1}
\end{figure*}

Fig.~\ref{kmax_data1} displays how the best fit parameters depend on the maximum scale for the four cosmologies: Planck13 (blue lines), H-Planck (red lines), L-Planck (green lines) and Mocks (black lines).  Dashed lines show the GR prediction for $f^{0.43}\sigma_8$ when a particular cosmological model is assumed.

 We conclude that there is no need to increase the errors estimated form the mocks on the quantity $f^{0.43}\sigma_8$ to account for uncertainty in the cosmological parameters.

\section{Tests on N-body simulations and survey mock catalogs}\label{section:systematics}
 We have performed extensive tests to check for systematic errors induced by our method  and to assess the performance of the different approximations we had to introduce. In particular we have  tested   the power spectrum and bispectrum  modelling on  dark matter particles, haloes and mock galaxy catalogs. 
 We also quantify the effects of the survey geometry and our approximation of these to match the FKP-estimator derived results.
 
\subsection{Tests on N-body dark matter particles}\label{test_dark_matter}
In order to test  the effect of our choice of triangle shapes on  the  best fit values and errors, we focus first on the simpler and cleaner  case of dark matter simulations. 

As described in \S~\ref{sec:bispectrummethod}, in the analysis of this paper we have chosen to use a subset of triangles where one of the ratios between two sides is fixed to equal $k_2 / k_1 = 1$ or $k_2 / k_1 = 2$. By doing so we are discarding information contained in the triangle shapes we do not use, but analytically estimating exactly how this affects the errors is difficult since different triangles are in general correlated.  Our kernel was calibrated on a slightly more extended set of shapes (see \citealt{HGMetal:2011,HGMetal:inprep}) by reducing the average differences from the simulations; this decision could hide  subtle cancellations that do not hold as well  when only a sub-set of shapes is considered. Thus, we need to check for possible shifts in the parameter estimates.

 One may instead choose to use all possible triangle configurations, varying all the three sides of the triangles with a step equal to the fundamental mode of the survey and imposing only that they form a closed triangle. This approach of course requires  significantly more computational power, especially since our estimate of the errors is done by analysing on hundreds of mocks, but it is, in principle, possible. When using all shapes one must extrapolate and interpolate the effective bispectrum kernel beyond the shapes for which it was calibrated,  and this can induce a systematic error.

\begin{figure*}
\centering
\includegraphics[scale=0.3]{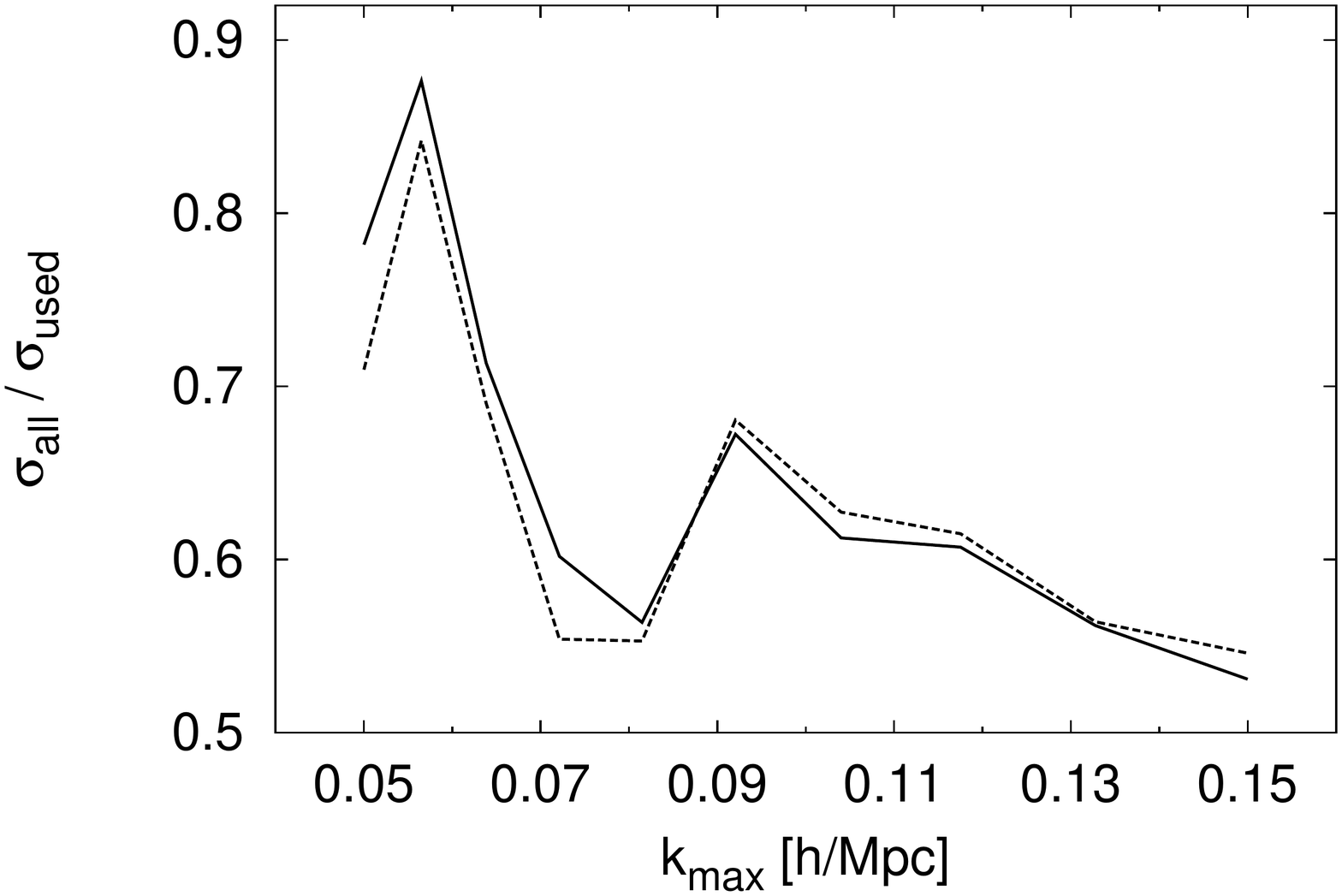}
\includegraphics[scale=0.3]{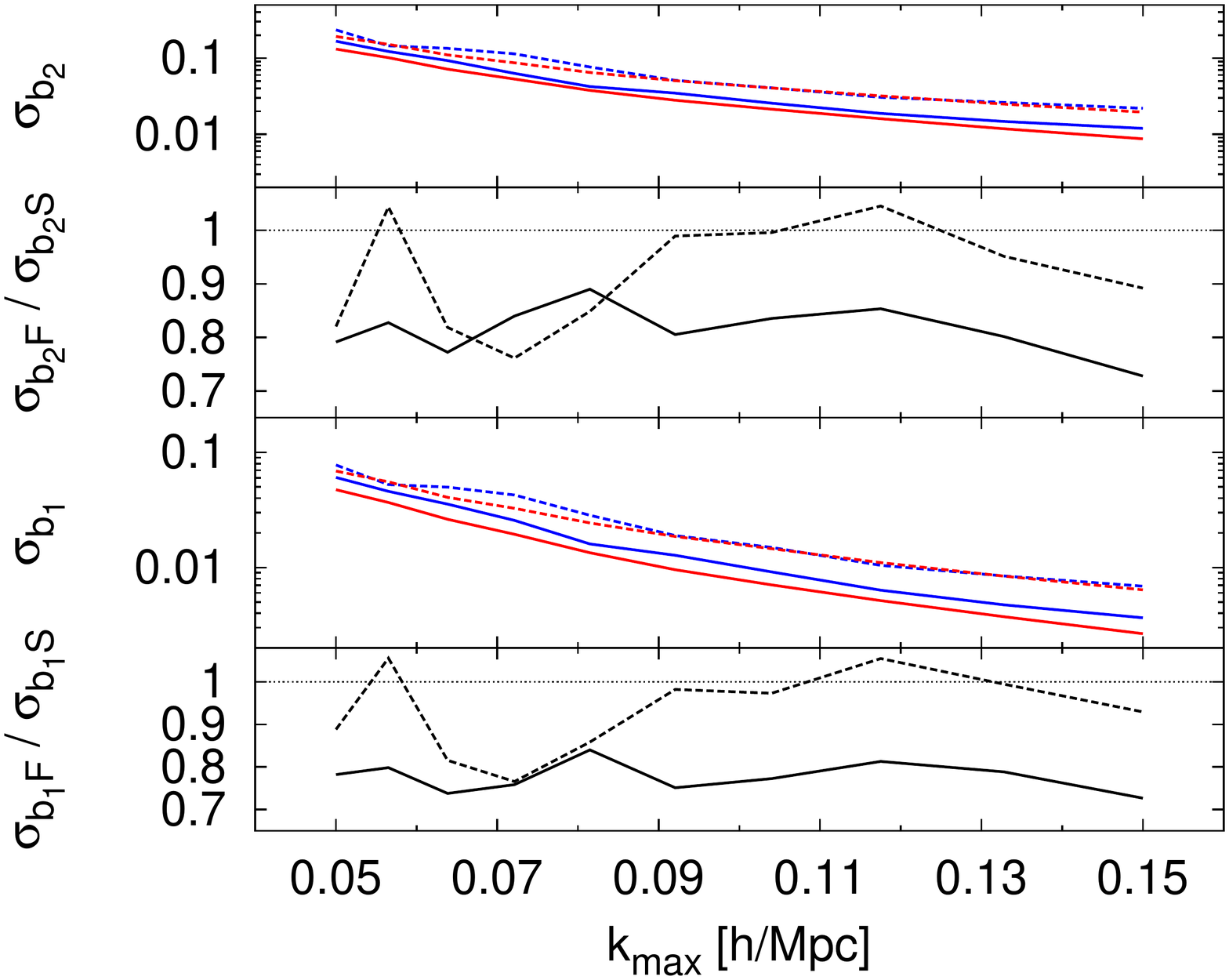}

\caption{Left Panel:  ratio between errors obtained using all possible triangles and only  $k_2/k_1=1,\, 2$ triangles. The solid line is for $b_1$ and dashed line is for $b_2$.  Errors are computed from the scatter of 60 realisations of dark matter. Right panel: Red lines correspond to the predictions of the errors of $b_1$ and $b_2$ using Fisher analysis, whereas blue lines when these errors are predicted from the scatter of best fit values of different realisations. Black lines correspond to the ratio between Fisher predictions (subscript F)  and scatter predictions (subscript S). Solid lines are the predictions when all the possible triangles are used, whereas dashed lines are for triangles with $k_2/k_1=1,\,2$. These plots indicate that the   the statistical errors  could potentially be reduced by using more shapes, although by doing this, the systematic effects would dominate the results and the full benefit of  shrinking the statistical errors will not be realised}
\label{all_triangles}
\end{figure*}

In order to tackle this issue  we apply our analysis to the simple case of dark matter in real space, for which we know that by definition $b_1 = 1$ and $b_2 = 0$, without complications due to halo bias, survey window etc. 
We use 60 N-body simulations among those used in \cite{HGMetal:2011} for an effective volume that is about 140 times larger than that of the survey.
Using only bispectrum measurements, we find that there is no significant bias in $b_1$ using either the two selected shapes or all shapes. For $b_2$ we find a hint of a possible $+0.05$   bias which is, however, at the $1.5\sigma$ level and thus completely negligible for our data set. Using all shapes  leads to reduced error-bars.
This result is shown in  the left panel of Fig.~\ref{all_triangles}.

The fractional difference in the errors indicates  there is  roughly a factor two improvement in using all the configurations. 

In the right panel of Fig.~\ref{all_triangles} we compare the errors obtained with a simple Fisher matrix estimate (following \citealt{SCFFHM98} Appendix A2 and \citealt{HGMetal:2011} Eq.  A.3). This figure indicates that  that one can take the --band-power-- bispectra to have a Gaussian distribution for this  volume and for the binning adopted here.

These findings demonstrate that in principle  the statistical errors  could be reduced by using more shapes. This approach, however,  will not be implemented here for several reasons: {\it i)} It is computationally extremely challenging {\it ii)} It requires an extrapolation/interpolation of kernels that have been calibrated on a subset of shapes. This extrapolation works fine for real space  but its effectiveness has not been explored in redshift space {\it iii)} Most importantly, in  the present  analysis, systematic errors are kept  (just) below the statistical errors, so the full benefit of  shrinking the statistical errors will not be realised.

\subsection{N-body haloes vs \textsc{PTHALOS} in real space and redshift space}\label{5_2_section}
The mock galaxy catalogs  are  based on    \textsc{PThalos}, which only provides an approximation to  fully non-linear dark matter halo distributions. Here we check  the differences at the level of the power spectrum and bispectrum between N-body haloes and \textsc{PThalos}.

\textsc{PThalos} and N-body haloes simulations  (\S~\ref{sec:sims-mocks})   have the  same underlying cosmology, but  different mass resolutions. The large scale power spectrum is therefore different for the two catalogs (there is a relative bias) because the minimum mass of the resolved  haloes is not identical. However, since the definition of halo cannot be the same for both (see \citealt{Maneraetal:2013} for a complete discussion on the differences between N-body-halo and \textsc{PThalos} mass), setting the mass threshold to be the same for the two catalogues does not completely solve this problem.

 Therefore we choose the minimum mass of the N-body catalogues so that the resulting halo  power spectrum matches the amplitude of \textsc{PThalos} power spectrum at large scales in real space.  This occurs at
 $\log_{10}(M_{\rm min}[M_\odot/h])=12.892$ where for the  \textsc{PThalos} catalogue the  minimum mass is $\log_{10}(M_{\rm min}[M_\odot/h])=12.700$. The \textsc{PThalos} mass we report, is the sum of the masses of the particles that form each \textsc{PThalo}. Hence, this is the halo mass {\it before} the re-assignment and should not be confused with the re-assigned mass that matches the mass function from N-body haloes.

\begin{figure*}
\centering
\includegraphics[clip=false, trim= 0mm 15mm 0mm 00mm,scale=0.3]{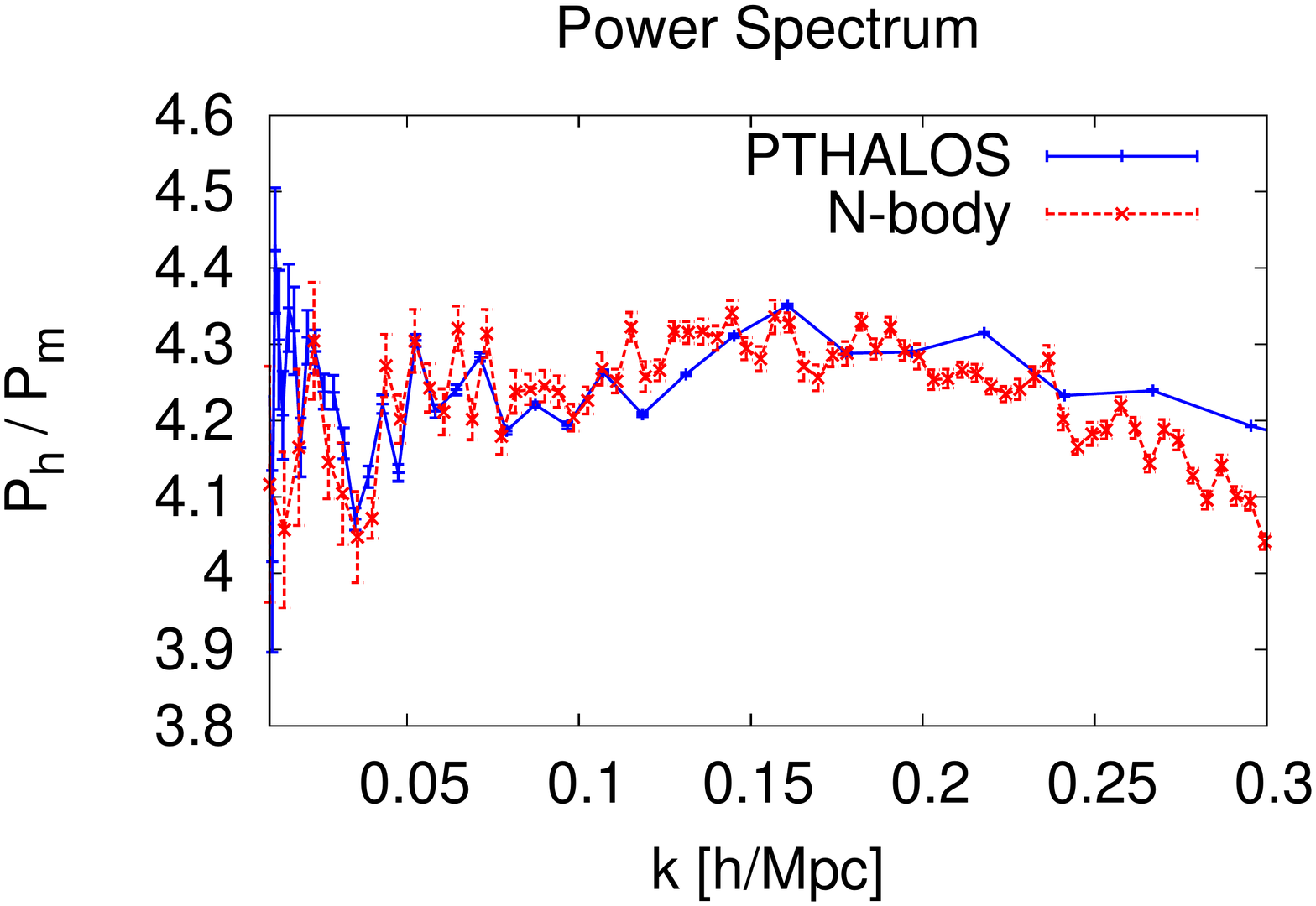}
\includegraphics[clip=false, trim= 0mm 15mm 0mm 00mm,scale=0.3]{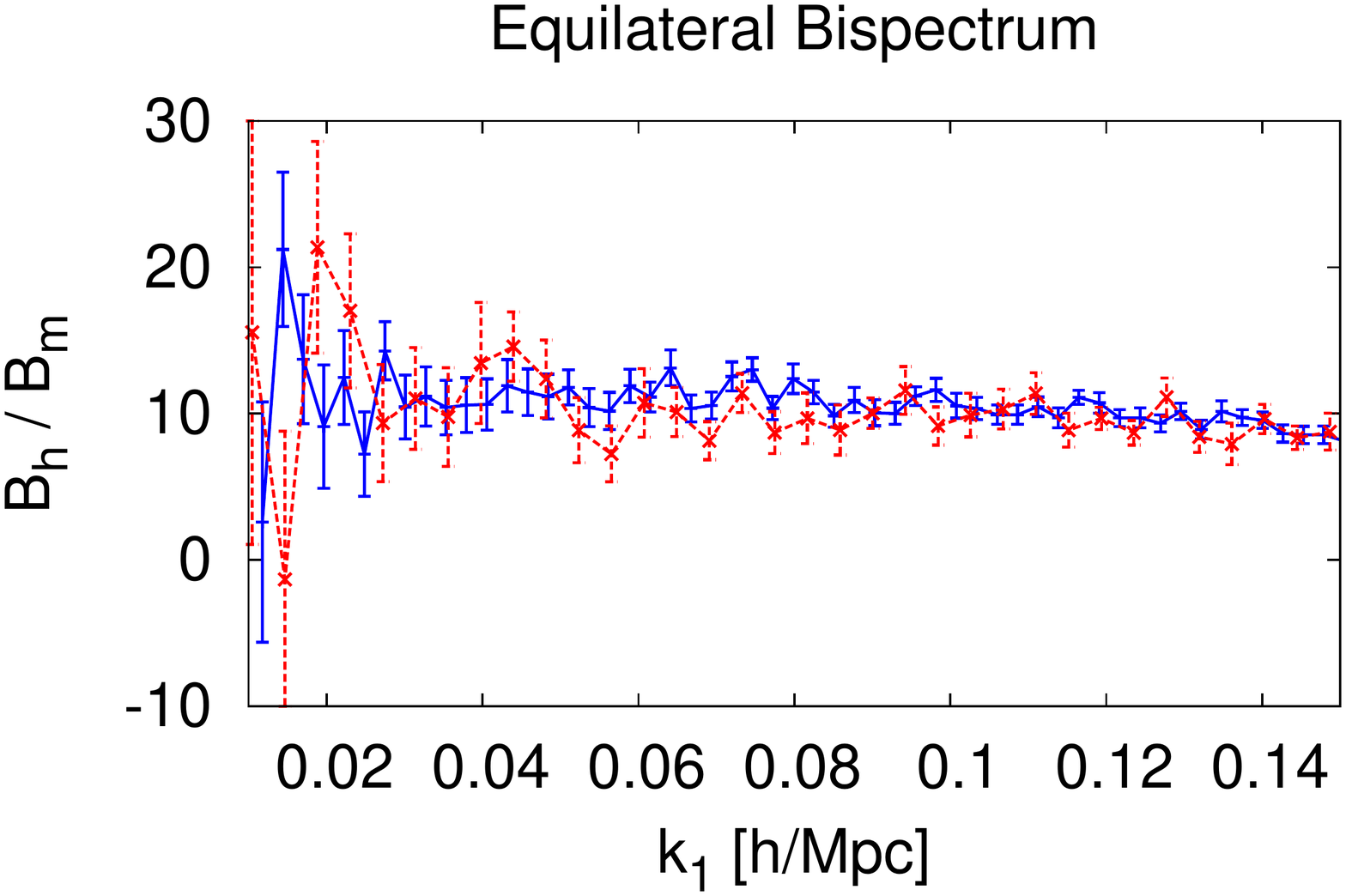}

\includegraphics[clip=false, trim= 0mm 10mm 0mm 30mm,scale=0.3]{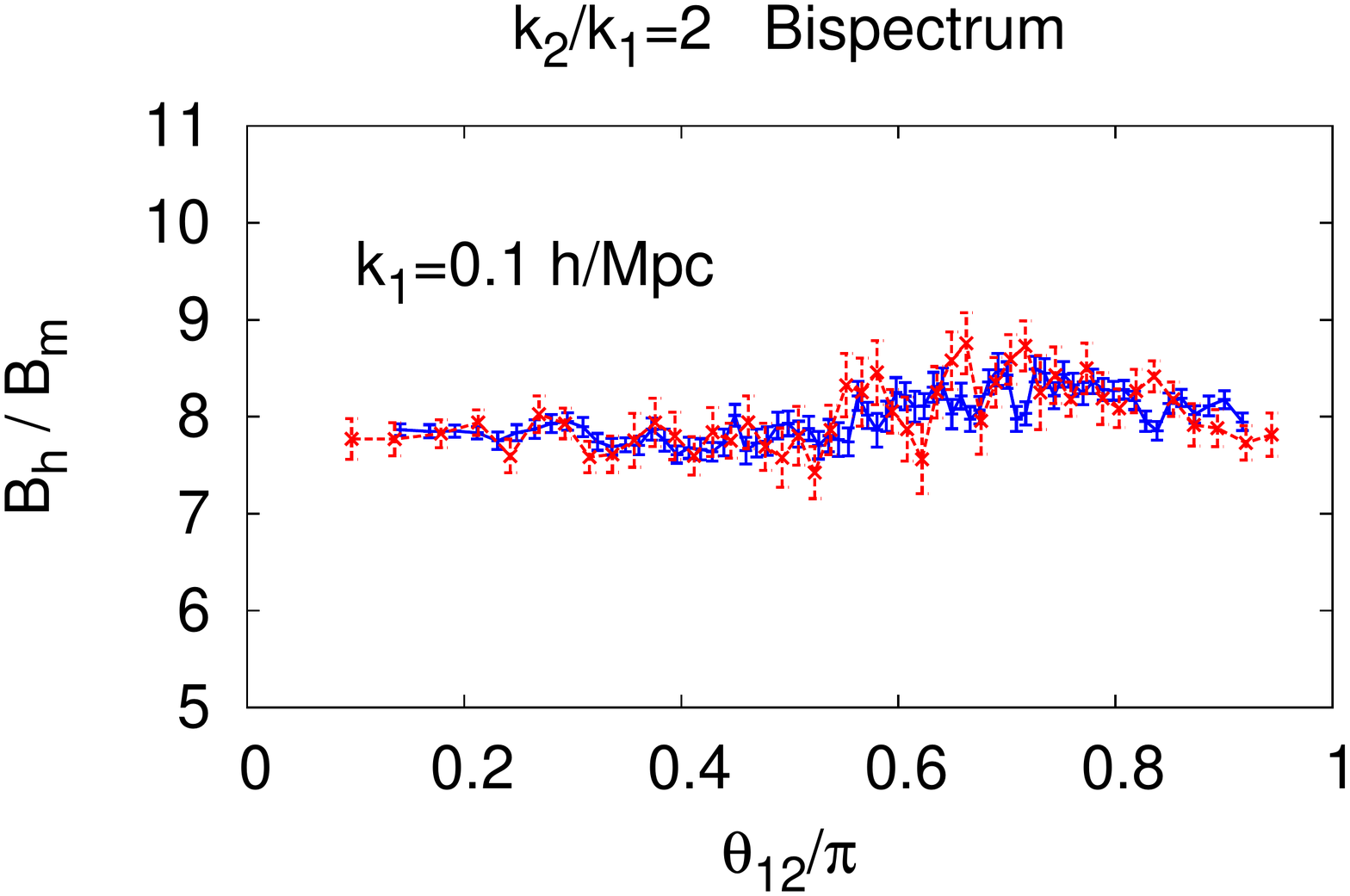}
\includegraphics[clip=false, trim= 0mm 10mm 0mm 30mm,scale=0.3]{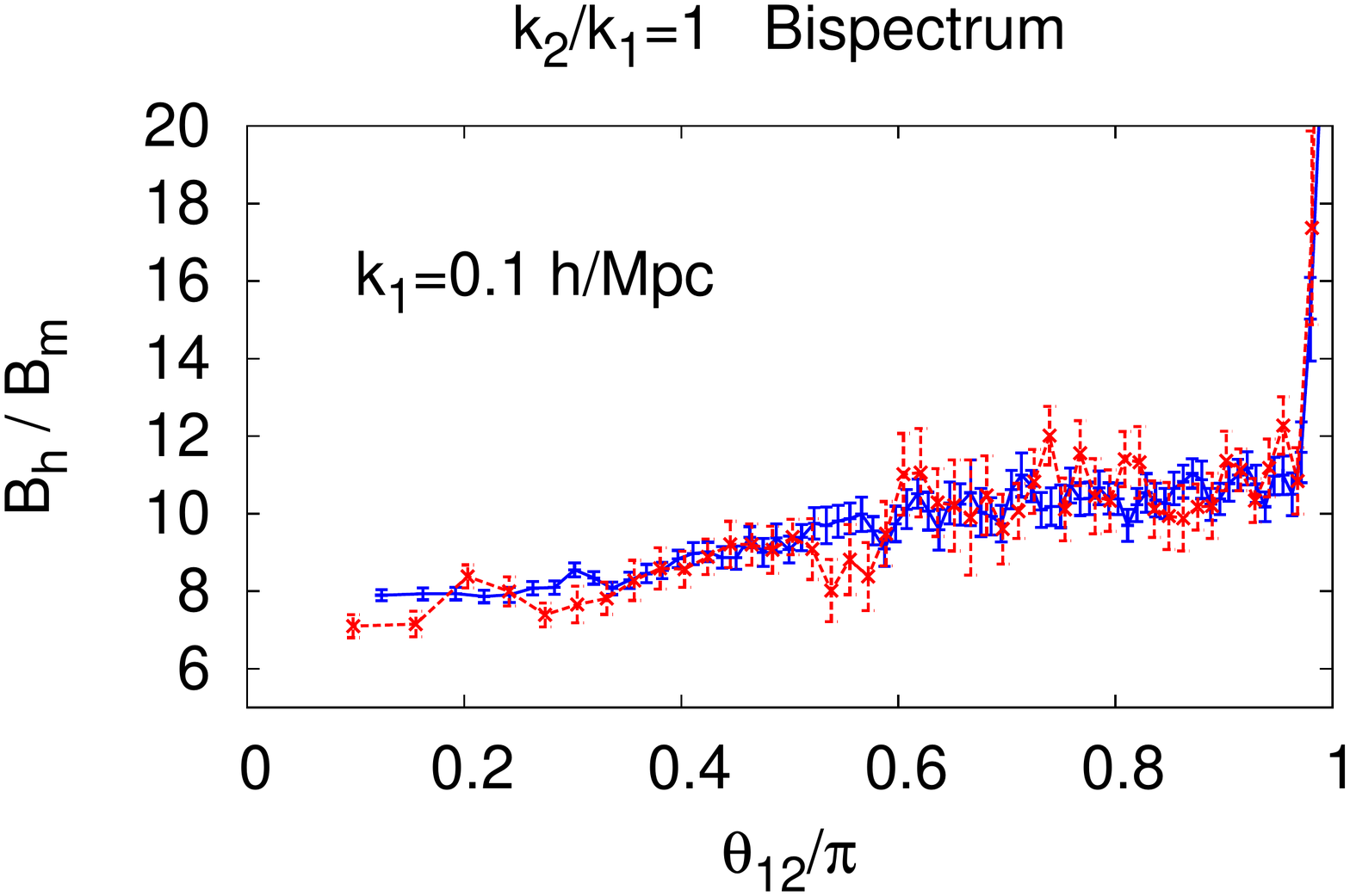}

\caption{Power spectra (top left panel) and bispectra (other panels) for N-body haloes (red lines) and \textsc{PThalos} (blue lines) both in real space normalised by $P_{\rm matter}$ and $B_{\rm matter}$, respectively. Poisson noise is assumed. There is good agreement for power spectrum and bispectrum of N-body haloes and \textsc{PThalos}  for $k\lesssim 0.2$ $h{\rm Mpc}^{-1}$. The halo mass cut has been $\log_{10}(M_{\rm min}[M_\odot h^{-1}])=12.892$.}
\label{plotRealspace}
\end{figure*}

Fig.~\ref{plotRealspace} presents the comparison between N-body haloes (red lines) and \textsc{PThalos} (blue lines). The top left panel shows the comparison between the power spectra in real space (normalised by the non-linear matter power spectrum prediction for clarity) and the others of the panels display the comparison between different shapes of the bispectrum in real space (also normalised by the non-linear matter prediction): equilateral triangles, $k_2/k_1=1$ and $k_2/k_1=2$ triangles, as indicated in each panel. In all the panels the symbols represent the mean value among 50 realisations for \textsc{PThalos} and 20 realisations for N-body haloes. The errors-bars correspond to the error of the mean. The error-bars for N-body haloes are slightly larger due to the difference in the number of realisations ($\sqrt{(50\times2.4)/(20\times1.5)}=2$), and therefore in the total volume. Note also that these error-bars do not take into account the uncertainty on the measurement of $P_m$ and $B_m$, which have been computed using 5 realizations, and therefore the displayed error-bars are slightly under-estimated.  The agreement between N-body haloes and \textsc{PThalos} is excellent at large scales for the power spectrum. At small scales, $k\geq0.2\,h{\rm Mpc}^{-1}$, the \textsc{PThalos} power spectrum overestimates the N-body prediction by few percent. The agreement is also good for the bispectrum. For the equilateral shape  both N-body and \textsc{PThalos} agree for $k\leq0.15\,h{\rm Mpc}^{-1}$. We do not go beyond this scale, given that our set of triangles with $k_1=k_2$ are limited to $k_1\leq0.15\,h{\rm Mpc}^{-1}$, as we have mentioned in \S\ref{sec:38}. Also for the scale of $k_1=0.1\,h{\rm Mpc}^{-1}$,  \textsc{PThalos} reproduces  the shape described by N-body haloes, for different values of $k_2/k_1$ ratio. Therefore we conclude that   \textsc{PThalos} is able to describe accurately the clustering predicted by N-body haloes for both the power spectrum and bispectrum up to mildly non-linear scales, typically $k_i\lesssim0.2$ at $z=0.55$ (recall that in deriving our main results we use $k_{\rm max}=0.17\,h{\rm Mpc}^{-1}$).

\begin{figure*}
\centering
\includegraphics[clip=false, trim= 0mm 15mm 0mm 00mm,scale=0.3]{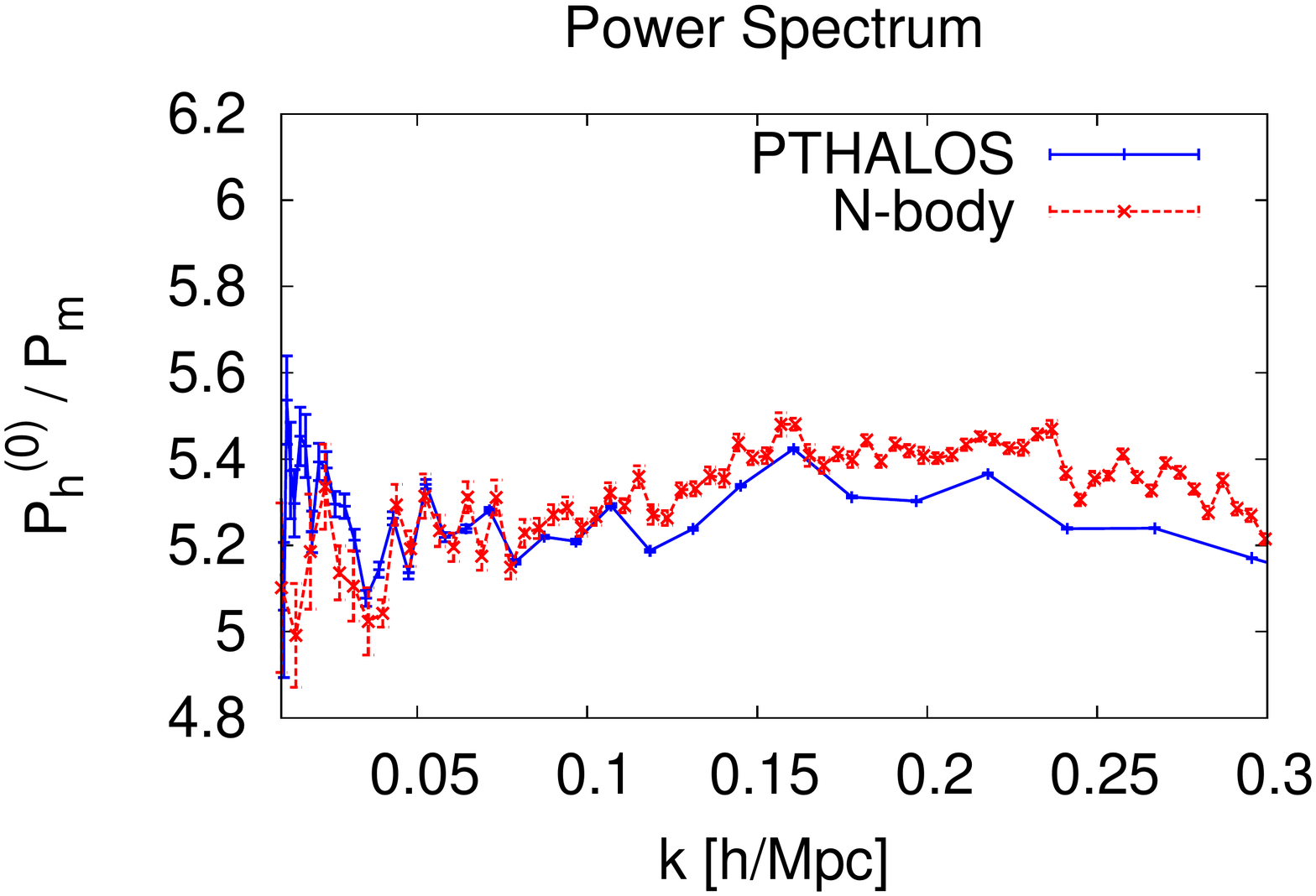}
\includegraphics[clip=false, trim= 0mm 15mm 0mm 00mm,scale=0.3]{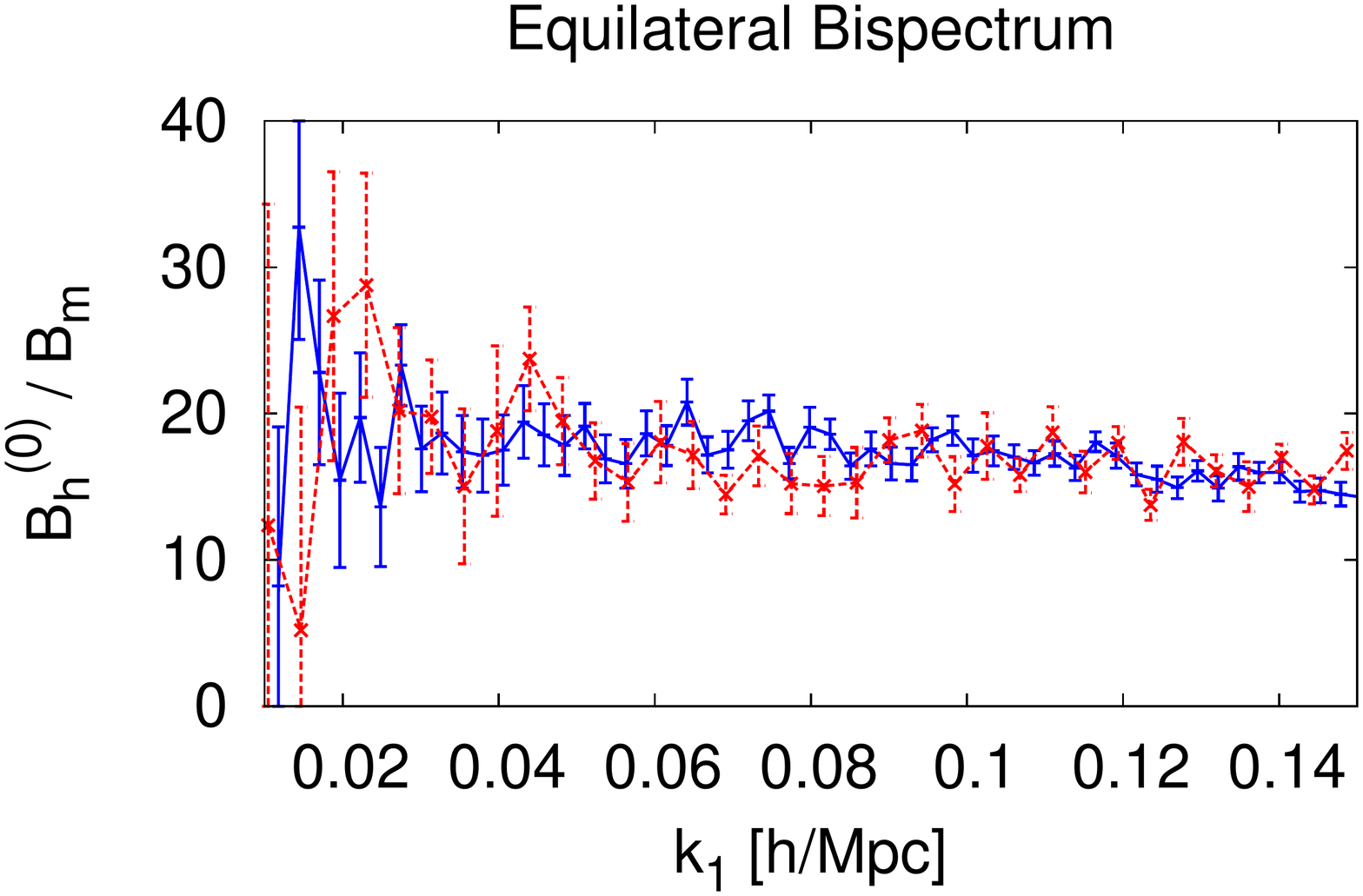}

\includegraphics[clip=false, trim= 0mm 10mm 0mm 30mm,scale=0.3]{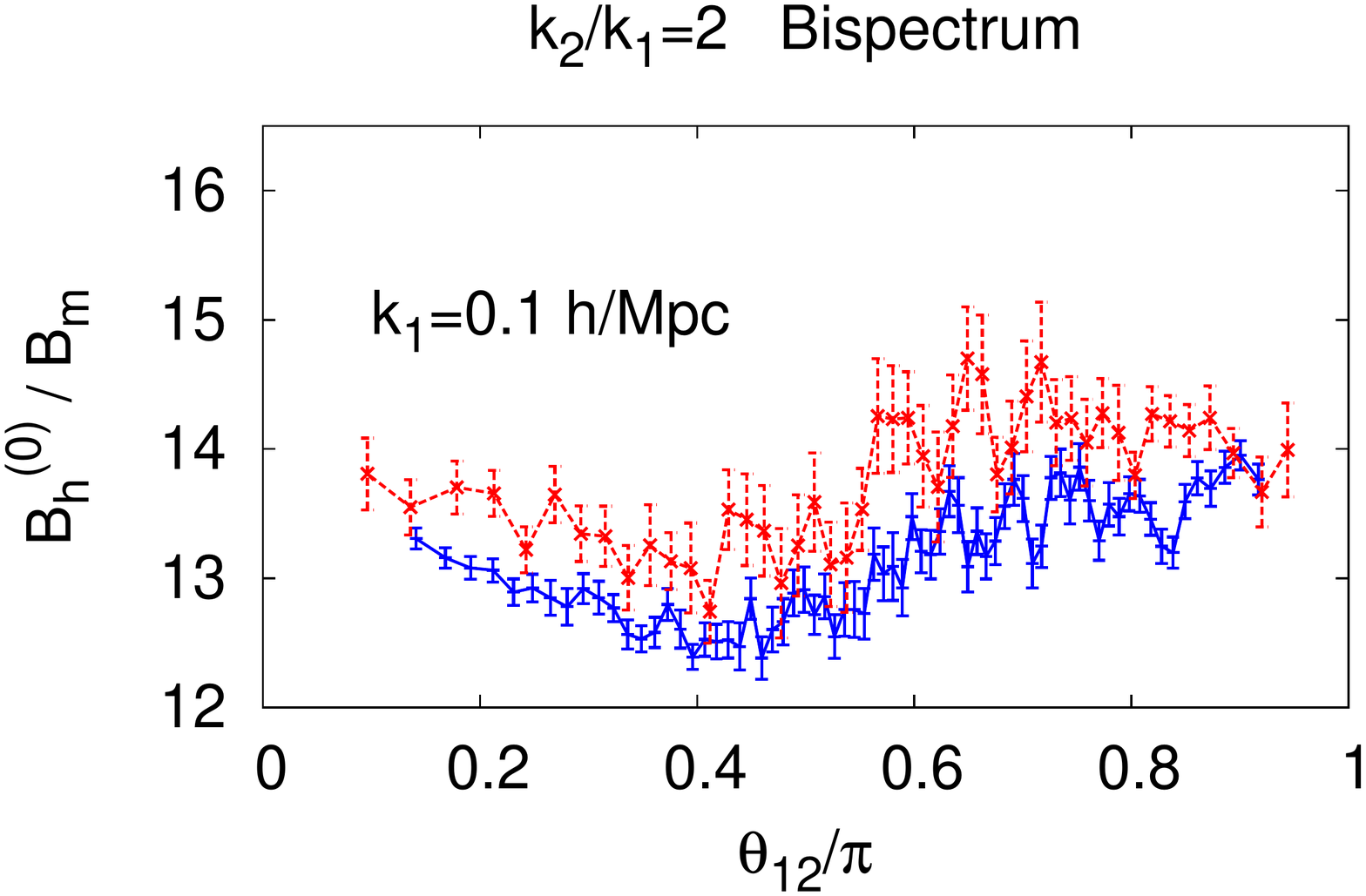}
\includegraphics[clip=false, trim= 0mm 10mm 0mm 30mm,scale=0.3]{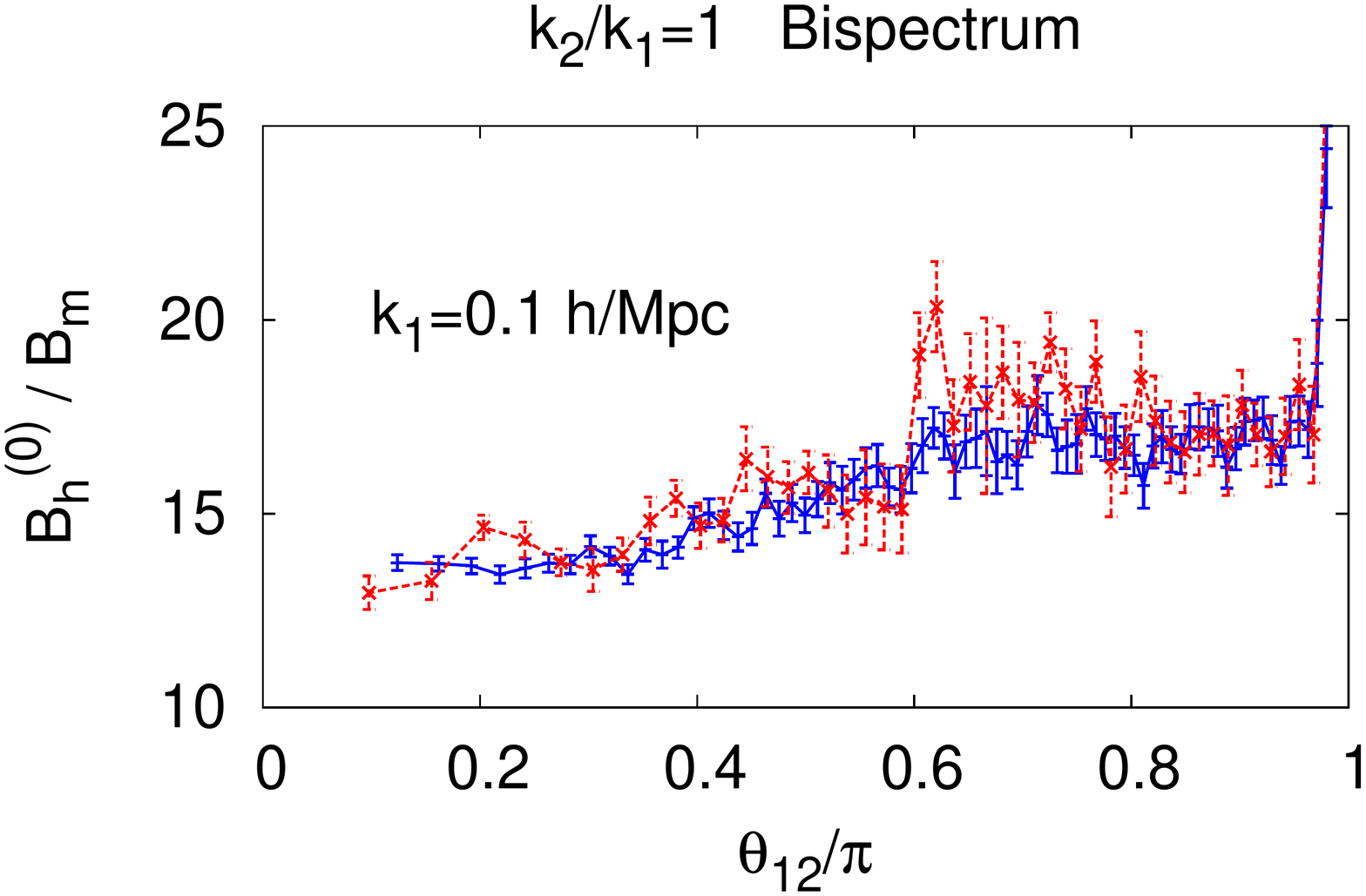}

\caption{ Same notation that in Fig.~\ref{plotRealspace} but for redshift space monopole statistics. \textsc{PThalos} tend to underestimate the monopole redshift space quantities, and this start to be significant ($\simeq5\%$ deviation) for the power spectrum at $k>0.10\, h{\rm Mpc}^{-1}$ and for the bispectrum shape where $k_2=2k_1$. The halo mass cut has been $\log_{10}(M_{\rm min}[M_\odot h^{-1}])=12.875$.}
\label{plotRSD}
\end{figure*}

 The panels of Fig.~\ref{plotRSD} use the same notation as Fig.~\ref{plotRealspace}  showing the  redshift space monopole for the power spectrum and bispectrum. In this case the halo mass cut for N-body has been set to $\log_{10}(M_{\rm min}[M_\odot h^{-1}])=12.875$. If the mass cut were maintained to the same value than in real space, we would not have obtained a good match between N-body and PTHAOS. Assuming  that N-body haloes are a better description of real haloes than \textsc{PThalos}, these discrepancies may indicate that  even large-scale redshift space distortions  are not well captured by \textsc{PThalos}.  However, we observe that these discrepancies can be mitigated rescaling slightly the mass cut for N-body halo catalogues. This does not  represent any practical problem, since for the final galaxy mocks the mass cut of the mocks is calibrated with observations, i.e. in redshift space.

From the panels of Fig.~\ref{plotRSD} we observe a very good match on the power spectrum monopole for $k\leq0.10\,h{\rm Mpc}^{-1}$. For large values of $k$ the differences slightly grow, but they are always below $5\%$. There are small differences between the bispectra of N-body and \textsc{PThalos} for the $k_1=k_2=1$ shape. There is a significant $\sim5\%$ offset for the $k_2/k_1=2$. From the plots of Fig.~\ref{plotRSD}, it is not clear how these offsets can affect to the parameter estimation. In order to check this, we compare the recovered the bias parameters from N-body and \textsc{PThalos} both in real and redshift space.

We start by  estimating  the bias parameters $b_1$ and $b_2$ for \textsc{PThalos} and N-body haloes assuming that the underlying cosmological parameters, such as $\sigma_8$ and $f$, are known. For simplicity (and speed)  we  also assume  no damping term is needed for the redshift space bispectrum monopole (i.e., Eq.~\ref{B_gggs} applies with $D_{\rm FoG}^B=1$). It is well known that no Finger-of-God-like velocity dispersion is expected when considering the clustering of haloes (mapped by their centre of mass  point).

 In order to estimate the bias parameters we follow the method described in \S~\ref{section_method}, in particular \S~\ref{method:powerspectrum} and \S~\ref{sec:bispectrummethod}, but  using only the bispectrum.  For the non-linear density dark matter power spectrum needed in the bispectrum model, we use the quantity directly estimated from dark matter simulations themselves. For this analysis, we have  only  three parameters: $b_1$, $b_2$ and $A_{\rm noise}$.
 
\begin{figure*}
\centering
\includegraphics[scale=0.31]{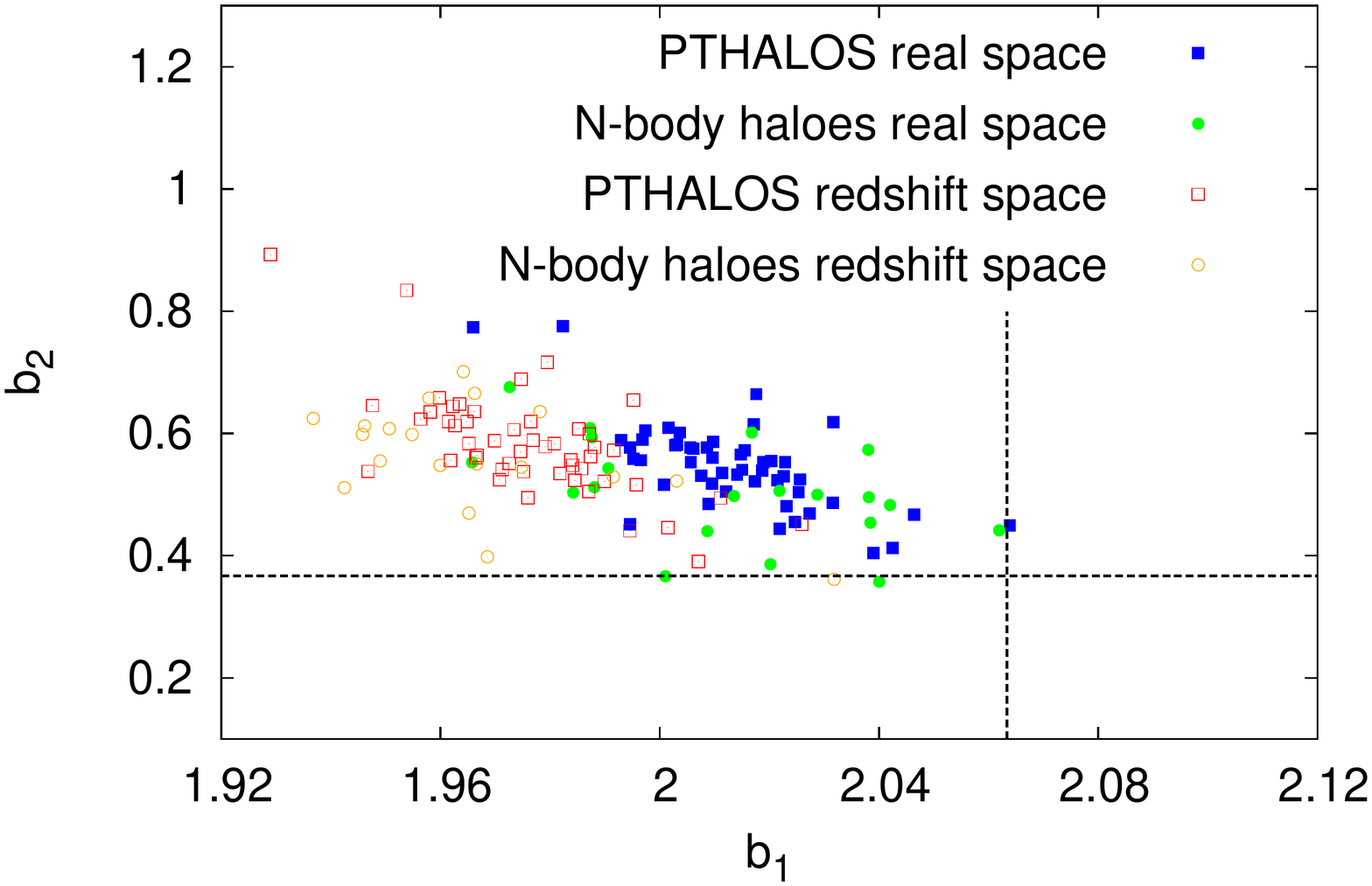}
\includegraphics[scale=0.3]{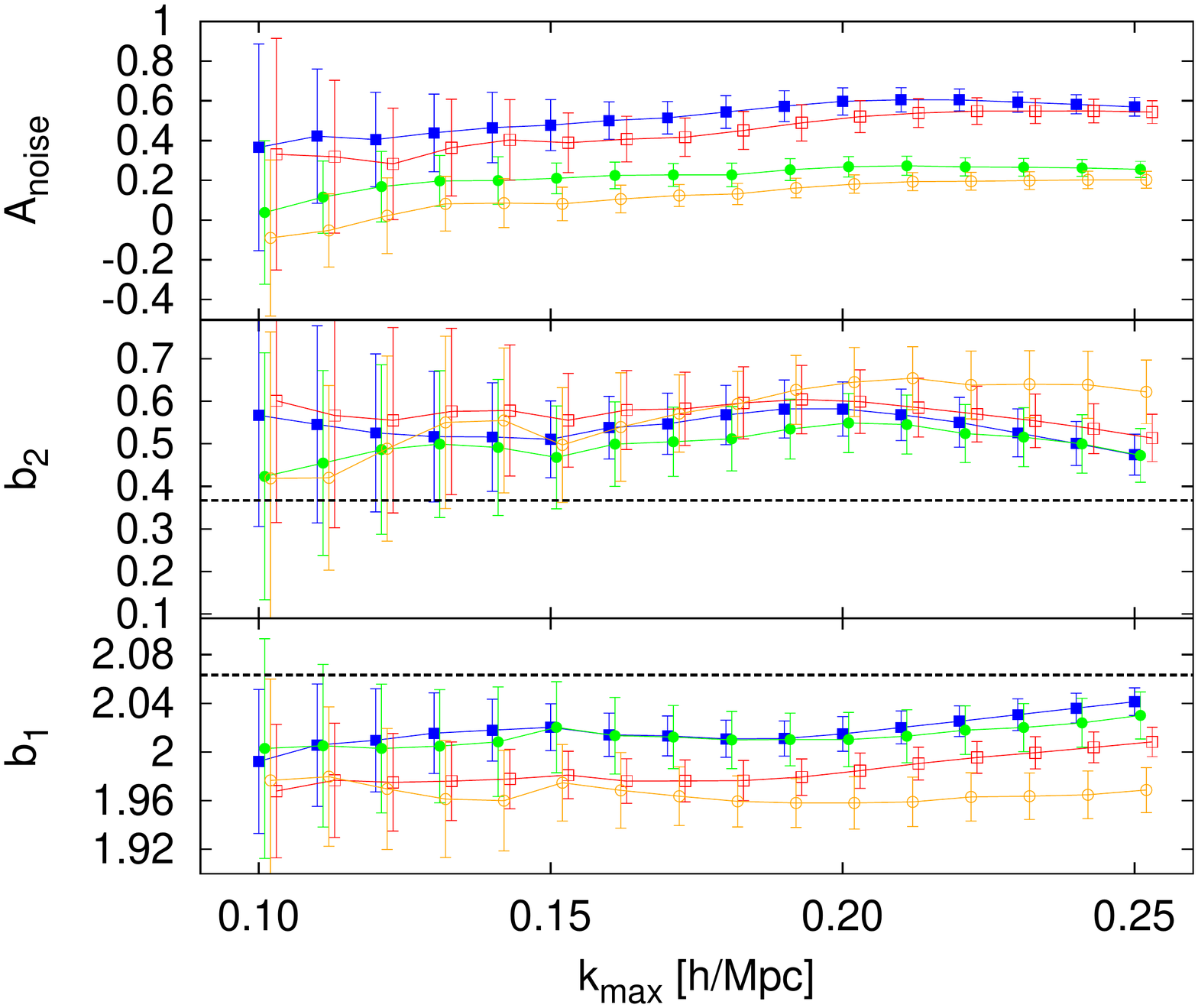}
\caption{{ Left Panel:} Best fit bias parameters for N-body haloes and \textsc{PThalos} estimated from their bispectrum only. Green (blue) symbols are N-body haloes (\textsc{PThalos}) best fit values from real space bispectrum. Red (orange) symbols are N-body haloes (\textsc{PThalos})  best fit values from redshift space monopole bispectrum. { Right Panel:} Best fit bias parameters and shot noise amplitude as a function of $k_{\rm max}$, using the same colour notation that in left panel. Error-bars correspond to the 1-$\sigma$ dispersion among the different realisations.  In both panels, black dashed lines represent the measured cross bias parameters as they are defined in Eq. \ref{b1_cross}-\ref{b2_cross}. This analysis assumes $k_{\rm max}=0.17\,h {\rm Mpc}^{-1}$ There are no significant differences in the bias parameters predicted from N-body haloes and \textsc{PThalos} catalogues.}
\label{plotbias1}
\end{figure*}

The left panel of Fig.~\ref{plotbias1} presents the best fit bias parameters,  $b_1$ and $b_2$, for the 20 (50) different  realisations for N-body haloes (\textsc{PThalos}) using the bispectrum triangles with $k_2/k_1=1$ and $2$. Blue filled squares show the estimate from \textsc{PThalos} in real space, green filled circles from N-body haloes in real space, red empty squares from \textsc{PThalos} in redshift space and orange empty circles N-body haloes in redshift space. All these estimates were made setting the maximum $k_i$ ($i=1,\,2,\,3$)  to $0.17\,h{\rm Mpc}^{-1}$. The right panel of Fig.~\ref{plotbias1} displays how the mean value of $b_1$, $b_2$ and $A_{\rm noise}$ changes with $k_{\rm max}$. The colour notation is the same in both panels. The error-bars in the right panel represent the 1$\sigma$ dispersion among all the realisations. 
 We also include the values of $b^{\rm cross}_1$ and $b^{\rm cross}_2$ measured from the cross halo-matter power spectrum, $P_{\rm hm}$ and the cross halo-matter-matter bispectrum for comparison in black dashed lines,

\begin{eqnarray}
\label{b1_cross}b^{\rm cross}_1&\equiv&\langle P_{\rm hm}(k)/P_{\rm mm}(k)\rangle_{k,\,{\rm realiz.}}, \\
\label{b2_cross} b^{\rm cross}_2&\equiv& \langle[B_{\rm hmm}(k_1,k_2,k_3)-b_1^{\rm cross}B_{\rm mmm}(k_1,k_2,k_3)]/[P_{\rm mm}(k_2) P_{\rm mm}(k_3)] +4/7(b^{\rm cross}_1-1)  S_2(k_1,k_2,k_3)\rangle_{k_i,\,{\rm realiz.}},
\end{eqnarray}
 where the average $\langle\ldots\rangle_{k,\,{\rm realiz.}}$ is taken among different $k$-modes and 70 different realizations of 2LPT dark matter and \textsc{PThalos}. For the $b^{\rm cross}_1$ we have considered $k$-bins with $k\leq0.03\,h{\rm Mpc}^{-1}$, and for $b_2^{\rm cross}$ we have taken into account the $k_1/k_2=1,\,2$ triangles with $0.01\leq k_i [h{\rm Mpc}^{-1}]\leq 0.03 $. The obtained values for the cross-bias parameters are: $b_1^{\rm cross}=2.063$ and $b_2^{\rm cross}=0.367$. The reason of using only such the large scale modes is because we have checked that 2LPT is not a good description of N-body dark-matter at smaller scales. 

In general we do not observe any significant differences for the bias parameters estimated from the  real space bispectrum: both \textsc{PThalos} (blue lines/symbols) and N-body haloes (green lines/symbols) present a similar distribution of $b_1$ and $b_2$ values over the entire $k$-range studied here. In redshift space,  there is  also in good agreement  for the  $b_1$ and $b_2$  between N-body (orange lines/symbols) and \textsc{PThalos} (red lines/symbols) prediction, with $\lesssim1\%$ deviation for $k\leq0.20\,h{\rm Mpc}^{-1}$. The bias parameters estimated from the bispectrum of N-body haloes and \textsc{PThalos} in real space present differences respect to the cross-bias parameters obtained from Eq. \ref{b1_cross}-\ref{b2_cross}: $b_1$ is underestimated by $\sim2.5\%$ respect to $b_1^{\rm cross}$ and $b_2$ is overestimated by $\sim50\%$ respect to $b_2^{\rm cross}$. However, these differences are considerably reduced when $P_{\rm hh}$ is combined with $B_{\rm hhh}$. As shown in \cite{HGMetal:inprep} (see table 4 and figure 9, where the same set of N-body haloes is used), the bias parameters estimated in real- and redshift-space from $P_{\rm hh}^{(0)}$ and $B_{\rm hhh}^{(0)}$ are $b_1=2.05$ and $b_2=0.47$, which represent a deviation of $\sim0.5\%$ and $\sim30\%$ respect to the $b_i^{\rm cross}$ values. Thus, we conclude that the linear bias parameter, $b_1$, obtained from $B_{\rm hhh}$ is biased by about $\sim2.5\%$ respect to $b_1^{\rm cross}$, but this difference is reduced to $0.5\%$ when $P_{\rm hh}$ is added to the analysis. On the other hand, $b_2$ is significantly biased respect to the $b_2^{\rm cross}$ value.

We also observe  differences in the $A_{\rm noise}$ parameter. First of all, redshift space quantities present a lower $A_{\rm noise}$ parameter than real space quantities, which means that the shot noise tends to be more super-Poisson in redshift space. This result can be perfectly understood if we recall that objects in redshift space present a higher clustering, which produce super-Poisson statistics. We will return to this point in \S~\ref{sec:real_redshift}. Conversely, N-body statistics presents a significant different noise than \textsc{PThalos} statistics: N-body haloes have a shot noise closer to the Poisson prediction, whereas \textsc{PThalos} statistics have sub-Poissonian shot noise.  The original differences observed in Fig. \ref{plotRSD} are somehow absorbed by the $A_{\rm noise}$ parameter,  and  the bias parameters are  relatively insensitive to these differences. The reason why these two simulations present different shot noise is unclear, but it may be related to the definition of halo, which varies from \textsc{PThalos} to N-body haloes.
 This issue should not concern us here, as we will  treat  $A_{\rm noise}$ as a nuisance parameter and marginalise over it. Moreover, we use the mocks to estimate error-bars not to model the signal directly.

We conclude that using \textsc{PThalos}  rather than N-body haloes for the mock survey catalogs does not introduce significant systematic biases in the determination of the $b_1$ and $b_2$ parameters at $k_{\rm max}<0.20\,h{\rm Mpc}^{-1}$. Smaller scales may introduce systematic errors, especially for the second-order bias, $b_2$. We also detect a small systematic in the estimation of $b_1$ between real and redshift spaces, which may arise from the halo bias and the modelling of redshift space distortions (see \citealt{HGMetal:inprep} for further discussion). Since this systematic is smaller than the statistical errors of this survey, we do not consider to correct for this effect.

The possible  bias introduced on the growth parameter $f$ is investigated in \S~\ref{sec:systematics}.

\subsection{Test of the effect of the survey geometry on dark matter haloes}\label{5_3_section}
In this section we test how the survey geometry, or mask, affects the power spectrum, and, more importantly, the bispectrum, and the performance of our approximations. In \S~\ref{sec:bispectrummethod} we saw how the fiducial statistics are related to the measured statistics through a convolution with the window mask (see Eq.~\ref{power_spectrum_FKP} and~\ref{Bispectrum_FKP}). In order to explore the effect of the mask we use the 50 realisations of \textsc{PThalos} used in \S~\ref{5_2_section}. These realisations are contained in a box with a constant mean density. Hereafter we will refer to them as {\it unmasked} \textsc{PThalos} realisations. On the other hand, we also have 50 realisations of \textsc{PThalos} with the the northern DR10 survey geometry. We will refer them as the {\it masked} \textsc{PThalos} realisations. By, computing the power spectrum and bispectrum for these two different sets of 50 \textsc{PThalos} realisations we can directly quantify the effect of the survey geometry.

\begin{figure*}
\centering
\includegraphics[clip=false, trim= 0mm 15mm 0mm 00mm,scale=0.30]{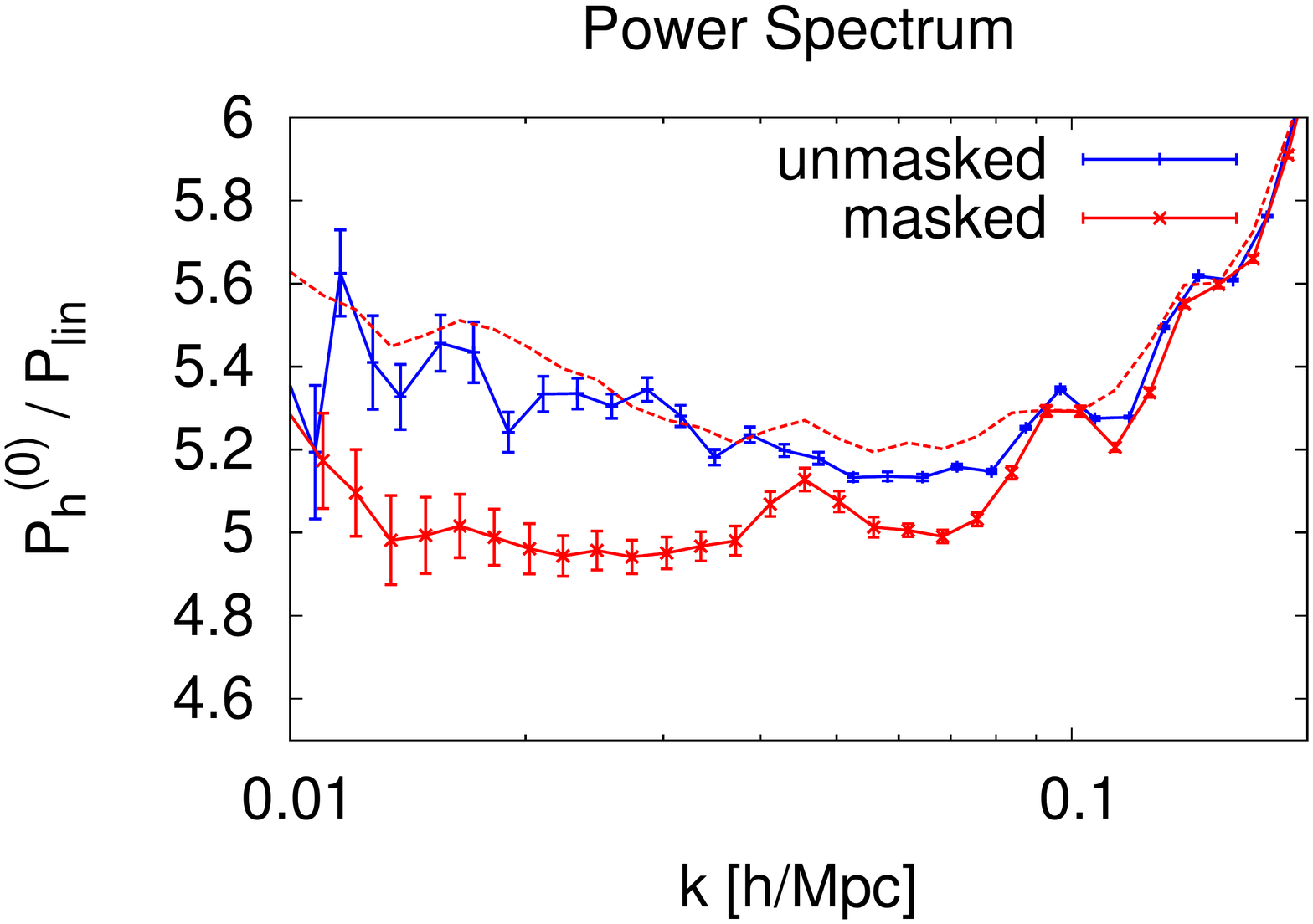}
\includegraphics[clip=false, trim= 0mm 15mm 0mm 00mm,scale=0.30]{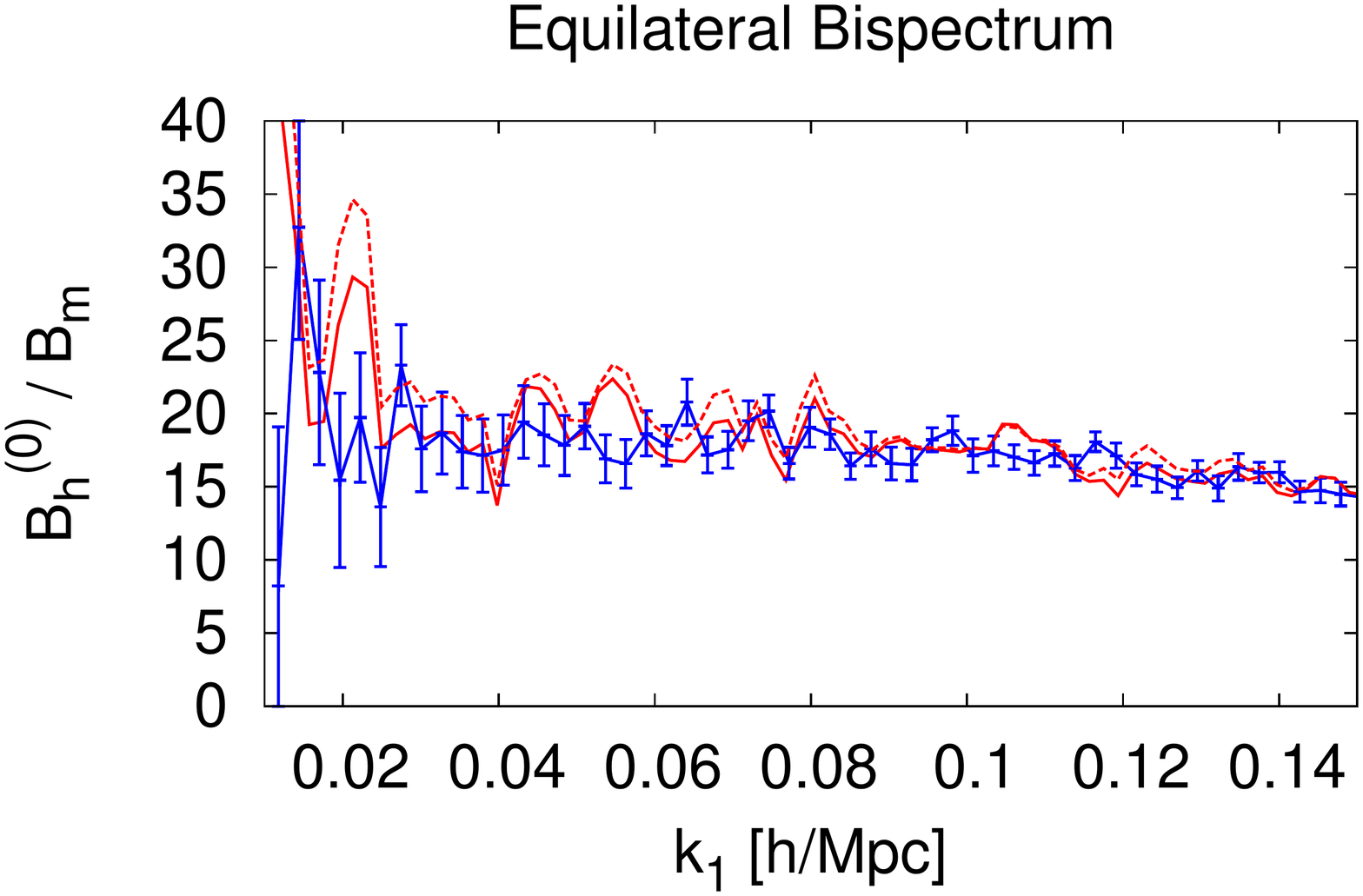}

\includegraphics[clip=false, trim= 0mm 10mm 0mm 30mm,scale=0.30]{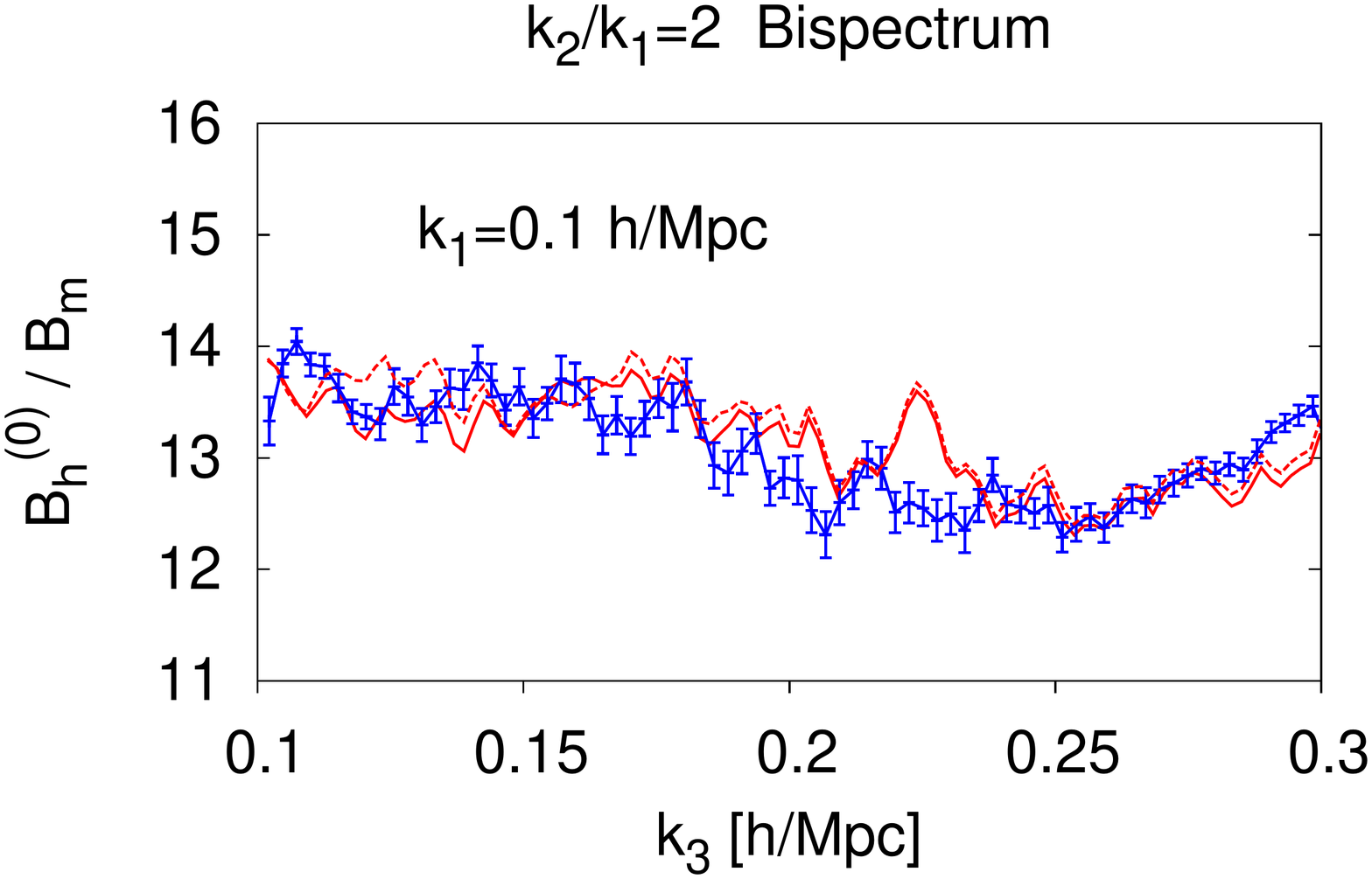}
\includegraphics[clip=false, trim= 0mm 10mm 0mm 30mm,scale=0.30]{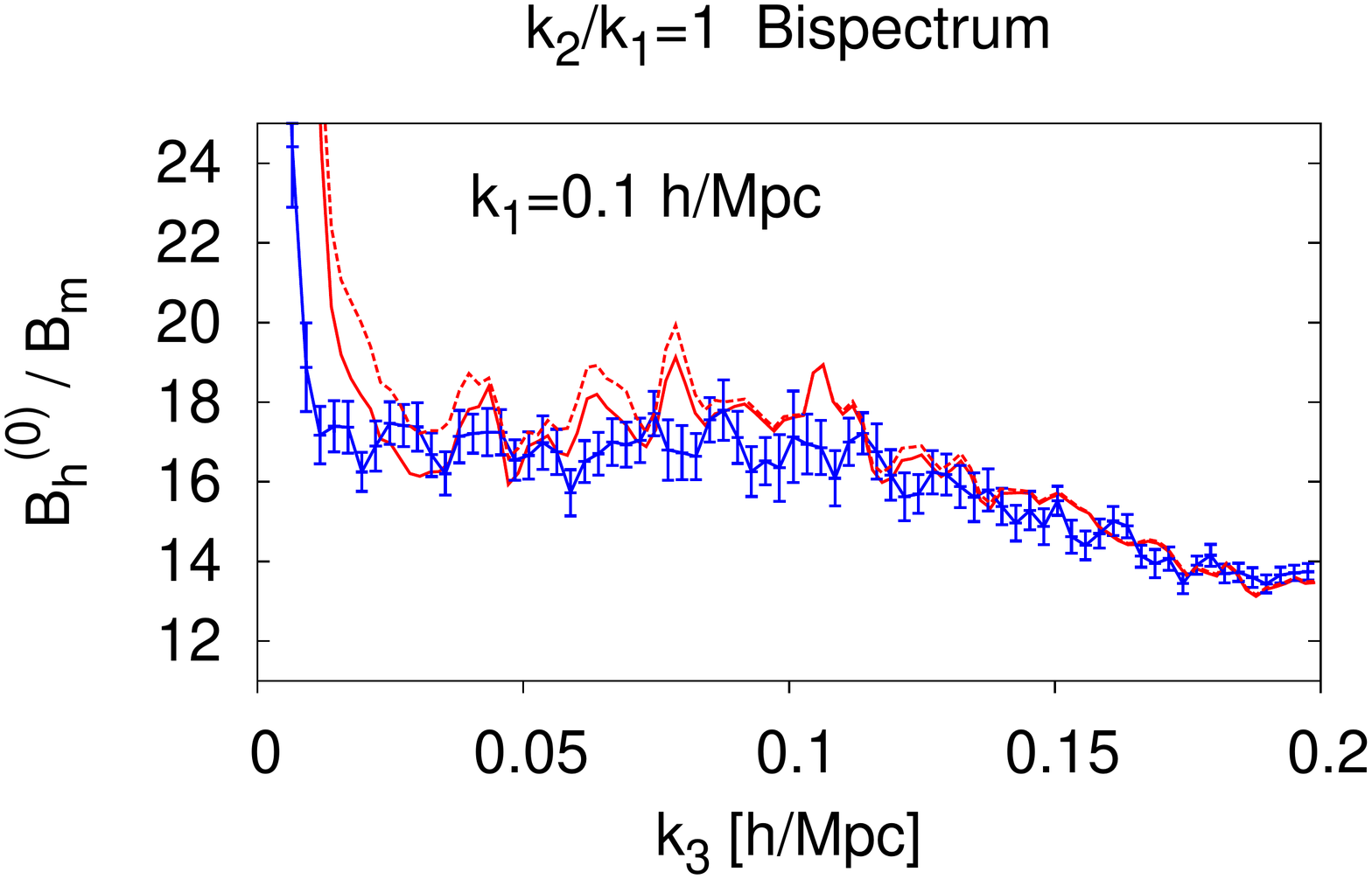}

\caption{Power spectra (top left panel) and bispectra (other panels) for \textsc{PThalos} in redshift space. The red (blue) solid lines are the measurements of the power spectrum and bispectrum from the masked (unmasked)  \textsc{PThalos} normalised by their linear power spectrum and matter bispectrum, respectively: $\langle F_2^2\rangle/P^{\rm lin}$ and $\langle F_3^3\rangle/B_{\rm matter}$. The red dashed lines are the measurement of power spectrum and bispectrum from the masked \textsc{PThalos} normalised by the convolution of the linear power spectrum and real space matter bispectrum, respectively, as it described in the right hand side of Eq.~\ref{power_spectrum_FKP} and the approximation described by Eq.~\ref{bispectrum_approximation}: $\langle F^2_2\rangle/(P^{\rm lin}\otimes W_2)$ and $\langle F^3_3\rangle/(B_{\rm matter}\otimes W_3)$.  Poisson noise is assumed. The effect of the mask is accurately modelled by the FKP-estimator described in \S~\ref{sec:estimator_ps} and \S~\ref{sec:bispectrumestimator}.}
\label{plotmask}
\end{figure*}

For the power spectrum, the effect of the survey geometry is described by Eq.~\ref{power_spectrum_FKP}, which is an exact relation between the fiducial power spectrum, $P_{\rm gal}$, and the measured one, $\langle F_2^2\rangle$. The top left panel of Fig.~\ref{plotmask} presents the redshift space power spectrum monopole from 50 unmasked realisations (blue symbols) and from the masked ones (red symbols). Both power spectra have been normalised by the linear power spectrum for clarity, therefore the plotted quantity is  the square of an effective bias parameter. Differences  are stronger  at large scales and unimportant at small scales: this result is expected, as discussed in  \S~\ref{sec:bispectrummethod}, where we argue that  the effect of the survey mask becomes  negligible at small scales. 
 
 To test the performance of the convolution described in Eq.~\ref{power_spectrum_FKP}, we divide the measured monopole power spectrum from the masked realisations, namely $\langle F_2^2\rangle$, by the linear power spectrum convolved with the window, as is described in the right hand side of Eq.~\ref{power_spectrum_FKP}.  This calculation is shown by  the dashed red line. The original difference between the masked and unmasked power spectra is now corrected. The different lines of Fig. \ref{plotmask} are summarised as follows.

\begin{itemize}
\item $\langle F^2_2\rangle/P^{\rm lin}$, where $\langle F^2_2\rangle$ is computed from the unmasked sample (blue solid lines).
\item  $\langle F^2_2\rangle/P^{\rm lin}$, where $\langle F^2_2\rangle$ is computed from the masked sample (red solid lines).
\item   $\langle F^2_2\rangle/(P^{\rm lin}\otimes W_2)$, where $\langle F^2_2\rangle$ is computed from the masked sample and $P^{\rm lin}\otimes W_2$ is the convolution of $P^{\rm lin}$ with the survey window according to Eq. \ref{combolucion_window}  (red dashed lines).
\end{itemize}

For the bispectrum, the effect of the mask is fully described by the Eq.~\ref{Bispectrum_FKP}. However, this equation involves a double convolution between the mask and the theoretical bispectrum formula. Since this calculation   is computationally too expensive to be viable in practice, we have introduced the approximation described by Eq.~\ref{bispectrum_approximation}, which splits the double convolution into two simple ones, i.e., the complexity of this computation is reduced to the same complexity used for the power spectrum. The remainder of the panels of Fig.~\ref{plotmask} display the redshift space bispectrum monopole measurement, $\langle F_3^3\rangle$ for the unmasked \textsc{PThalos} catalogue (blue lines) and for the masked dataset (red solid lines)\footnote{The errors on the masked measurements are not shown for clarity.}. Both the unmasked and masked bispectrum monopole are normalised by the real space matter prediction. The red dashed lines represent the masked bispectrum monopole normalised by the real space matter prediction convolved with the mask according the approximation described by Eq.~\ref{bispectrum_approximation}. The different cases can be summarised as,

\begin{itemize}
\item $\langle F^3_3\rangle/B_{\rm matter}$, where $\langle F^3_3\rangle$ is computed from the unmasked sample and $B_{\rm matter}$ is the tree-level matter bispectrum without any window effect (blue solid lines).
\item  $\langle F^3_3\rangle/B_{\rm matter}$, where $\langle F^3_3\rangle$ is computed from the masked sample and $B_{\rm matter}$ is the tree-level matter bispectrum without any window effect  (red solid lines).
\item   $\langle F^3_3\rangle/(B_{\rm matter}\otimes W_3)$, where $\langle F^3_3\rangle$ is computed from the masked sample and $B_{\rm matter}\otimes W_3$ is the convolution of $B_{\rm matter}$ with the survey window according to the approximation described in the second line of Eq. \ref{bispectrum_approximation}  (red dashed lines).
\end{itemize}

The difference between the the dark matter bispectrum and its convolution according to Eq.~\ref{bispectrum_approximation} are  small (red solid and dashed lines are  similar).   For the power spectrum, the effect of the mask is a clear broadband  suppression  of $\sim5\%$ level at scales of $k\sim0.03\,h{\rm Mpc}^{-1}$ (and even higher at larger scales) and therefore include the standard mask calculation where calculating models. For the bispectrum,  the effect of the mask is an enhancement of the bispectrum signal at $k_3\lesssim0.03\,h{\rm Mpc}^{-1}$.  At smaller scales the differences are always below $\sim10\%$ and we do not observe any clear systematic trend generated by the effect of the mask. However, we have checked that not including the mask in the bispectrum model (through the approximation described in Eq.~\ref{bispectrum_approximation}) leads to a systematic error in the estimation of the linear and nonlinear bias parameters by 1-2\%. Therefore, in this paper we will account the effect of the mask by correcting the bispectrum model using the approximation described in Eq.~\ref{bispectrum_approximation}.   In any case, since the bispectrum measurement presents a considerable scatter due to sample variance limitations (both for masked and unmasked) it is difficult to quantify exactly the accuracy of the approximation below $\sim10\%$. 

For most of  the shapes and scales of the bispectra compared here,  the differences between masked and unmasked are at the few percent level. However,  for very squeezed triangles, $k_3\lesssim k_1= k_2$, the bispectrum for masked \textsc{PThalos} over-predicts  the unmasked  one, even when the approximation of the mask correction is applied (Eq.~\ref{bispectrum_approximation}). We have determined that  this is a large-scale effect; for $k_i\gtrsim0.03\,h{\rm Mpc}^{-1}$, the masked and unmasked \textsc{PThalos} bispectrum agree, and the only discrepancies occur at large scales. Thus, in order to avoid spurious effects,  in this paper we only consider $k$-modes larger than $0.03\,h{\rm Mpc}^{-1}$ when estimating the bispectrum.

We conclude that the approximation of Eq.~\ref{bispectrum_approximation} introduces  a completely negligible  systematic error for  $k_i\gtrsim0.03\,h{\rm Mpc}^{-1}$: thus 
the effect of the mask can accurately described by Eq.~\ref{power_spectrum_FKP} and~\ref{bispectrum_approximation}.

\begin{figure*}
\centering
\includegraphics[scale=0.3]{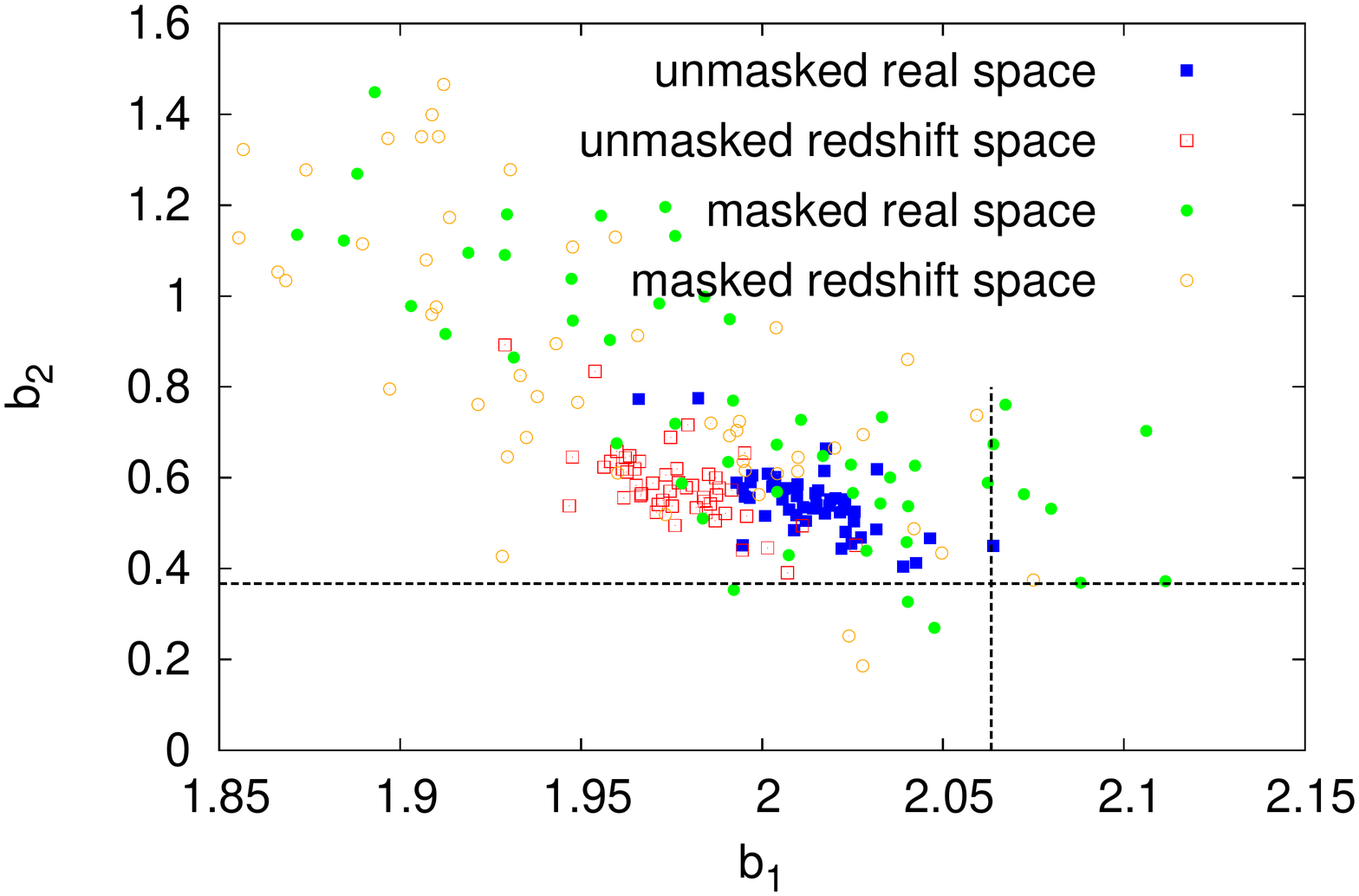}
\includegraphics[scale=0.3]{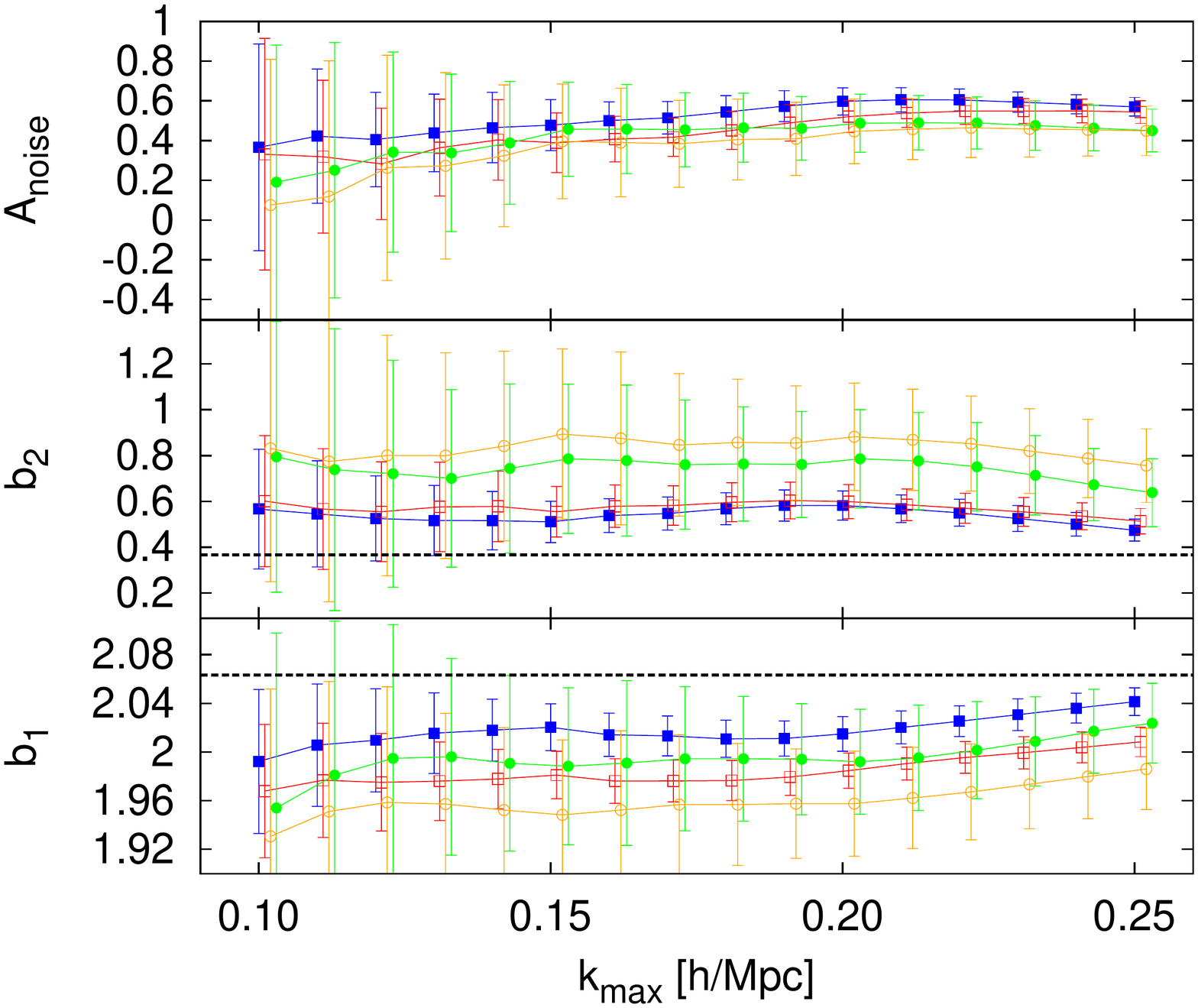}
\caption{{ Left Panel:} Best fit bias parameters for \textsc{PThalos} estimated from masked and unmasked realisations, from the real and redshift space monopole bispectra. Green (blue) symbols are the best fit values from real space bispectrum masked (unmasked) realisations. Red (orange) symbols  are best fit values from redshift space monopole bispectrum masked (unmasked) realisations using $k_{\rm max}=0.17\,h{\rm Mpc}^{-1}$. Right Panel: Best fit bias parameters and $A_{\rm noise}$ as a function of $k_{\rm max}$ using same colour notation that in the left panel. Error-bars correspond to the 1-$\sigma$ dispersion among the different realisations.   In both panels, black dashed lines represent the measured cross bias parameters defined in Eq. \ref{b1_cross}-\ref{b2_cross}. The observed differences between the masked and unmasked catalogues are significantly smaller than $1\sigma$ of the typical statistical errors obtained for the   CMASS galaxy survey. }
\label{biasmask}
\end{figure*}

In order to test the performance of the approximation of Eq.~\ref{bispectrum_approximation} in describing the mask,  we estimate $b_1$ and $b_2$ for the masked and unmasked \textsc{PThalos} using the bispectrum triangles with $k_2/k_1=1$ and $2$. As before, we follow the method of \S~\ref{sec:bispectrummethod} using the same model that in \S~\ref{5_2_section}. We set the cosmological parameters to their fiducial values and  set $A_{\rm noise}$ to be a free parameter in the fitting process. We adopt $k_{\rm min}$ to $0.03\,h{\rm Mpc}^{-1}$ to avoid the large scale mask effects that cannot be accounted by our approximation. The left panel of Fig.~\ref{biasmask} presents a  similar information to the one shown in Fig.~\ref{plotbias1} for $k_{\rm max}=0.17\,h{\rm Mpc}^{-1}$. In this case, blue (green) points refer to the best fit values $b_1$ and $b_2$ computed from the real space bispectrum monopole of unmasked (masked) \textsc{PThalos}, whereas red (orange) points are computed from the redshift space monopole bispectrum of unmasked (masked) \textsc{PThalos}. In black dashed lines the values of $b_1^{\rm cross}$ and $b_2^{\rm cross}$ measured according to Eq. \ref{b1_cross}-\ref{b2_cross} are shown. In both real and redshift space  the effect of the mask is to enhance the scatter. This effect is due to the differences in effective volumes between the masked and unmasked catalogues. Recalling that the masked catalogues have been generated from the unmasked ones by masking off haloes in order to match both the angular and the radial mask. The effective volume of the masked sample can be defined as \citep{Tegmark97}, 
\begin{equation}
V_{\rm mask}^{\rm eff}(k)\equiv\int \frac{\left[\bar{n}({\bf r})P(k)\right]^2}{\left[1+\bar{n}({\bf r})P(k)\right]^2}\,d^3{\bf r}.
\end{equation}
At $k=0.17\,h/{\rm Mpc}$, the amplitude of the power spectrum is about $8000\,[{\rm Mpc}/h]^3$ and the effective volume of the masked sample about $4.66\times10^8\,[{\rm Mpc}/h]^3$. For the unmasked sample, the volume is at any $k$,  $V_{\rm unmask}=2400^3\, [{\rm Mpc}/h]^3$
The effective volume has been reduced by $V^{\rm eff}_{\rm mask}/V_{\rm unmask}\simeq0.033$ at scales of $k\sim0.17\,h{\rm Mpc}^{-1}$; thus we expect that at these scales the $1\sigma$ dispersion is $\sqrt{V_{\rm unmask}/V^{\rm eff}_{\rm mask}}\simeq 5.4$ higher.
The right panel of Fig.~\ref{biasmask} displays the best-fit values for $b_1$, $b_2$ and $A_{\rm noise}$ as a function of $k_{\rm max}$.

In summary,  the recovered 
 $b_1$ tends to be smaller in the {\it masked} realizations than in the {\it unmasked} one, although the differences are smaller than the statistical errors. We observe these differences both in real and in redshift space, so they may be due to some residual effect of the mask.   We quantify these shifts to be  about $\sim1\%$ for  $b_1$, which represents a $\sim40\%$ shift of $1\sigma$ of the masked realizations. The effect of the mask is more important for  $b_2$: the {\it masked} realizations predict a $\sim0.2$ higher $b_2$ ($\sim30\%$)  than the {\it unmasked} realizations, which in this case represent $\sim80\%$ shift of the $1\sigma$ of the masked realizations. These differences are  within $1\sigma$ of the statistical errors.  In particular, this $+0.2$ shift for $b_2$ tends to {\it cancel}  the $-0.2$ shift  seen in \S~\ref{5_2_section} and \ref{sec:real_redshift}. Moreover,  in this paper we treat $b_2$ as a nuisance parameter that can absorb other systematic effects, such as the effect of truncation. We  therefore advocate not correcting the $b_2$ recovered values for a systematic shift.   The differences between the estimated bias parameters and the cross-bias parameters from Eq. 33-34 are similar  and fully consistent with the ones reported in \S\ref{5_2_section}.

Bear in mind that all the statistical $\sigma$-values reported in \S\ref{section:systematics} correspond to the marginal error distribution respect to $b_1$, $b_2$ and $A_{\rm noise}$; where $f$, $\sigma_8$, and $\sigma_B$ have been set to their fiducial values. When we have analysed the data in \S\ref{sec:results}, all the reported errors were marginalized with respect to all the parameters, i.e. $\{b_1,\, b_2,\, ,\sigma_8,\, f,\, A_{\rm noise},\,\sigma_{\rm FoG}^P,\,\sigma_{\rm FoG}^B\}$. Therefore, the statistical error values reported for the data in \S\ref{sec:results} are larger than the statistical errors reported in \S\ref{section:systematics}.

\subsection{Test: Is the measurement consistent across shapes?}

In this section we test how the choice of different triangle shapes affects the estimation of the bias parameters from the bispectrum. In the ideal case, we should always obtain the same bias parameters, whatever shapes are chosen. However, the bispectrum model may present different systematic errors that can vary from shape to shape as the  {\it anzatz} for effective the kernel  was set {\it a priori} and then the kernel was calibrated to reduce the  average  differences from the simulations. Moreover, the maximum $k$ at which the model is accurate might depend on the shape chosen. 

The main point of this sub-section is, therefore, to check whether the measurements of the bias parameters are consistent across shapes. Thus, the idea is to test  effects one by one, isolating each from all the other as much as possible in order to gain insight into each of the presented tests.  We tried to isolate this question from other factors such as survey effects or redshift space distortions. Therefore we think that  a  simple and clean way to approach this question is using unmasked boxes because they have larger volume and therefore  is easier to detect potential systematics. Since the effect of the mask is tested elsewhere, we  prefer not to re-introduce it  here. We could have done this test in redshift space. However, redshift space modelling adds and extra degree of complexness, which is addressed and discussed  (separately) later in \S\ref{sec:real_redshift}.

Here we consider separately the performance of the two shapes adopted: $k_2/k_1=1$ and $k_2/k_1=2$.
As we have said, for simplicity, we stay in real space and we use the unmasked realisations.  As the shot noise should not vary with the triangle shape, we assume that the shot noise is given by Poisson statistics. Any variation form the Poisson prediction will be the same for all triangles and we are only concerned with relative changes. The theoretical model is given by Eq.~\ref{B_ggg2}, and the cosmological parameters are set to their fiducial values. To estimate the bias parameters we use the bispectrum  applying the method described in \S~\ref{section_method}, as in \S~\ref{5_2_section} and \S~\ref{5_3_section}. We use the  (unmasked) \textsc{PThalos} realisations as this also tests the performance of the adopted bias model. As discussed in \S~\ref{sec:biasmethod}, this approach is a truncation of an expansion of the complex relationship between $\delta_m$ and $\delta_h$, and will have a limited regime of validity. 

\begin{figure*}
\centering
\includegraphics[scale=0.3]{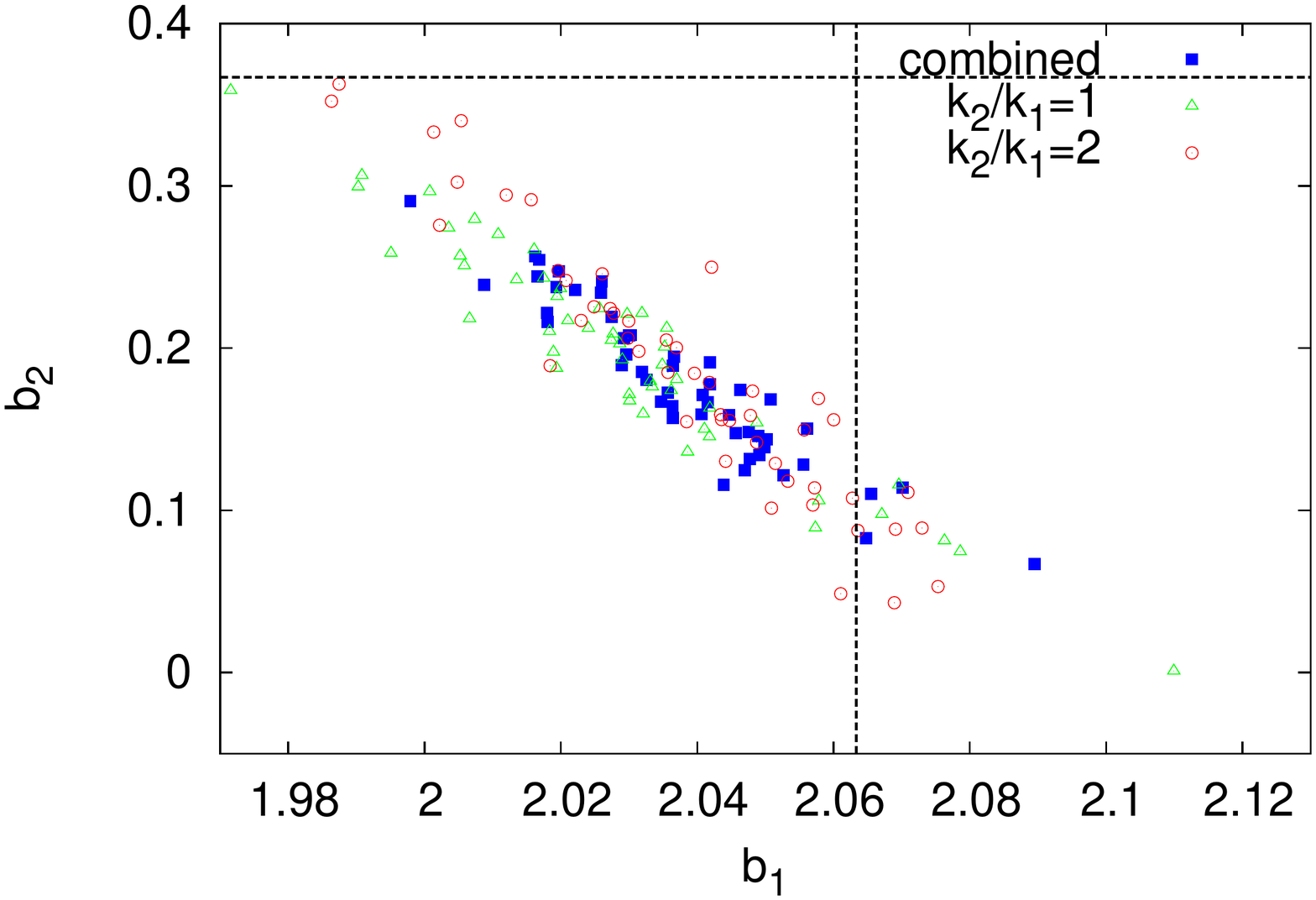}
\includegraphics[scale=0.3]{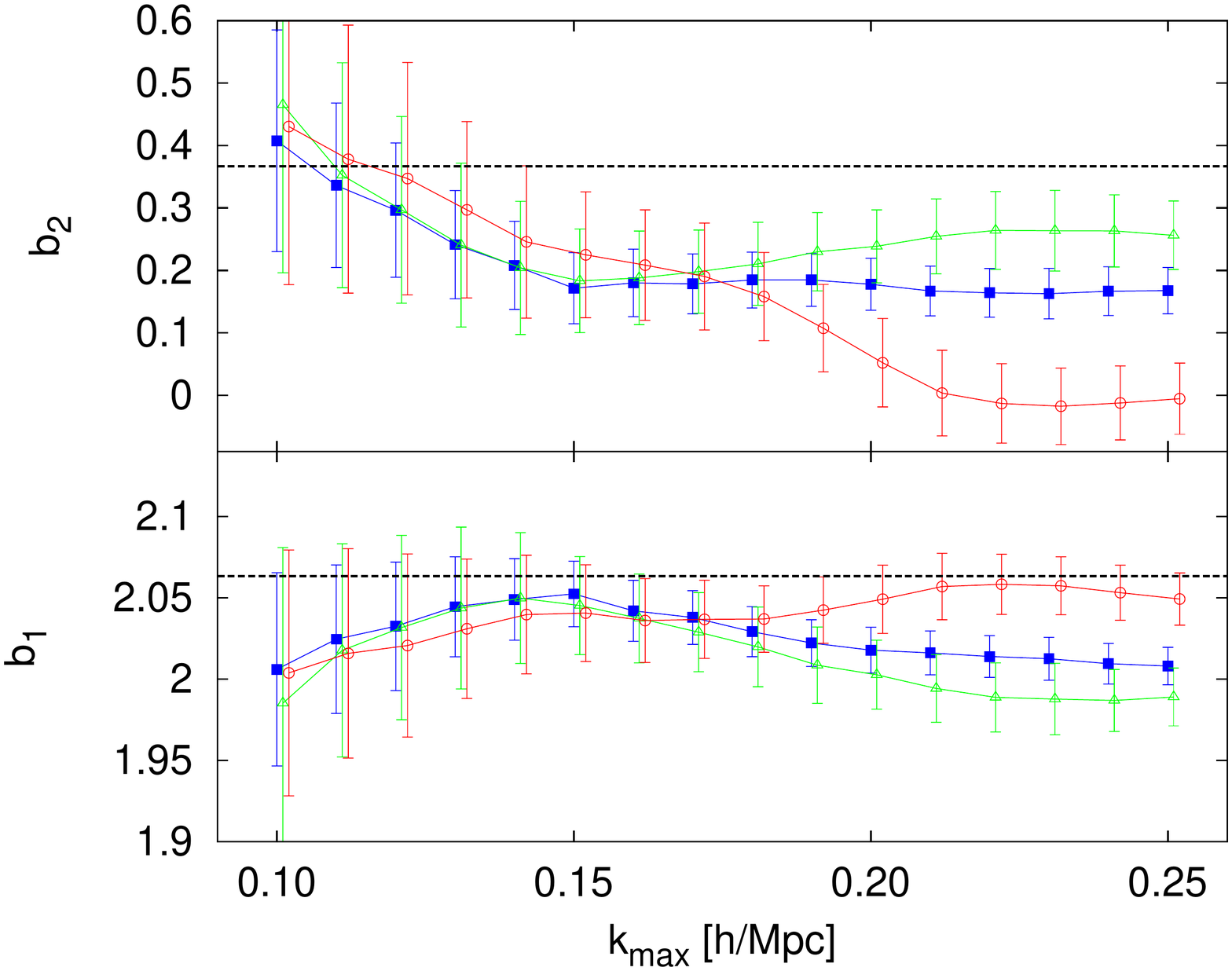}
\caption{{ Left panel:} Best fit bias parameter for \textsc{PThalos} from the real space bispectrum using different triangular shapes: $k_2/k_1=1$ (green points), $k_2/k_1=2$ (red points), and a combination of both (blue points), where $k_{\rm max}=0.17\,h{\rm Mpc}^{-1}$. { Right panel:} Best fit bias parameters as a function of $k_{\rm max}$. Same colour notation in both panels. There is no significant shape dependence on the bias parameters for $k_{\rm max}\leq0.17\,h{\rm Mpc}^{-1}$. In both panels, black dashed lines represent the measured cross bias parameters defined in Eq. \ref{b1_cross}-\ref{b2_cross}.}
\label{plotshape}
\end{figure*}

The left panel of Fig.~\ref{plotshape} presents the best fit  $b_1$ and $b_2$ parameters from the (unmasked) \textsc{PThalos} realisations. The red points show best fit parameters estimated from the bispectrum using the $k_2/k_1=1$ shape; the green points from $k_2/k_1=2$ shape; and the blue points both shapes combined. In this figure the maximum $k$ is set to $0.17\,h{\rm Mpc}^{-1}$. The right panel displays the best fit parameters as a function of $k_{\rm max}$ with the same colour notation in both panels. The errors are the 1$\sigma$ dispersion among the 50 \textsc{PThalos} realisations.  Black dashed lines show the measured cross-bias parameters as defined in Eq. \ref{b1_cross}-\ref{b2_cross}.

For $k_i\leq0.18\,h{\rm Mpc}^{-1}$, both shapes predict the same bias parameters. For $k>0.18\,h{\rm Mpc}^{-1}$ the $k_2/k_1=2$ shape tends to over-predict $b_1$ and under-predict $b_2$ with respect to the $k_2/k_1=1$ shape,  for which the  inferred  parameters  do not change significantly. In order to understand the behaviour of the $k_2/k_1=2$ triangles, one must recall that this shape is always limited by $k_1\leq0.1\,h{\rm Mpc}^{-1}$ and therefore by $k_2\leq0.2\, h{\rm Mpc}^{-1}$. So in the range $0.2\leq k\, [h{\rm Mpc}^{-1}]\leq 0.3$, this shape  only adds new scales through $k_3$, for those triangles with $k_1\simeq0.1\,h{\rm Mpc}^{-1}$. The decrease in recovered $b_2$ with $k_{\rm max}$ in Fig.~\ref{plotshape}, which matches the trend seen in the full fits, suggests that such triangles are responsible of misestimating the bias parameters at these scales. On  larger scales, the effect of these triangles is suppressed by other shapes, which also satisfy $k_2/k_1=2$. In fact, when we add both $k_2/k_1=1$ and 2 shapes, the bias parameters at the scales  $0.2\leq k\, [h{\rm Mpc}^{-1}]\leq 0.3$ have a consistent behaviour with larger scales. This analysis confirms two features: i) the responsibility for misestimating the bias parameters lies with the folded triangles with $k_1\simeq k_3 \simeq k_2/2$, and ii) the effect of these triangles is mitigated by including  other shapes.

 Comparing the real and redshift space measurements later in \S\ref{sec:real_redshift} we find no systematic offset for $b_1$. Since there are no systematics between real and redshift space for $b_1$ and there are no systematic across shapes in real space, it is reasonable to assume that there are not systematics between shapes in redshift space neither.

We conclude that for $k\leq0.18\, h{\rm Mpc}^{-1}$, the best fit bias parameters are robust to the choice of the bispectrum shape (at least in real space and for haloes). For smaller scales,  the behaviour of the  $k_2/k_1=2$ triangles is responsible for underestimating $b_2$.

 We observe that $b_1^{\rm cross}$ agrees better with the obtained $b_1$ from $B_{\rm hhh}$ than it does in Fig. \ref{plotbias1} and \ref{biasmask}. On the other hand, $b_2$ is underestimated respect to $b_2^{\rm cross}$. These differences are because in this section we have set $A_{\rm noise}$ to 0 for simplicity. However, as we reported in \S\ref{5_2_section}, when $P_{\rm hh}$ is added to the analysis and $A_{\rm noise}$ is set free, we are able to recover $b_1$ with almost no bias, and $b_2$ with a $\sim30\%$ bias.

\subsection{Tests on galaxy mocks.}\label{test_mocks}
In this section we perform a series of tests on the galaxy mocks used to estimate the errors of the data in \S~\ref{sec:results}. Since some tests have already been performed for the \textsc{PThalos} boxes they are not repeated for the mocks. By using mocks we include many real-world effects present in the survey and  test the performance of the adopted bias model, which was derived for haloes and not galaxies. In particular, we focus on three tests  for aspects that can produce the systematic errors. First, we check the consistency of the bias parameters estimated from the power spectrum and bispectrum. An inconsistency would indicate that  the bias model adopted  cannot describe the clustering of galaxies. 
Second, we check the effect of redshift space distortions on estimating the bias parameters when we combine the power spectrum and bispectrum. Finally, we investigate the possible systematic errors produced when we estimate the growth factor simultaneously  as the bias parameters and $\sigma_8$. In order to estimate the best fit parameters, for both power spectrum and bispectrum, we  use the same method applied to the data and  described in \S~\ref{section_method}. 
For the power spectrum we  use  Eq.~\ref{PS_real} for real space and~\ref{Psg} for redshift space, where the non-linear power spectrum terms $P_{\delta\delta}$, $P_{\delta\theta}$ and $P_{\theta\theta}$ are described by 2L-RPT (Eq.~\ref{Pab}). The bispectrum is given by Eq.~\ref{B_ggg2} (real space) and~\ref{B_gggs} (redshift space).
The {\it rms} scatter among the mocks provides our estimate of the 1-$\sigma$ uncertainty for the survey measurements. 

\subsubsection{Bias parameters from power spectrum \& bispectrum}\label{PS_BS_section}
We start by analyzing the power spectrum and bispectrum in redshift space for the CMASS DR11 NGC galaxy mocks. These mocks contain the same observational effects as the data, so for extracting the statistical moments we  use the FKP estimator as described in \S~\ref{sec:bispectrummethod}. We weight the galaxies according to the systematic weights described in \S~\ref{section_data}. The effect of the weights  on the shot noise term is described in Appendix~\ref{appendix_shot_noise}.

Our goal is to extract the bias parameters from different statistics and to check their consistency. Since we are considering galaxy clustering in redshift space, we expect a non-linear damping term due to the Fingers-of-God effect of the satellite galaxies inside the haloes. In total, the list of free parameters to be fitted: $b_1$, $b_2$, $A_{\rm noise}$, $\sigma_{\rm FoG}^P$ and $\sigma_{\rm FoG}^B$. In this section we set the cosmological parameters $f$ and $\sigma_8$ to their fiducial value, as well as fixing the shape of the linear matter power spectrum.

\begin{figure*}
\centering
\includegraphics[scale=0.3]{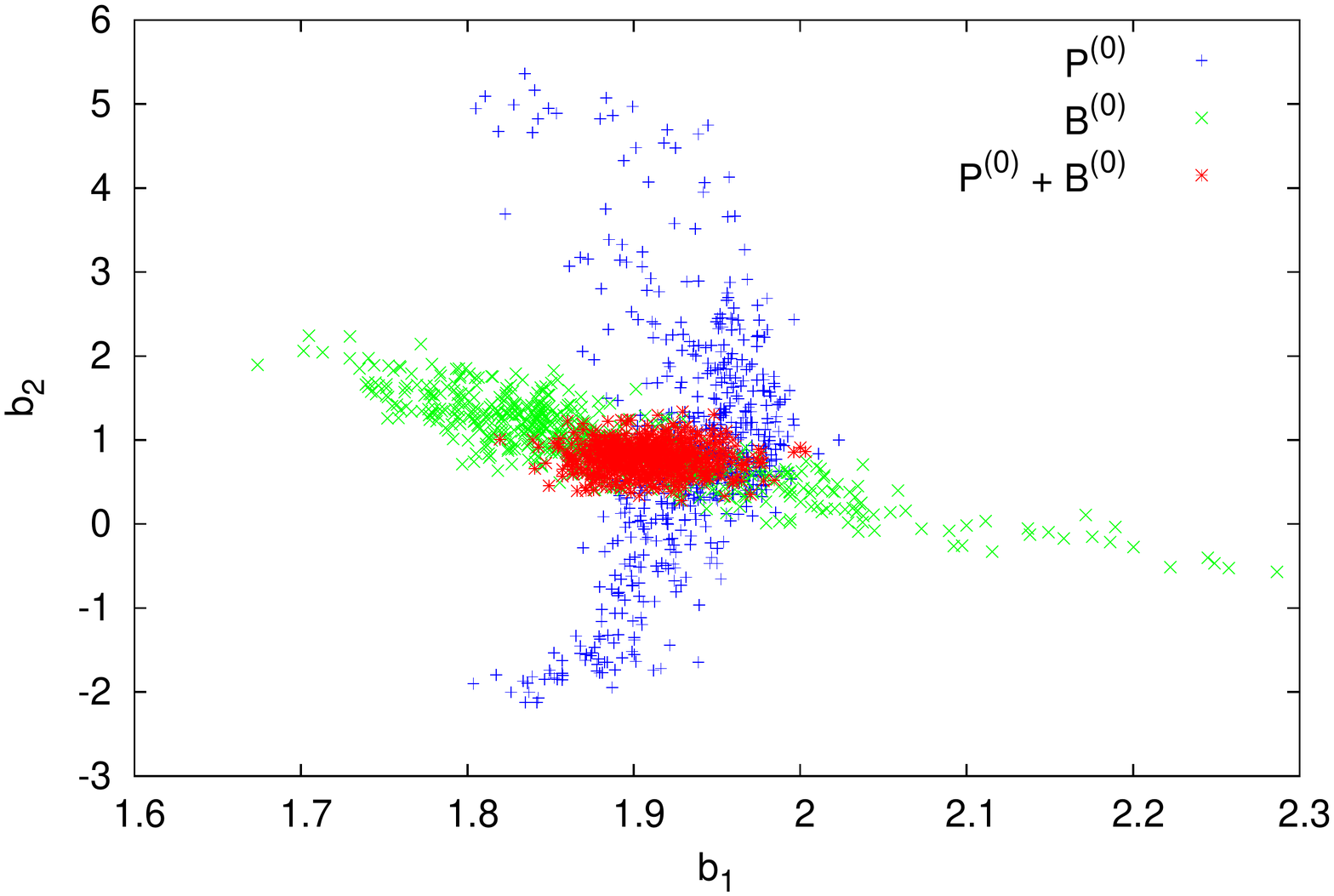}
\includegraphics[scale=0.3]{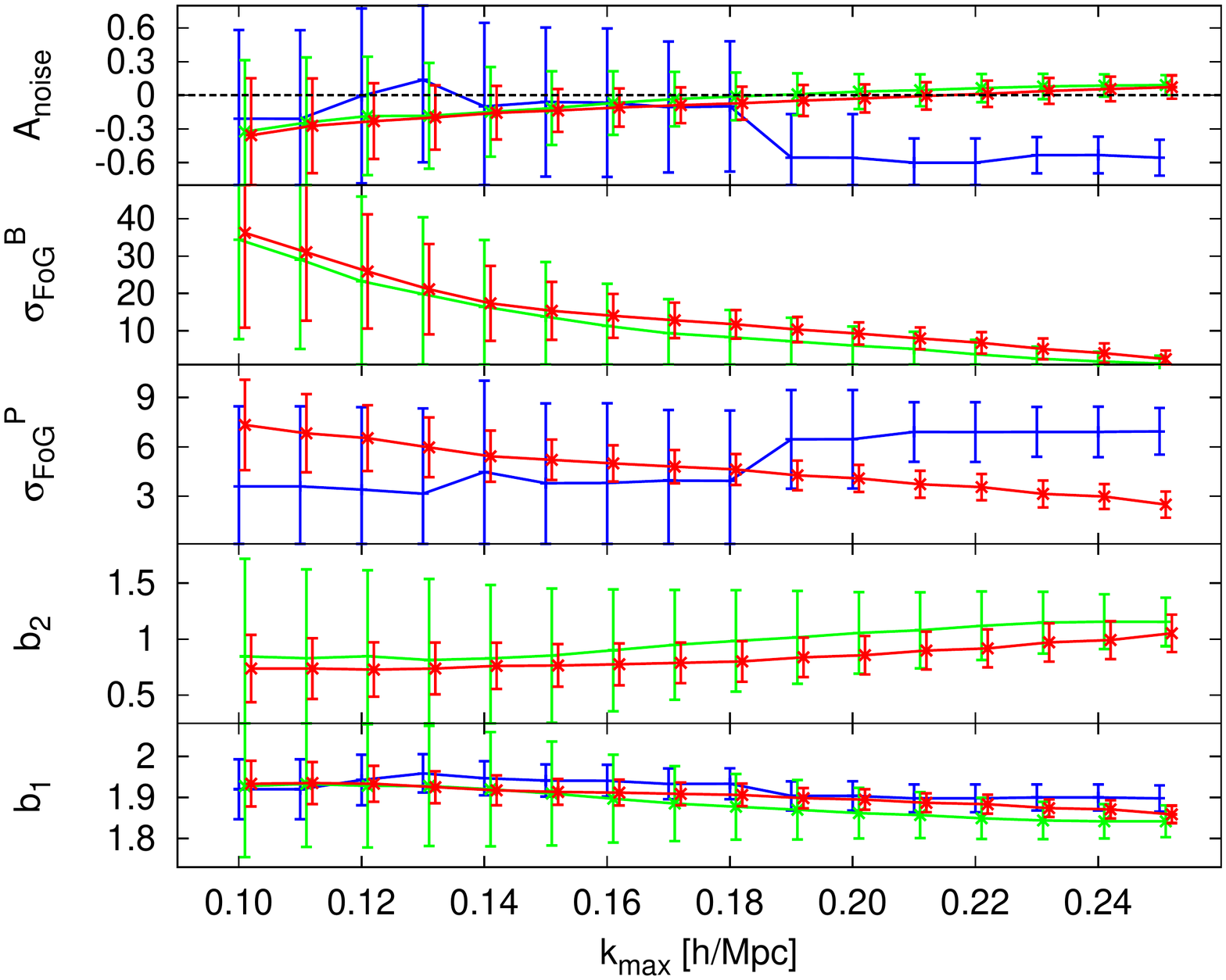}
\caption{{ Left panel}: Best fit $b_1$, $b_2$ and $A_{\rm noise}$ parameters for the galaxy mocks in redshift space, when the power spectrum monopole is used (blue points), when the bispectrum monopole is used (green points), and when both statistics are combined (red points). The quantities  $\sigma_{\rm FoG}^P$ and $\sigma_{\rm FoG}^B$ are varied but are  not shown for clarity. The maximum $k$ used for this fitting is $0.17\,h{\rm Mpc}^{-1}$. { Right panel}: Best fit parameters as a function of $k_{\rm max}$. The error-bars are the 1$\sigma$ dispersion for a single realisation. There is a good agreement in the bias parameters, $b_1$ and $b_2$, estimated form the power spectrum and bispectrum.}
\label{PS_BS_plot}
\end{figure*}

The left panel of Fig.~\ref{PS_BS_plot} presents the scatter of the 600 best fit values for the galaxy mocks with the CMASS DR11 NGC survey mask. The blue points are the constraints from the power spectrum monopole,  green points from the bispectrum monopole, and red points the combination of both statistics. The $k_{\rm max}$ used is $0.17\,h{\rm Mpc}^{-1}$.

When using only one statistic there are large degeneracies between parameters. In particular, for the power spectrum monopole, $b_2$ is poorly constrained  as it is highly degenerate with $A_{\rm noise}$ and $\sigma_{\rm FoG}^P$, whereas $b_1$ is relatively well constrained. Indeed $b_2$ only affects the  power spectrum amplitude at mildly non-linear scales, which is precisely  where the shot noise term  and  $\sigma_{\rm FoG}^P$ start to be relevant. On the other hand, the amplitude of the clustering at large scales is solely determined by $b_1$.

The constraints placed by the bispectrum on the bias parameters show a strong degeneracy between $b_1$ and $b_2$, and are consistent with the power spectrum predictions. The bispectrum  constrains $A_{\rm noise}$ much better than the power spectrum for two reasons, {\it i)} the shot noise is more important compared to the signal for the bispectrum and {\it ii)}  the shape dependence of this parameter is different from that of e.g., the  bias  parameters.
The strong degeneracy between $b_1$ and $b_2$ is well known;  at leading order in perturbation theory  for a power law power spectrum every shape can only constrain a linear combination of $b_1$ and $b_2$. The linear combination has a weak shape dependence, which is why   combining different shapes both parameters can be measured.

The right panel of Fig.~\ref{PS_BS_plot}, shows how the mean value of the best fit parameters estimated from the different statistics evolve with the variation of $k_{\rm max}$. The error-bars correspond to the $1\sigma$ dispersion among the different realisations.

For $k_{\rm max}\lesssim0.17\,h{\rm Mpc}^{-1}$, the bias parameters do not present a strong trend with the maximum scale used and the estimates obtained from power spectrum and bispectrum  agree.
However, as probe smaller scales,  there is a small tension for the best fit value of $b_1$ between the power spectrum and bispectrum predictions. For the noise parameter, $A_{\rm noise}$, there is a suggestion that, as we increase $k_{\rm max}$, $A_{\rm noise}$ moves from slightly super-Poisson values ($A_{\rm noise}<0$) to slightly sub-Poisson values ($A_{\rm noise}>0$). We do expect this parameter to change with the scale, due to the different clustering at different scales.

We also observe that the two FoG parameters, $\sigma_{\rm FoG}^P$ and $\sigma_{\rm FoG}^B$, clearly decrease with $k_{\rm max}$. These parameters aim to parametrise the internal dispersion of galaxies inside haloes, consistent with setting the constraints $\sigma_{\rm fog}^P>0$ and $\sigma_{\rm fog}^B>0$, and there  being low signal-to-noise ratio for small $k_{\rm max}$. In addition, we have argued previously  that  these parameters should be interpreted as nuisance rather than physical parameters.

  A comparison of the bias parameters we would get form the cross galaxy-matter power spectrum is not possible for galaxies. The reason is that we do not have realizations of galaxies without the survey mask geometry. This is because the galaxies were added to the halo and dark matter field at the end of the production of the galaxy mocks, after the survey geometry was applied. Thus, is not possible to compute a cross correlation between dark matter and galaxies in this case.
  
\subsubsection{Effect of redshift space distortions on the bias parameters}\label{sec:real_redshift}

In this section we test the differences between the bias parameters and shot noise obtained from real and redshift space power spectrum and bispectrum. Following  the same methodology as in \S~\ref{PS_BS_section}.  In this section we keep  $f$ and $\sigma_8$ fixed to their  fiducial values in order to isolate the effect of redshift space distortions into the bias parameters. Later in section \S\ref{sec:systematics} we check the effects of the survey mask and of the modelling on estimating these two parameters.

\begin{figure*}
\centering
\includegraphics[scale=0.3]{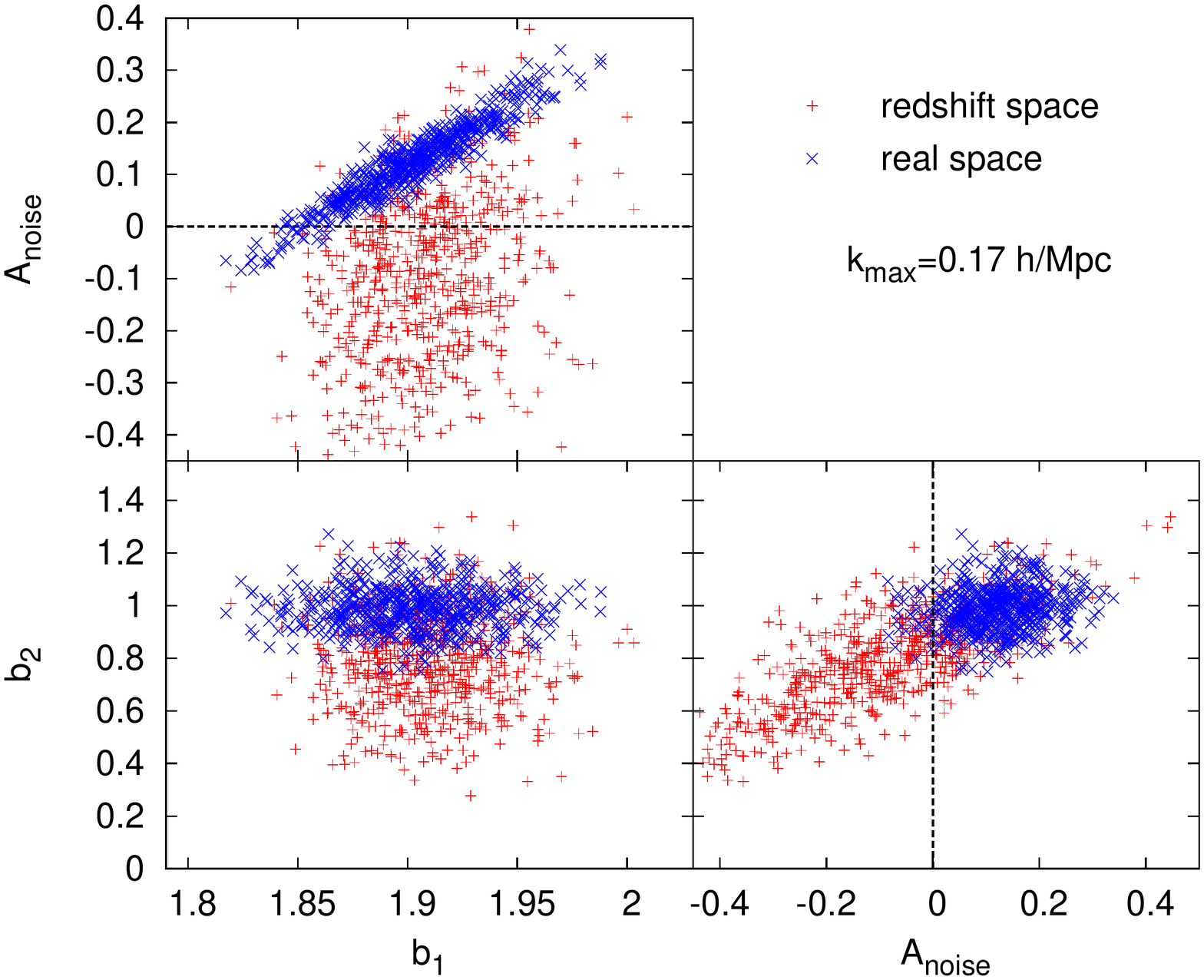}
\includegraphics[scale=0.3]{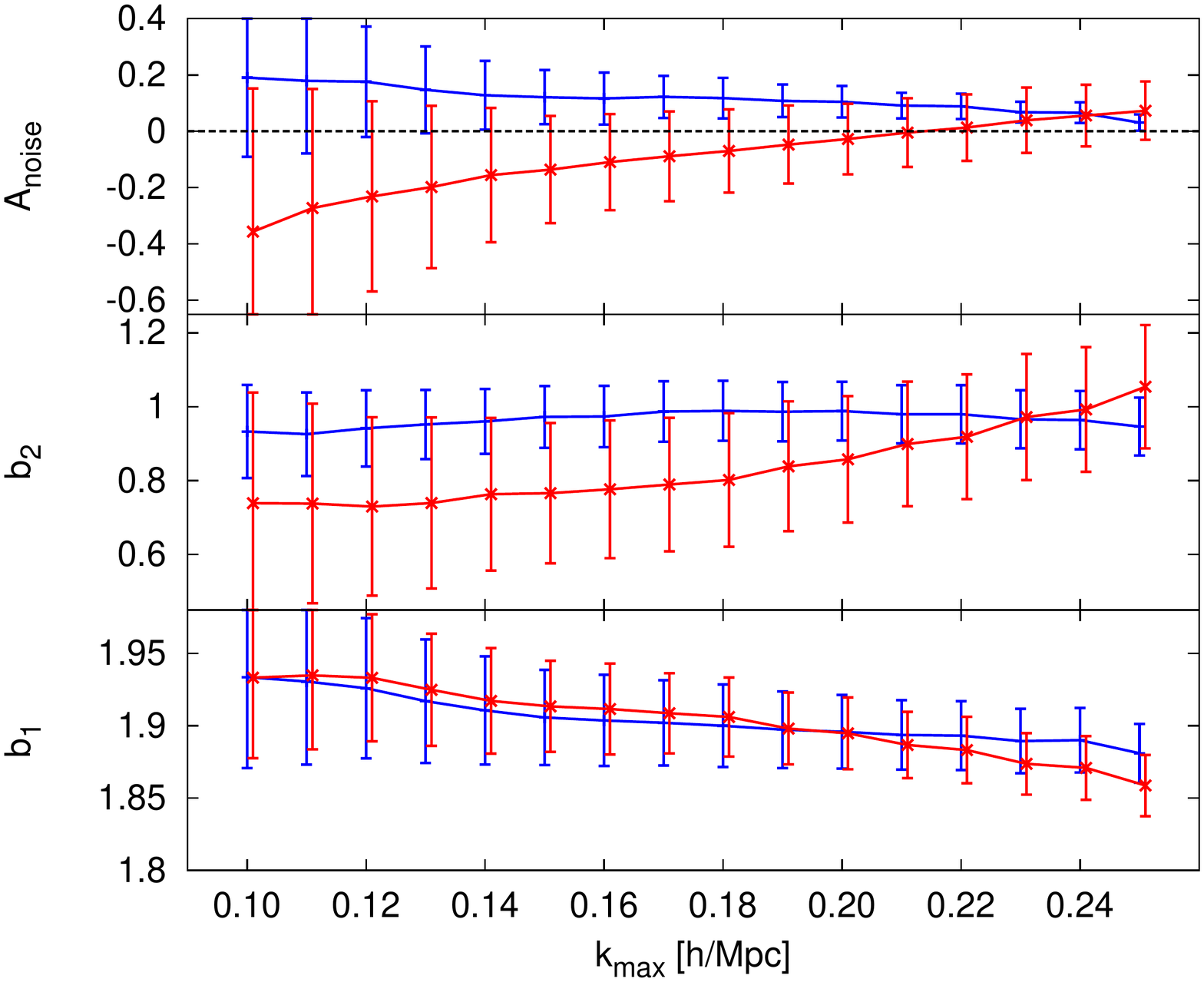}
\caption{{ Left panel}: Best fit parameters, $b_1$, $b_2$, $A_{\rm noise}$, for the galaxy mocks in real space (blue points) and in redshift space (red points). The maximum scale for the fitting is set to  $k_{\rm max}=0.17\,h{\rm Mpc}^{-1}$. {Right panel}: Best fit parameters as a function of $k_{\rm max}$. Same colour notation that in the left panel. The error-bars correspond to $1\sigma$ dispersion of the 600 realisations. There is a good agreement in the bias parameters, $b_1$ and $b_2$, estimated form the real and redshift space.}
\label{real_redshift_space_plot}
\end{figure*}
The left panel of Fig.~\ref{real_redshift_space_plot} displays the best fit parameters, $b_1$, $b_2$, and $A_{\rm noise}$ for the galaxy mocks in real space (blue points) and in redshift space (red points), where $k_{\rm max}$ is set to $0.17\,h{\rm Mpc}^{-1}$. 
The large scale bias parameter, $b_1$, is  consistent between real and redshift space statistics. 
Conversely,   the scatter of the $b_2$ parameter is larger for the redshift space statistics. This result is due to the fact that for redshift space there are two more free parameters that describe the FoG effect. We know that both $b_2$ and $\sigma_{\rm FoG}^{P}$ affect  the amplitude of the power spectrum at mildly non-linear scales: the two parameters are highly correlated, so by allowing $\sigma_{\rm FoG}^P$ to vary freely and then marginalising over it we  naturally add more uncertainty on $b_2$.
On the other hand, we observe a small tendency for $b_2$ to be underestimated by about $\sim 0.2$ in redshift space  with respect to real space, although the shift is within $1\sigma$. 

The best-fit value for $A_{\rm noise}$ is significantly different from real to redshift space. In real space we see that $A_{\rm noise}$ tends to be slightly  sub-Poisson, which is generally associated with halo-exclusion \citep{Casas-Mirandaetal:2002,Manera_Gazta:2011}. This result indicates that for this particular type of galaxies, the halo exclusion dominates over the clustering at the scales studied here. Recall that for the CMASS galaxy sample, most of the haloes are occupied only by a central galaxy. However, in redshift space there is more clustering at large scales due to the  Kaiser effect \citep{Kaiser:1987} which is not prevented by halo exclusion. This extra-clustering produces a higher  shot noise in redshift space than in real space.  In real space, halo exclusion is driving the shot noise towards the sub-Poisson region, whereas the redshift space extra-clustering drives it back towards the Poisson prediction and overtakes it slightly, making the final noise slightly super-Poisson. Since the extra-clustering in redshift space is scale dependent, we expect that the effective shot noise in redshift space possesses a scale dependence, from higher values at large scales to lower values at smaller scales. In the right panel of Fig.~\ref{real_redshift_space_plot} we see the dependence of the bias parameters and $A_{\rm noise}$ as a function of the maximum scale. The shot noise follows the expected trend:  in real space the shot noise is slightly sub-Poisson at all studied scales, whereas the shot noise in redshift space presents a scale dependence that moves from super-Poisson  at large scales towards a sub-Poisson  at smaller scales.

The right panel of Fig.~\ref{real_redshift_space_plot} demonstrates that the prediction for $b_1$ is consistent in real and redshift space and does not depend on the scale for $k_{\rm max}\lesssim0.17\,h{\rm Mpc}^{-1}$, which is the range of validity for the power spectrum model. It is also clear that $b_2$ has some scale dependence in redshift space (which becomes more significant for $k>k_{\rm max}$). This behaviour may be due to the fact that this parameter is highly correlated with  $\sigma_{\rm FoG}$, producing a parameter degeneracy in redshift space. 
Furthermore, the  adopted Finger-of-God model  is phenomenological and may not fully describe the  non-linearities in the power spectrum (and perhaps also in the bispectrum); other parameters sensitive to the same range of scales  may therefore be mis-estimated. However, given the size of the error-bars of this particular galaxy survey, the scale dependence of $b_2$ is negligible.

We conclude that, given the the typical errors of CMASS DR11 galaxy sample, the redshift space models for the power spectrum (Eq.~\ref{Psg}) and bispectrum (Eq.~\ref{B_gggs}) give a consistent description of the (mock) galaxy clustering for scales $k\leq 0.17\,h{\rm Mpc}^{-1}$.

\subsubsection{Constraining gravity and bias simultaneously}\label{sec:constraining_gravity}
In this section we drop  the assumption that the growth of structure is described by general relativity (GR)  and introduce two extra parameters: the linear growth rate $f$ and the linear  matter power spectrum amplitude parametrised by $\sigma_8$.
We constrain simultaneously  $b_1$, $b_2$, $A_{\rm noise}$, $\sigma_{\rm FoG}^P$, $\sigma_{\rm FoG}^B$, $f$ and $\sigma_8$ from the measurement of the power spectrum and bispectrum monopole. 
We still have to assume that the bispectrum kernels remain the same as those calibrated on GR-based N-body simulations and that the mildly non-linear evolution of the power spectrum is  well described by our model. We also assume that the initial linear power spectrum is given by GR.
However the analysis can be considered as a null hypothesis  test if no significant deviations from the GR-predicted values for $f$ are found. Moreover studies show that, at least for the $f(R)$ family of modified gravity theories, the GR-derived bispectrum kernel  is still a good description of the bispectrum \citep{HGMfR}. 

\begin{figure*}
\centering
\includegraphics[scale=0.5]{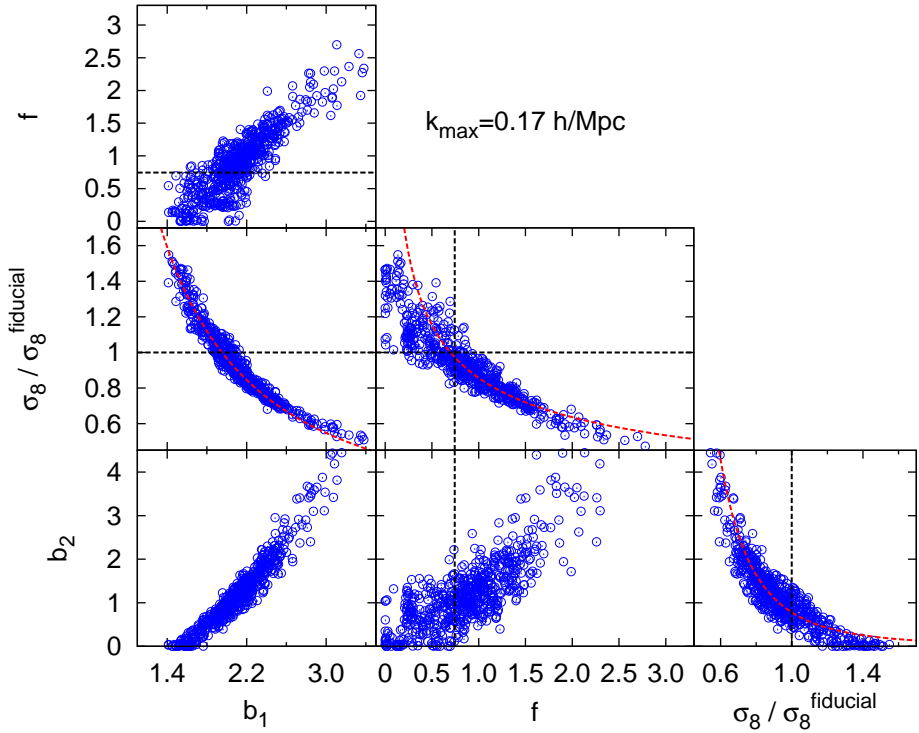}
\caption{Best fit parameters, $b_1$, $b_2$, $f$, $\sigma_8/\sigma_8^{\rm fiducial}$ for 600 realisations of NGC galaxy mocks in redshift space (blue points) when power spectrum and bispectrum monopole are used. The relations between the best fit parameters can be empirically modeled by power law relations. In particular,  red dashed lines represent the power-law relations for $\sigma_8-b_1$, $\sigma_8-b_2$ and $\sigma_8-f$ (see text for their exact values). Black dashed lines show the fiducial values for $f$ and $\sigma_8/\sigma_8^{\rm fiducial}$. The maximum scale for the analysis is set to  $k_{\rm max}=0.17\,h{\rm Mpc}^{-1}$.}
\label{plot_degenerations}
\end{figure*}

Fig.~\ref{plot_degenerations} displays the scatter for some of these parameters  from 600 realisations of the NGC galaxy mocks (blue symbols). The black dashed lines show the fiducial values for $f$ and $\sigma_8$.
 Since we are only using two statistics (power spectrum and bispectrum monopole), we cannot constrain efficiently both $\sigma_8$ and $f$. In a similar way, if we were using the power spectrum monopole and quadrupole, only the combination $f\sigma_8 $ would be suitable to be efficiently constrained. For the 
joint analysis of  power spectrum and bispectrum monopole,  a slightly different combination of  $f$ and $\sigma_8$ is measured efficiently. This creates the possibility of measuring both  $f$ and $\sigma_8$  from a combined analysis of power spectrum monopole and quadrupole and bispectrum monopole \citep{HGMetal:inprep}. 
 While in the case of the power spectrum monopole and quadrupole it is  clear from examining the large scale limit of the model that the relevant parameter combination is  $\sigma_8\sim f^{-1}$, this is not the case for the power spectrum and bispectrum monopole combination. The bias parameters are involved and even at large scales, the power spectrum has a non-negligible contribution of $b_2$.  
 Fig.~\ref{plot_degenerations} suggests that parameters are mostly distributed along one-to-one relations determined  directly from the distribution of the best fit parameters from the mocks.  Thus we can empirically determine  the degeneracy directions of importance.

We approximate these relations with power-law equations, which are the red dashed lines in Fig.~\ref{plot_degenerations}.
This information suggests that we can constrain  three combinations of the four parameters $b_1$, $b_2$, $f$ and $\sigma_8$.
 In particular, given the  {\it ansatz} relations $\sigma_8\sim f^{-n_1}$, $\sigma_8\sim b_1^{-n_2}$ and $\sigma_8\sim b_2^{-n_3}$, the best-fit to  the distributions  around the maximum are $n_1=0.43$, $n_2=1.40$, $n_3=0.30$. We recognise that these values do not correspond to  universal relations for these parameters, but are effective fits given a particular galaxy population. For other samples they may no longer be optimal.
 
\begin{figure*}
\centering
\includegraphics[scale=0.3]{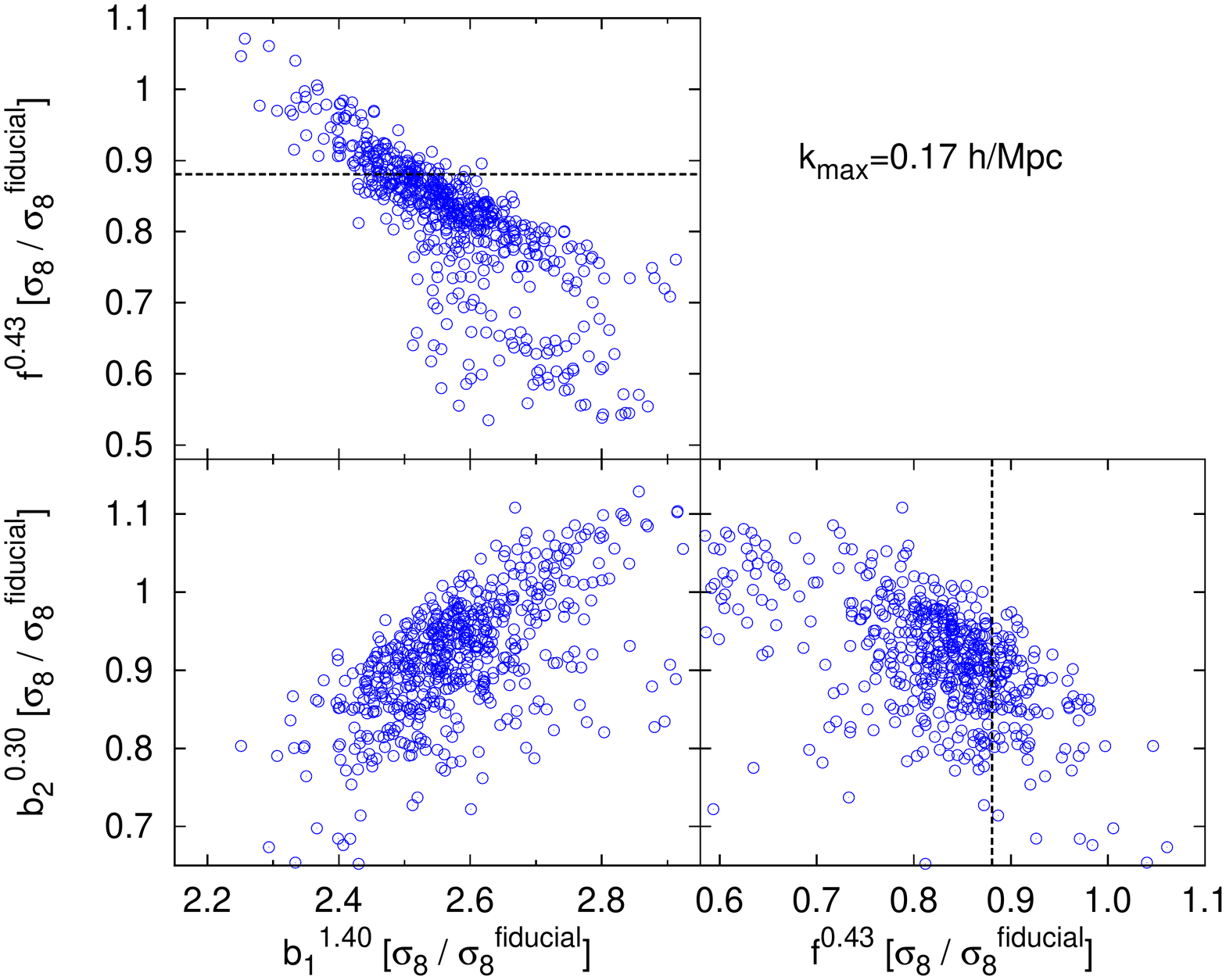}
\includegraphics[scale=0.3]{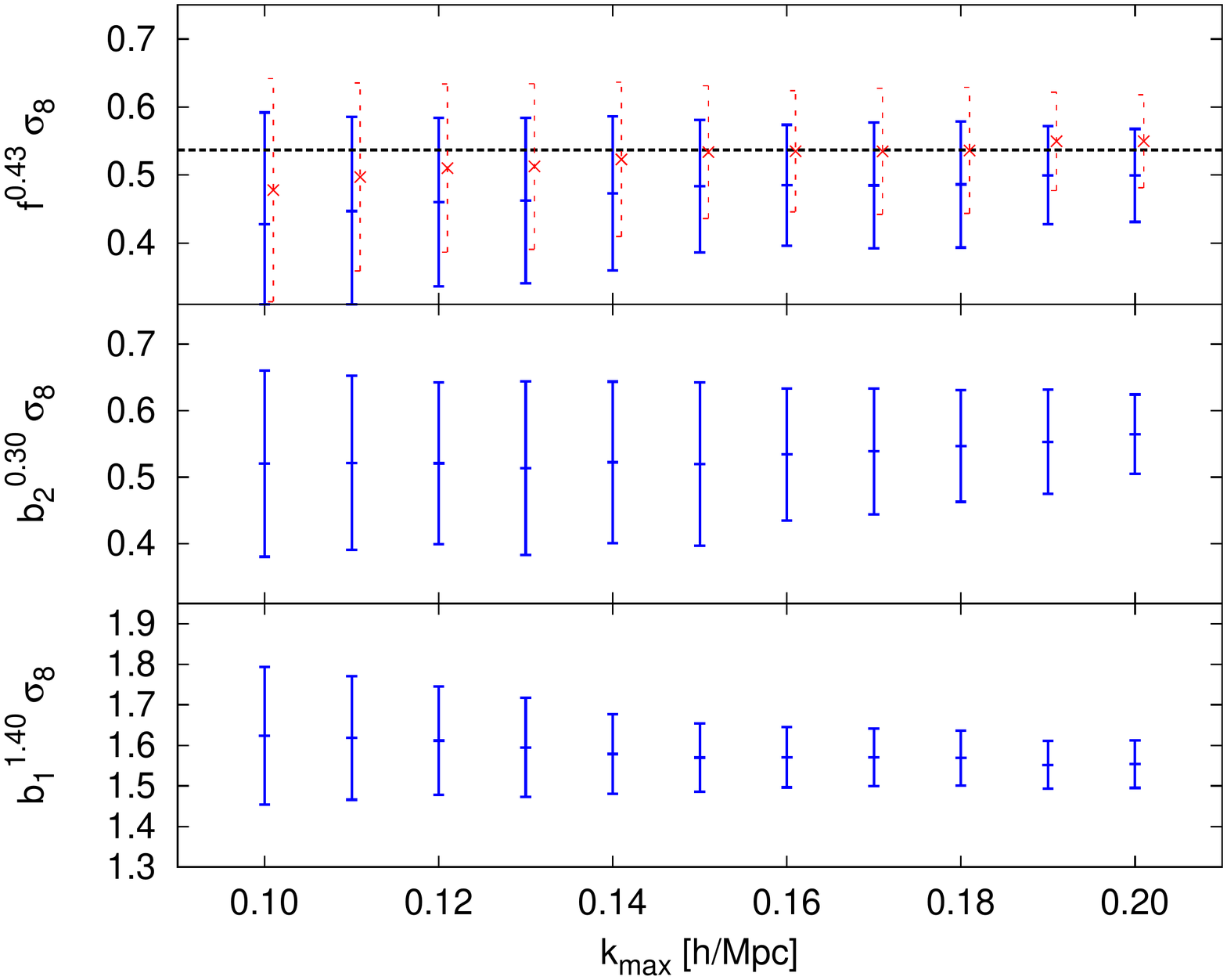}
\caption{ Left panel: Best fit parameters, $b_1^{1.40}\sigma_8$, $b_2^{0.30}\sigma_8$, $f^{0.43}\sigma_8$, for 600 realisations of galaxy mocks in redshift space (blue points) when power spectrum and bispectrum monopole are measured. When these new variables are used, the scatter distribution is more Gaussian and the errors can be estimated from the dispersion among the different realisations. Black dashed lines show the fiducial values for $f^{0.43}\sigma_8$. The maximum scale for the fitting is set to  $k_{\rm max}=0.17\,h{\rm Mpc}^{-1}$. Right panel:  single parameters estimate  as a function of $k_{\rm max}$. Blue error-bars correspond to $1\sigma$ dispersion. For the panel corresponding to $f^{0.43}\sigma_8$, the results corrected by a systematic offset of 0.05 are shown in red dashed lines. No significant $k_{\rm max}$-dependence is observed.}
\label{plot_degenerations2}
\end{figure*}

Results in the new combinations  $f^{0.43}\sigma_8$, $b_1^{1.40}\sigma_8$ and $b_2^{0.30}\sigma_8$ are shown in right panel of  Fig.~\ref{plot_degenerations2}. In these new variables, the distribution appears more Gaussian, and  it is more meaningful to estimate the error-bars from the dispersion of the distribution.

 In the right panel  the blue solid lines show the mean and the error-bars (computed from the distribution of the mocks best fit values) for these variables as a function of $k_{\rm max}$. The black dashed line in the panels of Fig.~\ref{plot_degenerations2} is the fiducial value for $f^{0.43}\sigma_8$. There is an offset between the mean of the galaxy mocks and the fiducial value, which is  constant with $k_{\rm max}$. This offset is at the 0.05 level,  below  $1\sigma$ statistical error for the survey, but the analysis tends to under-estimate the fiducial value of $f^{0.43}\sigma_8$. In red dashed lines the value of $f^{0.43}\sigma_8$ is corrected by this 0.05 offset. Recall that the error on the mean is  some 24 times smaller than the reported errors, so while the systematic shift is below the statistical error for the survey, it can be measured from the mocks with high statistical significance, and can also be  observed in Fig~\ref{plot_degenerations}.
In the next section we explore the source of this systematic error.

\subsection{Systematic errors on $f$ and $\sigma_8$}\label{sec:systematics}
There are several effects that could systematically shift in the combination $f^{0.43}\sigma_8$.
To assess  the treatment of the survey window and the fact that galaxy mocks are based on \textsc{PThalos} and not on N-body haloes, we  estimate $b_1$, $b_2$, $f$, $\sigma_8$, $A_{\rm noise}$ and $\sigma_{\rm FoG}^P$ from the 20 realisations of N-body haloes and from the 50 realisations of masked and unmasked \textsc{PThalos}.  Since we are considering the clustering of haloes  all the FoG contributions should vanish (i.e.,  we should strictly set $\sigma_{\rm FoG}^P$ and  $\sigma_{\rm FoG}^B$ to 0).  However, it has been shown \citep{Nishimichietal:2011} that at least for the power spectrum, it is necessary to incorporate a term of the form of $\sigma_{\rm FoG}^P$ in order to account for inaccuracies of the model, hence our inclusion of  $\sigma_{\rm FoG}^P$ as a free parameter.

\begin{figure*}
\centering
\includegraphics[scale=0.5]{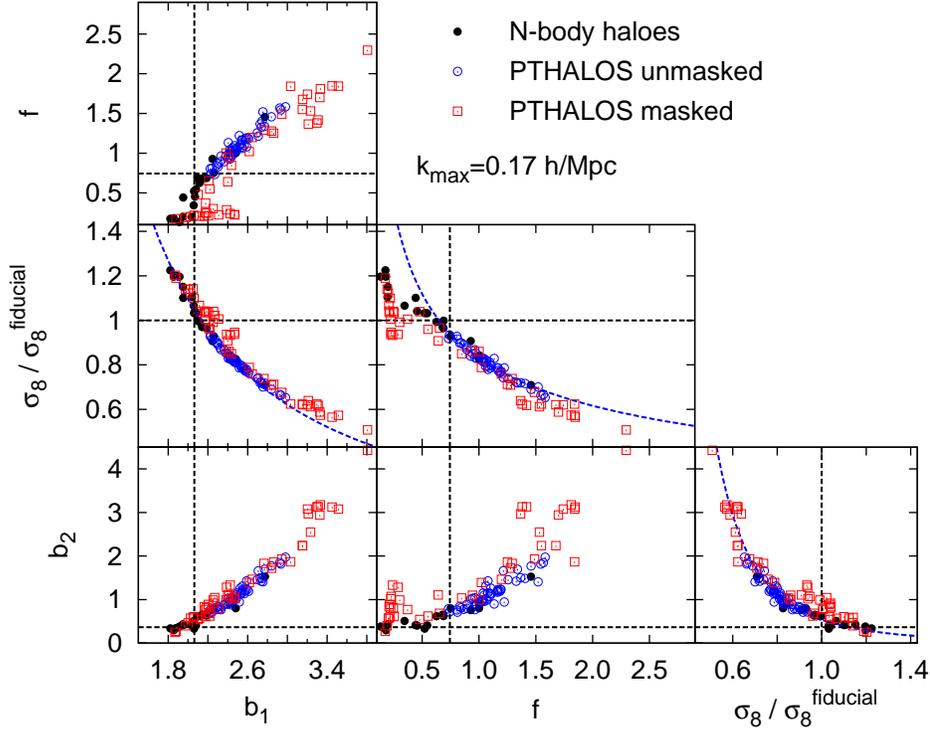}
\caption{Best fit parameters, $b_1$, $b_2$, $f$, $\sigma_8/\sigma_8^{\rm fiducial}$ for 20 realisations of N-body haloes in redshift space (black filled circles), for 50 realisations of masked (red empty squares) and unmasked (blue empty circles) \textsc{PThalos} when power spectrum and bispectrum monopole are measured. Black dashed lines show the fiducial values for $f$ and $\sigma_8/\sigma_8^{\rm fiducial}$ and the measured cross-bias parameters defined in  Eqs. \ref{b1_cross}-\ref{b2_cross}. Blue dashed lines show the power-law relations for some of these parameters (see text for their exact values). The maximum scale for the fitting is set to  $k_{\rm max}=0.17\,h{\rm Mpc}^{-1}$. The power-law relations observed in Fig.~\ref{plot_degenerations} for the galaxy mocks are very similar for N-body haloes and  \textsc{PThalos}, and therefore potentially applicable to the observed dataset.}
\label{plot_degenerations_haloes_1}
\end{figure*}

\begin{figure*}
\centering
\includegraphics[scale=0.3]{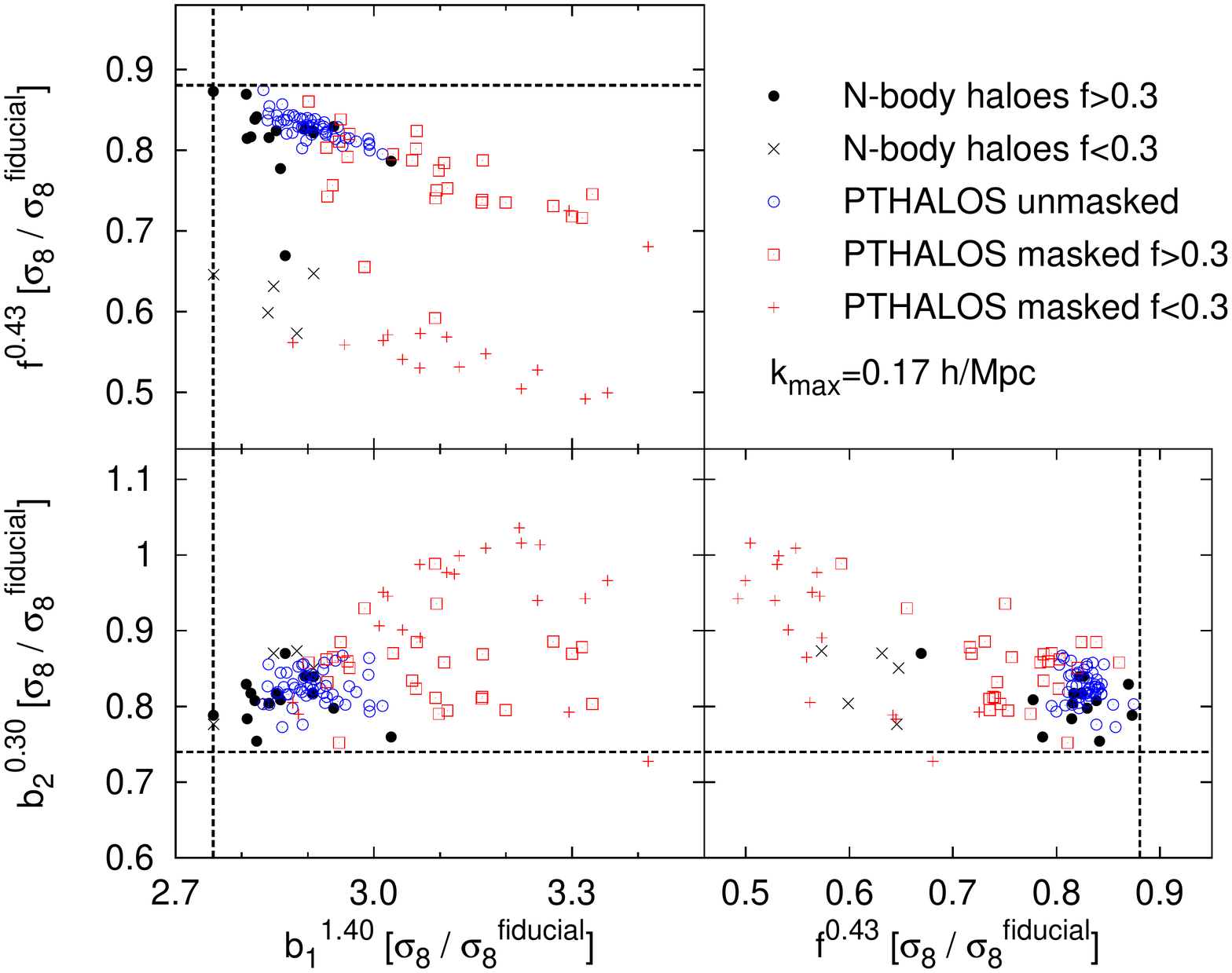}
\includegraphics[scale=0.3]{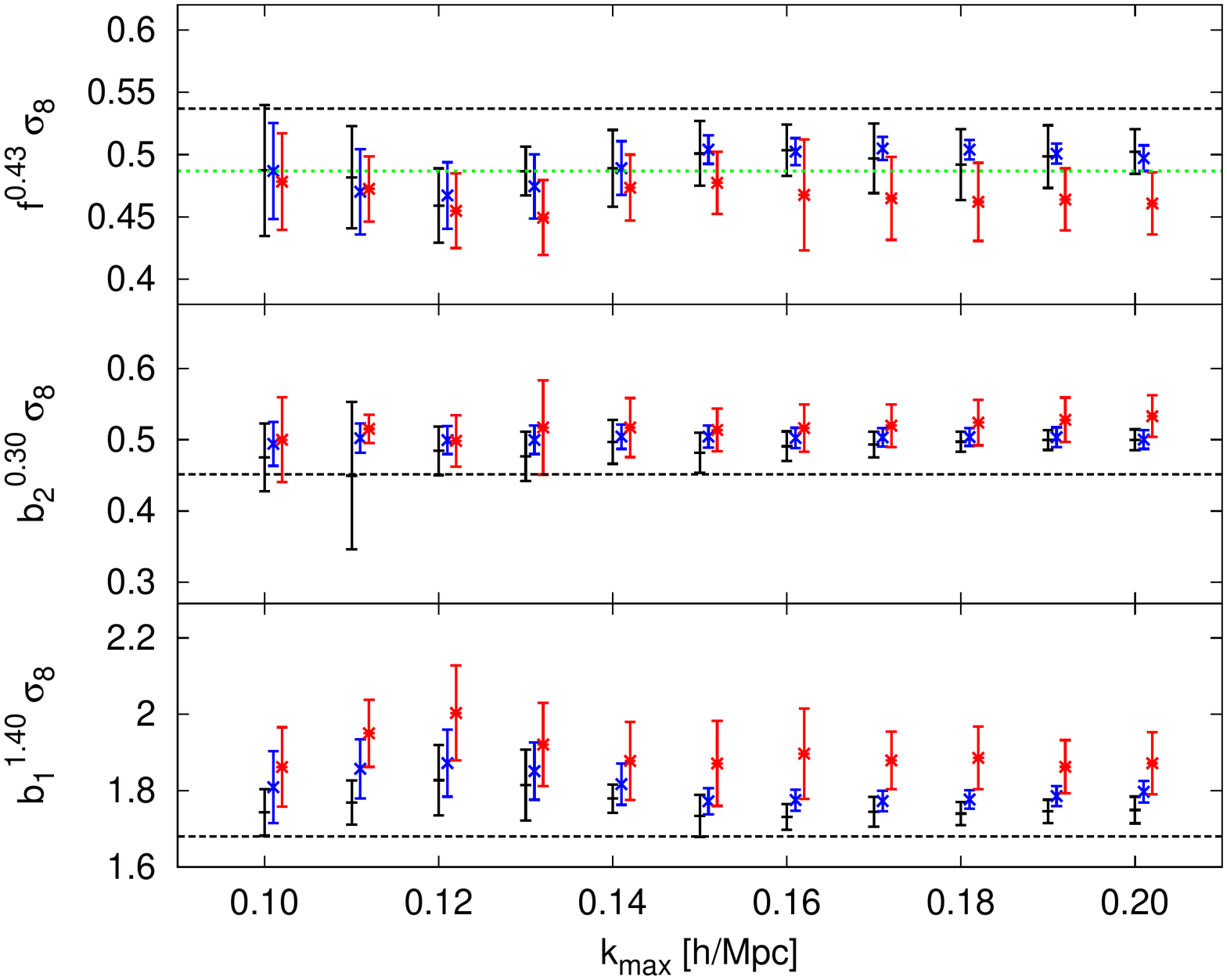}

\caption{Best fit parameters, $b_1^{1.40}\sigma_8$, $b_2^{0.30}\sigma_8$, $f^{0.43}\sigma_8$, for 20 realisations of N-body haloes, masked and unmasked \textsc{PThalos} (same colour notation that in Fig.~\ref{plot_degenerations_haloes_1}), when power spectrum and bispectrum monopole are measured.  For  \textsc{PThalos}  and N-body haloes, the crosses show those realizations whose best fit $f$ is below 0.3, whereas squares and circles above 0.3, respectively. Only those realizations whose $f>0.3$ have been included in the computation of mean values and error-bars of the right panel. Black dashed lines show the fiducial values for $f^{0.43}\sigma_8$ and for ${(b_1^{\rm cross})}^{1.40}\sigma_8$ and ${(b_2^{\rm cross})}^{0.30}\sigma_8$. The maximum scale for the fitting is set to  $k_{\rm max}=0.17\,h{\rm Mpc}^{-1}$. Green dotted line is the theoretical prediction reduced by a systematic offset of 0.05. When the new variables are used the original distributions of Fig. \ref{plot_degenerations_haloes_1} appears more Gaussian. However, the systematic shift on $f^{0.43}\sigma_8$ observed for the galaxy mocks, is also present for N-body haloes. This indicates that the systematic shift is not due to a limitation of the mocks, but a limitation in the theoretical description of the halo power spectrum and bispectrum in redshift space.}
\label{plot_degenerations_haloes_2}
\end{figure*}

Fig.~\ref{plot_degenerations_haloes_1} presents the distribution  of the best fit values for $b_1$, $b_2$, $f$ and $\sigma_8$ for N-body haloes (black filled circles), for unmasked \textsc{PThalos} (blue empty circles) and for masked \textsc{PThalos} (red empty squares) estimated from the power spectrum monopole and bispectrum. Recall that these three different halo catalogues have different effective volumes, so we expect  different magnitudes of the scatter for the estimated parameters. However, the best fit values should be the same for the three sets if there are no systematics related to the nature of the simulation or the window. We observe that there are no significant differences when comparing  masked and unmasked catalogs, indicating  (as already shown in \S~\ref{5_3_section}) that the  survey window is modelled correctly for both the power spectrum and bispectrum.  If we now compare  the N-body  and \textsc{PThalos}   results we notice  few differences. N-body haloes tend to have a smaller value for $b_1$, $b_2$  and $f$, but a higher value for $\sigma_8$, than \textsc{PThalos}. However, these differences are small and lie along the degeneracy direction  (blue dashed lines). As for galaxy mocks, we assume power-law relations between $b_1$, $b_2$ and $f$.  In black dashed lines, we show the cross-bias parameters reported in \S\ref{5_2_section} combined with $\sigma_8$. 

We assume that the values for the indices $n_1$, $n_2$ and $n_3$ are the same as those obtained from the galaxy mocks: $n_1=0.43$, $n_2=1.40$ and $n_3=0.30$. Independently of these relations, the parameter distributions for  N-body haloes and \textsc{PThalos} are slightly offset from  the fiducial value in the $f$-$\sigma_8$ panel of Fig.~\ref{plot_degenerations_haloes_1} in a similar way as observed for the galaxy mocks in Fig.~\ref{plot_degenerations}.

The relation between $f$ and $\sigma_8$ obtained (i.e. $f^{0.43}\sigma_8$) is not always perfect and does not hold for any value of $f$ or $\sigma_8$. This can be seen in the $\sigma_8-f$ panel in Fig. \ref{plot_degenerations_haloes_1}. Let us say that $f$ and $\sigma_8$ are correlated according to $f^{0.43}\sigma_8=\,{\rm constant}$, for $0.3\leq f $, which is a wide range for the possible values of $f$ (it is very unlikely that the observed galaxies have an $f$ value outside this range, but we could take it as a mild prior). We note that for the unmasked \textsc{PThalos} the volume of the boxes is large enough that $f$ is always inside this range, and the relation $f^{0.43}\sigma_8$ holds for all the mocks. When we reduce the volume (masking the boxes) the scatter increases and some realizations predict a best fit value of $f$ outside this range. Since for these points the $f^{0.43}\sigma_8$ relation does not hold anymore they seem to present a larger deviation.

The left panel of Fig.~\ref{plot_degenerations_haloes_2} displays the distribution of  these parameters combinations obtained from  the different realisations of N-body haloes, masked and unmasked \textsc{PThalos} with the same colour notation that in Fig.~\ref{plot_degenerations_haloes_1}. The fiducial value for $f^{n_1}\sigma_8$ is represented by black dotted line. For  \textsc{PThalos}  and N-body we have plotted the values whose $f<0.3$ as crosses and the values whose $f>0.3$ as squares and circles, respectively. We see clearly that the binomial distribution observed for  \textsc{PThalos}  and N-body in the $f^{0.43}\sigma_8-b_1^{1.40}\sigma_8$ panel is due to the fact that low $f$ values do not follow the $f^{0.43}\sigma_8$ relation. In this section we consider the mild prior $f>0.3$, which helps to hold the $f^{0.43}\sigma_8$ relation, when the total volume is small.  In these new variables and taking into account the mild prior on $f$,  is easy to appreciate the good agreement between masked and unmasked realisations and between \textsc{PThalos} and N-body haloes.
 The right panel of Fig.~\ref{plot_degenerations_haloes_2} shows how these parameters depend on  $k_{\rm max}$. Again the offset in $f^{0.43}\sigma_8$ is constant across $k_{\rm max}$ and also present at large scales.  For the  \textsc{PThalos}  and N-body haloes, the mild prior $f>0.3$ has been applied.

This feature indicates that the systematic offset observed in \S~\ref{sec:constraining_gravity} is  present in \textsc{PThalos}, with and without survey mask, and in N-body haloes.  It is therefore produced by a failure of the modelling of  the combination of redshift-space distortions and bias for haloes.  \cite{HGMetal:inprep} reports that the modelling of redshift space distortions adopted here works well and does not induce any bias for the (unbiased) dark matter distribution in  redshift space. When we examine (biased) haloes in redshift space, the adopted model seem to be insufficient to reach accuracy levels of few per cent. We believe we have reached the limitations of the currently available semi-analytic  modelling  of redshift-space clustering of  dark matter tracers: shrinking the statistical errors below this level is  not useful until these limitations can be overcome.

 We conclude that the method adopted here  to measure $f^{0.43}\sigma_8$ from the power spectrum monopole and bispectrum underestimate its fiducial value by about 0.05, which is a$~10\%$ effect. In the right panel of Fig.~\ref{plot_degenerations_haloes_2} this offset is shown by  the green dotted line, while  black dashed line corresponds to the fiducial value. When reporting our main results we will always apply a correction for this offset.  We also see that the values $b_1^{1.40}\sigma_8$ and $b_2^{0.30}\sigma_8$ are biased respect to the true values of $\sigma_8$ and the measured cross-bias parameters defined in Eqs. \ref{b1_cross}-\ref{b2_cross}: $b_1^{1.40}\sigma_8$ and $b_2^{0.30}\sigma_8$  are biased $\sim10\%$ higher respect to ${(b^{\rm cross}_1)}^{1.40}\sigma_8$ and ${(b^{\rm cross}_2)}^{0.30}\sigma_8$, respectively. In this paper we do not correct the bias parameter by the systematic shifts found respect to the cross-bias parameters.   The main result of this paper  is the  constraint on the combination of  $f$ and $\sigma_8$ from the CMASS galaxy data, but not the galaxy bias parameters, the shot noise properties or the  Fingers-of-God redshift space distortion  parameter,  which we treat as a nuisance parameters.

\section{Conclusions}\label{sec:conclusions}
We have presented a measurement of the bispectrum of the CMASS DR11 galaxy sample of the Baryon Oscillations Spectroscopic Survey of the Sloan Digital Sky Survey III.  This is the largest survey (in terms of volume and number of objects) to date where the bispectrum  has been measured, offering an unprecedented signal-to-noise ratio.

The bispectrum is the Fourier counterpart of the three point function, and as such encloses information about non-linear clustering, biasing and gravity.  Because  of the  complicated nature of redshift space distortions on  the triplets of  Fourier modes that create the bispectrum, we have only considered the bispectrum monopole (i.e., angle-averaged with respect to the line of sight direction).
The bispectrum  signal is detected  at high statistical significance, which enables its use  to measure cosmological parameters of interest. The bispectrum shows the characteristic shape dependence  induced by gravitational evolution in the mildly non-linear regime, indicating that  the large volume of the survey  allows us to discard  highly non-linear scales  and still have a useful signal-to-noise ratio.
We aim at measuring galaxy bias and the growth of structure. To reduce degeneracies among these quantities we jointly fit the power spectrum and bispectrum monopoles. 

In order to interpret this signal we have developed a description of the mildly non-linear power spectrum and bispectrum  for biased dark matter tracers in redshift space, which is presented in \S\ref{section_method}. The bias model is particularly important. The simple, local, quadratic bias expansion, which has been the workhorse to date to analyse the bispectrum from surveys and is widely used for forecasts, is not good enough for the precision offered by the CMASS DR11 survey. For instance, the bias parameters recovered from analysing the bispectrum are not consistent with those obtained from the power spectrum  adopting this  bias model. Similar problems were reported by \cite{Pollacketal:2013} when the local model is applied. Here, for CMASS galaxies,  we must move  beyond this simple model.
We  adopt for the bispectrum a nonlinear, nonlocal bias model that was originally developed for halos \citep{McDonaldRoy:2009,Baldaufetal:2012,Saitoetal:2014,HGMetal:inprep} and recently applied to power spectrum analyses \citep{Beutleretal:2013}.  This approach is still a two parameter bias model,  but $b_1$ and  $b_2$ do not have the same meaning as in the quadratic local bias model. Despite this bias model being strictly physically motivated for dark matter  halos,  we apply it to galaxies,  motivated by the fact that  CMASS galaxies  are believed to closely  trace  massive dark matter halos. Nevertheless the quadratic bias parameter $b_2$  should be treated as an effective parameter that absorbs  limitations of the adopted modelling. 

The mildly non-linear description of these statistics in redshift space is  also a crucial starting ingredient; because of the complicated formulae, the description and derivations are reported in the Appendices.  In brief we  use  the bispectrum kernel calibrated from N-body simulations in real and redshift space and include a suite of effective parameters  which, in principle, describe physical quantities such as non-linear incoherent velocity dispersion (Finger-of-God effects), and deviations from purely Poisson shot noise. In practice  we treat these quantities as nuisance parameters    to be marginalised over, and these parameters {\it absorb} several  of possible  inaccuracies of the modelling.
Even with this improvement, there are indications that  we have reached the limitations of the currently available  modelling  of redshift-space clustering of  dark matter tracers: shrinking the statistical errors below this level is  not useful until these limitations can be overcome.\\

Our measurements are  supported by an extensive series of tests performed on dark matter N-body simulations, halo catalogs (obtained both from N-body and \textsc{PThalos} simulations) and mock galaxy catalogs. These tests are also used to identify the regime of validity of the adopted modelling:  this regime occurs  when all $k$ modes of the bispectrum  triangles are larger than $0.03 h{\rm Mpc}^{-1}$ and less than $k_{\rm max}=0.17\, h{\rm Mpc}^{-1}$ being conservative or less than  $k_{\rm max}=0.20\, h{\rm Mpc}^{-1}$ being more optimistic.
We also account for real word effects such as survey windows and systematic weighting of objects.   We opt to add in quadrature the statistical error and half of the systematic shift to account for the uncertainty in the systematic correction.  

The bispectrum calculation is computationally intensive because of the number of bispectrum triplets, which increases as the number of $k$ modes in the survey to the third power. For this reason we only consider a subset of all possible bispectrum shapes. This is consistent with what has been done in previous literature; while it does not extract all the possible information from the survey it is a good compromise between  accuracy and computational feasibility. If  we were to use all possible shapes we could, in principle, almost halve the statistical error-bars. The price to pay, however, will be much less control over the theoretical  modelling,  and  the resulting measurements would become systematic-dominated.

An additional complication we had to overcome to perform the analysis is that there is no fully developed, tested and motivated estimator for the bispectrum or a quantity that depends on it (see e.g., \citealt{Verdeetal:2013}), whose probability distribution function is known, and none exist for the joint power spectrum and bispectrum analyses.  
We therefore had to resort to a sub-optimal but still unbiased  approach.  We ignore correlations between shapes in determining the parameters and then estimate the errors  from the distribution of the best parameters  values obtained from 600 mock galaxy surveys. 
Our cosmologically interesting parameters are two bias parameters $b_1$ and $b_2$, the linear  matter clustering amplitude $\sigma_8$ and the growth rate of fluctuations $f=d\ln \delta /d\ln a$, where $\delta$ denotes the dark matter over density and $a$ the scale factor. If gravity is described by general relativity  at  cosmological scales,  then $f$ is effectively  given by $\Omega_m$.

We find that even jointly,  the bispectrum and power spectrum monopole cannot measure all four parameter separately, but do constrains the following combinations: $f^{0.43}\sigma_8$, $b_1^{1.40}\sigma_8$ and $b_2^{0.30}\sigma_8$. In these variables the distribution of the best-fit parameters for the mock catalogs are much closer  a Gaussian distribution than in the original four parameters.
 When we set $k_{\rm max}=0.17\,h{\rm Mpc}^{-1}$ we obtain  $b_1(z_{\rm eff})^{1.40}\sigma_8(z_{\rm eff})=1.672\pm 0.060$ and  $b_2^{0.30}(z_{\rm eff})\sigma_8(z_{\rm eff})=0.579\pm0.082$ at the effective redshift of  the survey, $z_{\rm eff}=0.57$.  The main cosmological result in this case is the constraint on the combination $f^{0.43}(z_{\rm eff})\sigma_8(z_{\rm eff})=0.582\pm0.084$. Adopting a less conservative approach allow us to set $k_{\rm max}=0.20\,h{\rm Mpc}^{-1}$, which produces: $b_1(z_{\rm eff})^{1.40}\sigma_8(z_{\rm eff})=1.681\pm 0.046$, $b_2^{0.30}(z_{\rm eff})\sigma_8(z_{\rm eff})=0.571\pm0.043$ and $f^{0.43}(z_{\rm eff})\sigma_8(z_{\rm eff})=0.584\pm0.051$.

 The $f^{0.43}\sigma_8$ combination is affected by a 0.05 systematic error --extensively quantified and calibrated from simulations--  and this correction has been applied. This issue represents the main obstacle in further reducing the statistical errors.

The present analysis measures a combination of $f$-$\sigma_8$ that differs from that obtained from the combination of the  power spectrum monopole and quadrupole (which yields $f\sigma_8$). This creates the possibility of measuring both  $f$ and $\sigma_8$  from a combined analysis of power spectrum monopole and quadrupole and bispectrum monopole. The potential of this approach   is presented in the companion paper \citep{paper2}; a more detailed joint analysis is left to future work. 

The mock catalogues based in \textsc{PThalos} are adequate for performing the analysis described in this paper. In particular they are essential to extract the empirical relations between $b_1$, $b_2$, $\sigma_8$ and $f$, which are applied to the data, as well as to obtain a reliable estimation of the diagonal terms of the covariance matrix of the power spectrum and bispectrum. On the other hand, the limitation of the mocks for describing the observed clustering of the data at mildly non-linear scales suggests that there is space for improvement. Performing a similar bispectrum analysis on the next generation of surveys  will require more realistic mocks that better match the observations both of the mildly non-linear power spectrum and bispectrum for the adopted tracers. This will be an important ingredient to improve the modeling of the data to  significantly reduce the systematic errors  and keep them below the statistical ones.

The constraints on $f^{0.43}\sigma_8$ will be  useful in a joint analysis with other cosmological data sets (in particular CMB data)  for setting stringent constraints on on neutrino mass, gravity, curvature as well as number of neutrino species.  Further, the joint constraints on $f^{0.43}\sigma_8$, $b_1^{1.40}\sigma_8$, and $b_2^{0.30}\sigma_8$, can be used to include the broadband shape and amplitude  of the galaxy power spectrum when doing cosmological parameters estimation. These are presented in a companion paper \citep{paper2}.

\section*{Acknowledgements}
HGM thanks Florian Beutler for useful discussions about the survey mask and the power spectrum and bispectrum estimator. We thank Rom\'an Scoccimarro for useful comments on the final draft.  We also thank Beth Reid for providing the N-body haloes used to test the systematics of the power spectrum and bispectrum model.

HGM is grateful for support from the UK Science and Technology Facilities Council through the grant
ST/I001204/1.
JN is supported in part by ERC grant FP7-IDEAS-Phys.LSS.
LV   is supported by the European Research Council under the European Community's Seventh Framework Programme grant FP7-IDEAS-Phys.LSS and  acknowledges Mineco grant FPA2011-29678- C02-02.
WJP is grateful for support from the UK Science and Technology Facilities Research Council through the grant 
ST/I001204/1, and the European  Research Council through the ``Darksurvey" grant.

Funding for SDSS-III has been provided by the Alfred P. Sloan
Foundation, the Participating Institutions, the National Science
Foundation, and the U.S. Department of Energy Office of Science. The
SDSS-III web site is http://www.sdss3.org/.

SDSS-III is managed by the Astrophysical Research Consortium for the
Participating Institutions of the SDSS-III Collaboration including the
University of Arizona,
the Brazilian Participation Group,
Brookhaven National Laboratory,
University of Cambridge,
Carnegie Mellon University,
University of Florida,
the French Participation Group,
the German Participation Group,
Harvard University,
the Instituto de Astrofisica de Canarias,
the Michigan State/Notre Dame/JINA Participation Group,
Johns Hopkins University,
Lawrence Berkeley National Laboratory,
Max Planck Institute for Astrophysics,
Max Planck Institute for Extraterrestrial Physics,
New Mexico State University,
New York University,
Ohio State University,
Pennsylvania State University,
University of Portsmouth,
Princeton University,
the Spanish Participation Group,
University of Tokyo,
University of Utah,
Vanderbilt University,
University of Virginia,
University of Washington,
and Yale University.
This research used resources of the National Energy Research Scientific
Computing Center, which is supported by the Office of Science of the
U.S. Department of Energy under Contract No. DE-AC02-05CH11231.

Numerical computations were done on the Sciama High Performance Compute (HPC) cluster which is supported by the ICG, SEPNet and the University of Portsmouth and on Hipatia ICC-UB BULLx High Performance Computing Cluster at the University of Barcelona.

The simulations for N-body haloes used in this paper were analysed at the National Energy Research Scientific Computing Center, the Shared Research Computing Services Pilot of the University of California and the Laboratory Research Computing project at Lawrence Berkeley National Laboratory.

%
%  These Macros are taken from the AAS TeX macro package version 4.0.
%  Include this file in your LaTeX source only if you are not using
%  the AAS TeX macro package and need to resolve the macro definitions
%  in the BibTeX entries returned by the ADS abstract service.
%
%  For more information on the AASTeX macro package, please see the URL
%	http://www.aas.org/publications/aastex.html
%  For more information about ADS abstract server, please see the URL
%	http://adswww.harvard.edu/ads_abstracts.html
%

% Abbreviations for journals.  The object here is to provide authors
% with convenient shorthands for the most "popular" (often-cited)
% journals; the author can use these markup tags without being concerned
% about the exact form of the journal abbreviation, or its formatting.
% It is up to the keeper of the macros to make sure the macros expand
% to the proper text.  If macro package writers agree to all use the
% same TeX command name, authors only have to remember one thing, and
% the style file will take care of editorial preferences.  This also
% applies when a single journal decides to revamp its abbreviating
% scheme, as happened with the ApJ (Abt 1991).

\def\jnl@style{\it}
%commente par Seb
\def\aaref@jnl#1{{\jnl@style#1}}
%ref remplace par aaref pour eviter conflit...

\def\aaref@jnl#1{{\jnl@style#1}}

\def\aj{\aaref@jnl{AJ}}                   % Astronomical Journal
\def\araa{\aaref@jnl{ARA\&A}}             % Annual Review of Astron and Astrophys
\def\apj{\aaref@jnl{ApJ}}                 % Astrophysical Journal
\def\apjl{\aaref@jnl{ApJ}}                % Astrophysical Journal, Letters
\def\apjs{\aaref@jnl{ApJS}}               % Astrophysical Journal, Supplement
\def\ao{\aaref@jnl{Appl.~Opt.}}           % Applied Optics
\def\apss{\aaref@jnl{Ap\&SS}}             % Astrophysics and Space Science
\def\aap{\aaref@jnl{A\&A}}                % Astronomy and Astrophysics
\def\aapr{\aaref@jnl{A\&A~Rev.}}          % Astronomy and Astrophysics Reviews
\def\aaps{\aaref@jnl{A\&AS}}              % Astronomy and Astrophysics, Supplement
\def\azh{\aaref@jnl{AZh}}                 % Astronomicheskii Zhurnal
\def\baas{\aaref@jnl{BAAS}}               % Bulletin of the AAS
\def\jrasc{\aaref@jnl{JRASC}}             % Journal of the RAS of Canada
\def\memras{\aaref@jnl{MmRAS}}            % Memoirs of the RAS
\def\mnras{\aaref@jnl{MNRAS}}             % Monthly Notices of the RAS
\def\pra{\aaref@jnl{Phys.~Rev.~A}}        % Physical Review A: General Physics
\def\prb{\aaref@jnl{Phys.~Rev.~B}}        % Physical Review B: Solid State
\def\prc{\aaref@jnl{Phys.~Rev.~C}}        % Physical Review C
\def\prd{\aaref@jnl{Phys.~Rev.~D}}        % Physical Review D
\def\pre{\aaref@jnl{Phys.~Rev.~E}}        % Physical Review E
\def\prl{\aaref@jnl{Phys.~Rev.~Lett.}}    % Physical Review Letters
\def\pasp{\aaref@jnl{PASP}}               % Publications of the ASP
\def\pasj{\aaref@jnl{PASJ}}               % Publications of the ASJ
\def\qjras{\aaref@jnl{QJRAS}}             % Quarterly Journal of the RAS
\def\skytel{\aaref@jnl{S\&T}}             % Sky and Telescope
\def\solphys{\aaref@jnl{Sol.~Phys.}}      % Solar Physics
\def\sovast{\aaref@jnl{Soviet~Ast.}}      % Soviet Astronomy
\def\ssr{\aaref@jnl{Space~Sci.~Rev.}}     % Space Science Reviews
\def\zap{\aaref@jnl{ZAp}}                 % Zeitschrift fuer Astrophysik
\def\nat{\aaref@jnl{Nature}}              % Nature
\def\iaucirc{\aaref@jnl{IAU~Circ.}}       % IAU Cirulars
\def\aplett{\aaref@jnl{Astrophys.~Lett.}} % Astrophysics Letters
\def\apspr{\aaref@jnl{Astrophys.~Space~Phys.~Res.}}
                % Astrophysics Space Physics Research
\def\bain{\aaref@jnl{Bull.~Astron.~Inst.~Netherlands}} 
                % Bulletin Astronomical Institute of the Netherlands
\def\fcp{\aaref@jnl{Fund.~Cosmic~Phys.}}  % Fundamental Cosmic Physics
\def\gca{\aaref@jnl{Geochim.~Cosmochim.~Acta}}   % Geochimica Cosmochimica Acta
\def\grl{\aaref@jnl{Geophys.~Res.~Lett.}} % Geophysics Research Letters
\def\jcp{\aaref@jnl{J.~Chem.~Phys.}}      % Journal of Chemical Physics
\def\jgr{\aaref@jnl{J.~Geophys.~Res.}}    % Journal of Geophysics Research
\def\jqsrt{\aaref@jnl{J.~Quant.~Spec.~Radiat.~Transf.}}
                % Journal of Quantitiative Spectroscopy and Radiative Transfer
\def\memsai{\aaref@jnl{Mem.~Soc.~Astron.~Italiana}}
                % Mem. Societa Astronomica Italiana
\def\nphysa{\aaref@jnl{Nucl.~Phys.~A}}   % Nuclear Physics A
\def\physrep{\aaref@jnl{Phys.~Rep.}}   % Physics Reports
\def\physscr{\aaref@jnl{Phys.~Scr}}   % Physica Scripta
\def\planss{\aaref@jnl{Planet.~Space~Sci.}}   % Planetary Space Science
\def\procspie{\aaref@jnl{Proc.~SPIE}}   % Proceedings of the SPIE
\def\jcap{\aaref@jnl{J. Cosmology Astropart. Phys.}}
                % Journal of Cosmology and Astroparticle Physics

\let\astap=\aap
\let\apjlett=\apjl
\let\apjsupp=\apjs
\let\applopt=\ao

\newcommand{\etal}{et al.\ }

\newcommand{\mpc}{\, {\rm Mpc}}
\newcommand{\kpc}{\, {\rm kpc}}
\newcommand{\hmpc}{\, h^{-1} \mpc}
\newcommand{\ihmpc}{\, h\, {\rm Mpc}^{-1}}
\newcommand{\ikms}{\, {\rm s\, km}^{-1}}
\newcommand{\kms}{\, {\rm km\, s}^{-1}}
\newcommand{\hkpc}{\, h^{-1} \kpc}
\newcommand{\lya}{Ly$\alpha$\ }
\newcommand{\lyb}{Lyman-$\beta$\ }
\newcommand{\lyaf}{Ly$\alpha$ forest}
\newcommand{\lr}{\lambda_{{\rm rest}}}
\newcommand{\bF}{\bar{F}}
\newcommand{\bS}{\bar{S}}
\newcommand{\bC}{\bar{C}}
\newcommand{\bB}{\bar{B}}
\newcommand{\vdF}{{\mathbf \delta_F}}
\newcommand{\vdS}{{\mathbf \delta_S}}
\newcommand{\vdf}{{\mathbf \delta_f}}
\newcommand{\vdn}{{\mathbf \delta_n}}
\newcommand{\vdC}{{\mathbf \delta_C}}
\newcommand{\vdX}{{\mathbf \delta_X}}
\newcommand{\xrei}{x_{rei}}
\newcommand{\lrmin}{\lambda_{{\rm rest, min}}}
\newcommand{\lrmax}{\lambda_{{\rm rest, max}}}
\newcommand{\lmin}{\lambda_{{\rm min}}}
\newcommand{\lmax}{\lambda_{{\rm max}}}
\newcommand{\hi}{\mbox{H\,{\scriptsize I}\ }}
\newcommand{\heii}{\mbox{He\,{\scriptsize II}\ }}
\newcommand{\vp}{\mathbf{p}}
\newcommand{\vq}{\mathbf{q}}
\newcommand{\vxperp}{\mathbf{x_\perp}}
\newcommand{\vkperp}{\mathbf{k_\perp}}
\newcommand{\vrperp}{\mathbf{r_\perp}}
\newcommand{\vx}{\mathbf{x}}
\newcommand{\vy}{\mathbf{y}}
\newcommand{\vk}{\mathbf{k}}
\newcommand{\vR}{\mathbf{r}}
\newcommand{\tdtwo}{\tilde{b}_{\delta^2}}
\newcommand{\tstwo}{\tilde{b}_{s^2}}
\newcommand{\tbthree}{\tilde{b}_3}
\newcommand{\tadtwo}{\tilde{a}_{\delta^2}}
\newcommand{\tastwo}{\tilde{a}_{s^2}}
\newcommand{\tabthree}{\tilde{a}_3}
\newcommand{\vnabla}{\mathbf{\nabla}}
\newcommand{\tpsi}{\tilde{\psi}}
\newcommand{\vv}{\mathbf{v}}
\newcommand{\fnl}{{f_{\rm NL}}}
\newcommand{\tfnl}{{\tilde{f}_{\rm NL}}}
\newcommand{\gnl}{g_{\rm NL}}
\newcommand{\orderfour}{\mathcal{O}\left(\delta_1^4\right)}
\newcommand{\SDSSPF}{\cite{2006ApJS..163...80M}}
\newcommand{\PF}{$P_F^{\rm 1D}(k_\parallel,z)$}
\newcommand\ion[2]{#1$\;${\small \uppercase\expandafter{\romannumeral #2}}}%  
\newcommand\ionalt[2]{#1$\;${\scriptsize \uppercase\expandafter{\romannumeral #2}}}%  
\newcommand{\vxone}{\mathbf{x_1}}
\newcommand{\vxtwo}{\mathbf{x_2}}
\newcommand{\vRot}{\mathbf{r_{12}}}
\newcommand{\cm}{\, {\rm cm}}

\bibliographystyle{mn2e}
\bibliography{bisp.bib}

\appendix

\section{Shot noise for weighed galaxy mock catalogues}\label{appendix_shot_noise}
In this appendix we propose a formalism to incorporate the completeness weights in the Poisson shot noise terms of the FKP-estimator. The formalism itself is general enough that can be applied to any galaxy catalogue with completeness weights. However, the values for the $x_i$ parameters of Eq.~\ref{P_noise_eff},~\ref{B_noise_eff} and~\ref{Q_noise_eff} must be calibrated to match the unweighted galaxy power spectrum for the specific mocks (here we use \citealt{Maneraetal:2013}).

According to the FKP-estimator, the Poisson shot noise  contribution for the unweighted field $F_2$ (see Eq.~\ref{power_spectrum_FKP}) is given by 
\begin{eqnarray}
\label{P_noise0}P^{\rm nw}_{\rm noise}=I_2^{-1}\int d{\bf r}\, \langle n\rangle({\bf r})\left[ 1 + \alpha^{-1}  \right].
\end{eqnarray}
where $\alpha$ is the ratio between the number of galaxies in the survey and the number in the synthetic random catalog.
When the completeness weights $w_{\rm c}$ (and systematic weights $w_{\rm sys}$) are introduced into the formalism, the shot noise depends on them. In this appendix we assume that the systematic weights do not modify the shot noise when they are added. This behaviour is expected from the fact that, although the correction is not random, is related to a Poisson process, such as the presence of a galaxy around a star (see \S~\ref{sec:data} for details).
On the other hand, recall that the completeness weights are included to take into account galaxies whose radial position (redshift) is unknown.
For the CMASS DR11 sample this can arise, for instance, because of fiber collisions and redshift failures (see \S~\ref{sec:data} for a complete discussion). In the end, the effect of the  completeness weighting process is to remove the affected galaxy and to upweight a nearby one. The missing and the up-weighted galaxies are angularly close, but we do not know if they are a true pair or just a chance alignment.
If all of these angular pairs  were true pairs, the weighting process would not modify the large-scale shot noise, in the same way that a smoothing filter of the galaxy field does not change the large-scale shot noise. In this case, if we assume that the shot noise is Poisson, the correlation function of the weighted number density of galaxies would read,
\begin{eqnarray}
\langle w_c({\bf r}_1)n({\bf r}_1)w_c({\bf r}_2)n({\bf r}_2)\rangle=\langle w_c n \rangle({\bf r}_1)\langle w_c n \rangle({\bf r}_2)\left[1+\xi_{\rm gal}({\bf r}_1-{\bf r}_2)  \right]+w_{\rm sys}({\bf r}_2)\langle w_c n \rangle({\bf r}_1)\delta^D({\bf r}_1-{\bf r}_2),
\end{eqnarray}
and therefore the corresponding  shot noise term is, 
\begin{eqnarray}
\label{P_noise2}P^{\rm (true\, pairs)}_{\rm noise}=I_2^{-1}\int d{\bf r}\, w^2_{\rm FKP}({\bf r})\langle w_c n\rangle({\bf r})\left[ w_{\rm sys}({\bf r}) + \alpha  \right].
\end{eqnarray}
On the other hand, if all these angular pairs  were not true pairs, the process of removing one and  up-weighting the other introduces extra shot noise. In this case the correlation function of galaxies would read,
\begin{eqnarray}
\langle w_c({\bf r}_1)n({\bf r}_1)w_c({\bf r}_2)n({\bf r}_2)\rangle=\langle w_c n \rangle({\bf r}_1)\langle w_c n \rangle({\bf r}_2)\left[1+\xi_{\rm gal}({\bf r}_1-{\bf r}_2)  \right]+w_{\rm c}({\bf r}_2)\langle w_c n \rangle({\bf r}_1)\delta^D({\bf r}_1-{\bf r}_2),
\end{eqnarray}
and  therefore the shot noise term is, 
\begin{eqnarray}
\label{P_noise1}P^{\rm (false\, pairs)}_{\rm noise}=I_2^{-1}\int d{\bf r}\, w^2_{\rm FKP}({\bf r})\langle w_c n\rangle({\bf r})\left[ w_c({\bf r}) + \alpha \right].
\end{eqnarray}
We can write these two extreme cases in a more compact way,
\begin{eqnarray}
\label{P_noisei}P^{(i)}_{\rm noise}=I_2^{-1}\int d{\bf r}\, w^2_{\rm FKP}({\bf r})\langle w_c n\rangle({\bf r})\left[ w_{ i}({\bf r}) + \alpha  \right],
\end{eqnarray}
where $i$ can be ``true pairs'' and $w_{ i}$ is $w_{\rm sys}$, or  $i$ corresponds to ``false pairs" and  $w_{ i}$ is $w_{\rm c}$. Reality will be an intermediate case where a fraction of the missing  galaxies are true paris and the rest are chance alignments.
We propose a parametrisation of the effective shot noise as,
\begin{eqnarray}
\label{P_noise_eff}P_{\rm noise}=x_{\rm PS}P_{\rm noise}^{\rm(false\, pairs)}+(1-x_{\rm PS})P_{\rm noise}^{\rm (true\, pairs)}.
\end{eqnarray}
where $x_{\rm PS}$ is a free parameter 
between 0 and 1 to be fitted from the galaxy mocks.

\begin{figure*}
\centering
\includegraphics[clip=false, trim= 18mm 0mm 22mm 0mm,scale=0.24]{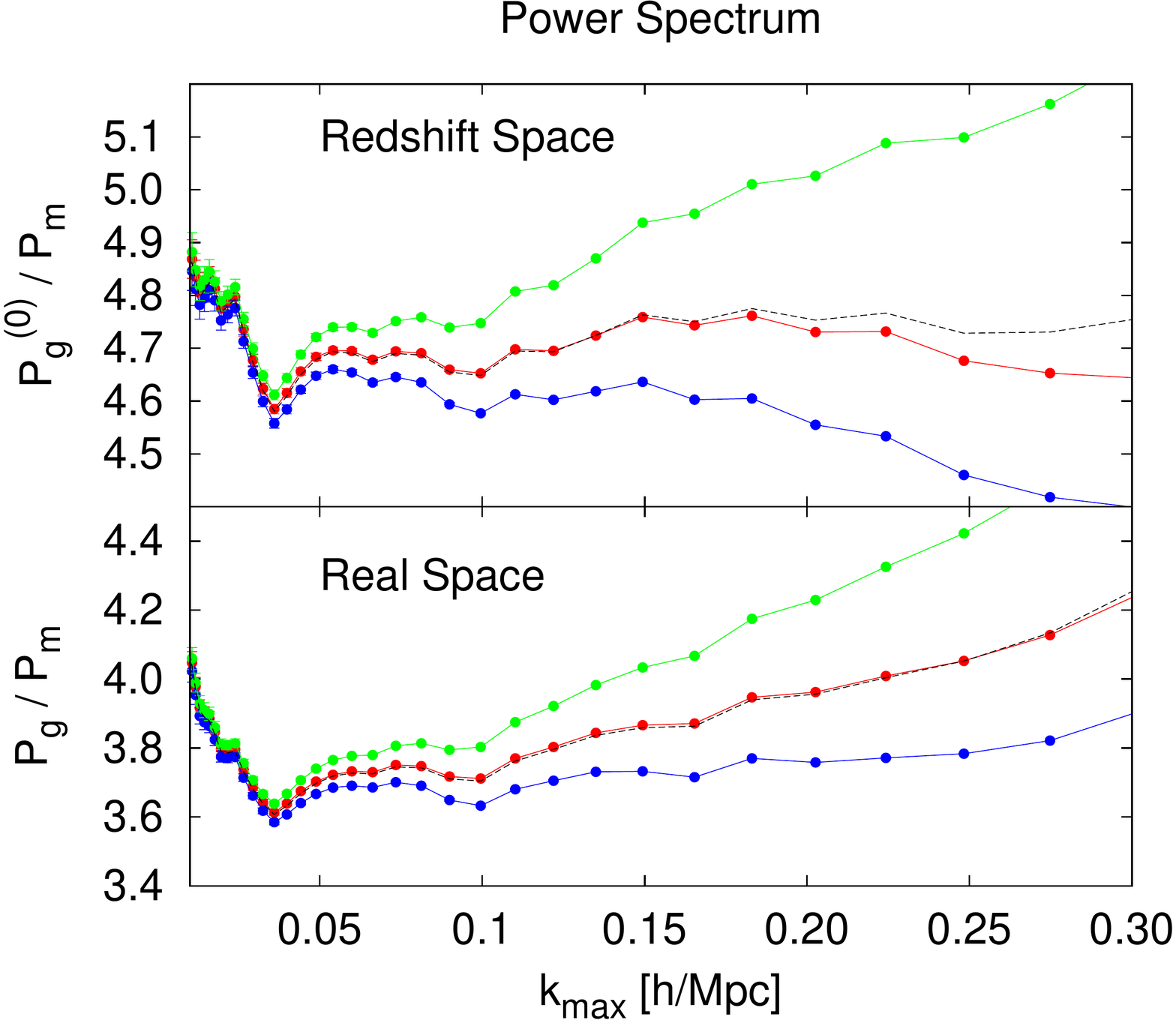}
\includegraphics[clip=false, trim= 18mm 0mm 22mm 0mm,scale=0.24]{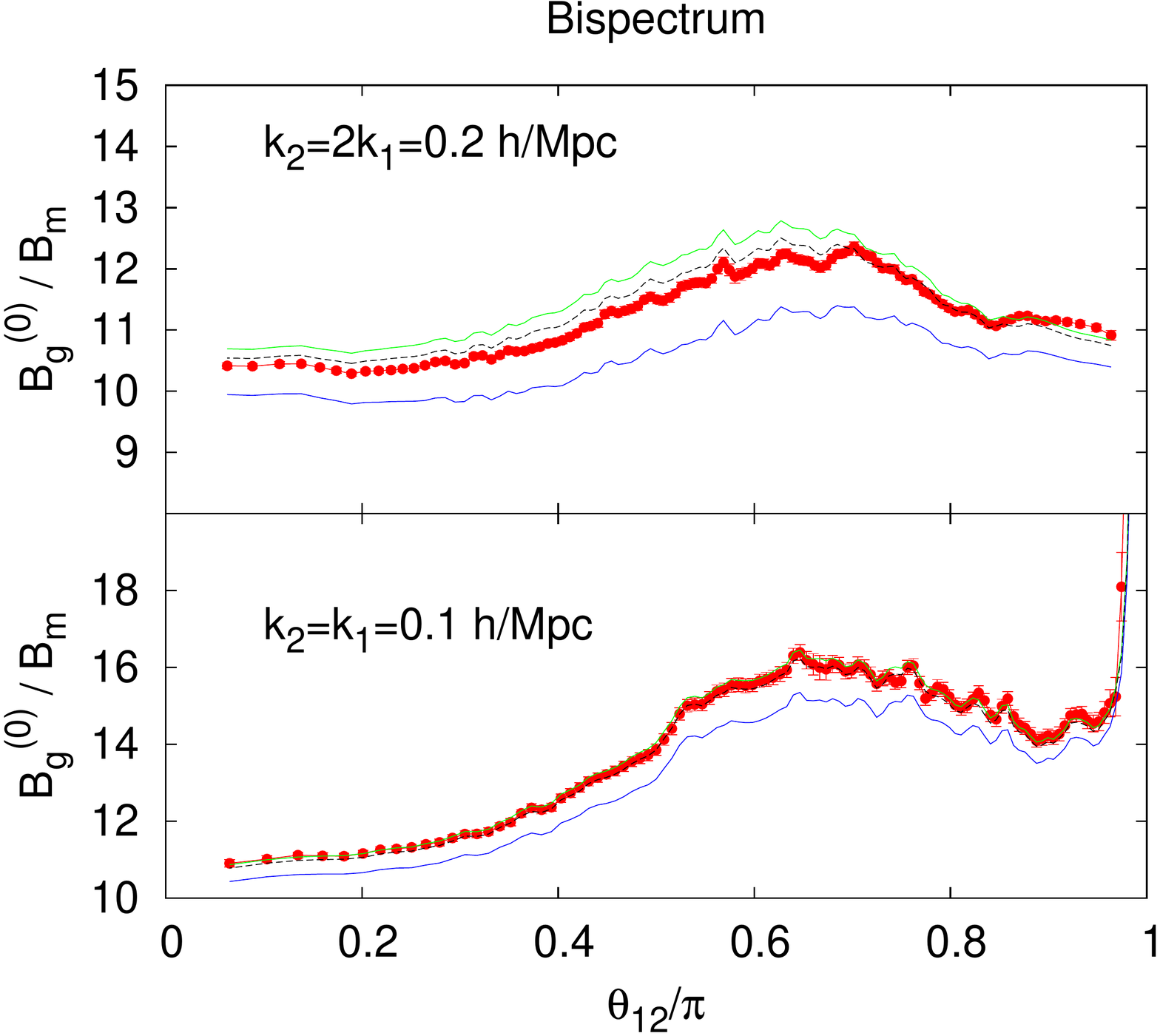}
\includegraphics[clip=false, trim= 18mm 0mm 22mm 0mm,scale=0.24]{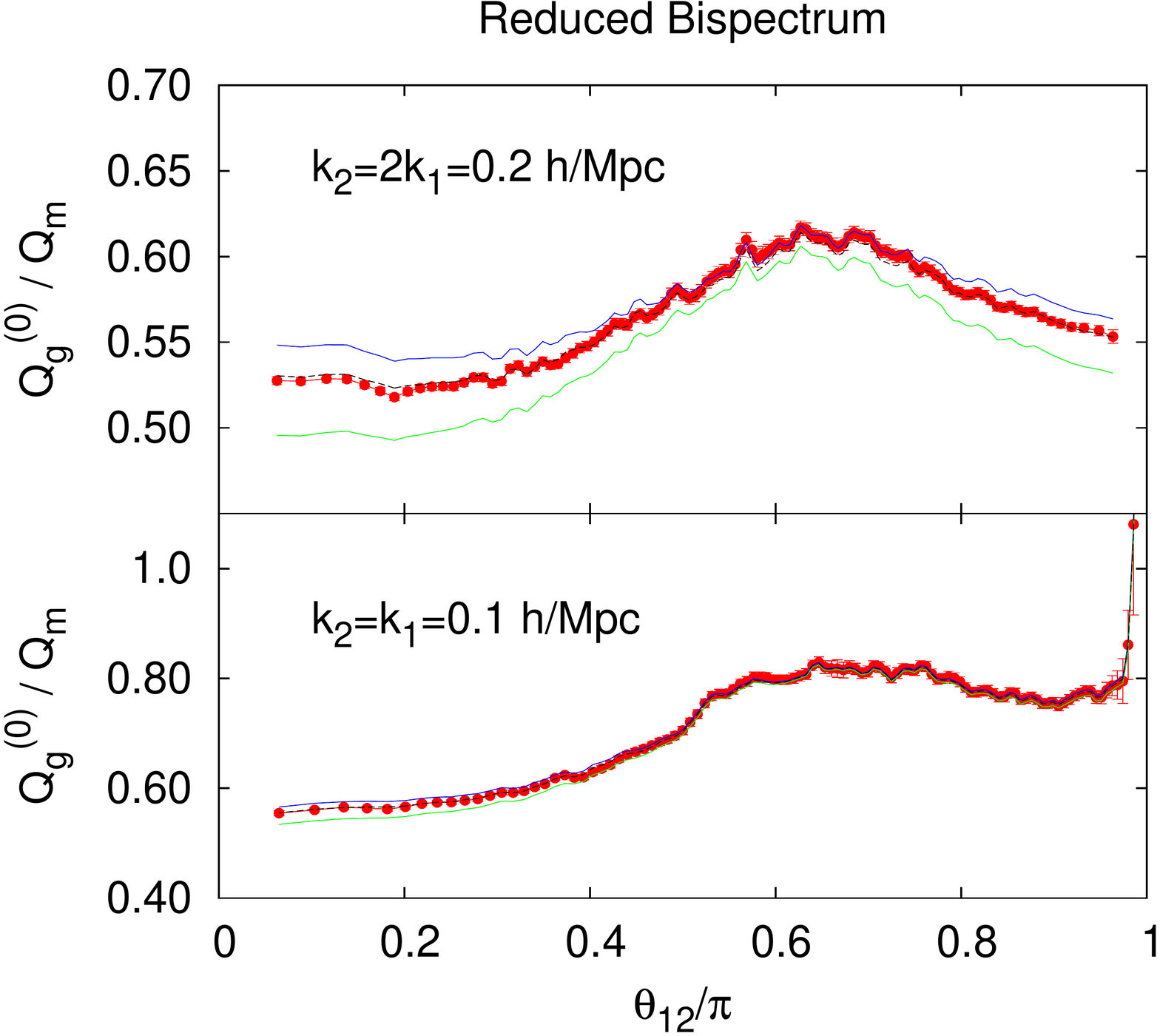}
\caption{Left panel:  the power spectrum normalised by the non-linear matter (convolved with the corresponding window)  for the unweighted galaxy mocks (red line) and for the weighted mocks with a subtraction according to $P_{\rm noise}^{({{\rm false\, pairs}})}$ (blue line) and $P_{\rm noise}^{({\rm true\, paris })}$ (green line). Our proposed  model of  Eq.~\ref{P_noise_eff} is shown in dashed black line for $x_{\rm PS}=0.58$ and is able to accurately describe the unweighted galaxy mocks for the $k\lesssim0.20\,h{\rm Mpc}^{-1}$. As labeled, the upper panel presents redshift space quantities and the lower panel the real space value. The central and right panels show the redshift space monopole of the bispectrum and reduced bispectrum, respectively, normalised by the non-linear matter bispectrum model of Eq.~\ref{B_ggg2}, for two different shapes, $k_1/k_2=1,\,2$, as labeled. The colour notation is the same as in the left panels. In this case the black line represents our proposed model of  Eq.~\ref{B_noise_eff} with $x_{\rm Bis}=0.2$ and Eq.~\ref{Q_noise_eff} with $x_{\rm Q}=0.66$ for the reduced bispectrum. Also for the bispectrum, our proposed model describe accurately the unweighted measurements.}
\label{noise_plot}
\end{figure*}

In the left panel of Fig.~\ref{noise_plot} we show the comparison of these two shot noise predictions 
 for the power spectrum of the galaxy mocks:  unweighted galaxy power spectrum (red line),  weighted galaxy power spectrum with the shot noise assumption of Eq.~\ref{P_noise1} (blue line), and weighted galaxy power spectrum with the shot noise assumption of Eq.~\ref{P_noise2} (green line). In this case the galaxy power spectrum has been normalised by the non-linear matter power spectrum for clarity. For $x_{\rm PS}=0.58$, our proposed {\it ansatz} of Eq.~\ref{P_noise_eff} produces  a good fit to the unweighted true distribution (black dotted lines) up to $k_{\rm max}\sim 0.18\, h{\rm Mpc}^{-1}$ in redshift space (much larger $k$ in real space). 
This result indicates that the maximum $k$ for our final joint power- and bi- spectra analysis  should be close to and  not be much larger than this value. 

The same argument used for the power spectrum can be applied to the bispectrum. The unweighted quantity for the shot noise when is assumed Poisson is given by,
\begin{eqnarray}
\nonumber B^{\rm nw}_{\rm noise}({\bf k}_1,{\bf k}_2)&=&\frac{I_2}{I_{3}} \int \frac{d{\bf k'}}{(2\pi)^3}\, P_{\rm gal}({\bf k}') |W^{\rm nw}_2({\bf k}_1-{\bf k}')|^2+\rm{cyc.}\\
&+&I_{3}^{-1}\int d{\bf r}\, \langle n \rangle({\bf r})\left[ 1 - \alpha^2 \right],
\label{B_noise_nw}
\end{eqnarray}
with 
\begin{eqnarray}
W_2^{\rm nw}({\bf k})\equiv I_{2}^{-1/2}\int d^3{\bf r}\, {\langle n\rangle}({\bf r})e^{+i{\bf k}\cdot{\bf r}}.
\end{eqnarray}
As before, the  (Poisson) shot noise contribution for the bispectrum depends on whether the angular  triplets are true triplets or not (note that for simplicity we do not consider mix triplets between true and false). Expanding this expression produces,
\begin{eqnarray}
\nonumber B^{(i)}_{\rm noise}({\bf k}_1,{\bf k}_2)&=&\frac{I_2}{I_{3}} \int \frac{d{\bf k'}}{(2\pi)^3}\, P_{\rm gal}({\bf k}') W_2^*({\bf k}_1-{\bf k}') \widetilde{W}_{2}^{(i)}({\bf k}_1-{\bf k}')+\rm{cyc.}\\
\label{B_noise_i}&+&I_{3}^{-1}\int d{\bf r}\, \langle w_c n \rangle({\bf r})w^3_{\rm FKP}({\bf r})\left[ w_i^2({\bf r}) - \alpha^2 \right],
\end{eqnarray}
where we have introduced $\widetilde W_2^{(i)}$ as,
\begin{eqnarray}
\widetilde W_2^{(i)}({\bf k})\equiv I_{2}^{-1/2}\int d^3{\bf r}\, w^2_{\rm FKP}({\bf r})w_i({\bf r}){\langle w_c n\rangle}({\bf r})e^{+i{\bf k}\cdot{\bf r}}.
\end{eqnarray}
and $W_2$ is the same as defined in Eq.~\ref{G_2_definition},
\begin{eqnarray}
W_2({\bf k})\equiv I_{2}^{-1/2}\int d^3{\bf r}\, w_{\rm FKP}({\bf r}){\langle w_c n\rangle}({\bf r})e^{+i{\bf k}\cdot{\bf r}}.
\end{eqnarray}
Our goal is to write Eq.~\ref{B_noise_i} as a function of the measured power spectrum. We define,
\begin{eqnarray}
\mathcal{A}^{(i)}\equiv\int d{\bf r}\, \langle w_i({\bf r})n_g({\bf r})\rangle^2({\bf r}) w_i({\bf r}) w^3_{\rm FKP},
\end{eqnarray}
which provides the normalization for the power spectrum convolution of Eq.~\ref{B_noise_i}. Thus, we can perform the approximation,
\begin{eqnarray}
\frac{I_{2}}{\mathcal{A}^{(i)}}\int \frac{d{\bf k'}}{(2\pi)^3}\, P_{\rm gal}({\bf k}') W_2^*({\bf k}-{\bf k}') \widetilde{W}_{2}^{(i)}({\bf k}-{\bf k}')\simeq\int \frac{d{\bf k'}}{(2\pi)^3}\, P_{\rm gal}({\bf k}') |W_2({\bf k}-{\bf k}') |^2=\langle |F_2({\bf k})|^2\rangle-P^{(i)}_{\rm noise},
\end{eqnarray}
which should be a accurate assumption, especially at small scales where the shot noise term is important.
Thus, finally we write Eq.~\ref{B_noise_i} in terms of the measured power spectrum $\langle |F_2({\bf k})|^2\rangle$,
\begin{eqnarray}
B^{(i)}_{\rm noise}({\bf k}_1,{\bf k}_2)&=&\frac{\mathcal{A}^{(i)}}{I_3}\left[ \langle |F_2({\bf k}_1)|^2\rangle+{\rm cyc.}-3P^{(i)}_{\rm noise}  \right]+I_3^{-1}\int d{\bf r}\, \langle w_c n \rangle({\bf r}) w^3_{\rm FKP}({\bf r})\left[ w_i^2({\bf r}) - \alpha^2  \right].
\end{eqnarray}
In a similar approach as was used for the power spectrum, we can approximate the effective (Poisson) shot noise term for the bispectrum as,
\begin{equation}
\label{B_noise_eff}B_{\rm noise}({\bf k}_1,{\bf k}_2)=x_{\rm Bis}B_{\rm noise}^{\rm(false\,triplets)}({\bf k}_1,{\bf k}_2)+(1-x_{\rm Bis})B_{\rm noise}^{\rm(true\, triplets)}({\bf k}_1,{\bf k}_2).
\end{equation}
Finally, combining the shot noise terms obtained for the power spectrum and bispectrum, we can write the  (Poisson)   shot noise terms for the reduced bispectrum $Q$ as,
\begin{eqnarray}
Q_{\rm noise}^{\rm nw}(k_1,k_2,k_3)=\frac{B_{\rm noise}^{\rm nw}(k_1,k_2,k_3)}{[P(k_1)-P_{\rm noise}^{\rm nw}][P(k_2)-P_{\rm noise}^{\rm nw}]+\rm{cyc.}},
\end{eqnarray}

\begin{eqnarray}
Q_{\rm noise}^{(i)}(k_1,k_2,k_3)=\frac{B_{\rm noise}^{(i)}(k_1,k_2,k_3)}{[P(k_1)-P_{\rm noise}^{(i)}][P(k_2)-P_{\rm noise}^{(i)}]+\rm{cyc.}},
\end{eqnarray}
and therefore the effective term of the Poisson  shot noise for the reduced bispectrum is,
\begin{equation}
\label{Q_noise_eff}Q_{\rm noise}({\bf k}_1,{\bf k}_2)=x_{\rm Q}Q_{\rm noise}^{\rm(false\,triplets)}({\bf k}_1,{\bf k}_2)+(1-x_{\rm Q})Q_{\rm noise}^{\rm(true\,triplets)}({\bf k}_1,{\bf k}_2)
\end{equation}
In the central panel of Fig.~\ref{noise_plot} we show the redshift space monopole galaxy bispectrum (normalised by the corresponding non-linear matter bispectrum) of the unweighed galaxy catalogue with the shot noise subtraction of Eq.~\ref{B_noise_nw} (red), and for the weighed galaxy catalogue when  the shot noise term subtracted is $B_{\rm noise}^{({\rm false\, triplets})}$ (blue lines) and $B_{\rm noise}^{({\rm true\,triplets})}$ (green lines). The black dashed lines display the interpolated model of Eq.~\ref{B_noise_eff} with the fitted value $x_{\rm Bis}=0.20$. The top panel presents the bispectrum for the shape $k_2/k_1=2$, whereas the bottom panel for $k_2/k_1=1$ as indicated. In the right panel the same formalism applied to the reduced bispectrum $Q$. In this case, the interpolation parameter has been set to $x_{\rm Q}=0.66$.
Note that $x_Q$ could be in principle related to $x_P$ and $x_B$, since $Q$, $P$ and $B$ are related. However, this relation is far from being simple as Eq. \ref{Q_noise_eff} and the functional between $Q_{\rm noise}$, $Q_{\rm noise}^{\rm(false\,triplets)}$ and $Q_{\rm noise}^{\rm(true\,triplets)}$ is not linear. In this paper, we have tried a linear relation, treating $x_Q$ as a free parameter. Given that the performance of $x_Q$ in Fig. A1 seems pretty similar to $x_B$, we assume that Eq. A19 is a good approximation of the full relation given by $x_B$ and $x_P$.

To conclude, in this paper we  always  assume that the Poisson shot noise prediction of the weighted galaxy catalogues by \cite{Maneraetal:2013} is given by Eqs.~\ref{P_noise_eff},~\ref{B_noise_eff} and~\ref{Q_noise_eff} with the values summarised in Table~\ref{x_table}.
\begin{table}
\begin{center}
\begin{tabular}{|c|c|c|}
\hline
 & $x_{(P,B,Q)}$ & Eq. \\
 \hline
 \hline
 $P$ & 0.58 &~\ref{P_noise_eff}\\
 \hline
  $B$ & 0.20 &~\ref{B_noise_eff}\\
\hline
 $Q$ & 0.66 &~\ref{Q_noise_eff}\\
\end{tabular}
\end{center}
\caption{Interpolation values $x_{P},x_{B},x_{Q}$ used for the shot noise weighted statistics.}
\label{x_table}
\end{table}%

\section{Power spectrum in redshift space}\label{appendix_power_spectrum}
\begin{figure*}
\centering
\includegraphics[scale=0.32]{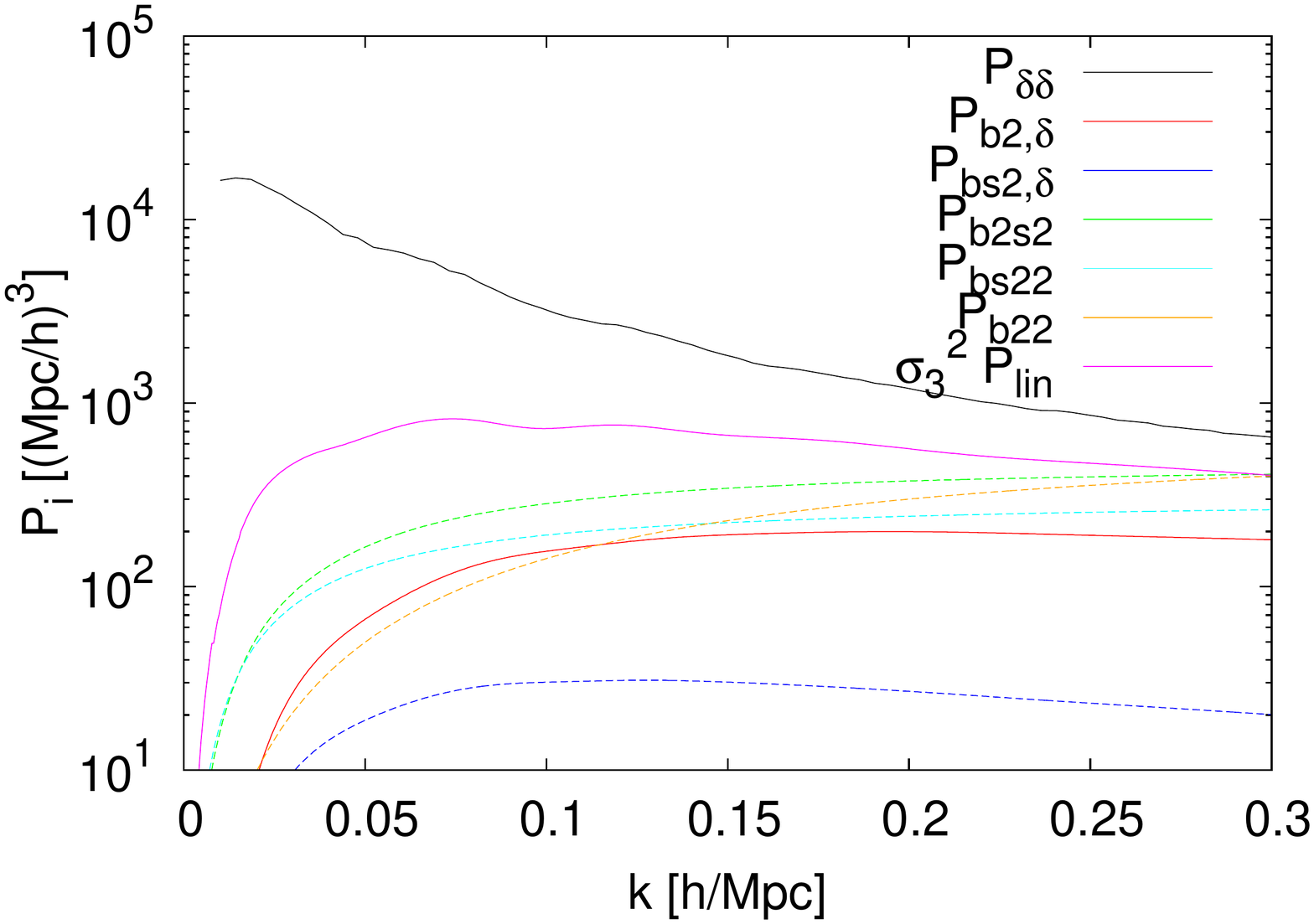}
\includegraphics[scale=0.29]{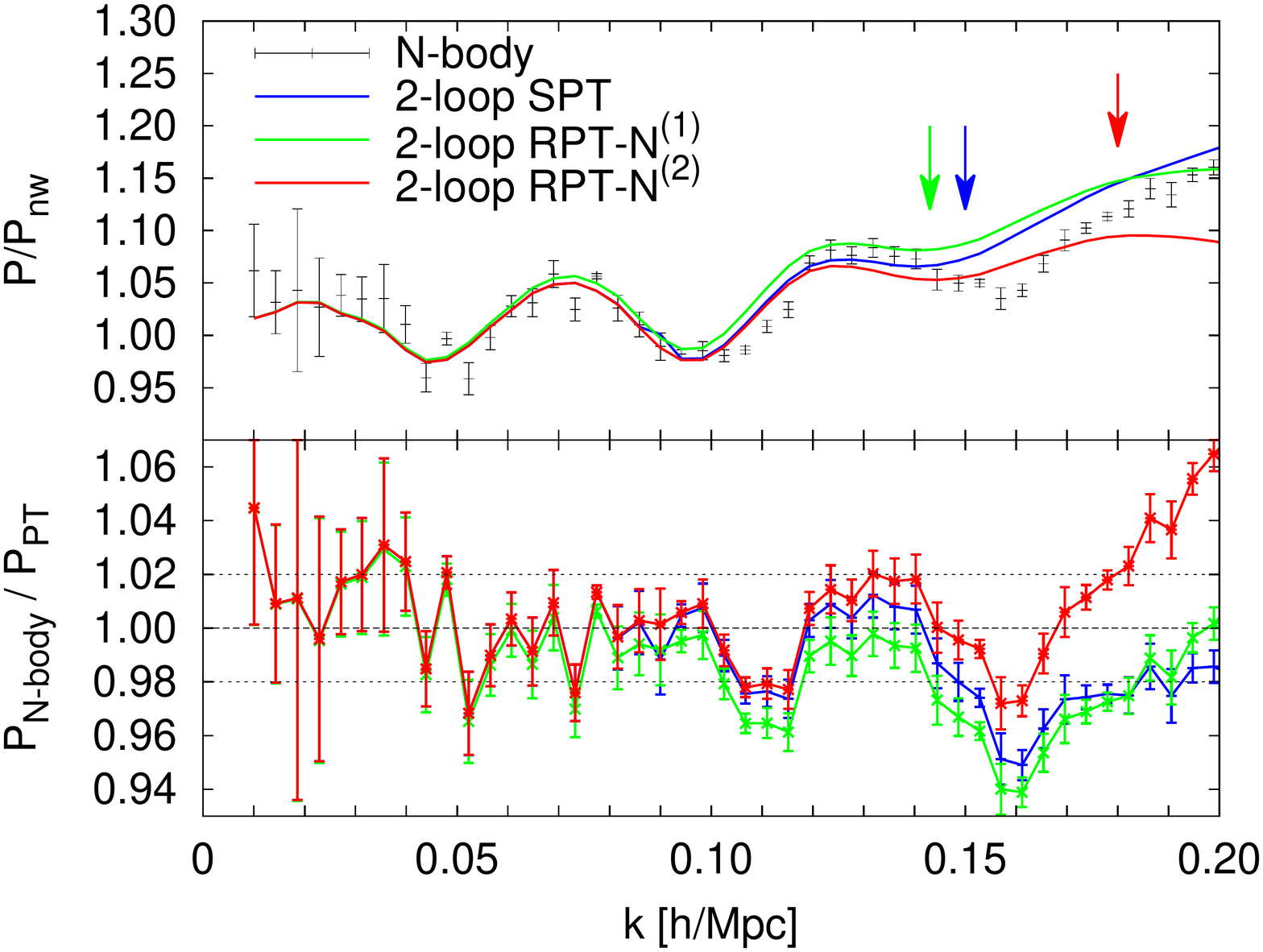}
\caption{{ Left Panel}: The different contributions of Eq.~\ref{b2_terms1}-~\ref{b2_terms2}. { Right panel}: Perturbation theory and N-body simulation predictions for the dark matter power spectrum $P_{\delta\delta}$. The top panel displays the actual power spectrum normalised by a non-wiggle linear model for clarity. Bottom panel shows the relative difference of each PT model to the N-body simulations. Blue lines correspond to SPT, green lines to RPT-${\mathcal N}^{(1)}$ and red lines to RPT-${\mathcal N}^{(2)}$. The arrows indicate where each model starts to deviate with respect to N-body mocks higher than $2\%$. The cosmology chosen is the same of the galaxy mocks described in \S~\ref{sec:sims-mocks} at $z=0.55$. The errors of N-body correspond to the error of the mean among five different realisations, with a total effective volume of $V_{\rm eff}=16.875\,{\rm Mpc}h^{-1}$.}
\label{appendixb_plot}
\end{figure*}

In this appendix we specify  the formulae we use to compute the galaxy power spectrum in redshift space. The full formulae derivation can be found in the papers cited by the equations.
The starting point is the non-local bias model given in Eq.~\ref{deltah}. From there we  obtain  the real space power spectrum,
\begin{eqnarray}
P_{{g},\delta\delta}(k)&\!\!\!=\!\!\!&b_1^2 P_{\delta\delta}(k)+2b_2b_1P_{b2,\delta}(k)+2b_{s2}b_1P_{bs2,\delta}(k)+b_2^2P_{b22}(k)+2b_2b_{s2}P_{b2s2}(k)+b_{s2}^2P_{bs22}(k)+2b_1b_{3\rm nl}\sigma_3^2(k)P^{\rm lin}(k)
\end{eqnarray}
where $P_{\delta\delta}$ and $P^{\rm lin}$ are the non-linear and linear matter power spectra. The power spectra that multiply the bias parameters $b_2$ and $b_{s}$ can be given by the following 1-loop integrals  \citep{McDonaldRoy:2009,Beutleretal:2013},
\begin{eqnarray}
\label{b2_terms1}P_{b2,\delta}&=&\int \frac{d^3q}{(2\pi)^3}\, P^{\rm lin}(q)P^{\rm lin}(|{\bf k}-{\bf q}|) {\cal F}_2^{\rm SPT}({\bf q},{\bf k-q}),\\
P_{bs2,\delta}&=& \int \frac{d^3q}{(2\pi)^3}\, P^{\rm lin}(q)P^{\rm lin}(|{\bf k}-{\bf q}|) {\cal F}_2^{\rm SPT}({\bf q},{\bf k-q}) S_2({\bf q},{\bf k-q}),  \\
P_{b2s2}&=& -\frac{1}{2}\int \frac{d^3q}{(2\pi)^3}\, P^{\rm lin}(q)\left[ \frac{2}{3}P^{\rm lin}(q) - P^{\rm lin}(|{\bf q}-{\bf k}|)S_2({\bf q},{\bf k}-{\bf q}) \right],  \\
P_{bs22}&=& -\frac{1}{2} \int \frac{d^3q}{(2\pi)^3}\, P^{\rm lin}(q)\left[  \frac{4}{9}P^{\rm lin}(q) - P^{\rm lin}(|{\bf k}-{\bf q}|) S_2({\bf q},{\bf k}-{\bf q})^2\right], \\
P_{b22}&=& -\frac{1}{2} \int \frac{d^3q}{(2\pi)^3}\, P^{\rm lin}(q)\left[ P^{\rm lin}(q)-P^{\rm lin}({\bf k}-{\bf q}|)\right],\\
\label{b2_terms2}\sigma^2_3(k)&=&\int\frac{d^3{\bf q}}{(2\pi)^3}\,P^{\rm lin}(q)\left[ \frac{5}{6}+\frac{15}{8}S_2({\bf q},{\bf k}-{\bf q})S_2(-{\bf q},{\bf k})-\frac{5}{4}S_2({\bf q},{\bf k}-{\bf q})  \right].
\end{eqnarray} 
The $S_2$ kernel is given in Eq.~\ref{eq:S2kernel} and the ${\cal F}_2^{\rm SPT}$ kernel (e.g.,  \citealt{Goroff:1986, CM94a, CM94b} and \citealt{Bernardeauetal:2002} for a review)  is given by,
\begin{equation}
\label{eq:F2_kernel}
 {\cal F}_2^{\rm SPT}({\bf k}_i,{\bf k}_j)= \frac{5}{7}+\frac{1}{2}\frac{{\bf k}_i\cdot{\bf k}_j}{k_ik_j}\left(\frac{k_i}{k_j}+\frac{k_j}{k_i}\right)+\frac{2}{7}\left[ \frac{{\bf k}_i\cdot{\bf k}_j}{k_ik_j} \right]^2\,.
\end{equation}
These integrals can be reduced to 2-dimensional integrals due to rotational invariance of the linear power spectrum.
These contributions are illustrated in the left panel of Fig.~\ref{appendixb_plot}.

 To  obtain the redshift space power spectrum we also need the terms $P_{g\theta}$ and $P_{\theta\theta}$. Since we  assume no velocity bias, $P_{\theta\theta}$ is the same for non-linear matter and galaxies,
\begin{eqnarray}
P_{g\theta}(k)=b_1P_{\delta\theta}(k)+b_2P_{b2,\theta}(k)+b_{s2}P_{bs2,\theta}(k)+b_{3\rm nl}\sigma_3^2(k)P^{\rm lin}(k),
\end{eqnarray}
where $P_{\delta\theta}$ is the matter density-velocity non-linear power spectrum, and the other two terms are given by 1-loop integrals,
\begin{eqnarray}
P_{b2,\theta}(k)&=&\int\frac{d^3q}{(2\pi)^3}\,P^{\rm lin}(q)P^{\rm lin}(|{\bf k}-{\bf q}|){\cal G}_2^{\rm SPT}({\bf q},{\bf k}-{\bf q}),\\
P_{bs2,\theta}(k)&=&\int\frac{d^3q}{(2\pi)^3}\,P^{\rm lin}(q)P^{\rm lin}(|{\bf k}-{\bf q}|){\cal G}_2^{\rm SPT}({\bf q},{\bf k}-{\bf q})S_2({\bf q},{\bf k}-{\bf q}).\\
\end{eqnarray}
The ${\cal G}_2^{\rm SPT}$ kernels are \citep{Goroff:1986, CM94a, CM94b},
\begin{equation} 
\label{G_kernel}{\cal G}_2^{\rm SPT}({\bf k}_i,{\bf k}_j)= \frac{3}{7}+\frac{1}{2}\frac{{\bf k}_i\cdot{\bf k}_j}{k_ik_j}\left(\frac{k_i}{k_j}+\frac{k_j}{k_i}\right)+\frac{4}{7}\left[ \frac{{\bf k}_i\cdot{\bf k}_j}{k_ik_j} \right]^2.
\end{equation}

The kernels  ${\cal F}_2^{\rm SPT}$ and ${\cal G}_2^{\rm SPT}$ have only weak cosmology dependence  \citep{Bouchetetal:1992, CLMM95, Bernardeau94b, Eis97,MVH97,KamionBuchalter99}.
Once we have the real space quantities, $P_{gg}$, $P_{g\theta}$ and $P_{\theta\theta}$, the redshift space power spectrum can be written using the mapping provided by \cite{Taruyaetal:2010,Nishimichietal:2011},
\begin{eqnarray}
P_{g}^{(s)}(k,\mu)=D^P_{\rm FoG}(k,\mu,\sigma_{\rm FoG}^P[z])\left[ P_{g,\delta\delta}(k)+2f\mu^2P_{g,\delta\theta}(k)+f^2\mu^4P_{\theta\theta}(k)+b_1^3A^{\rm TNS}(k,\mu,f/b_1)+b_1^4B^{\rm TNS}(k,\mu,f/b_1)  \right]
\end{eqnarray}
where  $A^{\rm TNS}$ and $B^{\rm TNS}$ are correction terms  arising from  the coupling between  the Kaiser and the Fingers-of-God effects. 
The expression of these terms  (to leading order  for the bias) is given in \cite{Taruyaetal:2010},
\begin{eqnarray}
\label{taruya_a}A^{\rm TNS}(k,\mu,b)&=& (k\mu f)\int \frac{d^3{\bf q}}{(2\pi)^3}\frac{q_z}{q^2}\left\{B_\sigma({\bf q}, {\bf k}-{\bf q}, -{\bf k})-B_\sigma({\bf q},{\bf k}, -{\bf k}-{\bf q})\right\}, \\
\label{taruya_b}B^{\rm TNS}(k,\mu,b)&=&(k\mu f)^2\int\frac{d^3{\bf q}}{(2\pi)^3}F^{\rm TNS}({\bf q})F^{\rm TNS}({\bf k}-{\bf q}),
\end{eqnarray}
where,
\begin{equation}
F^{\rm TNS}({\bf q})\equiv\frac{q_z}{q^2}\left\{b_1P_{\delta\theta}(q)+f\frac{q_z^2}{q^2}P_{\theta\theta}(q)\right\},
\end{equation}
and
\begin{eqnarray}
\nonumber(2\pi)^3\delta_D({\bf k}_{123})B_\sigma({\bf k}_1,{\bf k}_2,{\bf k}_3)\equiv\left\langle \theta({\bf k}_1)\left\{b_1\delta({\bf k}_2)+f\frac{k_{2z}^2}{k_2^2}\theta({\bf k}_2)\right\}\left\{b_1\delta({\bf k}_3)+f\frac{k_{3z}^2}{k_3^2}\theta({\bf k}_3)\right\} \right\rangle,\\
\end{eqnarray}
with ${\bf k}_{123}\equiv{\bf k}_1+{\bf k}_2+{\bf k}_3$. Since we expect $A^{\rm TNS}$ and $B^{\rm TNS}$ to be small compared to $P_{\delta\delta}$, $P_{\delta\theta}$ and $P_{\theta\theta}$,  we have assumed that only the leading terms for the galaxy power spectrum and bispectrum contribute in the integrals of Eq.~\ref{taruya_a} and~\ref{taruya_b}. In other words,  $P_{\delta\delta}$, $P_{\delta\theta}$ and $P_{\theta\theta}$ are approximated by $P^{\rm lin}$ and $B_{\delta\delta\theta}$, $B_{\delta\theta\theta}$, $B_{\theta\theta\theta}$ by the corresponding tree level quantities in Eqs.~\ref{taruya_a} and~\ref{taruya_b}. 

The function $D^P_{\rm FoG}$ accounts for the the fully non-linear damping due to the velocity dispersion of satellite galaxies inside the host halo which we parametrise through a one-free parameter  Lorentzian  distribution,
\begin{eqnarray}
\label{DfogP}D^P_{\rm FoG}(k,\mu,\sigma^P_{\rm FoG}[z])=\left( 1+k^2\mu^2\sigma_{\rm FoG}^P[z]^2/2  \right)^{-2}.
\end{eqnarray}

We must specify the procedure to  compute the non-linear matter power spectra, $P_{\delta\delta}$, $P_{\delta\theta}$ and $P_{\theta\theta}$. One option would be to run a suite of N-body simulations and measure these quantities {\it in situ}. However, if we want to change the cosmology we would need to re-run  simulations with the new cosmological parameters, which would be prohibitively expensive. More importantly,  the quantities that involves the $\theta$-field need are delicate to compute as there are many  grid-cells with no particles (Voronoi tessellation methods can be used to address this issue). Here we adopt the approach of using analytical expressions  based on  perturbation theory. According to standard perturbation theory (SPT), the 2-loop prediction for the power spectrum reads 
(e.g., see \citealt{JainBertschinger94,MakinoSasakiSuto92} for the first pioneering studies)
\begin{eqnarray}
P^{\rm SPT}_{ij}(k)&=&P^{\rm lin}(k)+2P_{ij}^{(13)}(k)+P_{ij}^{(22)}(k)+2P_{ij}^{(15)}(k)+2P_{ij}^{(24)}(k)+P_{ij}^{(33)}(k)
\end{eqnarray}
with the compact notation where $i$ and $j$ can be $\delta$ and $\theta$. The terms $P_{ij}^{(13)}$, $P_{ij}^{(22)}$, $P_{ij}^{(15)}(k)$, $P_{ij}^{(24)}$ and $P_{ij}^{(33)}$ can be found in the references above and any perturbation theory review (see for e.g. \citealt{Bernardeauetal:2002} among many others).
\cite{CrocceScoccimarro:2006} proposed a reorganization of the infinite terms of the SPT series and a resummation of part of them in what  is called the {\it resummed propagator}. In this this formalism, the behaviour when truncating the infinite series at certain loop improves moderately with respect to SPT. According to this resummed perturbation theory (hereafter RPT), at 2-loop truncation the power spectrum reads,
\begin{eqnarray}
\label{Pab}P^{\rm RPT}_{ij}(k)=[P^{\rm lin}(k,z)+P_{ij}^{(22)}(k,z)+P_{ij}^{(33)\,-2{\rm L}}(k,z)]\,\mathcal{N}^{(i)}_{ij},
\end{eqnarray}
where $P_{ij}^{(33)\,-2{\rm L}}$ is the part of $P_{ij}^{(33)}$ that accounts for the full 2-loop coupling, and $\mathcal{N}_{ij}$ is the resummed propagator. The full expression of the resummed propagator $\mathcal N$ depends on how all the infinite terms of the series have been approximated just before the resummation. When these terms are resummed using 1-loop kernels we refer to the resummed propagator as ${\mathcal N}^{(1)}$. However,  the propagator can also be resummed using higher order loop kernels. In general we refer to the resummed propagator using $\ell$-loop kernels as, ${\mathcal N}^{(\ell)}$. The expressions for 1- and 2-loop can be found respectively in \cite{CrocceScoccimarro:2006} and \cite{HGMetal:2012} and read,
\begin{eqnarray}
{\mathcal N}^{(1)}_{ij}(k)&\equiv&\exp\left[P^{(13)}_{ij}(k)/P^{\rm lin}(k)\right],\\
\mathcal{N}^{(2)}_{ij}(k)&\equiv& \cosh\left[  \sqrt{ \frac{2P_{ij}^{(15)}(k)}{P^{\rm lin}(k)} }     \right] + \frac{P_{ij}^{(13)}(k)}{P^{\rm lin}(k)}\sqrt{\frac{P^{\rm lin}(k)}{2P_{ij}^{(15)}(k)}}\sinh\left[  \sqrt{ \frac{2P_{ij}^{(15)}(k)}{P^{\rm lin}(k)} }     \right].
\end{eqnarray}
The order at which we approximate the resummed propagator has nothing to do with the order of truncation of the infinite series of the remaining (non-resummed) terms, which is something done after the resummation process.

In Fig.~\ref{appendixb_plot} we show the performance of these different approximation schemes for the matter power spectrum: 2-loop SPT (blue lines), 2-loop RPT-${\mathcal N}^{(1)}$ (green lines) and 2-loop RPT-${\mathcal N}^{(2)}$ (red lines). The matter power spectrum at $z=0.55$ from N-body simulations (described in \S~\ref{sec:sims-mocks}) is indicated by the black symbols;  the cosmology is the same as the mock catalogs. The top panel displays the different power spectra normalised by a non-wiggle linear power spectrum for clarity. The bottom panel presents the relative difference to N-body predictions. The arrows indicate where every model starts to deviate more than $2\%$ with respect to N-body simulation measurements. For SPT and RPT-${\mathcal N}^{(1)}$, this happens at about $k\simeq 0.15\,h{\rm Mpc}^{-1}$, whereas RPT-${\mathcal N}^{(2)}$ is able to describe N-body result up to $k\simeq0.18\,h{\rm Mpc}^{-1}$, within $2\%$ errors. Because of this effect, in this paper we choose RPT-${\mathcal N}^{(2)}$ to compute $P_{ij}$. The observed behaviour in Fig.~\ref{appendixb_plot} indicates that our maximum $k$ for the analysis might not be much larger than the values pointed by the arrows, as our description starts breaking down. For simplicity, in the rest of the paper we refer to RPT-${\mathcal N}^{(2)}$ as 2L-RPT.

The  redshift space power spectrum  depends on the angle with respect to the line of sight and thus can be expressed in the Legendre polynomials base,
\begin{equation}
\label{Legendre}P^{(s)}(k,\mu)=\sum_{\ell=0}^{\infty} P^{(\ell)}(k)L_{\ell}(\mu),
\end{equation}
where $P_{\ell}$ are the $\ell$-order multipoles and $L_{\ell}$ are the Legendre polynomials. Most of the signal of the original $P^{(s)}$ function is contained in the first non-zero multipoles. In particular, at large scales, the only multipoles that are non-zero are $\ell=0$ (monopole), $\ell=2$ (quadrupole) and $\ell=4$ (hexadecapole), but almost all the signal is contained in the first two terms.  In this paper, we focus on the monopole. This is the only multipole whose Legendre polynomial is unitary, $L_{0}(\mu)=1$, and therefore it does not depend on the orientation of the line of sight. Because of this, we can safely apply the FKP-estimator to  measure it  from the galaxy survey. Inverting Eq.~\ref{Legendre}, we can express the multipoles as a function of $P^{(s)}$, 
\begin{equation}
P_{g}^{(\ell)}(k)=\frac{2\ell+1}{2}\int_{-1}^{+1} d\mu\,P^{(s)}_{g}(k,\mu)L_\ell(\mu).
\end{equation}
For $\ell=0$ we obtain the monopole, $P_g^{(0)}$.

\section{Bispectrum in  real and redshift space}\label{appendix_bispectrum}

According to perturbation theory, the leading order correction for the dark matter density- and velocity-bispectrum can be expressed as a function of the linear power spectrum and the symmetrised 2-point kernel  \citep{Fry:1994, HMV98,VHMM98,SCFFHM98} ,
\begin{eqnarray}
\label{matter_bispectrum} B_{\delta}({\bf k}_1,{\bf k}_2)&=&2P^{\rm lin}(k_1)P^{\rm lin}(k_2){\cal F}_2^{\rm SPT}({\bf k}_1,{\bf k}_2)+{\rm cyc.}\\
 \label{theta_bispectrum}B_{\theta}({\bf k}_1,{\bf k}_2)&=&2P^{\rm lin}(k_1)P^{\rm lin}(k_2){\cal G}_2^{\rm SPT}({\bf k}_1,{\bf k}_2)+{\rm cyc.},
\end{eqnarray}
with the two-point kernel ${\cal F}_2^{\rm SPT}$  given by Eq.~\ref{eq:F2_kernel} and  the ${\cal G}_2^{\rm SPT}$ kernel  by Eq.~\ref{G_kernel}.
These formulae only reproduce the N-body predictions at the largest scales. The tree-level model of Eq.~\ref{matter_bispectrum} can be improved by substituting the linear power spectra by the non-linear correction and the ${\cal F}_2^{\rm SPT}$ kernel by an effective kernel ${\cal F}_2^{\rm eff}$, as initially  proposed in \cite{Scoccimarro_Couchman},  improved in \cite{HGMetal:2011} and reported in Eq.~\ref{F2effective_kernel} in Appendix~\ref{fit_kernels} .
Similarly the tree-level Eq.~\ref{theta_bispectrum} can be improved and made valid into the (mildly)non-linear regime by  introducing an effective  kernel ${\cal G}_2^{\rm eff}$, described in \cite{HGMetal:inprep} reported in Eq.~\ref{G2effective_kernel} in  Appendix~\ref{fit_kernels}.

These effective kernels  do not show a strong dependence with cosmology or  with $z$: the dependence  of the bispectrum on cosmology and redshift is dominated by that of the power spectra. This model has shown a better description of the matter density-bispectrum up to mildly non-linear regime ($k\lesssim0.2\,h{\rm Mpc}^{-1}$) at low redshifts ($z\leq1.5$). 

The galaxy-bispectrum can be written according to the bias model of Eq.~\ref{deltah} as, 
\begin{eqnarray}
\label{B_ggg1_app}B_{g}(k_1,k_2,k_3)=b_1^3 B(k_1,k_2,k_3) +b_1^2\left[ b_2 P(k_1)P(k_2) + b_{s^2} P(k_1)P(k_2) S_2({\bf k}_1,{\bf k}_2) + {\rm cyc.} \right]
\end{eqnarray}
where $P$ and $B$ are the non-linear matter power spectrum and bispectrum, respectively. We neglect the terms proportional to $b_2^2$, $b_{s2}^2$, which belong to higher-order contributions. Applying the tree-level form for the matter bispectrum, we can write the real space galaxy bispectrum as a function of the non-linear matter power spectrum and the effective kernel,
\begin{eqnarray}
\label{B_ggg2_app}B_{g}(k_1,k_2,k_3)= 2P(k_1)P(k_2)\left[ b_1^3 {\cal F}^{\rm eff}_2({\bf k}_1,{\bf k}_2) + \frac{b_1^2b_2}{2}  + \frac{b_1^2b_{s^2}}{2}S_2({\bf k}_1,{\bf k}_2) \right] + {\rm cyc.}
\end{eqnarray}
In this case, the non-local bias $b_{s2}$ contributes to the leading order and introduces a new shape dependence through $S_2$, which was not present in the matter bispectrum. We do not consider the contribution of $b_{3\rm nl}$ because for the bispectrum (in contrast to the power spectrum) it only appears in fourth and higher order corrections in $\delta_g$.

We next derive the  expression for the galaxy bispectrum in redshift space.

The mapping from the real space radial coordinate to the redshift space radial coordinate depends on the Hubble flow and the Doppler effect due to the peculiar motions of particles, namely peculiar velocities $\bf v$. Under the distant observer approximation, the redshift space coordinate $\bf s$ reads as,
\begin{equation}
s=x+\frac{v_z({\bf x})}{H(a) a}\hat{\bf x}_z,
\end{equation}
where $v_z$ is the radial component of the velocity, $a$ the scale factor and $H$ the Hubble parameter. Using the scaled velocity field, ${\bf u}\equiv-{\bf v}/[H(a) a f(a)]$, where $f$ is the logarithmic grow factor, we write the mapping as,
\begin{equation}
{\bf s}={\bf x}-fu_z({\bf x})\hat{\bf x}_z.
\end{equation}
According to this expression, we can express the Fourier space density contrast in redshift space as a function of the real space density contrast as, 
\begin{eqnarray}
\delta^{(s)}({\bf k})=\int\frac{d^3{\bf x}}{(2\pi)^3}\,e^{-i{\bf k}{\bf x}}e^{ifk_zv_z({\bf x})}\left[ \delta({\bf x})+f\nabla_zv_z({\bf x})  \right],
\end{eqnarray}
where just those points with $f\nabla_z u_z({\bf x})<1$ have been taken into account. Expanding the second exponential in power series we can write the galaxy density contrast in redshift space as,
\begin{eqnarray}
\delta^{(s)}_g({\bf x})=\sum_{i=1}^\infty\int d^3{\bf k}_1\ldots d^3{\bf k}_n \delta^D({\bf k}-{\bf k}_1-\ldots{\bf k}_n) \delta_g({\bf k}_1)+ f\mu_1^2\theta({\bf k}_1)  \frac{(f\mu k)^{n-1}}{(n-1)!}\frac{\mu_2}{k_2}\theta({\bf k}_2)\ldots\frac{\mu_n}{k_n}\theta({\bf k}_n),
\end{eqnarray}
where we have defined $\theta({\bf k})\equiv [-i{\bf k}\cdot{\bf v}]/[a f(a) H(a)]$. We also have assumed an unbiased velocity bias relation for galaxies $\theta_g({\bf k})=\theta({\bf k})$. Plugging the  bias model of Eq.~\ref{deltah}  and expanding perturbatively the dark matter and $k$-velocities over-densities we can re-write,
\begin{eqnarray}
\delta^s_g({\bf k})=\sum_{i=1}^{\infty}\int d^3{\bf k}_1\ldots d^3{\bf k}_n\delta^D({\bf k}-{\bf k}_1-\ldots{\bf k}_n) Z_n({\bf k}_1,\ldots,{\bf k}_n)\delta^{(1)}({\bf k}_1)\ldots\delta^{(1)}({\bf k}_n),
\end{eqnarray}
where the $Z_i$ are the redshift space $i$-loop kernels. The first two kernels read as,
\begin{eqnarray}
\label{Zkernel1}Z_1({\bf k}_i)&\!\!\!\equiv\!\!\!& (b_1+f\mu_i^2), \\
\label{Zkernel2}Z_2({\bf k}_1, {\bf k}_2)&\!\!\!\equiv\!\!\!& b_1\left[ {\cal F}_2^{\rm SPT}({\bf k}_1, {\bf k}_2)+\frac{f\mu k}{2}\left(\frac{\mu_1}{k_1}+\frac{\mu_2}{k_2}\right)\right]+f\mu^2{\cal G}_2^{\rm SPT}({\bf k}_1, {\bf k}_2)+ \frac{f^2\mu k}{2}\mu_1\mu_2\left( \frac{\mu_2}{k_1}+\frac{\mu_1}{k_2} \right)+\frac{b_2}{2}+\frac{b_{s^2}}{2}S_2({\bf k}_1,{\bf k}_2),
\end{eqnarray}
with $\mu_i\equiv{\bf k}_i\cdot {\hat{x}_z}/k_i$, $\mu\equiv(\mu_1 k_1+\mu_2k_2)/k$, $k^2=({\bf k}_1+{\bf k}_2)^2$; ${\cal F}_2^{\rm SPT}$ and ${\cal G}_2^{\rm SPT}$ are the second order kernels of the densities and velocities, respectively (see Eqs.~\ref{eq:F2_kernel} and~\ref{G_kernel}).

The $Z_i$ kernels play the same role as ${\cal F}_i$ but now in redshift space. Thus, the redshift space galaxy bispectrum becomes,
\begin{equation}
B_{g}^{(s)}({\bf k}_1,{\bf k}_2)=2P(k_1)\,Z_1({\bf k}_1)\,P(k_2)\,Z_1({\bf k}_2)\,Z_2({\bf k}_1,{\bf k}_2)+\mbox{cyc.}
\end{equation}
For the unbiased case of dark matter without radial peculiar velocities ($b_1=1$ and $f=0$), $Z_1\rightarrow1$ and $Z_2\rightarrow {\cal F}_2^{\rm SPT}$, and we recover the tree-level expression in real space.

To extend this description more into the (mildly) non-linear regime  a Fingers-of-God term can be added,
\begin{equation}
\label{B_gggsbis}B_{g}^{(s)}({\bf k}_1,{\bf k}_2)=D^B_{\rm FoG}(k_1,k_2,k_3,\sigma_{\rm FoG}^B[z])\left[2P(k_1)\,Z_1({\bf k}_1)\,P(k_2)\,Z_1({\bf k}_2)\,Z^{\rm eff}_2({\bf k}_1,{\bf k}_2)+\mbox{cyc.}\right]\,.
\end{equation}
Similarly to what is done in real space, the redshift space kernel $Z_2$ has been substituted by an effective kernel $Z^{\rm eff}_2$ of the form \citep{HGMetal:inprep},

\begin{equation}
\label{Zkernel_eff}Z_2^{\rm eff}({\bf k}_1, {\bf k}_2)\equiv b_1\left[ {\cal F}_2^{\rm eff}({\bf k}_1, {\bf k}_2)+\frac{f\mu k}{2}\left(\frac{\mu_1}{k_1}+\frac{\mu_2}{k_2}\right)\right]+f\mu^2{\cal G}_2^{\rm eff}({\bf k}_1, {\bf k}_2)+
\frac{f^3\mu k}{2}\mu_1\mu_2\left( \frac{\mu_2}{k_1}+\frac{\mu_1}{k_2} \right)+\frac{b_2}{2}+\frac{b_{s^2}}{2}S_2({\bf k}_1,{\bf k}_2),
\end{equation}
where ${\cal F}_2^{\rm eff}$ is given by Eq.~\ref{F2effective_kernel} and ${\cal G}_2^{\rm eff}$ by Eq.~\ref{G2effective_kernel}.

In Eq.~\ref{B_gggsbis}, analogous to what was done for the power spectrum, we included  $D_{\rm FoG}^B$: a damping term that aims to describe the Fingers-of-God effect due to velocity dispersion inside virialised structures through 1-free parameter, $\sigma_{\rm FoG}^B$. For the bispectrum we parametrise this term as (see e.g., \citealt{VHMM98,SCF99})
\begin{eqnarray}
\label{D_fog_sc}D^B_{\rm FoG}(k_1,k_2,k_3,\sigma_{\rm FoG}^B[z])&=&\left( 1+[k_1^2\mu_1^2+k_2^2\mu_2^2+k_3^2\mu_3^2]^2\sigma_{\rm FoG}^B[z]^2/2  \right)^{-2},
\end{eqnarray}
where $\sigma_{\rm FoG}^B$ is a different parameter than $\sigma_{\rm FoG}^P$ in Eq.~\ref{Psg}. In this paper we treat $\sigma_{\rm FoG}^P$ and $\sigma_{\rm FoG}^B$ as independent parameters, although they may be weakly correlated.

As it is done for the power spectrum, we can express the redshift space bispectrum in spherical harmonics,
\begin{equation}
B^{(s)}({\bf k}_1,{\bf k}_2)=\sum_{\ell=0}^{\infty}\sum_{m=-{\ell}}^{\ell} B^{\ell}_m(k_1,k_2,k_3)Y_{\ell}^m(\mu_1,\mu_2).
\end{equation}
The original signal of $B^{(s)}$ is now spread along the different multipoles $B_{\ell,m}$. However, most of the signal is contained in those multipoles with lower values of $\ell$ and $m$. As for the power spectrum only the first multipole with $\ell=0$ and $m=0$ (monopole) can be extracted using the FKP estimator, since $Y_0^0=1$. The bispectrum monopole can be written as a function of the bispectrum in redshift space as\footnote{Since for $\ell=0$ there is only one possible $m$, we ignore this last parameter in the notation of the bispectrum monopole: ${B_{(0)}^{(0)}}_g\equiv B_g^{(0)}$}, 
\begin{eqnarray}
B_g^{(0)}(k_1,k_2,k_3)=\int d\mu_1d\mu_2 B_g^{(s)}({\bf k}_1, {\bf k}_2)\equiv\int_{-1}^{+1}d\mu_1\int_0^{2\pi}d\varphi\, B_g^{(s)}({\bf k}_1, {\bf k}_2),
\end{eqnarray}
where $\varphi$ has been defined to be $\mu_2\equiv\mu_1x_{12}-\sqrt{1-\mu_1^2}\sqrt{1-x_{12}^2}\cos\varphi$, where $x_{12}\equiv({\bf k}_1\cdot{\bf k}_2)/(k_1 k_2)$.

Integrating over the line of sight of the two vectors we obtain an expression for the monopole,
\begin{eqnarray}
B_g^{(0)}({\bf k}_1,{\bf k}_2)=\int d\mu_1d\mu_2 B_g^{(s)}({\bf k}_1, {\bf k}_2)\,.
\end{eqnarray}
 An expression for $B_g^{(0)}$  can be analytically written only when $D^B_{\rm FoG}=1$. This is not the case in general (only when we describe halo without substructure). However, even in this simplified case, having an analytical expression helps understanding the behaviour of the different terms,
\begin{eqnarray}
\label{Bhhh0_real}B_{g}^{(0)}({\bf k}_1, {\bf k}_2)&=&P(k_1)P(k_2)b_1^4\left\{ \frac{1}{b_1}{\cal F}_2(k_1,k_2,\cos\theta_{12}){ \mathcal D}_{\rm SQ1}^{(0)}+\frac{1}{b_1}{\cal G}_2(k_1,k_2,\cos\theta_{12}){\mathcal D}_{\rm SQ2}^{(0)} \right.\\ 
\nonumber&+&\left[\frac{b_2}{b_1^2}+\frac{b_{s^2}}{b_1^2}S_2({\bf k}_1,{\bf k}_2)\right] \left.{\mathcal D}_{\rm NLB}^{(0)}+{\mathcal D}_{\rm FoG}^{(0)}  \right\}+{\mbox{cyc.}}
\end{eqnarray}
where ${ \mathcal D}_{\rm SQ1}^{(0)}$ and ${ \mathcal D}_{\rm SQ2}^{(0)}$ are the first and second order contribution for the large scale squashing (Kaiser effect or pancakes-of-God), ${ \mathcal D}_{\rm NLB}^{(0)}$ is the non-linear bias contribution and ${ \mathcal D}_{\rm FoG}^{(0)}$ is the damping effect due to the velocity dispersion (linear part of of Fingers-of-God). The $\cal F$ and $\cal G$ terms can either be SPT or effective. All these terms depends on $x_{ij}$, $y_{ij}$ and $\beta\equiv f/b_1$: ${\mathcal D}_l^{(0)}(x_{ij},y_{ij};\beta)$,
\begin{eqnarray}
\label{DSQ1}{\mathcal D}^{(0)}_{\rm SQ1}&=&\frac{2(15+10\beta+\beta^2+2\beta^2x_{12}^2)}{15},\\
{\mathcal D}^{(0)}_{\rm SQ2}&=& 2\beta\left(35y_{12}^2 +28\beta y_{12}^2 +3\beta^2y_{12}^2 +35 +28\beta  +\right.\\
\nonumber&+&3\beta^2 +70y_{12}  x_{12}+84\beta y_{12}  x_{12}+18\beta^2y_{12}x_{12}+14\beta y_{12}^2x_{12}^2 +12\beta^2y_{12}^2x_{12}^2 +\\
\nonumber&+&\left.+14\beta x_{12}^2 +12\beta^2x_{12}^2+12\beta^2y_{12}x_{12}^3\right)/[105 (1+y^2_{12}+2x_{12}y_{12}) ],\\
{\mathcal D}^{(0)}_{\rm NLB}&=& \frac{(15+10\beta+\beta^2+2\beta^2 x_{12}^2)}{15},\\
\label{Dfog}{\mathcal D}^{(0)}_{\rm FoG}&=& \beta\left(210 + 210\beta  + 54\beta^2  + 6\beta^3  + 105 y_{12} x +189\beta y_{12} x_{12} +\right.\\
\nonumber&+&99\beta^2 y_{12} x_{12} + 15\beta^3 y_{12} x_{12} +105 y_{12}^{-1} x_{12} +189 \beta y_{12}^{-1} x + 99 \beta^2 y_{12}^{-1} x_{12} +15 \beta^3 y_{12}^{-1} x_{12} +\\
\nonumber&+&168\beta x_{12}^2 + 216 \beta^2  x_{12}^2 + 48 \beta^3  x_{12}^2 + 36 \beta^2 y_{12} x_{12}^3 + 20\beta^3 y_{12}^{-1} x_{12}^3 +\\
\nonumber&+&\left.36\beta^2 y_{12}^{-1} x_{12}^3 + 20 \beta^3 y_{12} x_{12}^3 + 16\beta^3 x_{12}^4\right)/315,
\end{eqnarray}
where $\beta\equiv f/b_1$, $x_{ij}\equiv{\bf k}_i\cdot{\bf k}_j/(k_ik_j)$, $y_{ij}\equiv k_i/k_j$.

\section{Explicit expressions for effective kernels}
\label{fit_kernels}
The performance of the tree-level form of the matter bispectrum  can be improved substantially at small scales by substituting the ${\cal F}_2^{\rm SPT}$ and ${\cal G}_2^{\rm SPT}$ kernels by effective analogues with  free fitting parameters that can be calibrated using N-body simulations \citep{HGMetal:2011,HGMetal:inprep},
\begin{eqnarray}
\label{F2effective_kernel}
{\cal F}_2^{\rm eff}({\bf k}_i,{\bf k}_j)&=&\frac{5}{7}a(n_i,k_i; {\bf a}^F)a(n_j,k_j; {\bf a}^F) +\frac{1}{2}\cos(\theta_{ij})  \left(\frac{k_i}{k_j}+\frac{k_j}{k_i}\right)b(n_i,k_i;{\bf a}^F)b(n_j,k_j ;{\bf a}^F) \\
&+&\frac{2}{7} \cos^2(\theta_{ij})c(n_i,k_i;{\bf a}^F)c(n_j,k_j;{\bf a}^F),\\
\label{G2effective_kernel}
{\cal G}_2^{\rm eff}({\bf k}_i,{\bf k}_j)&=&\frac{3}{7}a(n_i,k_i; {\bf a}^G)a(n_j,k_j; {\bf a}^G) +\frac{1}{2}\cos(\theta_{ij})  \left(\frac{k_i}{k_j}+\frac{k_j}{k_i}\right)b(n_i,k_i;{\bf a}^G)b(n_j,k_j ;{\bf a}^G) \\
&+&\frac{4}{7} \cos^2(\theta_{ij})c(n_i,k_i;{\bf a}^G)c(n_j,k_j;{\bf a}^G),\\
\end{eqnarray}
with the functions $a$, $b$ and $c$ defined as,
\begin{eqnarray}
 \nonumber \label{abc_new} {a}(n,k,{\bf a})&=&\frac{1+\sigma_8^{a_6}(z)[0.7Q_3(n)]^{1/2}(q a_1)^{n+a_2}}{1+(q a_1)^{n+a_2}}, \\
 {b}(n,k,{\bf a})&=&\frac{1+0.2a_3(n+3)(q a_7)^{n+3+a_8}}{1+(q a_7)^{n+3.5+a_8}}, \\
\nonumber {c}(n,k,{\bf a})&=&\frac{1+4.5a_4/[1.5+(n+3)^4](q a_5)^{n+3+a_9}}{1+(q a_5)^{n+3.5+a_9}}.
\end{eqnarray}
where $q\equiv k/k_{\rm nl}$ with $k_{\rm nl}(z)$ a characteristic scale defined as,
\begin{eqnarray}
 \frac{k_{\rm nl}(z)^3P^{\rm lin}(k_{\rm nl},z)}{2\pi^2}\equiv1;
\end{eqnarray}
 $n$ is the slope of the smoothed linear power spectrum,
\begin{eqnarray}
 n(k)\equiv\frac{d\log P_{\rm nw}^{\rm lin}(k)}{d\log k},
 \end{eqnarray}
 $Q_3(n)$ is defined as,
 \begin{eqnarray}
 Q_3(n)\equiv\frac{4-2^n}{1+2^{n+1}}
 \end{eqnarray}
and ${\bf a}=\{a_1,\ldots,a_9\}$, is a set of nine free parameters to be fit by comparison to N-body simulations. For the ${\cal F}_2^{\rm eff}$ these parameters are \citep{HGMetal:2011},
\begin{eqnarray}
\nonumber a^{\cal F}_1 &=& 0.484\quad\,\,\,\, a^{\cal F}_4 = 0.392\quad\,\,\,\, a^{\cal F}_7 = 0.128\\
\nonumber a^{\cal F}_2 &=& 3.740\quad\,\,\,\, a^{\cal F}_5 = 1.013\quad\,\,\,\, a^{\cal F}_8 = -0.722\\
\nonumber a^{\cal F}_3 &=& -0.849\quad a^{\cal F}_6 = -0.575\quad a^{\cal F}_9 = -0.926
\end{eqnarray}
and for the ${\cal G}_2^{\rm eff}$ kernel are \citep{HGMetal:inprep},
\begin{eqnarray}
\nonumber a^{\cal G}_1 &=& 3.599\quad\,\,\,\, a^{\cal G}_4 = -3.588\quad a^{\cal G}_7 = 5.022\\
\nonumber a^{\cal G}_2 &=&-3.879\quad a^{\cal G}_5 = 0.336\quad\,\,\,\, a^{\cal G}_8 = -3.104\\
\nonumber a^{\cal G}_3 &=& 0.518\quad\,\,\,\, a^{\cal G}_6 = 7.431\quad\,\,\,\,\, a^{\cal G}_9 = -0.484
\end{eqnarray}
These new kernels have been shown to improve the behaviour of bispectrum both in real and redshift space up to scales of $k\simeq0.2$ for a wide range of redshifts, $z\leq1.5$.

Both ${\cal F}_2^{\rm eff}$ and ${\cal G}^{\rm eff}_2$ have a similar dependence on $a$, $b$ and $c$ functions. However, the parameters on  which these functions depend, namely ${\bf a}^{\cal F}$ and ${\bf a}^{\cal G}$, are different for $\cal F$ and $\cal G$.

   \end{document}